**REPORT, ver. 3.02**

# Analysis of the possibility for time optimal control of the scanning system

Author:     Assoc. Prof. Dr Borislav Penev

Department of Optoelectronics and Laser Engineering

TU-Sofia, Branch Plovdiv

Work within     Work Package 4 - 3D Imaging UV Doppler LIDAR System

Task 4.4 - Development of Scanning/Imaging System

# Contents

# 1. Introduction

The scanning system is an essential part of the Green-Wake project's lidar. This system provides the appropriate way of scanning the space in front of the aircraft by a laser beam in order to obtain the right picture of the atmospheric turbulence. The scanning system shown in Figure 1 and designed by Sula Systems Ltd. comprises two light mirror actuators for rotational movement with perpendicular to each other axles of rotation. The light mirror actuators are built on the basis of a limited angle torquer. Each actuator includes both a rotary high precision encoder and a decoder to measure quickly the absolute rotary angle of each actuator with high accuracy. The load of each actuator represents the respective light weight mirror mounted on specific pivots. By the movement of each mirror actuator the system provides a specific reflection of a specially formed laser beam, so that the laser beam describes a respective trajectory in front of the scanning system. The projection of this laser beam in a vertical scan plane at a respective range forms the scan pattern needed to obtain the picture of the atmospheric turbulence at the demanded range in front of the aircraft.

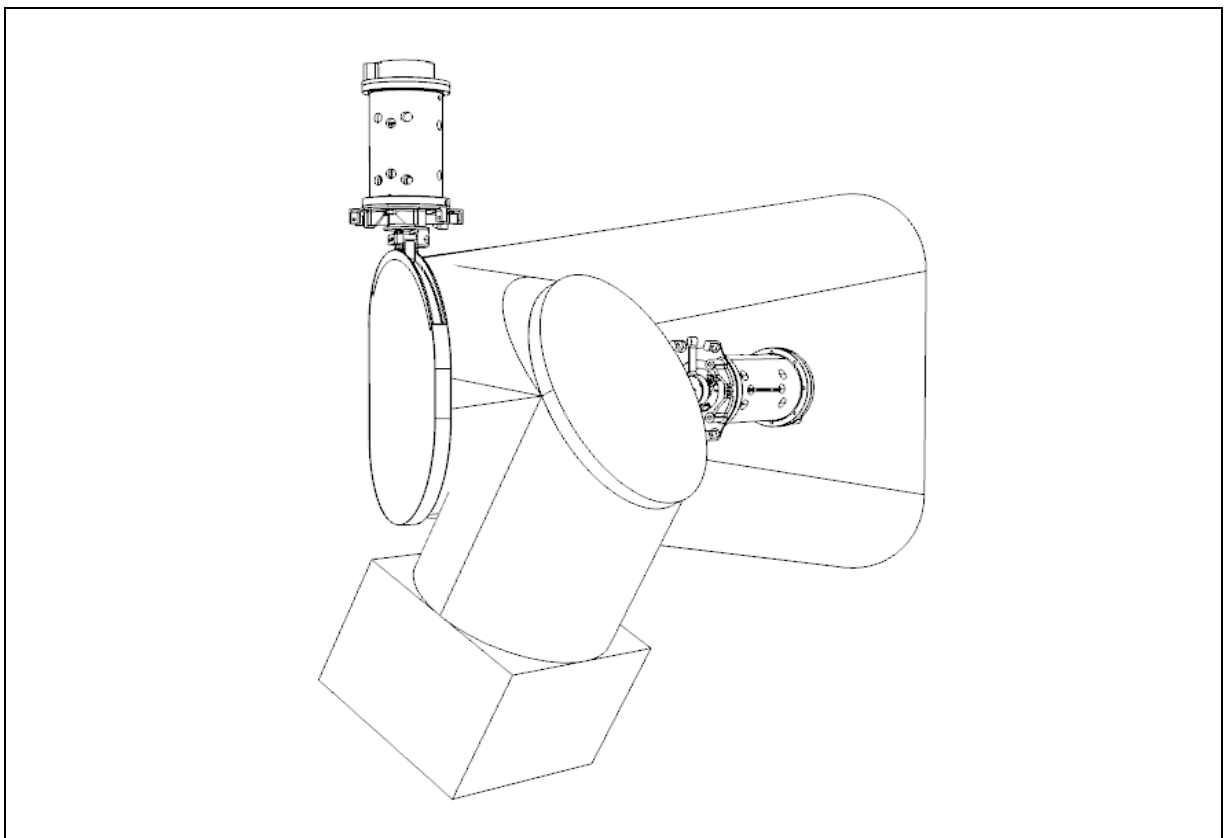

**Figure 1. View of the scanning system with two light mirror actuators.**

According to the objectives of Task 4.4 - Development of Scanning/Imaging System - within the Work Package 4 of the Green-Wake project, this scanning/imaging system must be scalable – enabling the system to be adapted and utilised in all aircraft of all categories – light, medium and large. A possible way to meet these requirements is developing a scanning system comprising a tracking control system for each mirror actuator having the best dynamic performance within the resources this system could provide. In other words, this approach leads to the problem of analyzing the possibility of time optimal control of the developed scanning system.

## 1.1 *On the linear time optimal control problem*

The linear time optimal control problem was formulated about 60 years ago. Feldbaum introduced first the concept of the optimal process in 1949 at the investigation of a linear system consisting of two integrators [ 7 ]. His papers [ 8 ] - [ 11 ] laid the basics of the theory of the optimal systems. Pavlov says in [ 5 ] that "academician Pontryagin and his associates Boltyanskii, Gamkrelidze and Mischenko went from the time optimal processes to development of the general theory of the optimal processes". Their formulation of the time optimal control problem [ 6 ] (§ 17) and [ 2 ] (§ 5, Section 18) is the following. The controlled system of order $n$ is described by the equations

( 1 ) $$\frac{d\boldsymbol{x}}{dt} = A\boldsymbol{x} + B\boldsymbol{u} \,,$$

where $\boldsymbol{x} = [x_1 \quad x_2 \quad ... \quad x_n]^{\mathrm{T}}$, $\boldsymbol{u} = [u_1 \quad u_2 \quad ... \quad u_r]^{\mathrm{T}}$, the matrices $A$ and $B$ are respectively of range $n \times n$ and $n \times r$.

It is assumed that the admissible control $\boldsymbol{u} = [u_1 \quad u_2 \quad ... \quad u_r]^{\mathrm{T}}$ is defined in the closed area $U$, representing a convex polyhedron in the r-dimensional space of $u_1$, $u_2$, ..., $u_r$; the origin of that space belongs to that area but does not appear its peak; each component $u_i$, $i = \overline{1, r}$, is a piecewise continuous function; in the points of discontinuity it is assumed

( 2 ) $u_i(\tau) = u_i(\tau + 0) \,;$

The initial and the target states of the system are

( 3 ) $$\boldsymbol{x}(t_0) = \boldsymbol{x}_0 \,,$$

( 4 ) $$\boldsymbol{x}(t_1) = \boldsymbol{x}_1 \,,$$

where $t_1$ is unspecified.

The time optimal control problem is to find an admissible control $\boldsymbol{u} = u(t)$ that transfers the system ( 1 ) from its initial state ( 3 ) to the target state ( 4 ) in minimum time, i.e. minimizing the performance index

( 5 ) $$J = \int_{t_0}^{t_1} dt = t_1 - t_0 \,.$$

In order to formulate the maximum principle the following Hamiltonian $H$ is introduced, which is a function of $\boldsymbol{x} = [x_1 \quad x_2 \quad ... \quad x_n]^T$, $\boldsymbol{u} = [u_1 \quad u_2 \quad ... \quad u_r]^T$ and the so called conjugate variables $\boldsymbol{\psi} = [\psi_1 \quad \psi_2 \quad ... \quad \psi_n]^T$,

( 6 )  $\qquad H(\boldsymbol{\psi}, \boldsymbol{x}, \boldsymbol{u}) = \boldsymbol{\psi}^T A \boldsymbol{x} + \boldsymbol{\psi}^T B \boldsymbol{u}$ .

In case $\boldsymbol{\psi}$ and $\boldsymbol{x}$ are fixed, the function $H$ becomes a function of $\boldsymbol{u}$; the upper boundary of that function is represented by $M(\boldsymbol{\psi}, \boldsymbol{x})$:

( 7 )  $\qquad M(\boldsymbol{\psi}, \boldsymbol{x}) = \sup_{\boldsymbol{u} \, \in \, U} H(\boldsymbol{\psi}, \boldsymbol{x}, \boldsymbol{u})$ .

By the help of $H$ the conjugate system of equations regarding $\boldsymbol{\psi} = [\psi_1 \quad \psi_2 \quad ... \quad \psi_n]^T$ is introduced

( 8 )  $\qquad \dfrac{d\psi_i}{dt} = -\dfrac{dH}{dx_i}, \quad i = \overline{1, n}$ ,

or

( 9 )  $\qquad \dfrac{d\boldsymbol{\psi}}{dt} = -A^T \boldsymbol{\psi}$ .

So the maximum principle of Pontryagin with respect to the considered time optimal control problem ( 1 ) - ( 5 ) is formulated by the following theorem [ 6 ] (Theorem 2, § 3, Chapter 1):

*Let $\boldsymbol{u}(t)$, $t_0 \leq t \leq t_1$, be an admissible control, which transfers the plant from its initial state $\boldsymbol{x}_0$ in to the state $\boldsymbol{x}_1$, and $\boldsymbol{x}(t)$ be the respective trajectory so that $\boldsymbol{x}(t_0) = \boldsymbol{x}_0$, $\boldsymbol{x}(t_1) = \boldsymbol{x}_1$. For the optimality (with respect to minimum time) of the control $\boldsymbol{u}(t)$ and the trajectory $\boldsymbol{x}(t)$ the existence of such non-trivial solution $\boldsymbol{\psi}(t) = [\psi_1(t) \quad \psi_2(t) \quad ... \quad \psi_n(t)]^T$, $t_0 \leq t \leq t_1$, of ( 8 ) corresponding to $\boldsymbol{u}(t)$ and $\boldsymbol{x}(t)$ so that:*

$\qquad$ *1.  for every one moment $t$, $t_0 \leq t \leq t_1$, the condition of the maximum is valid*

( 10 )  $\qquad H(\boldsymbol{\psi}(t), \boldsymbol{x}(t), \boldsymbol{u}(t)) = M(\boldsymbol{\psi}(t), \boldsymbol{x}(t))$ ;

$\qquad$ *2.  in the final moment $t_1$ it is valid*

( 11 )  $\qquad M(\boldsymbol{\psi}(t_1), \boldsymbol{x}(t_1)) \geq 0$ ;

*is necessary.*

*It turns out that, if $\boldsymbol{\psi}(t)$, $\boldsymbol{u}(t)$, $\boldsymbol{x}(t)$ satisfy ( 1 ), ( 8 ) and condition 1, then $M(\boldsymbol{\psi}(t), \boldsymbol{x}(t))$ is constant and the examination of ( 11 ) could be done for an arbitrary chosen moment $t$, $t_0 \leq t \leq t_1$.*

It is supposed for the linear time optimal control problem ( 1 ) - ( 5 ) that it satisfies the condition of normality [ 1 ] (§ 4.21), [ 2 ] (§ 5, Section 22), [ 6 ] (§ 17). The fundamental properties of the problem have been long investigated and summarized [ 1 ], [ 2 ], [ 4 ], [ 6 ], [ 13 ], [ 14 ], [ 20 ]: the maximum principle appears not only a necessary but also sufficient condition; the theorems regarding the existence and uniqueness are proved; the time optimal control represents piecewise constant function taking its values in the peaks of the polyhedron $U$ having finite number of switching over – the so called theorem of the finite number of switching over [ 2 ] (§ 6, Section 26) etc.

The solution of the time optimal control problem could be searched in two forms: form of program control, when the optimal control is presented as a function of the time $\boldsymbol{u} = u(t)$; form of synthesis, when the optimal control is presented as a function of the state $\boldsymbol{x}$ of the controlled system $\boldsymbol{u} = u(\boldsymbol{x})$.

A very important fact regarding the time optimal control in form of synthesis is the so called theorem of the $n$-th intervals [ 2 ] (Theorem 2.11, § 6), [ 6 ] (Theorem 10, § 17), which was first formulated and proved by Feldbaum ([ 8 ] - [ 10 ]):

*If it is supposed that the convex polyhedron $U$ represents a parallelepiped in the r-dimensional space, defined by inequalities on each one component of the admissible control $\boldsymbol{u} = [u_1 \quad u_2 \quad ... \quad u_r]^{\mathrm{T}}$ as*

**( 12 )**  $\quad \alpha_i \leq u_i \leq \beta_i, \quad i = \overline{1, r}$ ,

*and the eigenvalues of the matrix $A$ ( 1 ) are real, then each one function $u_i(t)$, $i = \overline{1, r}$, of the optimal control $\boldsymbol{u}(t) = [u_1(t) \quad u_2(t) \quad ... \quad u_r(t)]^{\mathrm{T}}$ represents a piecewise constant function, taking only values $\alpha_i$ and $\beta_i$ and having no more than $(n-1)$ switching over, i.e. maximum $n$ intervals of constancy.*

This property gives the opportunity for a differential approach to resolving the synthesis problem considering the properties of the controlled system. The presented estimation allows clear and simple, seemingly on the surface of it, application of the state space method for its solution especially in the particular case of one-dimensional control. We underline that our investigation here is focused on that very case.

The interest in the linear time optimal control problem has been considerably decreased for the last decades. Among the reasons for this are as it follows: the main approaches are known; within each one approach the main theoretical issues are resolved; the advantages and difficulties of the application of the techniques are known and so on. However in the field of the synthesis versus the approaches, which lead to program control, regardless of the fact that the synthesis is described by Pontryagin, Boltyanskii, Gamkrelidze and Mischenko in [ 6 ] (Chapter 3, § 20, §21, Example 3) it is mentioned by Gabassov, Cirillova and Prischepov in [ 3 ] that "the method of constructing  the hyper-surfaces of switching over could not be reputed to be the solution of the synthesis problem because of the excessive requirements to the amount of information stored. There is little hope for success at the direct solution of the problem by the method of dynamic programming". So despite many years of progress the time optimal control problem remains one of the most difficult of the optimal control theory.

A possible direction for finding the time optimal control solution is the approach based on the state space method but taking into account the factors that limit its application only to systems of low order. A state of the art method for synthesis of time optimal control of any order for a class of linear time optimal control problems is developed by the author [ 15] - [ 18]. The method avoids the well known synthesis techniques. It is based on the relations and recently discovered properties of the examined class of linear time optimal control problems and comprises two stages: analysis of the state-space properties of the class of problems and synthesis of time optimal control by using a multi-step procedure avoiding the switching hyper surface description. Based on that method synthesis of a number of closed loop systems including high order systems of 7th order [ 15], [ 18] and synthesis of time optimal control for pneumatic servo systems are presented in [ 15] and [ 19].

## *1.2 Phases of the analysis of the possibility for time optimal control of the scanning system*

The analysis of possibility for time optimal control of the scanning system requires solving consequently a set of problems, which can be divided into several groups. The first group of problems is connected with the modelling of the scanning system as a whole, which includes spatial modelling of the system, modelling of both the large and the small mirror actuators etc., so all the obtained mathematical and simulation models could serve the next group of problems connected with the control theory techniques for synthesis and design of tracking control systems for each one mirror actuator based on the author's method for synthesis of time optimal control for a class of linear systems.

# 2. Initial spatial modelling of the scanning system

A spatial simulation model of the scanning system has been developed. Figure 2 shows a Simulink block diagram of the model. The angle positions of both mirror actuators of the scanning system are considered here as model's inputs. The laser beam reflected by the mirror surfaces having a respective angular position in every moment is directed in front of the scanning system. The XY position of the beam's projection in the vertical scan plane at given range in front of the aircraft is considered as output of the scanning. The model is based on a specially developed function, ssx5y5f02, which actually provides the conversion of the mirrors' angle positions to XY position of the laser beam in the scan plane. Here at the initial spatial modelling the model's inputs are pure sinusoidal signals, given the fact that the mirror actuators represent ideal elements.

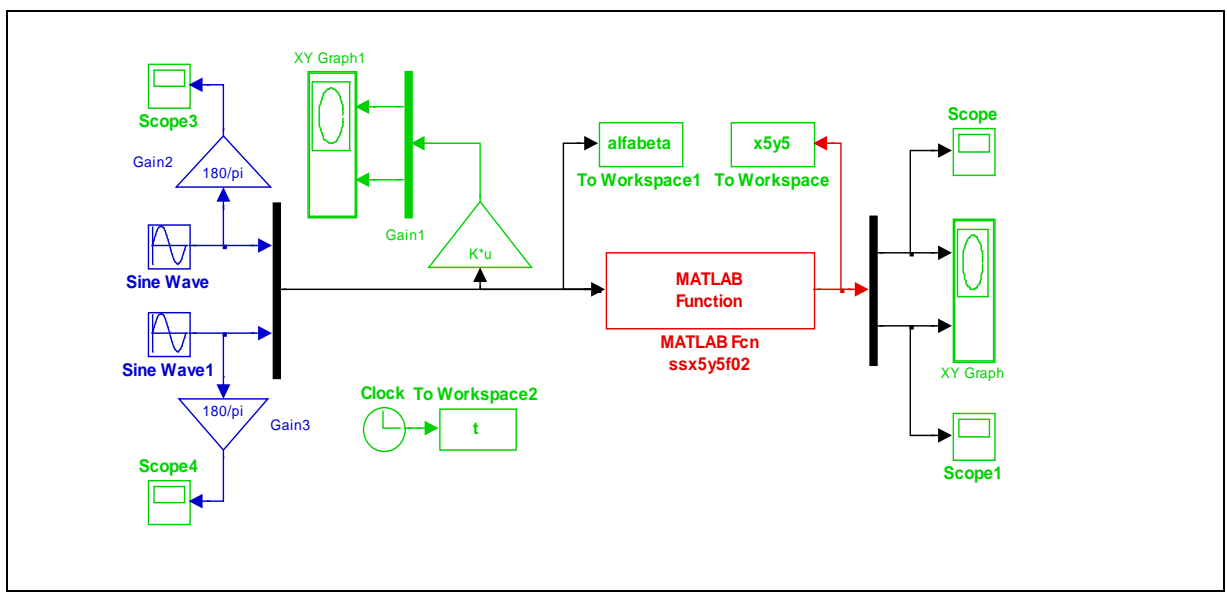

**Figure 2. A spatial simulation model of the scanning system.**

## 2.1 Results at the initial data provided by SULA Systems Ltd

The model in Figure 2 has been simulated by initial data given in Table 1. Figure 3 represents the scan pattern of the mirrors. Figure 4 represents the scan pattern in the vertical plane at range of 200 m. Table 2 contains the input and output data versus the simulation time points, respectively the angular position of the large and the small mirrors and the horizontal and vertical positions of the beam projection in the scan plane at the considered range of 200 m.

**Table 1. Initial data for the oscillations of the mirror actuators in the ideal case (provided by Sula Systems Ltd.)**

| | |
|---|---|
| Oscillation amplitude of the small mirror: | 3.57 degrees |
| Oscillation amplitude of the large mirror: | 8.35 degrees |
| Oscillation frequency of the small mirror: | 20 [Hz] |
| Oscillation frequency of the large mirror: | 2.5 [Hz] |
| Range: | 200 [m] |

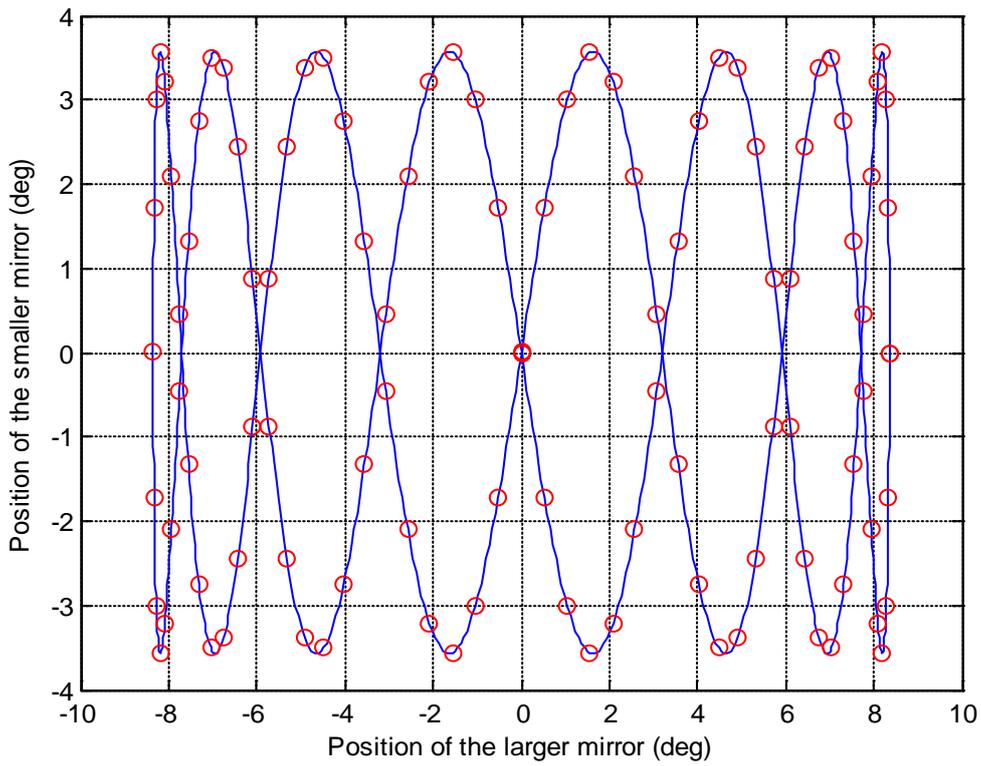

**Figure 3. Scan pattern of the mirrors.**

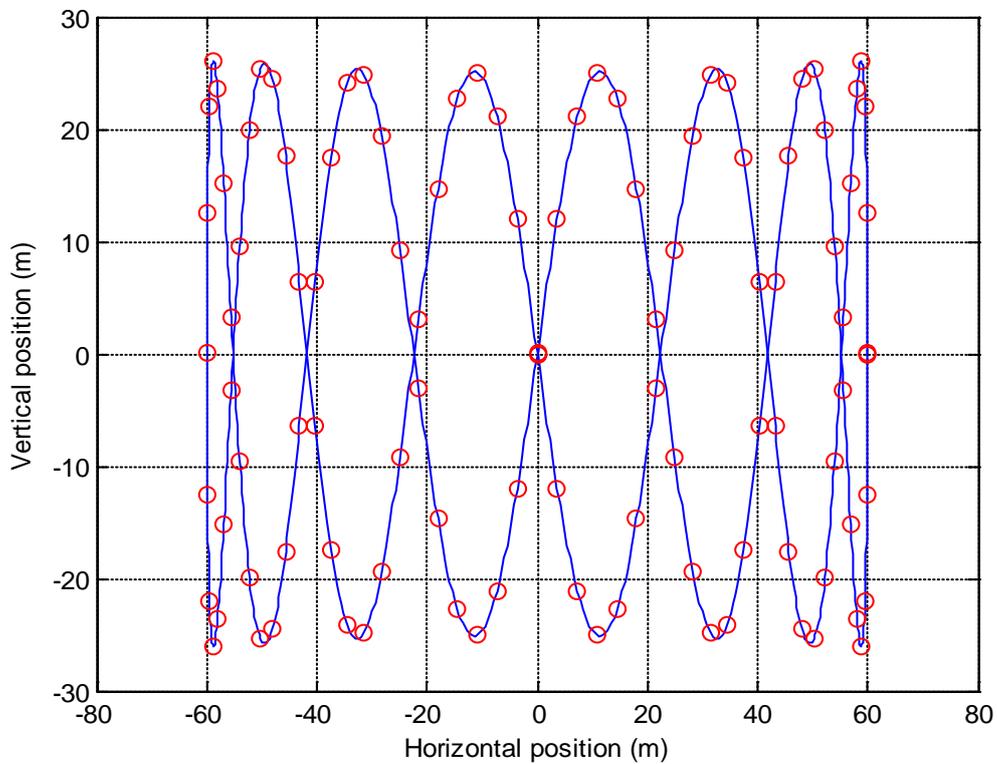

**Figure 4. The scan pattern in the vertical plane at range 200 m.**

**Table 2. Numerical results of the simulation of the ideal work of the scanning system.**

| Time (s) | Large mirror Position (deg) | Small mirror Position (deg) | Scan plane at range 200 m. Horizontal position (m) | Scan plane at range 200 m. Vertical position (m) |
|---|---|---|---|---|
| 0.000 | 8.35 | 0.00 | 60.00 | 0.00 |
| 0.004 | 8.33 | -1.72 | 59.88 | -12.57 |
| 0.008 | 8.28 | -3.01 | 59.50 | -22.07 |
| 0.012 | 8.20 | -3.56 | 58.88 | -26.10 |
| 0.016 | 8.09 | -3.23 | 58.01 | -23.61 |
| 0.020 | 7.94 | -2.10 | 56.91 | -15.28 |
| 0.024 | 7.76 | -0.45 | 55.57 | -3.25 |
| 0.028 | 7.56 | 1.31 | 54.00 | 9.52 |
| 0.032 | 7.32 | 2.75 | 52.22 | 19.94 |
| 0.036 | 7.05 | 3.51 | 50.24 | 25.41 |
| 0.040 | 6.76 | 3.40 | 48.05 | 24.53 |
| 0.044 | 6.43 | 2.44 | 45.69 | 17.57 |
| 0.048 | 6.09 | 0.89 | 43.15 | 6.35 |
| 0.052 | 5.72 | -0.89 | 40.44 | -6.33 |
| 0.056 | 5.32 | -2.44 | 37.59 | -17.43 |
| 0.060 | 4.91 | -3.40 | 34.60 | -24.20 |
| 0.064 | 4.47 | -3.51 | 31.49 | -24.94 |
| 0.068 | 4.02 | -2.75 | 28.27 | -19.48 |
| 0.072 | 3.56 | -1.31 | 24.95 | -9.27 |
| 0.076 | 3.07 | 0.45 | 21.54 | 3.15 |
| 0.080 | 2.58 | 2.10 | 18.06 | 14.76 |
| 0.084 | 2.08 | 3.23 | 14.52 | 22.74 |
| 0.088 | 1.56 | 3.56 | 10.93 | 25.08 |
| 0.092 | 1.05 | 3.01 | 7.31 | 21.17 |
| 0.096 | 0.52 | 1.72 | 3.66 | 12.04 |
| 0.100 | 0.00 | 0.00 | 0.00 | 0.00 |
| 0.104 | -0.52 | -1.72 | -3.66 | -12.04 |
| 0.108 | -1.05 | -3.01 | -7.31 | -21.17 |
| 0.112 | -1.56 | -3.56 | -10.93 | -25.08 |
| 0.116 | -2.08 | -3.23 | -14.52 | -22.74 |
| 0.120 | -2.58 | -2.10 | -18.06 | -14.76 |
| 0.124 | -3.07 | -0.45 | -21.54 | -3.15 |
| 0.128 | -3.56 | 1.31 | -24.95 | 9.27 |
| 0.132 | -4.02 | 2.75 | -28.27 | 19.48 |
| 0.136 | -4.47 | 3.51 | -31.49 | 24.94 |
| 0.140 | -4.91 | 3.40 | -34.60 | 24.20 |
| 0.144 | -5.32 | 2.44 | -37.59 | 17.43 |
| 0.148 | -5.72 | 0.89 | -40.44 | 6.33 |
| 0.152 | -6.09 | -0.89 | -43.15 | -6.35 |
| 0.156 | -6.43 | -2.44 | -45.69 | -17.57 |
| 0.160 | -6.76 | -3.40 | -48.05 | -24.53 |
| 0.164 | -7.05 | -3.51 | -50.24 | -25.41 |
| 0.168 | -7.32 | -2.75 | -52.22 | -19.94 |
| 0.172 | -7.56 | -1.31 | -54.00 | -9.52 |
| 0.176 | -7.76 | 0.45 | -55.57 | 3.25 |
| 0.180 | -7.94 | 2.10 | -56.91 | 15.28 |
| 0.184 | -8.09 | 3.23 | -58.01 | 23.61 |
| 0.188 | -8.20 | 3.56 | -58.88 | 26.10 |
| 0.192 | -8.28 | 3.01 | -59.50 | 22.07 |
| 0.196 | -8.33 | 1.72 | -59.88 | 12.57 |
| 0.200 | -8.35 | 0.00 | -60.00 | 0.00 |
| 0.204 | -8.33 | -1.72 | -59.88 | -12.57 |
| 0.208 | -8.28 | -3.01 | -59.50 | -22.07 |
| 0.212 | -8.20 | -3.56 | -58.88 | -26.10 |
| 0.216 | -8.09 | -3.23 | -58.01 | -23.61 |
| 0.220 | -7.94 | -2.10 | -56.91 | -15.28 |
| 0.224 | -7.76 | -0.45 | -55.57 | -3.25 |
| 0.228 | -7.56 | 1.31 | -54.00 | 9.52 |
| 0.232 | -7.32 | 2.75 | -52.22 | 19.94 |
| 0.236 | -7.05 | 3.51 | -50.24 | 25.41 |
| 0.240 | -6.76 | 3.40 | -48.05 | 24.53 |
| 0.244 | -6.43 | 2.44 | -45.69 | 17.57 |
| 0.248 | -6.09 | 0.89 | -43.15 | 6.35 |
| 0.252 | -5.72 | -0.89 | -40.44 | -6.33 |
| 0.256 | -5.32 | -2.44 | -37.59 | -17.43 |
| 0.260 | -4.91 | -3.40 | -34.60 | -24.20 |
| 0.264 | -4.47 | -3.51 | -31.49 | -24.94 |
| 0.268 | -4.02 | -2.75 | -28.27 | -19.48 |
| 0.272 | -3.56 | -1.31 | -24.95 | -9.27 |
| 0.276 | -3.07 | 0.45 | -21.54 | 3.15 |
| 0.280 | -2.58 | 2.10 | -18.06 | 14.76 |
| 0.284 | -2.08 | 3.23 | -14.52 | 22.74 |
| 0.288 | -1.56 | 3.56 | -10.93 | 25.08 |
| 0.292 | -1.05 | 3.01 | -7.31 | 21.17 |
| 0.296 | -0.52 | 1.72 | -3.66 | 12.04 |
| 0.300 | 0.00 | 0.00 | 0.00 | 0.00 |
| 0.304 | 0.52 | -1.72 | 3.66 | -12.04 |
| 0.308 | 1.05 | -3.01 | 7.31 | -21.17 |
| 0.312 | 1.56 | -3.56 | 10.93 | -25.08 |
| 0.316 | 2.08 | -3.23 | 14.52 | -22.74 |
| 0.320 | 2.58 | -2.10 | 18.06 | -14.76 |
| 0.324 | 3.07 | -0.45 | 21.54 | -3.15 |
| 0.328 | 3.56 | 1.31 | 24.95 | 9.27 |
| 0.332 | 4.02 | 2.75 | 28.27 | 19.48 |
| 0.336 | 4.47 | 3.51 | 31.49 | 24.94 |
| 0.340 | 4.91 | 3.40 | 34.60 | 24.20 |
| 0.344 | 5.32 | 2.44 | 37.59 | 17.43 |
| 0.348 | 5.72 | 0.89 | 40.44 | 6.33 |
| 0.352 | 6.09 | -0.89 | 43.15 | -6.35 |
| 0.356 | 6.43 | -2.44 | 45.69 | -17.57 |
| 0.360 | 6.76 | -3.40 | 48.05 | -24.53 |
| 0.364 | 7.05 | -3.51 | 50.24 | -25.41 |
| 0.368 | 7.32 | -2.75 | 52.22 | -19.94 |
| 0.372 | 7.56 | -1.31 | 54.00 | -9.52 |
| 0.376 | 7.76 | 0.45 | 55.57 | 3.25 |
| 0.380 | 7.94 | 2.10 | 56.91 | 15.28 |
| 0.384 | 8.09 | 3.23 | 58.01 | 23.61 |
| 0.388 | 8.20 | 3.56 | 58.88 | 26.10 |
| 0.392 | 8.28 | 3.01 | 59.50 | 22.07 |
| 0.396 | 8.33 | 1.72 | 59.88 | 12.57 |
| 0.400 | 8.35 | 0.00 | 60.00 | 0.00 |

**End of Table 2.**

# 3. Modelling of the mirror actuators

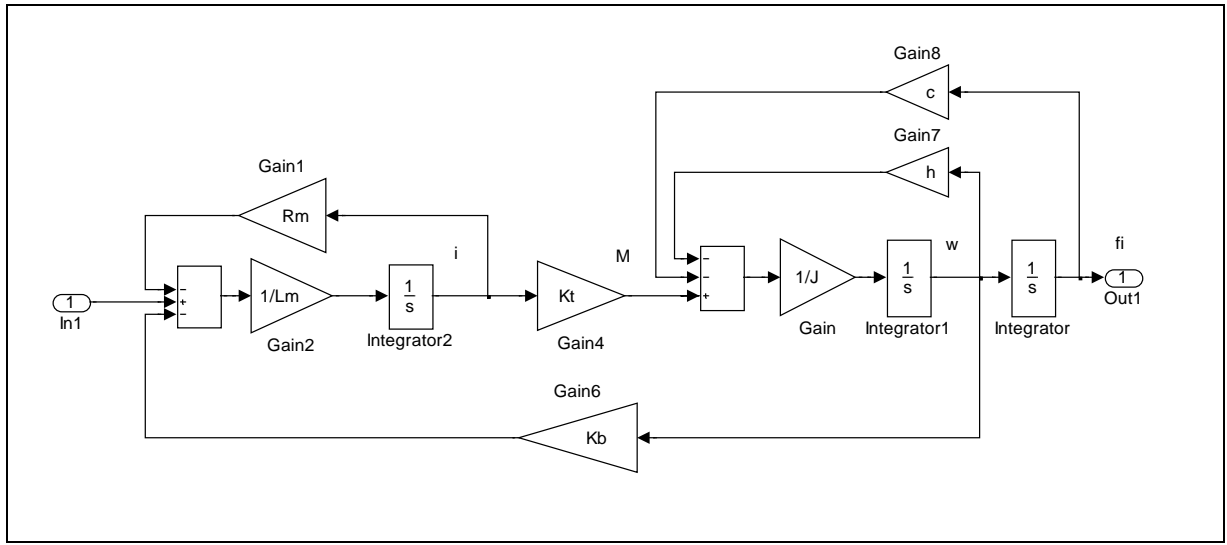

**Figure 5. A block diagram of the linear model of the mirror actuator.**

## 3.1 State space model

$\dot{x} = Ax + Bu$ ,

$y = Cx + Du$ ,

$$x = \begin{bmatrix} \varphi \\ \omega \\ i \end{bmatrix}, \quad A = \begin{bmatrix} 0 & 1 & 0 \\ -c/J & -h/J & Kt/J \\ 0 & -Kb/Lm & -Rm/Lm \end{bmatrix}, \quad B = \begin{bmatrix} 0 \\ 0 \\ 1/Lm \end{bmatrix},$$

$C = \begin{bmatrix} 1 & 0 & 0 \end{bmatrix}, \quad D = 0$ ,

where:

- $\varphi$ is the output angle (rad), $\omega$ is the angle velocity (rad/sec) and $i$ is the LAT current (Amperes);

- c is the Pivot Stiffness (N.m/rad), h is the Estimated Damping (N.m.sec/rad), J is the Inertia (Kg.m²);

- Lm is LAT Inductance (H), Rm is the LAT Resistance (Ohms), Kb is the LAT Back EMF Constant (V per rad/sec), Kt is the LAT Torque Sensitivity (N.m/Amp)

## 3.2 Input output model by transfer function

$$W_{\varphi u}(p) = \frac{K}{a_0 p^3 + a_1 p^2 + a_2 p + a_3} ,$$

where

$$K = \frac{K_t}{R_m c} \quad [rad/V],$$

$$a_0 = (\frac{L_m}{R_m})\frac{J}{c} \quad [s^3], \quad a_1 = (\frac{L_m}{R_m})\frac{h}{c} + \frac{J}{c} \quad [s^2], \quad a_2 = (\frac{L_m}{R_m}) + \frac{h}{c} + \frac{KtKb}{R_m c} \quad [s], \quad a_3 = 1.$$

## 3.3  Representation of the small mirror actuator

We use the following initial data based on the LAT Size and Winding Constants (2011_10_18 1819 F D_Bamford - (At3) LAT-1503.pdf ) and the GreenWake − Key data for performance modelling- Issue 2 (2011_10_27 1810 F D_Bamford - (At) G-W Parameters (4).doc):

```
c_sm=12.3        % N*m/rad
h_sm=0.03        % N*m*sec/rad
J_sm=0.7e-3      % kg*m^2

Rm=7.5           % Om
Kt=283e-3        % N*m/A
Kb=0.283         % V/(rad/sec)
Lm=4.5e-3        % H
```

So the state space model represents

$$\dot{x}_{sm} = A_{sm} x_{sm} + B_{sm} u_{sm},$$

where

$$A_{sm} = \begin{bmatrix} 0 & 1 & 0 \\ -17571 & -42.857 & 404.29 \\ 0 & -62.889 & -1666.7 \end{bmatrix}, \quad B_{sm} = \begin{bmatrix} 0 \\ 0 \\ 222.22 \end{bmatrix}.$$

The transfer function of the small mirror actuator represents

$$W_{\varphi_{sm}u_{sm}}(p) = \frac{K_{sm}}{a_{0sm}p^3 + a_{1sm}p^2 + a_{2sm}p + a_{3sm}},$$

where

$$K_{sm} = 0.0030678 \quad [rad/V],$$

$$a_{0sm} = 3.4146\text{e-}008 \quad [s^3], \quad a_{1sm} = 5.8374\text{e-}005 \quad [s^2], \quad a_{2sm} = 0.0039072 \quad [s], \quad a_{3sm} = 1.$$

The correspondent step responses, $u_{sm}(t) = 1(t)$, based on the above state space model, the above transfer function and the simulation of the model of Figure 5 at the initial data for the small mirror actuator are shown in Figure 6 and they are identical. The correspondent Bode diagram of the actuator is given in Figure 7.

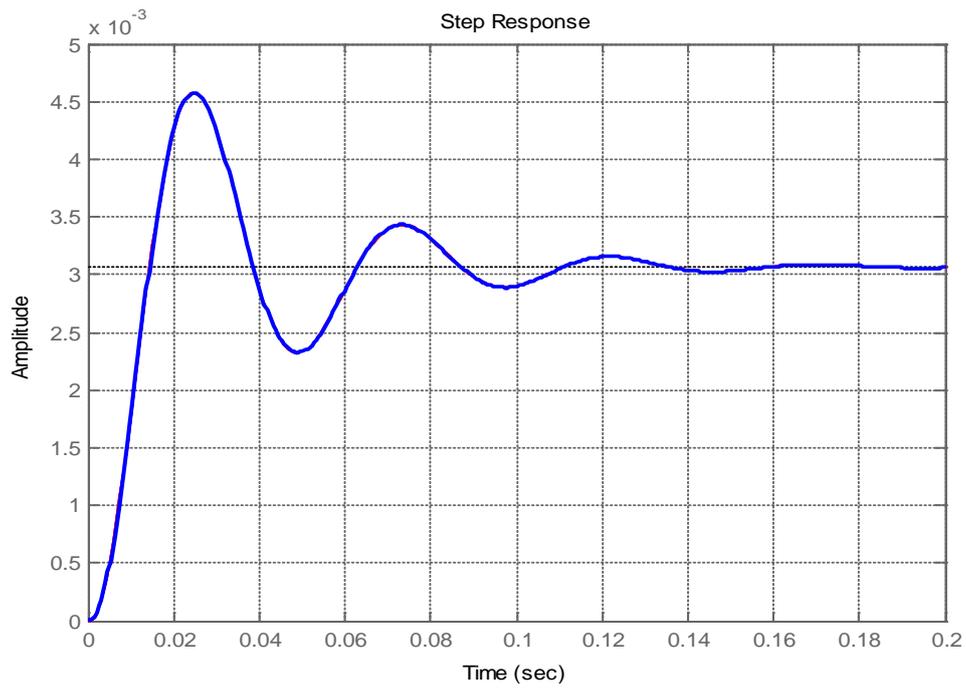

**Figure 6. Step responses of the state space, transfer function and block diagram models of the small mirror drive.**

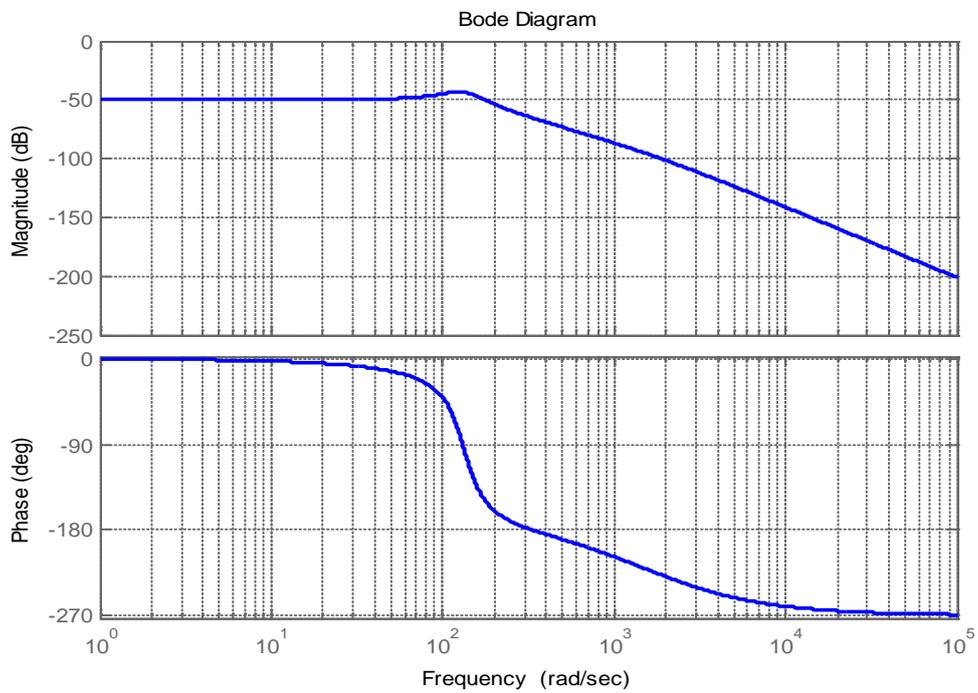

**Figure 7. Bode diagram of the linear small mirror actuator model.**

### 3.3.1 Simplified linear model

The Time Constant $T_e = \dfrac{L_m}{R_m} = 0.0006\,[s]$ is more than ten times smaller than the Time

Constant $T_{M,sm} = \sqrt{\dfrac{J_{sm}}{c_{sm}}} = 0.0075\,[s]$, so the inertia produced by $T_e$ in the inner closed loop on $i$ in Figure 5 could be ignored to simplify the model. The simplified model could be presented by the diagram in Figure 8.

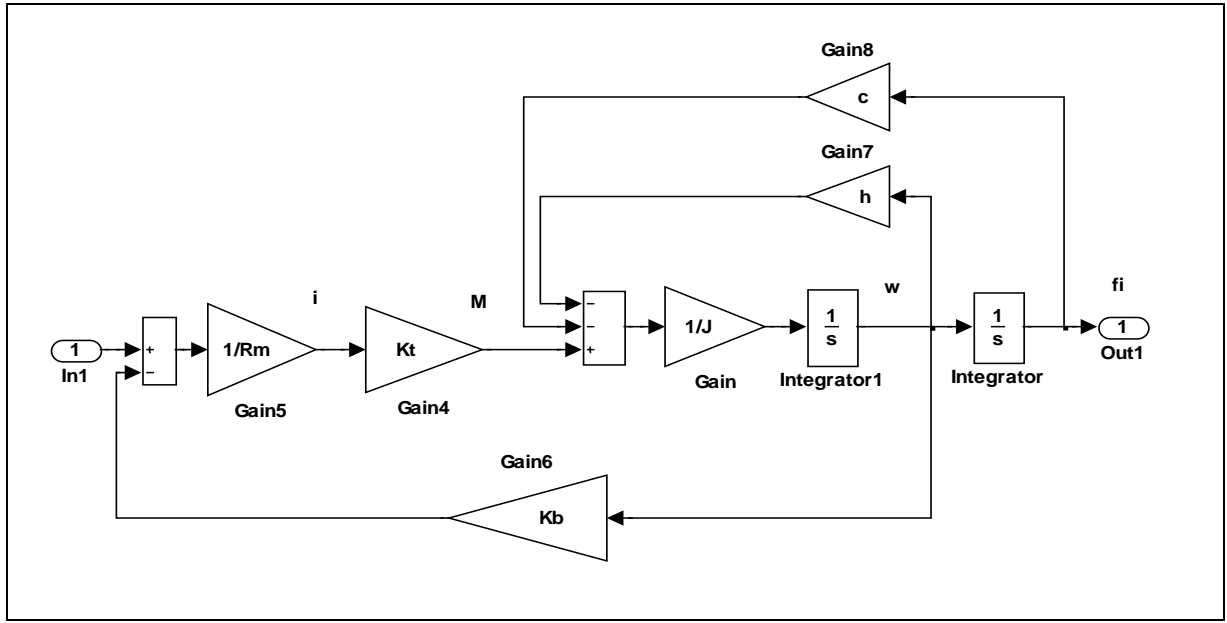

**Figure 8. A block diagram of the simplified linear model of the mirror actuator.**

This simplification reduces the order of the model. The correspondent state space representation is the following:

$$\dot{\boldsymbol{x}}_{sms} = A_{sms}\boldsymbol{x}_{sms} + B_{sms}u_{sm}\,,$$

$$y_{sms} = C_{sms}\boldsymbol{x}_{sms} + D_{sms}u_{sm}\,,$$

$$\boldsymbol{x}_{sms} = \begin{bmatrix} \varphi_{sm} \\ \omega_{sm} \end{bmatrix},\quad A_{sms} = \begin{bmatrix} 0 & 1 \\ -c_{sm}/J_{sm} & (-(K_bK_t)/(R_mJ_{sm})-h_{sm}/J_{sm}) \end{bmatrix},\quad B_{sms} = \begin{bmatrix} 0 \\ K_t/(R_mJ_{sm}) \end{bmatrix},$$

$$C_{sms} = \begin{bmatrix} 1 & 0 \end{bmatrix},\quad D_{sms} = 0\,.$$

The simplified transfer function becomes

$$W^{s}_{\varphi_{sm}u_{sm}}(p) = \frac{K_{sm}}{a_{0sms}p^2 + a_{1sms}p + a_{2sms}}\,,$$

where

$$a_{0sms} = \frac{J_{sm}}{c_{sm}} \quad [s^2], \quad a_{1sms} = \frac{h_{sm}}{c_{sm}} + \frac{K_t K_b}{R_m c_{sm}} \quad [s], \quad a_{2sms} = 1.$$

Based on the initial data for the parameters, the numerical representation of the simplified model is:

$$A_{sms} = \begin{bmatrix} 0 & 1 \\ -17571 & -58.11 \end{bmatrix}, \quad B_{sms} = \begin{bmatrix} 0 \\ 53.90 \end{bmatrix},$$

$$K_{sm} = 0.0030678 \quad [rad/V],$$

$$a_{0sms} = 5.6911e-005 \quad [s^2], \quad a_{1sms} = 0.0033072 \quad [s], \quad a_{2sms} = 1.$$

The correspondent step responses of the simplified model, $u_{sm}(t) = 1(t)$, based on the above state space model, the above transfer function and the simulation of the model of Figure 8 at the initial data for the small mirror drive are shown in the following Figure 9 and they are identical.

### 3.3.2 Comparison between the linear model and its simplified one

The step responses of both the models are presented in Figure 11, the correspondent Bode diagrams – in Figure 12. The vectors of the eigenvalues of the linear model and its simplified one are

$$\lambda_{sm} = \begin{bmatrix} -29.282 + 129.93j \\ -29.282 - 129.93j \\ -1651 \end{bmatrix}, \quad \lambda_{sms} = \begin{bmatrix} -29.056 + 129.33j \\ -29.056 - 129.33j \end{bmatrix}.$$

The eigenvalue $\lambda_{sm}(3) = -1651$, neglected at the simplified model, causes the differences in the step responses and the Bode diagrams. This eigenvalue has a corresponded part of the transfer function of the linear model $\frac{1}{(T_{3sm}p+1)}$, $T_{3sm} = \frac{-1}{\lambda_{sm}(3)}$, which part, missed in the transfer function of the simplified model, causes the extra phase delay in the phase diagram of the linear model in comparison to the phase diagram of the simplified model, 45 degrees at $\omega = -\lambda_{sm}(3) = 1651 \, [rad/s]$ and causes also the extra slope with -20 db/decade of the magnitude starting at this frequency.

The transfer function of the simplified model

$$W_{\varphi_{sm}u_{sm}}^{s}(p) = \frac{K_{sm}}{a_{0sms}p^2 + a_{1sms}p + a_{2sms}}$$

could be presented as

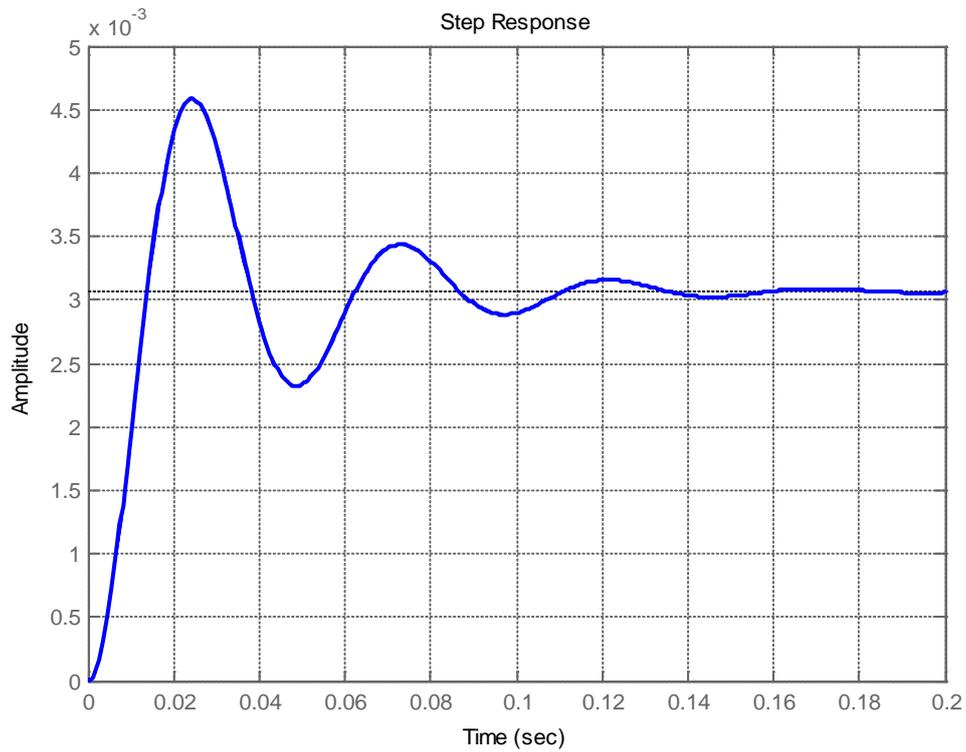

**Figure 9. Step responses of the state space, transfer function and block diagram models of the simplified small mirror drive.**

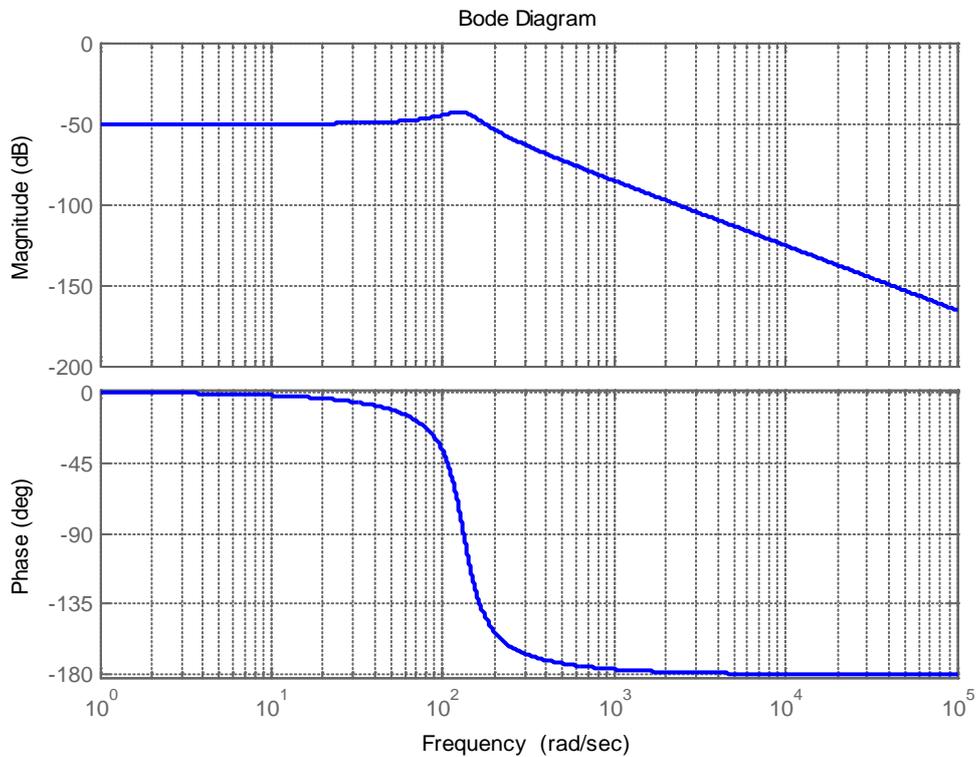

**Figure 10. Bode diagram of the simplified linear small mirror drive model.**

$$W_{\varphi_{sm}u_{sm}}^{s}(p) = \frac{K_{dsm}}{T_{sms}^{2}p^{2} + 2T_{sms}\xi_{sms}p + 1} = \frac{K_{dsm}\omega_{sms}^{2}}{p^{2} + 2\omega_{sms}\xi_{sms}p + \omega_{sms}^{2}},$$

where

$$K_{sm} = 0.0030678 \quad [rad/V],$$

$$T_{sms} = \sqrt{a_{0sms}} = 0.0075439\,[s], \quad \xi_{sms} = a_{1sms}/(2T_{sms}) = 0.2192, \quad \omega_{sms} = 1/T_{sms} = 132.56\,[rad/s].$$

Because of $\xi_{sms} = 0.2192 < \dfrac{1}{\sqrt{2}} = 0.707$, there is a resonance at the frequency

$$\omega_{smsR} = \omega_{sms}\sqrt{1 - 2\xi_{sms}^{2}} = 126.03\,[rad/s] \text{ or } f_{smsR} = \frac{\omega_{smsR}}{2\pi} = 20.058\,[Hz].$$

That corresponds very well to the fact shared in the end of the e-mail from 18th October: The mirror axles are mounted on flex pivots. These allow the mirrors to move through the limited angle required and allow the system to run at a pre-set resonant frequency with very little power.
…
The system can actually be run open loop or semi-open loop by simply provising a small sinusoidal signal of the right frequency. Using this approach, the small fast mirror can be run over the full range (+/- 7.13 degrees) at full speed (approximately 20 Hz) using a driving signal of just 5v and 40mA. ($\omega_{driv} = 2\pi 20 = 125.66\,[rad/s]$)

## 3.4  Linear representation of the large mirror actuator

The initial data for the linear model are based on the LAT Size and Winding Constants (2011_10_18 1819 F D_Bamford - (At3) LAT-1503.pdf ) and the GreenWake – Key data for performance modelling-Issue 2 (2011_10_27 1810 F D_Bamford - (At) G-W Parameters (4).doc):

### 3.4.1  State space representation

c_lm=1.54        % N*m/rad
h_lm=0.02        % N*m*sec/rad
J_lm=4.9e-3      % kg*m^2

Rm=7.5           % Om
Kt=283e-3        % N*m/A
Kb=0.283         % V/(rad/sec)
Lm=4.5e-3        % H

The state space model represents

$$\dot{x}_{lm} = A_{lm}x_{lm} + B_{lm}u_{lm},$$

$$y_{lm} = C_{lm}x_{lm} + D_{lm}u_{lm},$$

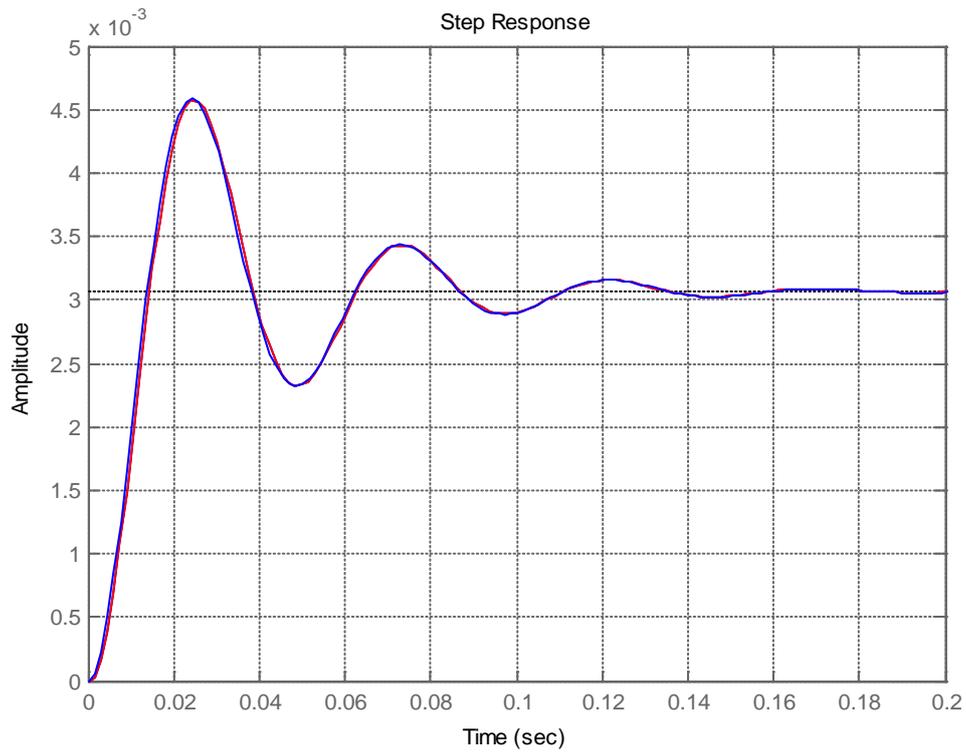

**Figure 11. Comparison of the step responses of the linear model (-) and the simplified linear one (-).**

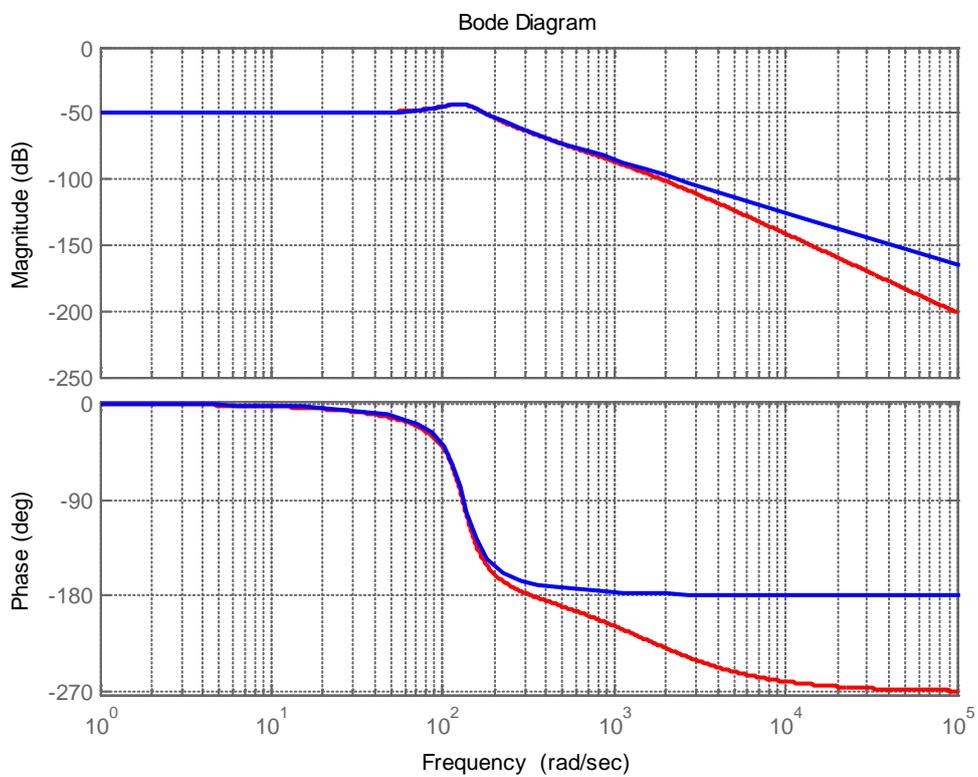

**Figure 12. Comparison of Bode diagrams of the linear model (-) and the simplified linear one (-).**

$$\boldsymbol{x}_{lm} = \begin{bmatrix} \varphi_{lm} \\ \omega_{lm} \\ i_{lm} \end{bmatrix}, \quad A_{lm} = \begin{bmatrix} 0 & 1 & 0 \\ -c_{lm}/J_{lm} & -h_{lm}/J_{lm} & Kt/J_{lm} \\ 0 & -Kb/Lm & -Rm/Lm \end{bmatrix}, \quad B_{lm} = \begin{bmatrix} 0 \\ 0 \\ 1/Lm \end{bmatrix},$$

$$C_{lm} = \begin{bmatrix} 1 & 0 & 0 \end{bmatrix}, \quad D_{lm} = 0,$$

where

$$A_{lm} = \begin{bmatrix} 0 & 1 & 0 \\ -314.29 & -4.0816 & 57.755 \\ 0 & -62.889 & -1666.7 \end{bmatrix}, \quad B_{lm} = \begin{bmatrix} 0 \\ 0 \\ 222.22 \end{bmatrix}.$$

### 3.4.2 Presentation by transfer function

$$W_{\varphi_{lm} u_{lm}}(p) = \frac{K_{lm}}{a_{0lm}p^3 + a_{1lm}p^2 + a_{2lm}p + a_{3lm}},$$

$$K_{lm} = \frac{K_t}{R_m c_{lm}} \quad [rad/V],$$

$$a_{0lm} = (\frac{L_m}{R_m})\frac{J_{lm}}{c_{lm}} \quad [s^3], \quad a_{1lm} = (\frac{L_m}{R_m})\frac{h_{lm}}{c_{lm}} + \frac{J_{lm}}{c_{lm}} \quad [s^2], \quad a_{2lm} = (\frac{L_m}{R_m}) + \frac{h_{lm}}{c_{lm}} + \frac{K_t K_b}{R_m c_{lm}} \quad [s], \quad a_{3lm} = 1.$$

where

$$K_{lm} = 0.0245 \quad [rad/V],$$

$$a_{0lm} = 1.9091\text{e-}006 \quad [s^3], \quad a_{1lm} = 0.00319 \quad [s^2], \quad a_{2lm} = 0.0205 \quad [s], \quad a_{3lm} = 1.$$

The correspondent step responses, $u_{sm}(t) = 1(t)$, based on the above state space model, the above transfer function and the simulation of the model of Figure 6 at the initial data for the large mirror actuator are shown in the following Figure 13 and they are identical. The correspondent Bode diagram of the large mirror actuator's model is given in Figure 14.

### 3.4.3 Simplified linear model of the large mirror actuator

Here the Time Constant $T_{M,lm} = \sqrt{\frac{J_{lm}}{c_{lm}}} = 0.056408\,[s]$ is 94 times greater than the Time Constant $T_e = \frac{L_m}{R_m} = 0.0006\,[s]$. The simplified model could be presented by analogy with the simplification of the small mirror actuator's model by the diagram in Figure 8. The respective simplified state space model is:

$$\dot{\boldsymbol{x}}_{lms} = A_{lms}\boldsymbol{x}_{lms} + B_{lms}u_{lm},$$

$$y_{lms} = C_{lms}\boldsymbol{x}_{lms} + D_{lms}u_{lm},$$

$$\boldsymbol{x}_{lms} = \begin{bmatrix} \varphi_{lm} \\ \omega_{lm} \end{bmatrix}, \quad A_{lms} = \begin{bmatrix} 0 & 1 \\ -c_{lm}/J_{lm} & (-(K_bK_t)/(R_mJ_{lm})-h_{lm}/J_{lm}) \end{bmatrix}, \quad B_{lms} = \begin{bmatrix} 0 \\ K_t/(R_mJ_{lm}) \end{bmatrix},$$

$$C_{lms} = \begin{bmatrix} 1 & 0 \end{bmatrix}, \quad D_{lms} = 0.$$

The respective numerical representation of the matrices is:

$$A_{lms} = \begin{bmatrix} 0 & 1 \\ -314.29 & -6.261 \end{bmatrix}, \quad B_{lms} = \begin{bmatrix} 0 \\ 7.7 \end{bmatrix}.$$

By analogy with the small mirror actuator's model the simplified transfer function of the large mirror actuator becomes

$$W^s_{\varphi_{lm}u_{lm}}(p) = \frac{K_{lm}}{a_{0lms}p^2 + a_{1lms}p + a_{2lms}},$$

$$a_{0lms} = \frac{J_{lm}}{c_{lm}} \quad [s^2], \quad a_{1lms} = \frac{h_{lm}}{c_{lm}} + \frac{K_tK_b}{R_mc_{lm}} \quad [s], \quad a_{2lms} = 1,$$

with the following numerical values of the coefficients of the denominator

$$a_{0lms} = 0.003182 \quad [s^2], \quad a_{1lms} = 0.01992 \quad [s], \quad a_{2lms} = 1.$$

The correspondent step responses of the simplified model, $u_{sm}(t) = 1(t)$, based on the above state space model, the above transfer function and the simulation of the model of Figure 8 at the initial data for the large mirror actuator are shown in Figure 15 and they are identical. The Bode diagram of the simplified actuator's model is shown in Figure 16.

The transfer function of the simplified model could be also presented as

$$W^s_{\varphi_{lm}u_{lm}}(p) = \frac{K_{lm}}{T^2_{lms}p^2 + 2T_{lms}\xi_{lms}p + 1} = \frac{K_{lm}\omega^2_{lms}}{p^2 + 2\omega_{lms}\xi_{lms}p + \omega^2_{lms}},$$

where

$$T_{lms} = \sqrt{a_{0lms}} = 0.056408 \, [s], \quad \xi_{lms} = a_{1lms}/(2T_{lms}) = 0.17658, \quad \omega_{lms} = 1/T_{lms} = 17.728 \, [rad/s].$$

Because of $\xi_{lms} = 0.17658 < \dfrac{1}{\sqrt{2}} = 0.707$, there is a resonance at the frequency

$$\omega_{lmsR} = \omega_{lms}\sqrt{1-2\xi^2_{lms}} = 17.166 \, [rad/s] \text{ or } f_{lmsR} = \frac{\omega_{lmsR}}{2\pi} = 2.7321 \, [Hz].$$

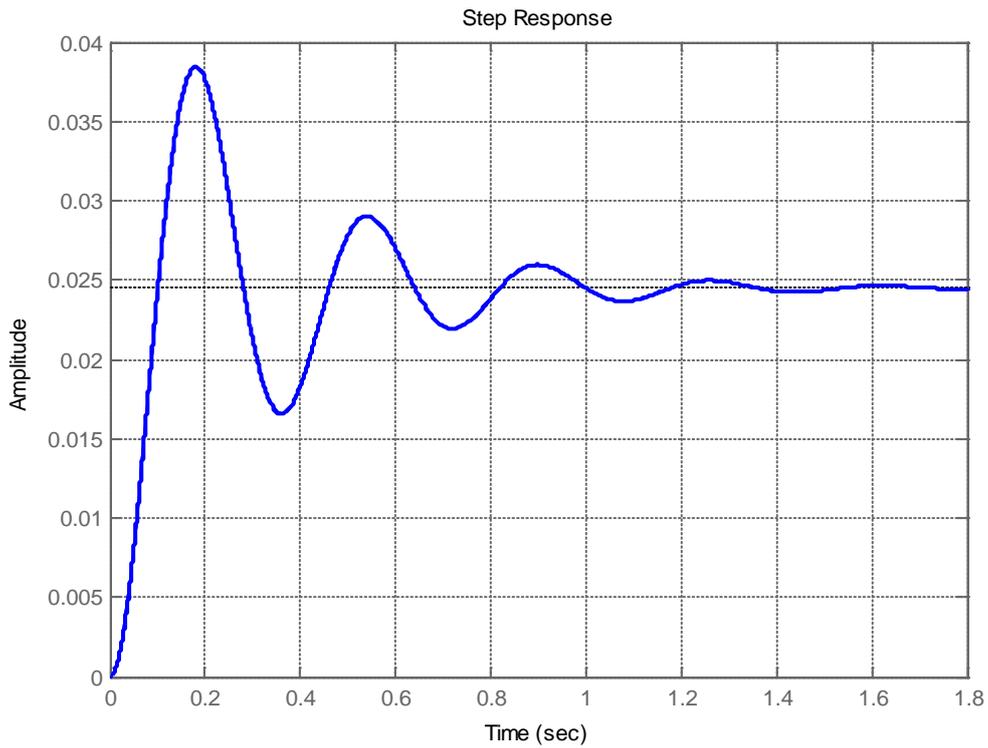

**Figure 13. Step responses of the state space, transfer function and block diagram models of the large mirror drive.**

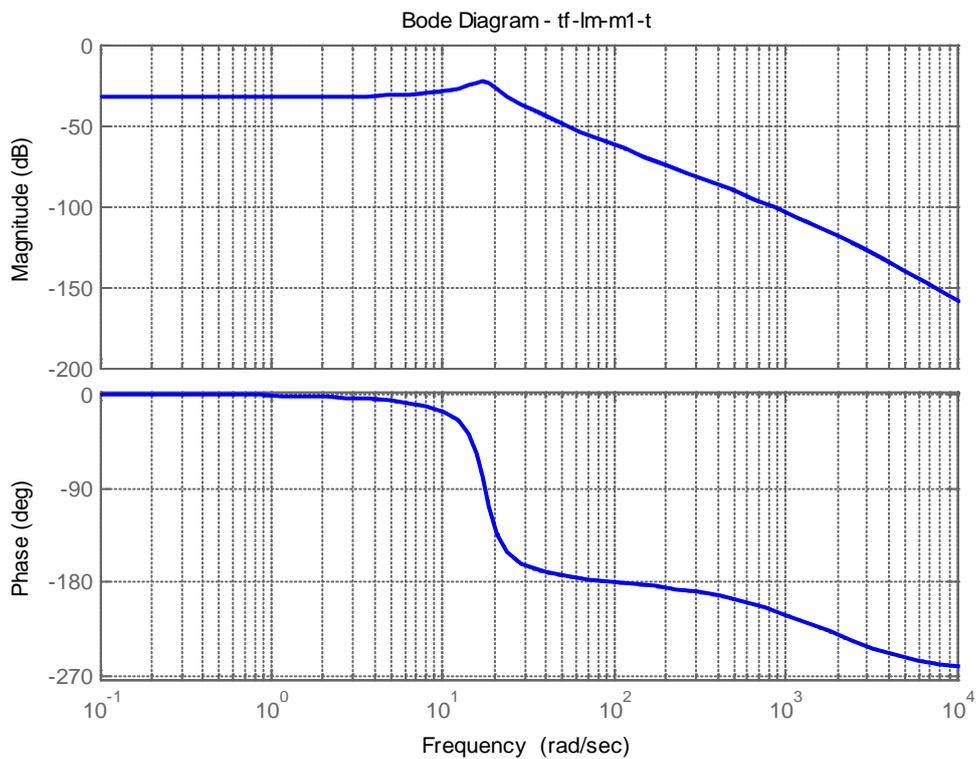

**Figure 14. Bode diagram of the linear large mirror actuator's model.**

### 3.4.4 Comparison of the linear large mirror actuator's models

Bode diagrams of both the linear large mirror actuator's model and the simplified one are shown in the following Figure 17. The eigenvalues of the linear model and the simplified one are respectively

$$\lambda_{lm} = \begin{bmatrix} -3.1345 + 17.46\,j \\ -3.1345 - 17.46\,j \\ -1664.5 \end{bmatrix}, \quad \lambda_{lms} = \begin{bmatrix} -3.1305 + 17.45\,j \\ -3.1305 - 17.45\,j \end{bmatrix}.$$

By analogy with the comparison of the small mirror actuator's models the eigenvalue $\lambda_{lm}(3) = -1664.5$, neglected at the simplified model, causes the differences in the Bode diagrams. This eigenvalue has a corresponded part of the transfer function of the linear model $\dfrac{1}{(T_{3lm}p + 1)}$, $T_{3lm} = \dfrac{-1}{\lambda_{lm}(3)}$, which part, missed in the transfer function of the simplified model, causes the extra phase delay in the phase diagram of the linear model in comparison to the phase diagram of the simplified model, 45 degrees at $\omega = -\lambda_{lm}(3) = 1664.5\,[rad/s]$ and causes also the extra slope with -20 db/decade of the magnitude starting at this frequency.

According to Bode diagrams, the amplitude of the frequency response at $\omega_{driv,lm} = 2\pi 2.5 = 15.708\,[rad/s]$ is $A_{lm}(\omega_{driv,lm}) = 0.064444$ ( $A_{lms}(\omega_{driv,lm}) = 0.064544$ respectively according to the simplified model). In case the amplitude of the sinusoidal control signal applied to LAT is $5\,[V]$, the respective amplitude of the steady sinusoidal reaction of the output angle $\varphi_{lm}$ becomes $\left(5A_{lm}(\omega_{driv,lm})\right)*180/\pi = 18.5$ degrees, which result is confirmed also by simulations.

Considering the small mirror actuator's model in the same manner, the amplitude of frequency response at $\omega_{driv,sm} = 2\pi 20 = 125.66\,[rad/s]$ is $A_{sm}(\omega_{driv,sm}) = 0.00717$. Respectively the amplitude of the steady reaction to the sinusoidal control signal at this frequency with amplitude of $5\,[V]$ is $\left(5A_{sm}(\omega_{driv,sm})\right)*180/\pi = 2.05$ degrees.

This result confirmed also by simulations is slightly different from the result shared in correspondence by email - the amplitude of the small mirror actuator is **(7.13/2)=3.57** degrees: **Using this approach, the small fast mirror can be run over the full range (+/- 7.13 degrees) at full speed (approximately 20 Hz) using a driving signal of just 5v and 40mA.**

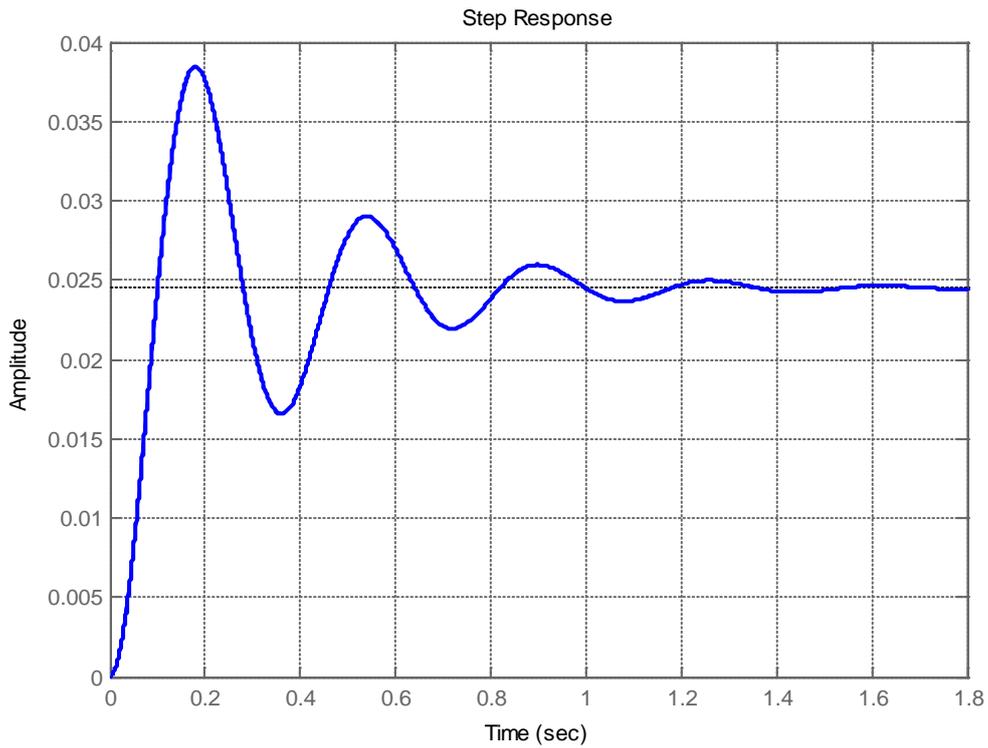

**Figure 15. Step responses of the state space, transfer function and block diagram models of the simplified large mirror actuator's model.**

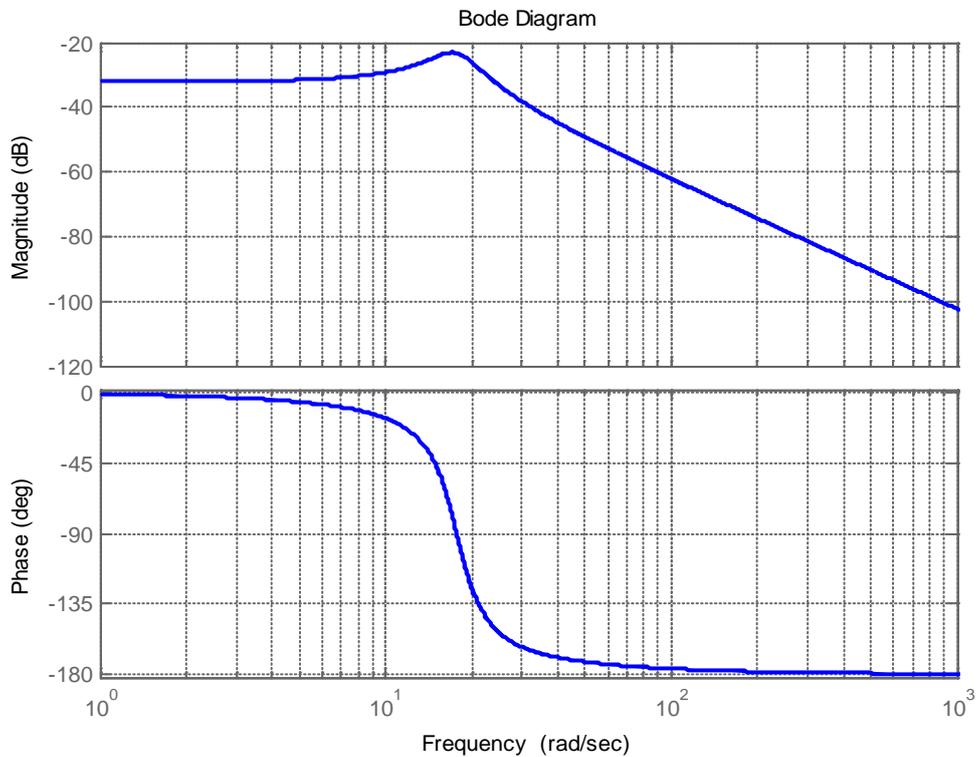

**Figure 16. Bode diagram of the simplified linear large mirror actuator's model.**

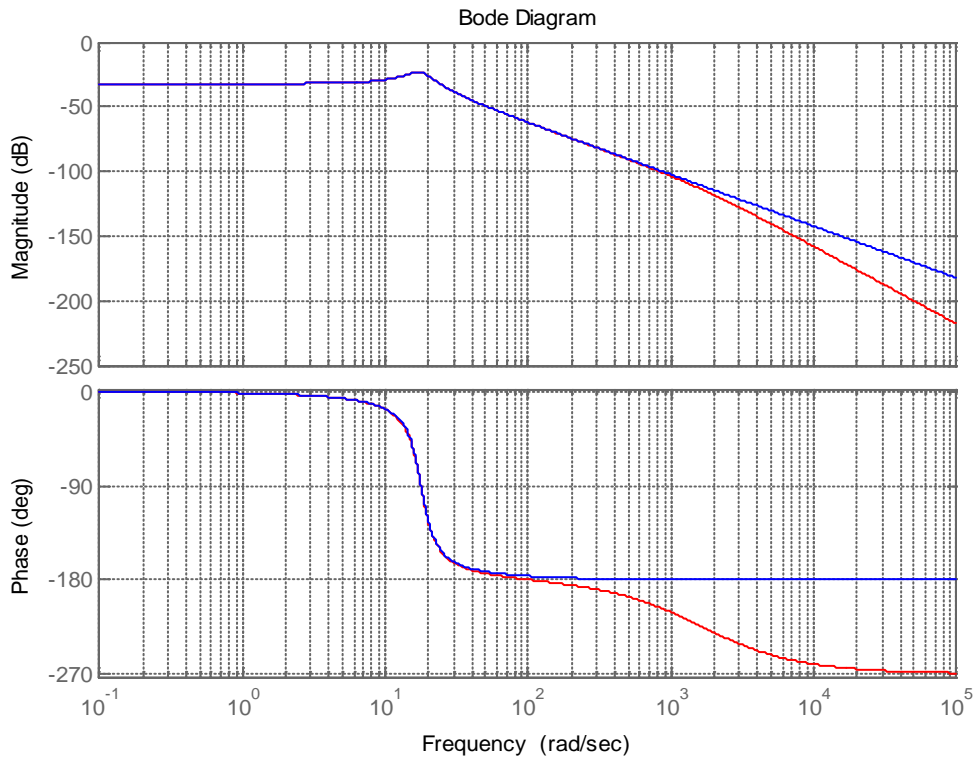

**Figure 17. Comparison between Bode diagrams of the linear large mirror actuator's model (-) and the simplified linear one (-).**

## 3.5  Inclusion of Coulomb's friction model into the linear models of the mirror actuators

At the linear modelling of the mirror actuators the resultant torque acting on the moment of inertia, assuming that there is no dry friction torque, is described by the expression

$$T_{RL}(t) = K_t i(t) - c\varphi(t) - h\omega(t)$$

and the respective equation of motion is

$$J\dot{\omega} = T_{RL}.$$

The Coulomb's perfect dry contact friction at rotation could be described by the following expression:

$$T_{CF} = \begin{cases} T_C & \text{if} \quad \omega > 0 \\ \in [-T_C, T_C] & \text{if} \quad \omega = 0, \\ -T_C & \text{if} \quad \omega < 0 \end{cases}$$

where $T_{CF}$ is the friction torque acting in a direction opposite to the direction of the rotating motion.

Taking into account the Coulomb's perfect dry contact friction at rotation, the equation of motion becomes

$$J\dot{\omega} = T_{RL} - T_{CF} = T_R \,.$$

A simulation model based on the above equation has been developed and included into the linear models of the mirror actuators.

The diagrams in Figure 18 represents the response of the large mirror actuator's linear model to the sinusoidal control signal $u_{lm}(t) = 5\cos(\omega_{driv,lm}t)$ $[V]$, $\omega_{driv,lm} = 2\pi 2.5 = 15.708$ $[rad/s]$.

The diagrams in Figure 19 represents the response of the large mirror actuator's model with included friction model to the same sinusoidal control signal as in the previous case.

The next three figures, Figure 20a, Figure 20b and Figure 20c, show collectively the diagrams of the linear component of the resultant torque $T_{RL}$ ('–'), of Coulomb's friction torque $T_{CF}$ ('–'), of the resultant torque $T_R$ ('–'), of the angular velocity $\omega_{lm}$ ('–') and of a flag of sticking ('–'), which becomes 1 in case of sticking and 0 in case of sliding, in the time interval [0, 5] seconds and zoomed – in the time intervals [0, 0.002] and [0, 0.35] seconds respectively.

There are three sticking phases in the beginning of the process in Figure 20, the process starts with a sticking phase, while at the steady oscillations the velocity crosses the zero with sliding without interlaced sticking phase.

The Step Response of the large mirror actuator's model with included friction model to the control signal $u_{lm}(t) = 5*1(t)$ $[V]$ is presented in Figure 21 and Figure 22. Figure 22 shows collectively the diagrams of the linear component of the resultant torque $T_{RL}$ ('–'), of Coulomb's friction torque $T_{CF}$ ('–'), of the resultant torque $T_R$ ('–'), of the angular velocity $\omega_{lm}$ ('–') and of a flag of sticking ('–').

Figure 23 shows the comparison between the step responses of the large mirror actuator's linear model ('–') and the large mirror actuator's model with included friction model ('–') to the control signal $u_{lm}(t) = 5*1(t)$ $[V]$.

Figure 24 shows collectively the step responses of the large mirror actuator's model with included friction model to the control signals $u_{lm}(t) = \left(\left(\dfrac{T_C R_{lm}}{K_t}\right) + u_0\right)*1(t)$ $[V]$, where $u_0 = 1, 2, 3, 4$ respectively.

The following expression describes the relation between the parameter $u_0$ and the steady output of the step responses in degrees:

$$\varphi_{lm}(\infty) = \begin{cases} 2.2026u_0 & \text{if} & u_{lm}(t) = \left(\left(\dfrac{T_C R_m}{K_t}\right) + u_0\right) * 1(t), & u_0 > 0 \\[3mm] 0 & \text{if} & u_{lm}(t) = \left(\left(\dfrac{T_C R_m}{K_t}\right) + u_0\right) * 1(t), & u_0 \in \left[-\left(\dfrac{T_C R_m}{K_t}\right), 0\right] \\[3mm] 0 & \text{if} & u_{lm}(t) = \left(-\left(\dfrac{T_C R_m}{K_t}\right) + u_0\right) * 1(t), & u_0 \in \left[0, \left(\dfrac{T_C R_m}{K_t}\right)\right] \\[3mm] 2.2026u_0 & \text{if} & u_{lm}(t) = \left(-\left(\dfrac{T_C R_m}{K_t}\right) + u_0\right) * 1(t), & u_0 < 0 \end{cases} \quad .$$

### 3.6 Conclusion regarding the scanning system's modelling

This part of the investigation provides mathematical and simulation models of the spatial movement of the scanning system as well as the linear and non-linear models of the mirror actuators. The modelling confirms the experimental results with the large and small mirror actuators as well as the whole scanning system. The elaborated linear and non-linear models of the actuators are an appropriate basis for the next stage of synthesizing time optimal control of the scanning system in order to achieve the best tracking control systems within the actual system's resources.

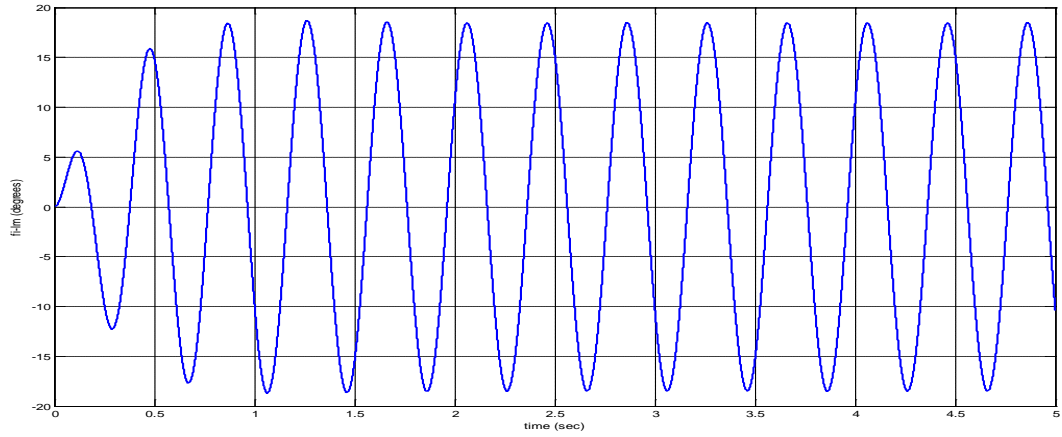

a) Output angle $\varphi_{lm}$ [deg]

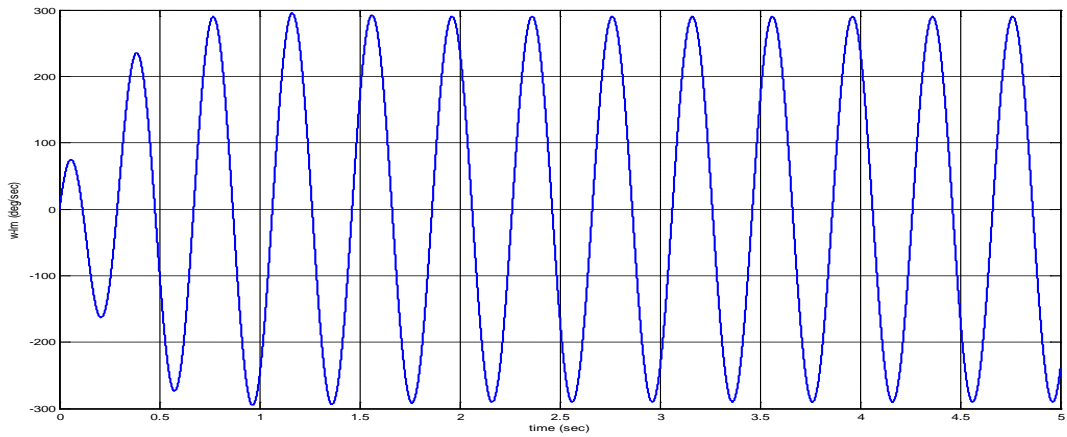

b) Angular velocity $\omega_{lm}$ [deg/sec]

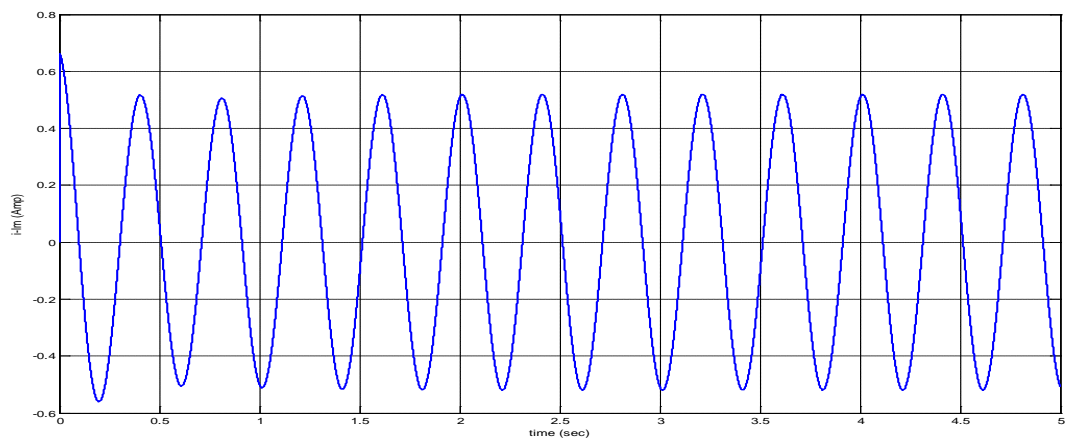

c) LAT current $i_{lm}$ [Amp]

**Figure 18. Response of the large mirror actuator's linear model to the sinusoidal control signal of 2.5 Hz and amplitude of 5 [V]** $u_{lm}(t) = 5\cos(\omega_{driv,lm}t)$ $[V]$.

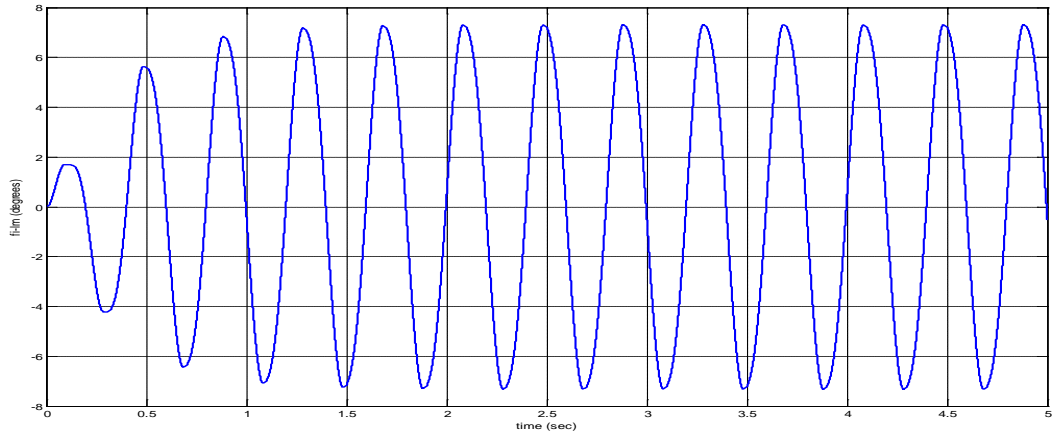

a)  Output angle $\varphi_{lm}$ [deg]

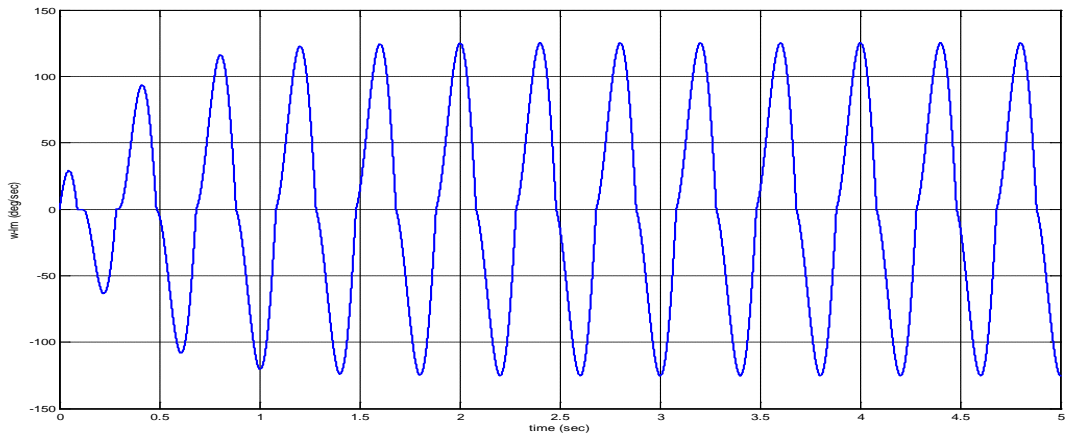

b)  Angular velocity $\omega_{lm}$ [deg/sec]

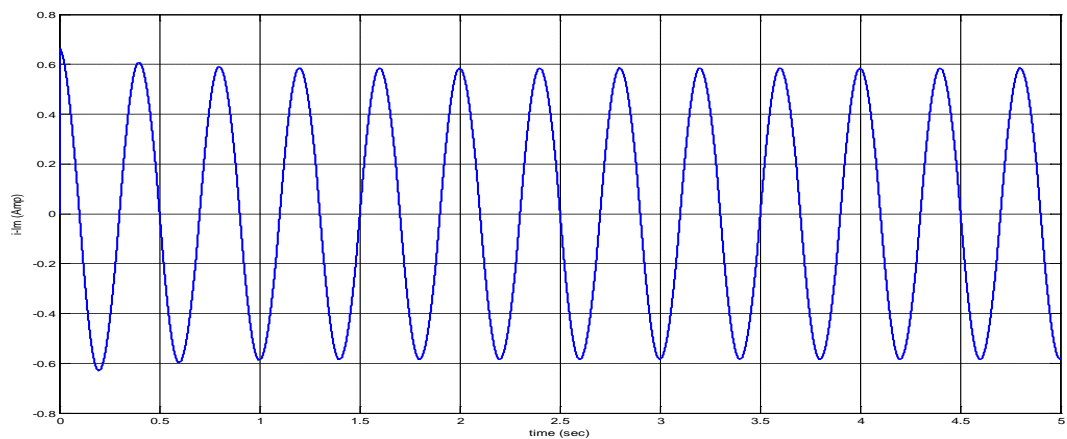

c)  LAT current $i_{lm}$ [Amp]

**Figure 19. Response of the large mirror actuator's model with included friction model to the sinusoidal control signal of 2.5 Hz and amplitude of 5 [V]** $u_{lm}(t) = 5\cos(\omega_{driv,lm}t)$ [V].

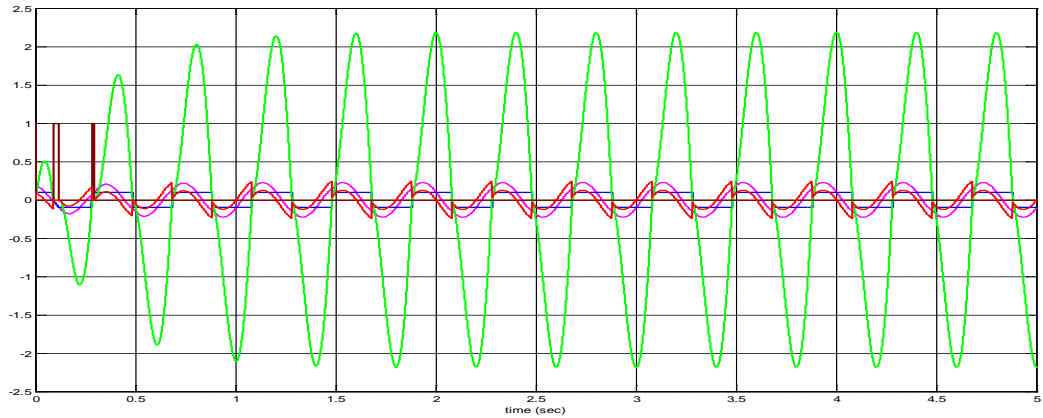

a) Time interval [0, 5] seconds

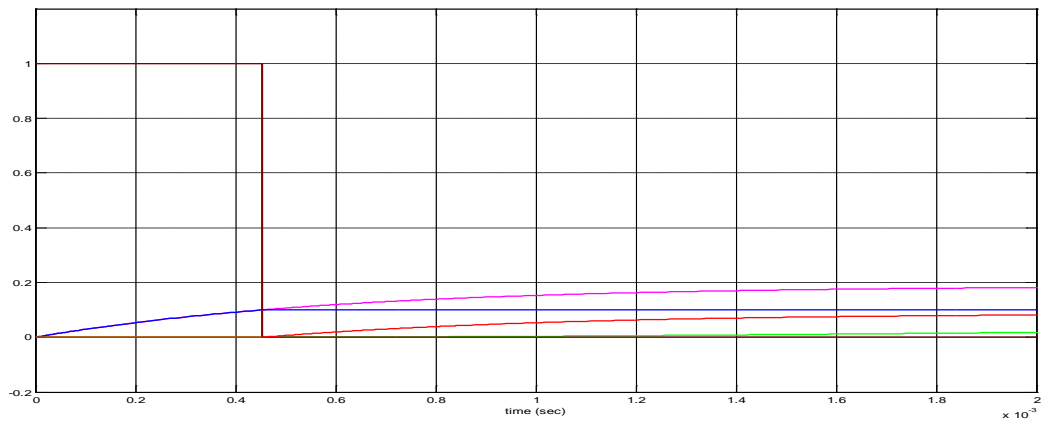

b) Time interval [0, 0.002] seconds

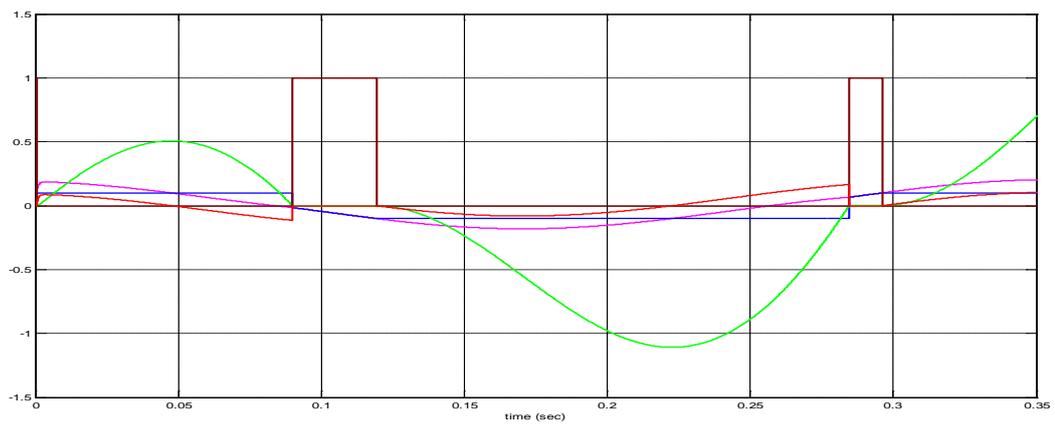

c) Time interval [0, 0.35] seconds

**Figure 20. Diagrams of the linear component of the resultant torque** $T_{RL}$ **[N.m] ('–'), of Coulomb's friction torque** $T_{CF}$ **[N.m] ('–'), of the resultant torque** $T_R$ **[N.m] ('–'), of the angular velocity** $\omega_{lm}$ **[rad/sec] ('–') and of a flag of sticking ('–') at sinusoidal control signal** $u_{lm}(t) = 5\cos(\omega_{driv,lm}t)\ [V]$.

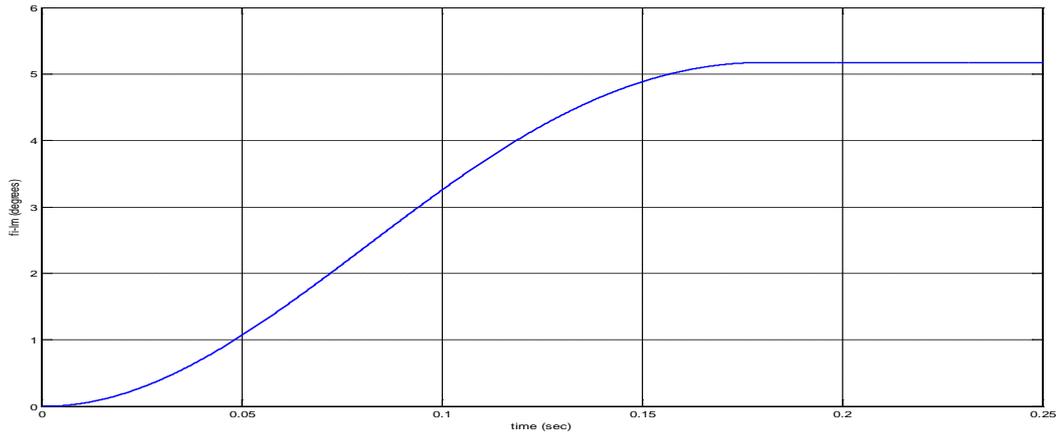

a)   Output angle  $\varphi_{lm}$  [deg]

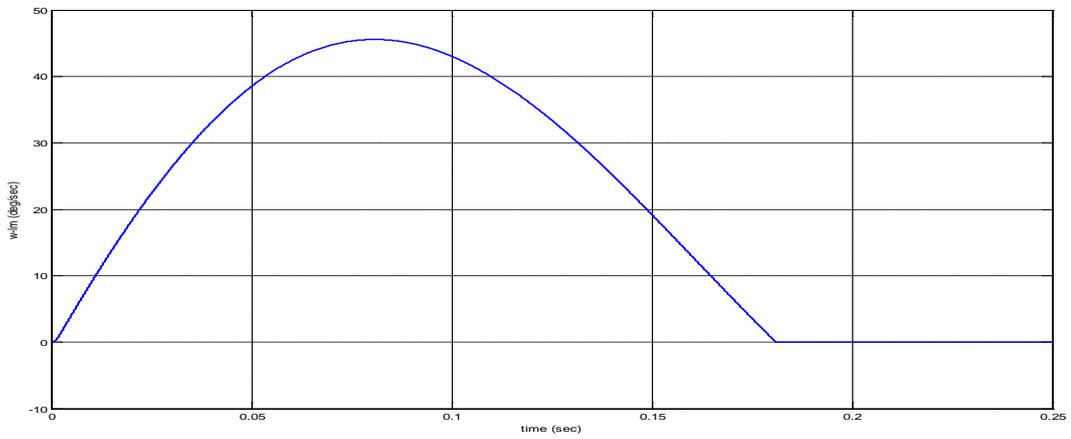

b)   Angular velocity  $\omega_{lm}$  [deg/sec]

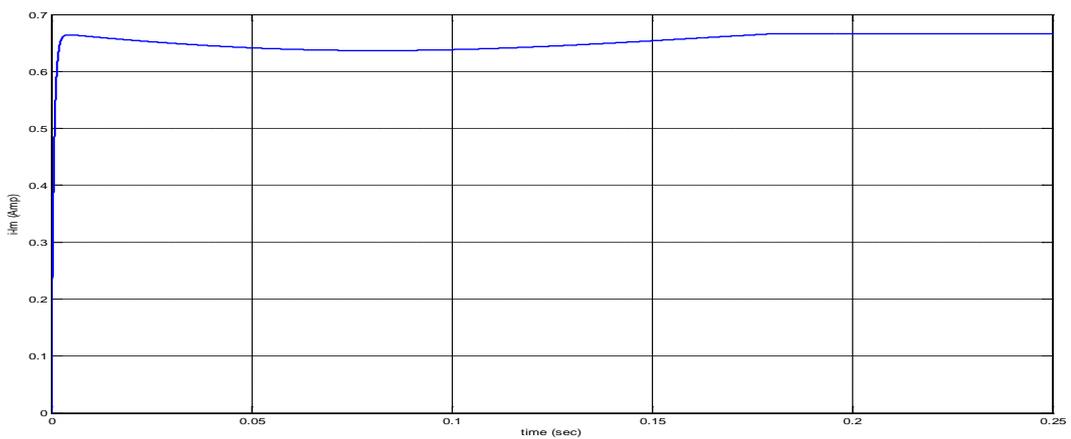

c)   LAT current  $i_{lm}$  [Amp]

**Figure 21. Step Response of the large mirror actuator's model with included friction model to the control signal** $u_{lm}(t) = 5*1(t)$ $[V]$.

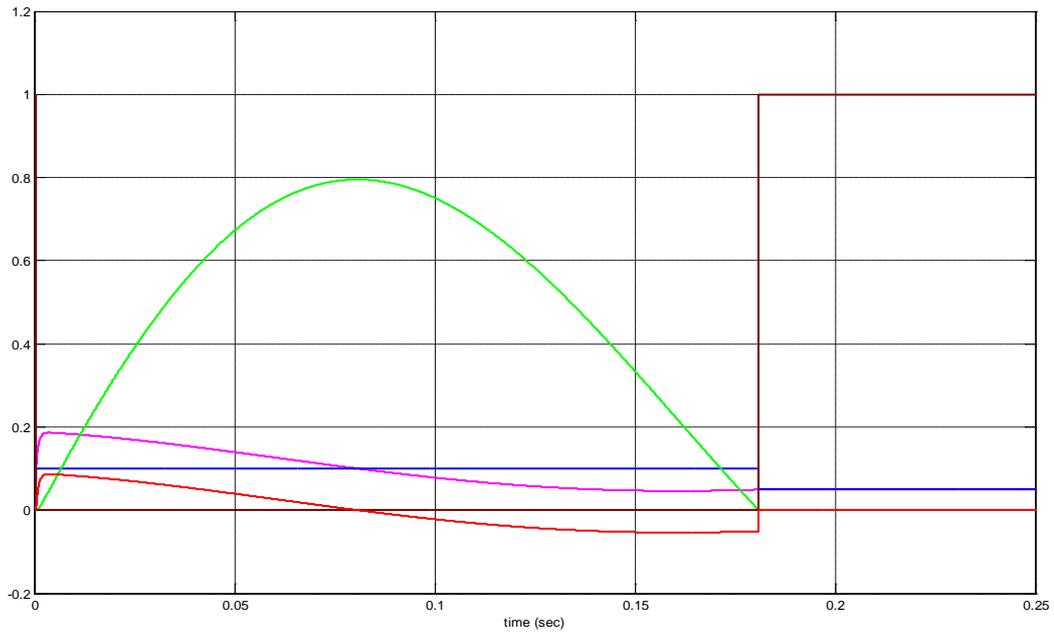

**Figure 22. Diagrams of the linear component of the resultant torque $T_{RL}$ [N.m] ('−'), of Coulomb's friction torque $T_{CF}$ [N.m] ('−'), of the resultant torque $T_R$ [N.m] ('−'), of the angular velocity $\omega_{lm}$ [rad/sec] ('−') and of a flag of sticking ('−') at control signal $u_{lm}(t) = 5*1(t)$ [V].**

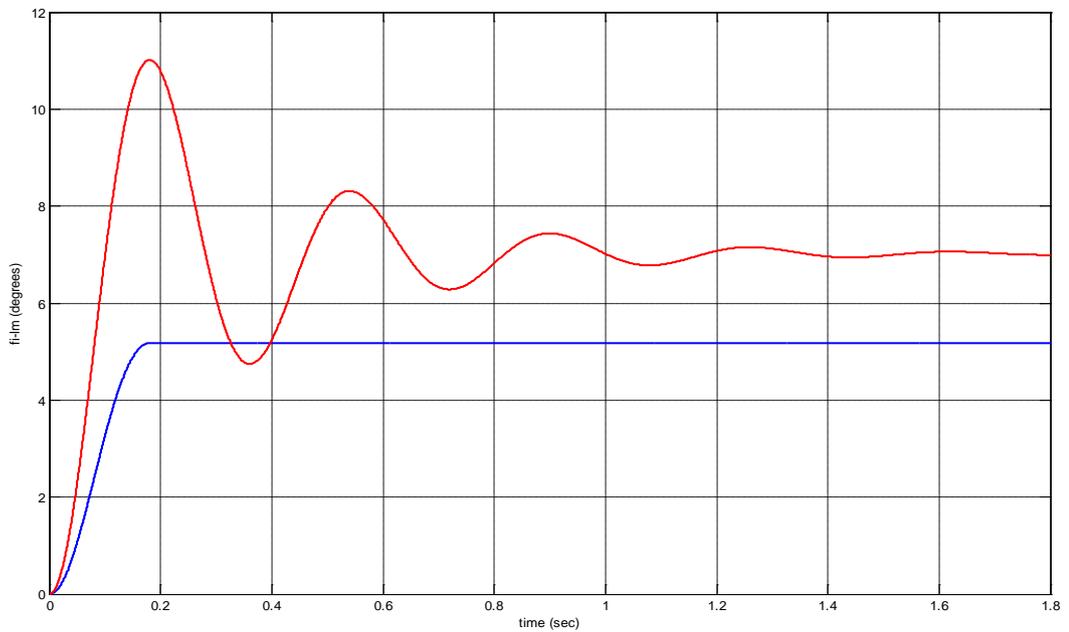

**Figure 23. Comparison between the step responses of the large mirror actuator's linear model ('−') and the large mirror actuator's model with included friction model ('−') to the control signal $u_{lm}(t) = 5*1(t)$ [V].**

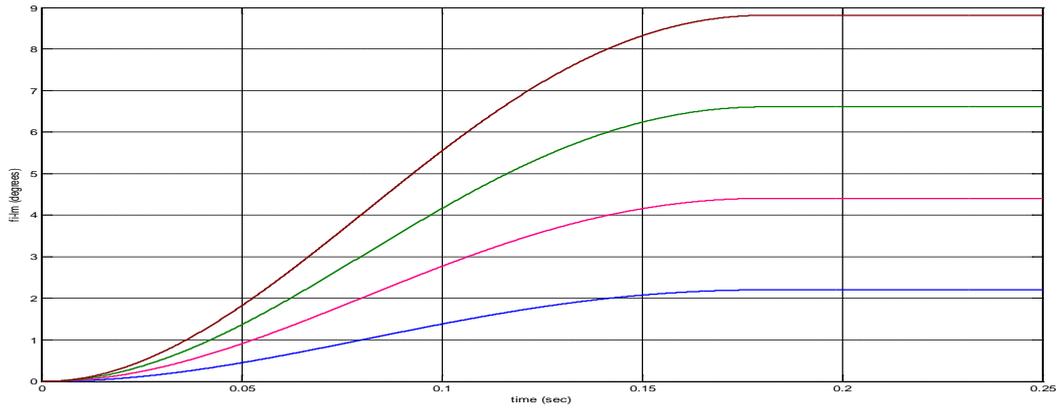

a) Output angle $\varphi_{lm}$ [deg]

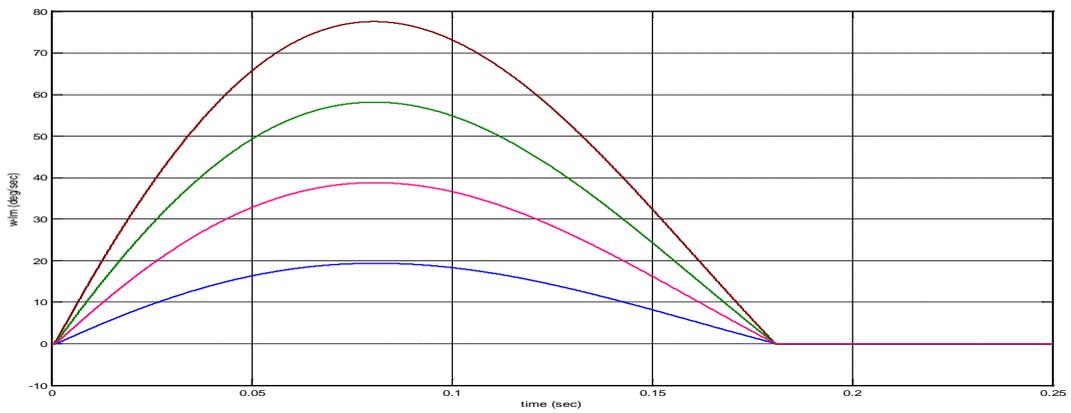

b) Angular velocity $\omega_{lm}$ [deg/sec]

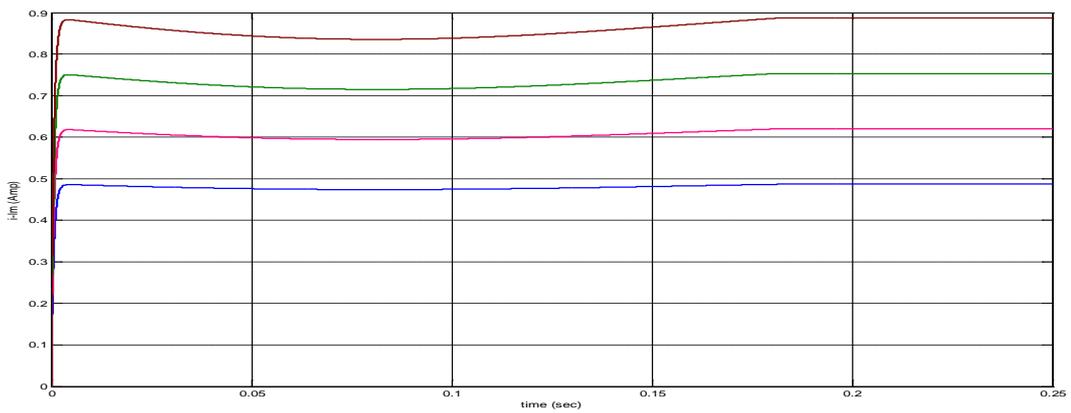

c) LAT current $i_{lm}$ [Amp]

**Figure 24. Comparison between the step responses of the large mirror actuator's model with included friction model to the control signals** $u_{lm}(t) = \left( \left( \dfrac{T_c R_m}{K_t} \right) + u_0 \right) * 1(t) \;\; [V]$ **, where** $u_0 = $ **1 ('−'), 2 ('−'), 3 ('−'), 4 ('−') respectively.**

# 4. Closed loop systems for the large mirror actuator based on the synthesis of time optimal control

Let us consider once again the simplified model of the large mirror actuator.

$$\dot{\boldsymbol{x}}_{lms} = A_{lms}\boldsymbol{x}_{lms} + B_{lms}u_{lm},$$

$$y_{lms} = C_{lms}\boldsymbol{x}_{lms} + D_{lms}u_{lm},$$

$$\boldsymbol{x}_{lms} = \begin{bmatrix} \varphi_{lm} \\ \omega_{lm} \end{bmatrix}, \quad A_{lms} = \begin{bmatrix} 0 & 1 \\ -c_{lm}/J_{lm} & (-(K_bK_t)/(R_mJ_{lm}) - h_{lm}/J_{lm}) \end{bmatrix}, \quad B_{lms} = \begin{bmatrix} 0 \\ K_t/(R_mJ_{lm}) \end{bmatrix},$$

$$C_{lms} = \begin{bmatrix} 1 & 0 \end{bmatrix}, \quad D_{lms} = 0.$$

At the initial data the respective numerical representation of the matrices is:

$$A_{lms} = \begin{bmatrix} 0 & 1 \\ -314.29 & -6.261 \end{bmatrix}, \quad B_{lms} = \begin{bmatrix} 0 \\ 7.7 \end{bmatrix}.$$

The eigenvalues of $A_{lms}$ are

$$\lambda_{lms} = \begin{bmatrix} -3.1305 + 17.45j \\ -3.1305 - 17.45j \end{bmatrix}.$$

They determine the properties, mentioned at considering the model, so that the actuator works well as an open loop system driven by a sinusoidal signal at a frequency, very close to the resonant frequency of the system.

The developed idea here outlines a tracking control system, where the demand signal represents a periodic signal with a frequency 2.5 Hz, where the controlled system – the mirror actuator – has no oscillating properties.

The approaches to achieve non-oscillating properties of the actuator without changes on the LAT and the moment of inertia involve a change of the pivot stiffness $c_{lm}$ or/and damping $h_{lm}$ of the actuator's mechanical subsystem.

## 4.1 Approach based on the time optimal control of a large mirror actuator with increased damping $h_{lm}$ of the mechanical subsystem

By increasing the damping $h_{lm}$ ten times, from $h_{lm} = 0.02$ to $h_{lm} = 0.2$ [ N*m*sec/rad], we obtain a new numerical representation of the system matrix $A_{lms}$

$$A_{lms} = \begin{bmatrix} 0 & 1 \\ -314.29 & -42.996 \end{bmatrix}$$

with respective eigenvalues

$$\boldsymbol{\lambda}_{lms} = \begin{bmatrix} -9.3376 \\ -33.658 \end{bmatrix}.$$

This change doesn't affect the coefficient $K_{lm}$, $K_{lm} = \dfrac{K_t}{R_m c_{lm}} = 0.0245 \quad [rad/V]$, of the large mirror actuator's transfer function, but drastically changes the properties of the step response as shown in the next Figure 25.

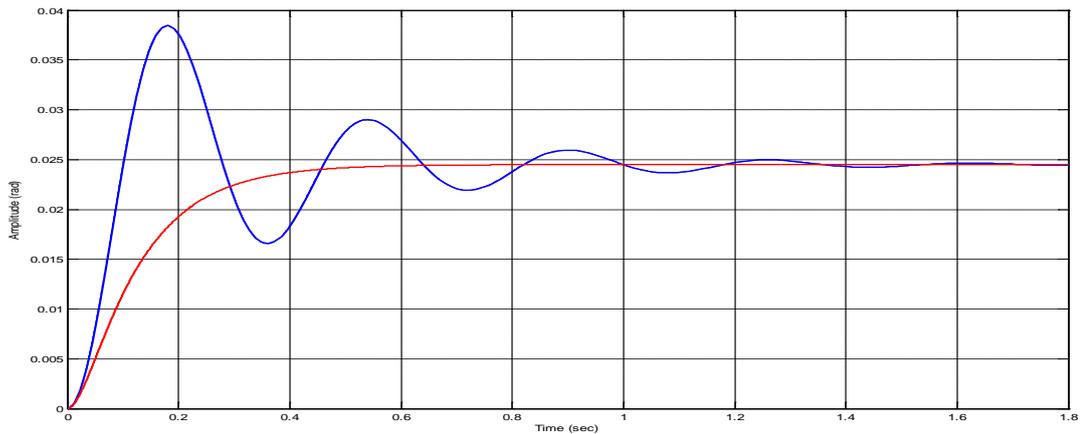

**Figure 25. Comparison between the step responses of the initial simplified linear model ('–') and corrected model by increasing the damping ten times ('–'). The control signal is** $u_{lm}(t) = 1(t) \quad [V]$ **.**

### 4.1.1  Time optimal control of the corrected large mirror actuator's models

Let us deal first with the solution of the time optimal control problem for the corrected linear model of the actuator. We solve the problem by the author's method. Let us assume the constraints for the control signal $u_{lm}(t)$ are

$$-u_0 \le u_{lm}(t) \le u_0 .$$

Figure 26 shows the time-diagrams on $\varphi_{lm}$, $\omega_{lm}$, and the control signal $u_{lm}$ for the transition from zero initial state to the final state

$$\boldsymbol{x}_{lms}(t_f) = \begin{bmatrix} \varphi_{lm}(t_f) \\ \omega_{lm}(t_f) \end{bmatrix} = \begin{bmatrix} 8.35 \\ 0 \end{bmatrix} \quad \begin{pmatrix} \text{deg} \\ \text{deg/sec} \end{pmatrix}$$

with an accuracy of $3.521\mathrm{e}\text{-}005\,(\mathrm{deg})$ on $\varphi_{lm}$ and $0.0012\,(\mathrm{deg/sec})$ on $\omega_{lm}$ in case $u_0 = 10\,[V]$. The solution represents a piece-wise constant function with two intervals of constancy and an amplitude $u_0 = 10\,[V]$. The length of the first interval is $0.12713\,(\mathrm{sec})$, and of the second one is $0.00652\,(\text{sec})$. The near minimum time for this transition is $t_f^{\tilde{a}} = 0.13365\,(\text{sec})$.

Figure 27 shows a near time optimal control solution for the transition from zero initial state to the final state as in the previous case, but with constraint on the control signal $u_0 = 20\,[V]$. The length of the first interval is $0.061457\,(\text{sec})$, the length of the second one is $0.010919\,(\text{sec})$. The near minimum time for this transition is $t_f^{\tilde{a}} = 0.072377\,(\text{sec})$.

The correction of damping causes only the numerical representation of the system matrix $A_{lm}$ of the linear model of the large mirror actuator of 3$^{\text{rd}}$ order.

$$\dot{\boldsymbol{x}}_{lm} = A_{lm}\boldsymbol{x}_{sm} + B_{lm}u_{lm}\,,$$

$$y_{lm} = C_{lm}\boldsymbol{x}_{lm} + D_{lm}u_{lm}\,,$$

$$\boldsymbol{x}_{lm} = \begin{bmatrix} \varphi_{lm} \\ \omega_{lm} \\ i_{lm} \end{bmatrix}, \quad A_{lm} = \begin{bmatrix} 0 & 1 & 0 \\ -c_{lm}/J_{lm} & -h_{lm}/J_{lm} & Kt/J_{lm} \\ 0 & -Kb/Lm & -Rm/Lm \end{bmatrix}, \quad B_{lm} = \begin{bmatrix} 0 \\ 0 \\ 1/Lm \end{bmatrix},$$

$$C_{lm} = \begin{bmatrix} 1 & 0 & 0 \end{bmatrix}, \quad D_{lm} = 0\,.$$

The initial representation of $A_{lm}$

$$A_{lm} = \begin{bmatrix} 0 & 1 & 0 \\ -314.29 & -4.0816 & 57.755 \\ 0 & -62.889 & -1666.7 \end{bmatrix},$$

having eigenvalues

$$\lambda_{lm} = \begin{bmatrix} -3.1345 + 17.46\,j \\ -3.1345 - 17.46\,j \\ -1664.5 \end{bmatrix},$$

at this correction changes to

$$A_{lm} = \begin{bmatrix} 0 & 1 & 0 \\ -314.29 & -40.816 & 57.755 \\ 0 & -62.889 & -1666.7 \end{bmatrix}$$

with eigenvalues

$$\lambda_{lm} = \begin{bmatrix} -9.3329 \\ -33.72 \\ -1664.4 \end{bmatrix}.$$

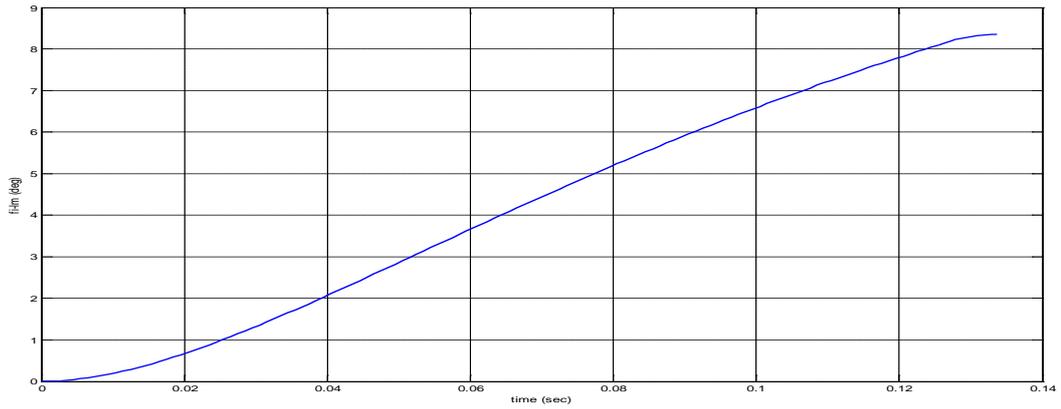

a) Output angle $\varphi_{lm}$ [deg]

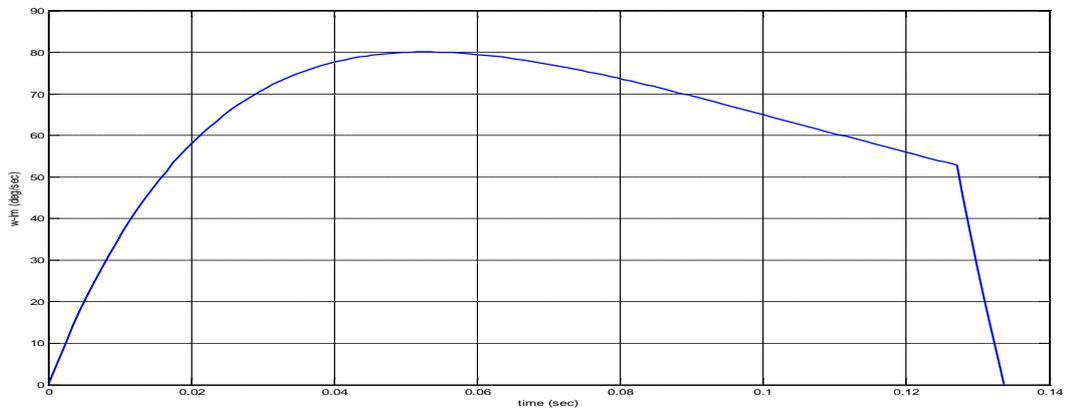

b) Angular velocity $\omega_{lm}$ [deg/sec]

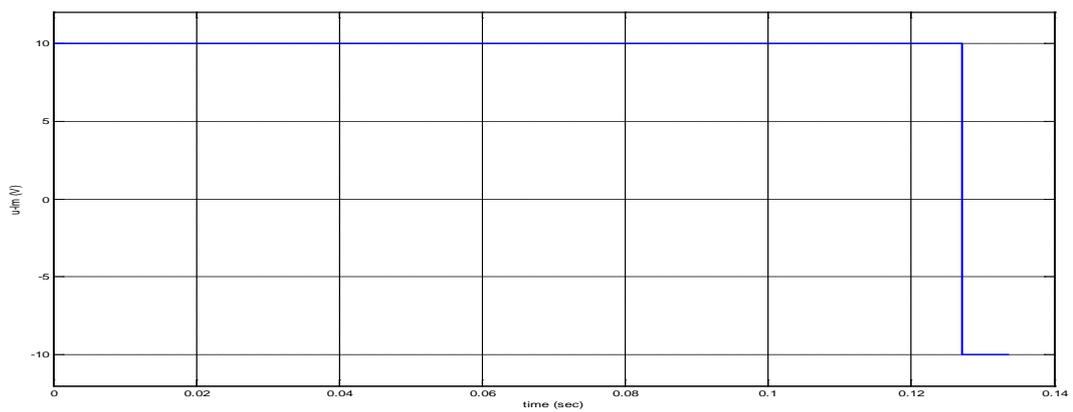

c) Near time optimal control signal $u_{lm}^{\tilde{o}}$ [V]

**Figure 26. Time-diagrams on the output angle $\varphi_{lm}$ [deg], angular velocity $\omega_{lm}$ [deg/sec] and near time optimal control signal $u_{lm}^{\tilde{o}}$ [V] in case $u_0 = 10\,[V]$ and demand position on $\varphi_{lm} - 8.35\,[\text{deg}]$.**

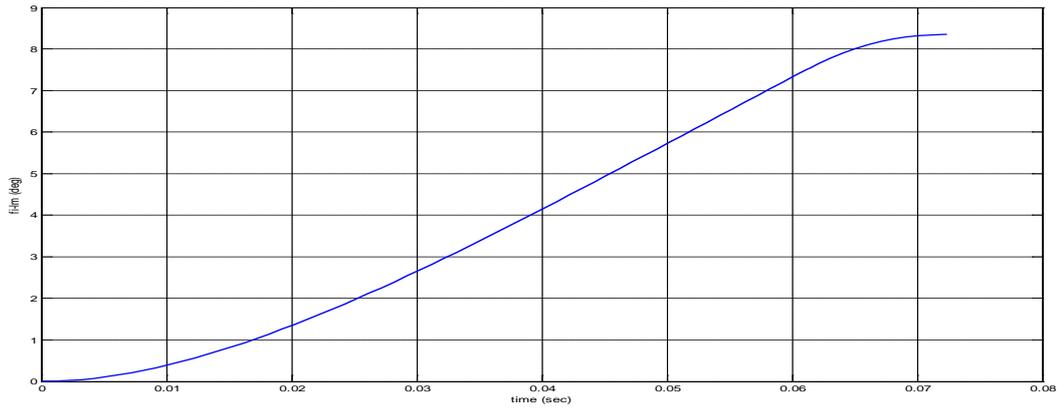

a)  Output angle  $\varphi_{lm}$  [deg]

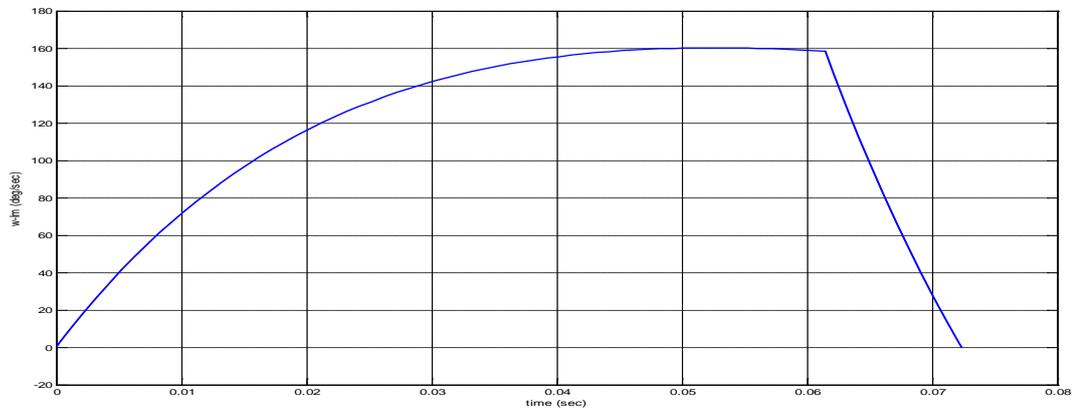

b)  Angular velocity  $\omega_{lm}$  [deg/sec]

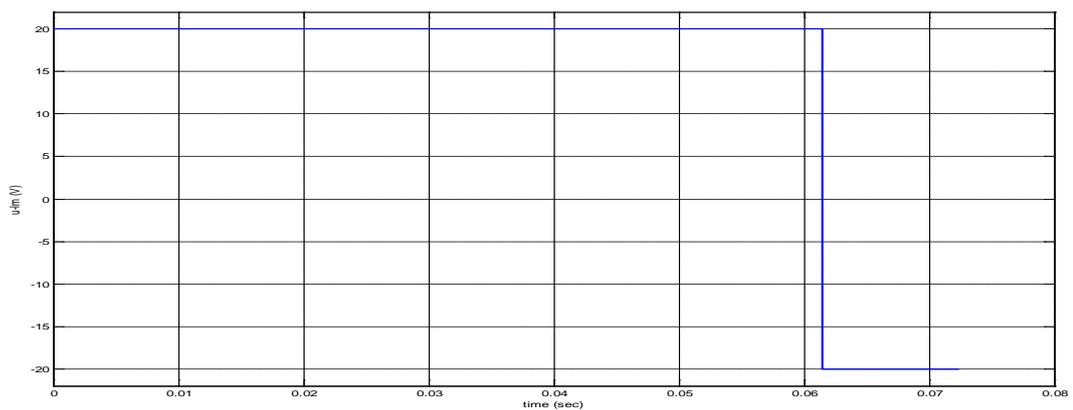

c)  Near time optimal control signal  $u_{lm}^{\tilde{o}}$  [V]

**Figure 27. Time-diagrams on the output angle  $\varphi_{lm}$  [deg], angular velocity  $\omega_{lm}$  [deg/sec] and near time optimal control signal  $u_{lm}^{\tilde{o}}$  [V] in case  $u_0 = 20\,[V]$  and demand position on  $\varphi_{lm} - 8.35\,[\mathrm{deg}]$.**

Let us deal now with the solution of the time optimal control problem for the corrected linear model of $3^{rd}$ order. Figure 28 shows the time-diagrams of a near the time optimal control solution for the transition from zero initial state to the final state

$$\boldsymbol{x}_{lm}(t_f) = \begin{bmatrix} \varphi_{lm}(t_f) \\ \omega_{lm}(t_f) \\ i_{lm}(t_f) \end{bmatrix} = \begin{bmatrix} 8.35 \\ 0 \\ 0 \end{bmatrix} \quad \begin{pmatrix} \text{deg} \\ \text{deg/sec} \\ \text{Amp} \end{pmatrix}$$

in case $u_0 = 10\,[V]$. The solution represents a piece-wise constant function with three intervals of constancy and an amplitude $u_0 = 10\,[V]$. The length of the first interval is $0.12715\,(\text{sec})$, of the second one is $0.006846\,(\text{sec})$ and of the third interval - $0.000417\,(\text{sec})$. The near minimum time for this transition is $t_f^{\tilde{o}} = 0.13441\,(\text{sec})$. The accuracy on $\varphi_{lm}$ is $7.0\text{e-}8\,[\text{deg}]$, on $\omega_{lm} - 2.0\text{e-}6\,[\text{deg/sec}]$ and on $i_{lm} - 2.0\text{e-}6\,[A]$.

Figure 29 shows a near time optimal control solution for the transition from zero initial state to the final state as in the previous case, but with constraint on the control signal $u_0 = 20\,[V]$. The length of the first interval is $0.061453\,(\text{sec})$, the length of the second one is $0.011276\,(\text{sec})$ and the length of the third one is $0.000417\,(\text{sec})$. The near minimum time for this transition is $t_f^{\tilde{o}} = 0.073145\,(\text{sec})$. The accuracy on $\varphi_{lm}$ is $2.0\text{e-}8\,[\text{deg}]$, on $\omega_{lm} - 5.0\text{e-}6\,[\text{deg/sec}]$ and on $i_{lm} - 3.0\text{e-}6\,[A]$.

Based on the comparison between the solutions for the simplified linear model of $2^{nd}$ order and the solutions for the linear model of $3^{rd}$ order, shown in Table 3 and Table 4 in case $u_0 = 10\,[V]$ and $u_0 = 20\,[V]$ respectively, we accept first the approach of controlling the large mirror actuator's models in the closed loop system on the time optimal control synthesis algorithm for the simplified model of the actuator.

**Table 3. Comparison between the solutions for the linear models of the $2^{nd}$ and $3^{rd}$ order in case $u_0 = 10\,[V]$.**

| Near time optimal control solution in case $u_0 = 10\,[V]$ | Simplified linear model of $2^{nd}$ order | Linear model of $3^{rd}$ order |
|---|---|---|
| Length of the first interval [s] | 0.12713 | 0.12715 |
| Length of the second interval [s] | 0.00652 | 0.00685 |
| Length of the third interval [s] | - | 0.00042 |
| Near minimum time for the transition $t_f^{\tilde{o}}$ [s] | 0.13365 | 0.13441 |

**Table 4.** Comparison between the solutions for the linear models of the 2$^{nd}$ and 3$^{rd}$ order in case $u_0 = 20\,[V]$.

| Near time optimal control solution in case $u_0 = 20\,[V]$ | Simplified linear model of 2$^{nd}$ order | Linear model of 3$^{rd}$ order |
|---|---|---|
| Length of the first interval [s] | 0.061457 | 0.061453 |
| Length of the second interval [s] | 0.010919 | 0.011276 |
| Length of the third interval [s] | - | 0.000417 |
| Near minimum time for the transition $t_f^{\tilde{a}}$ [s] | 0.072377 | 0.073145 |

Consequently have been investigated the systems with the following models of the actuator: the simplified linear model of 2$^{nd}$ order, the linear one of 3$^{rd}$ order, the non-linear model of 2$^{nd}$ order with included Coulomb's friction model, the non-linear model of 3$^{rd}$ order with included Coulomb's friction model.

Alongside with increasing the complexity of the controlled system's model, the impact of the sampling time has been investigated. One millisecond could be accepted as an upper limit for the sampling time. The inclusion of a special one-step prediction mechanism could reduce nearly twice the amplitude of the steady oscillations at tracking.

Figure 30 shows a comparison between the near time optimal control solution ('-'), the process of the closed loop system "Time optimal controller with sampling time of 1 millisecond − Large mirror actuator's linear model of 3$^{rd}$ order" ('-') and the closed loop system "Time optimal controller with sampling time of 1 millisecond and the prediction mechanism included − Large mirror actuator's linear model of 3$^{rd}$ order" ('-'). The demand position on $\varphi_{lm}$ is $8.35\,[\mathrm{deg}]$ and $u_0 = 10\,[V]$. The oscillations' amplitude around the demand position as mentioned is reduced by the prediction mechanism.

Figure 31 shows the time-diagrams in the closed loop system "Time optimal controller with sampling time of 1 millisecond and the prediction mechanism included − Large mirror actuator's non-linear model of 3$^{rd}$ order with included Coulomb's friction model" in case the demand position on $\varphi_{lm}$ is $8.35\,[\mathrm{deg}]$ and $u_0 = 10\,[V]$. In Figure 31a and Figure 31b by ('-') are shown also the near time optimal solution on $\varphi_{lm}$ and $\omega_{lm}$ for the linear actuator's model. As it is shown in Figure 31c, there is only one short sticking phase in the beginning of the process. After this phase the process continues with sliding.

Figure 32 shows the time-diagrams in the closed loop system "Time optimal controller with sampling time of 1 millisecond and the prediction mechanism included − Large mirror actuator's non-linear model of 3$^{rd}$ order with included Coulomb's friction model" in the same manner as Figure 31 for the same demand position on $\varphi_{lm}$ - $8.35\,[\mathrm{deg}]$, but with constraint on $u$ - $u_0 = 20\,[V]$. In Figure 32a and Figure 32b by ('$-$') are shown also the near time optimal solution on $\varphi_{lm}$ and $\omega_{lm}$ for the linear actuator's model in case $u_0 = 20\,[V]$.

Dealing with the closed loop system "Time optimal controller with sampling time of 1 millisecond and the prediction mechanism included – Large mirror actuator's non-linear model of 3$^{rd}$ order with included Coulomb's friction model", Figure 33 shows the time-diagrams in the closed loop system at periodic demand signal, representing a square wave with an amplitude of $8.35\,[\deg]$ and frequency 2.5 Hz, in case $u_0 = 10\,[V]$. Figure 34 shows the processes in the same system but with constraint on $u$ - $u_0 = 20\,[V]$. In case $u_0 = 10\,[V]$ the output angle $\varphi_{lm}$ can't reach the demand position of $8.35\,[\deg]$ in $0.2\,[\sec]$ representing the half of the demand signal's period. This fact could be seen already in Figure 31. But the system responses very well to the periodic square wave demand signal as shown in Figure 34 in case $u_0 = 20\,[V]$.

Figure 35 and Figure 36 show the tracking of the sinusoidal demand signal with an amplitude of $8.35\,[\deg]$ and frequency 2.5 Hz in case $u_0 = 10\,[V]$ and $u_0 = 20\,[V]$ respectively. At $u_0 = 10\,[V]$ the system can't follow in the demand harmonic signal with an amplitude of $8.35\,[\deg]$, but at $u_0 = 20\,[V]$ the system manages well with tracking. The system after the first control signal's switching from $+20\,[V]$ to $-20\,[V]$, Figure 36e, i.e. after the first two intervals of constancy, continues with tracking the demand signal in sliding mode.

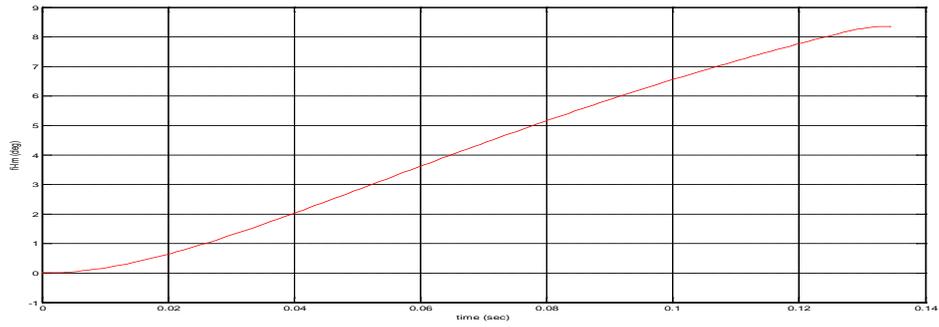

a)  Output angle $\varphi_{lm}$ [deg]

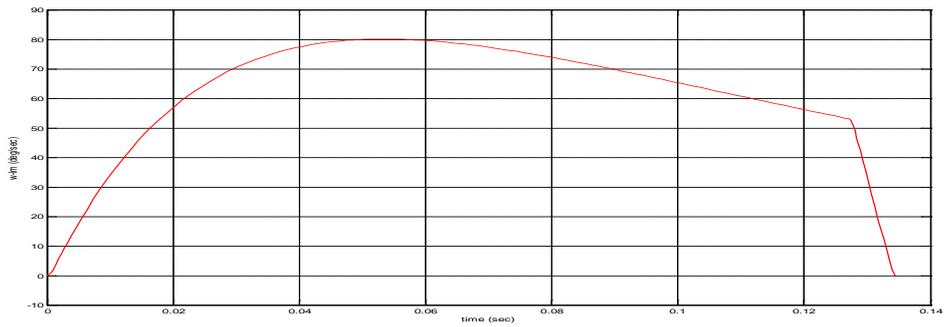

b)  Angular velocity $\omega_{lm}$ [deg/sec]

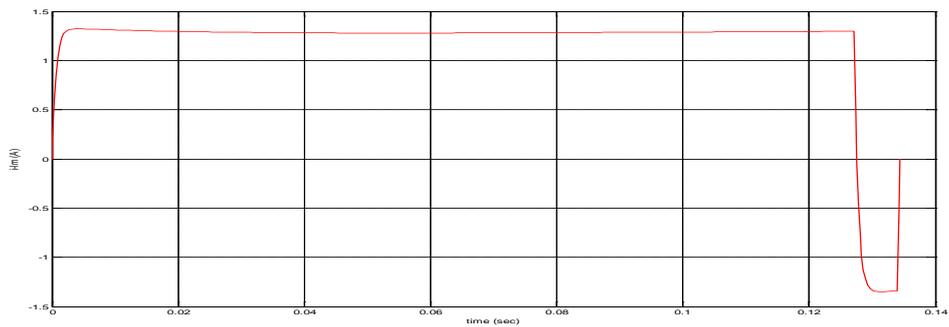

c)  LAT current $i_{lm}$ [A]

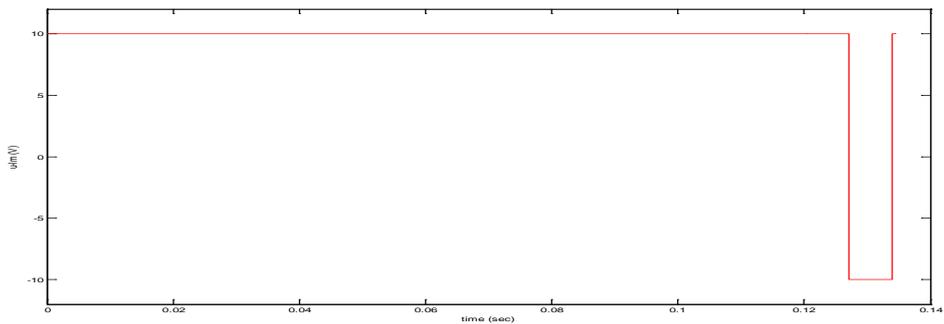

d)  Near time optimal control $u_{lm}^{\tilde{o}}$ [V]

**Figure 28. A Near time optimal solution for the transition from zero initial state to demand position on** $\varphi_{lm}$ $8.35\,[\text{deg}]$ **in case** $u_0 = 10\,[V]$ **with accuracy on** $\varphi_{lm} - 7.0\text{e-}8\,[\text{deg}]$**, on** $\omega_{lm} - 2.0\text{e-}6\,[\text{deg/sec}]$ **and on** $i_{lm} - 2.0\text{e-}6\,[A]$**.**

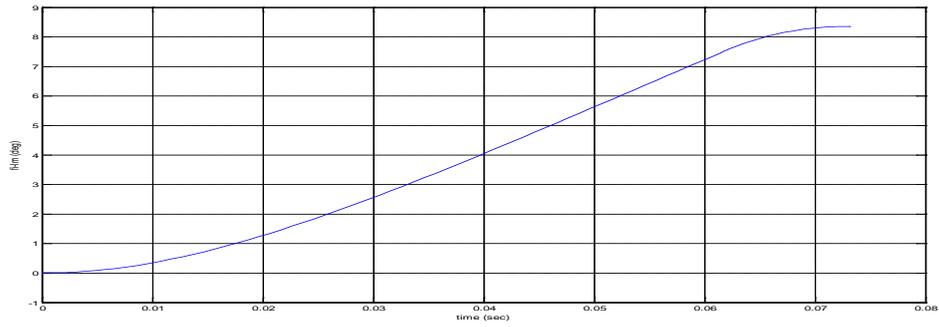

a) Output angle $\varphi_{lm}$ [deg]

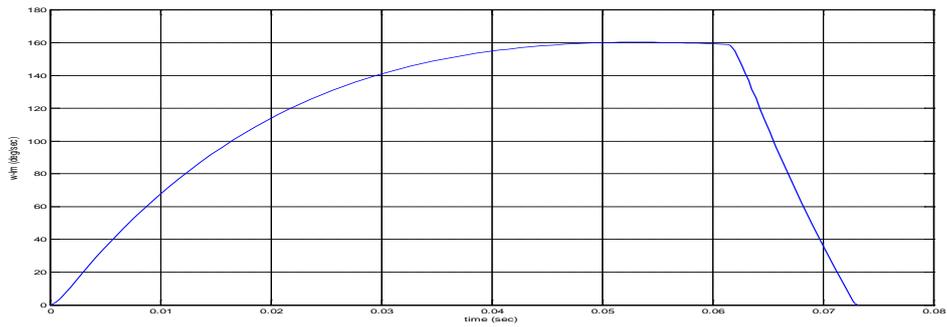

b) Angular velocity $\omega_{lm}$ [deg/sec]

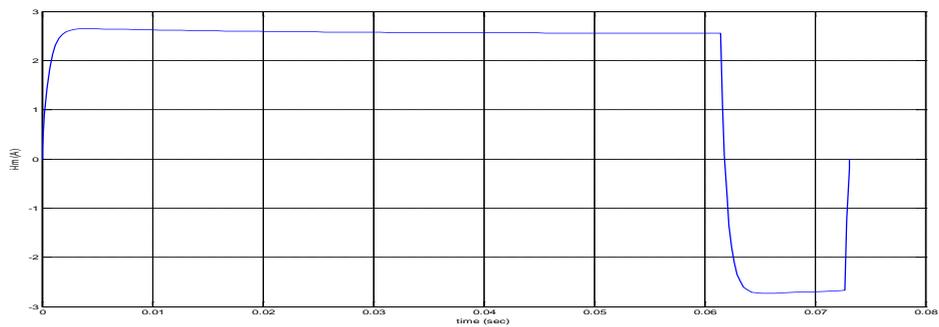

c) LAT current $i_{lm}$ [A]

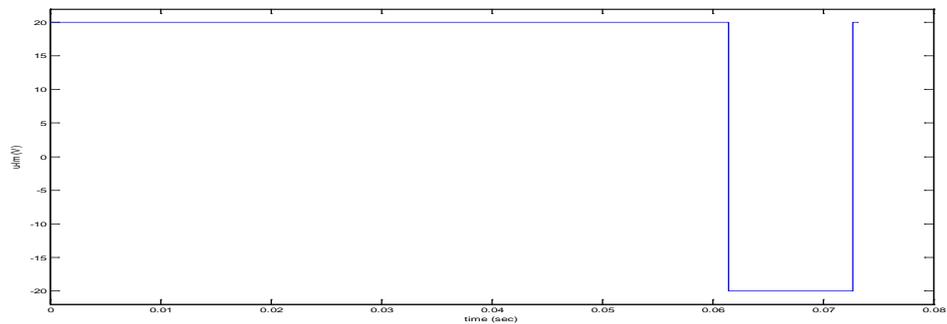

d) Near time optimal control $u_{lm}^{\tilde{o}}$ [V]

**Figure 29. A Near time optimal solution for the transition from zero initial state to demand position on** $\varphi_{lm}$ $8.35\,[\text{deg}]$ **in case** $u_0 = 20\,[V]$ **with accuracy on** $\varphi_{lm} - 2.0\text{e-}8\,[\text{deg}]$**, on** $\omega_{lm} - 5.0\text{e-}6\,[\text{deg/sec}]$ **and on** $i_{lm} - 3.0\text{e-}6\,[A]$ **.**

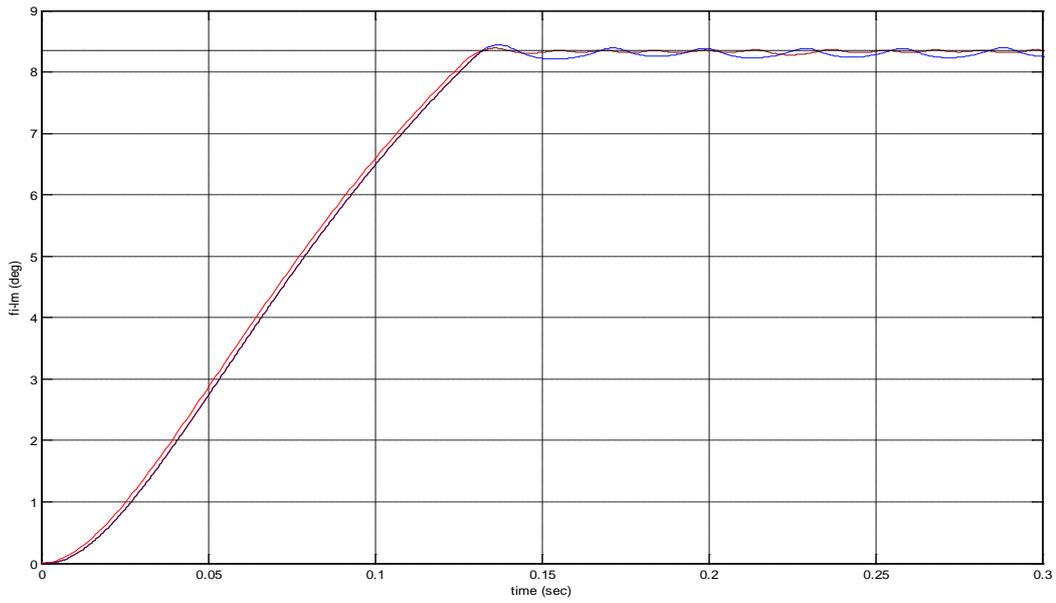

a) Output angle $\varphi_{lm}$ [deg]

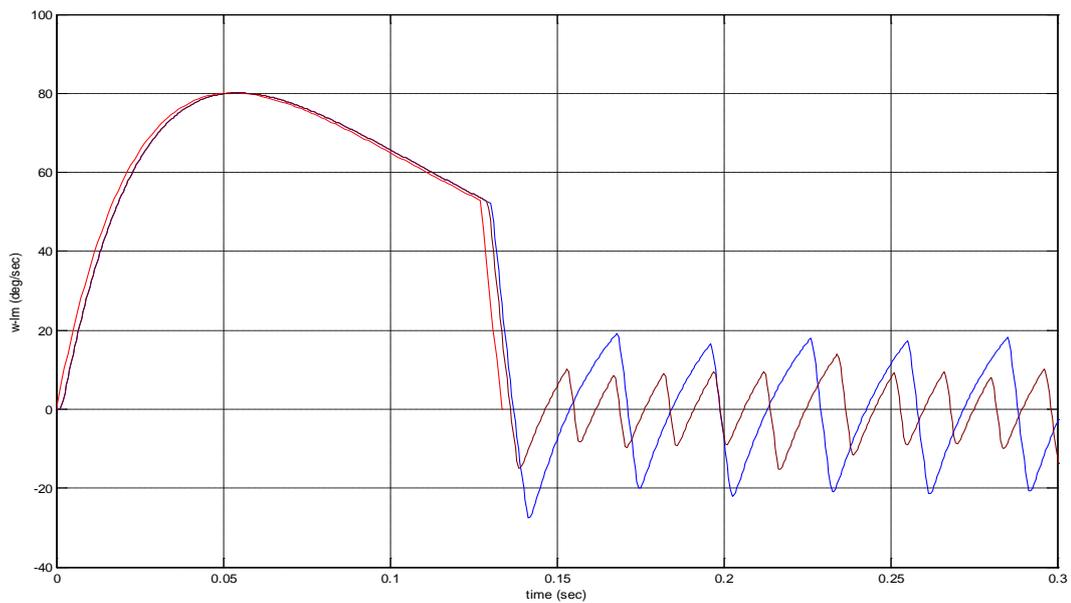

b) Angular velocity $\omega_{lm}$ [deg/sec]

**Figure 30. Time-diagrams on the output angle $\varphi_{lm}$ [deg] and angular velocity $\omega_{lm}$ [deg/sec] in case $u_0 = 10\,[V]$ and demand position on $\varphi_{lm} - 8.35\,[\text{deg}]$ of the near time optimal control solution ('–'), of the closed loop system "Time optimal controller with sampling time of 1 ms – Large mirror actuator's linear model" ('–') and of the closed loop system "Time optimal controller with sampling time of 1 ms and the prediction mechanism included – Large mirror actuator's linear model" ('–').**

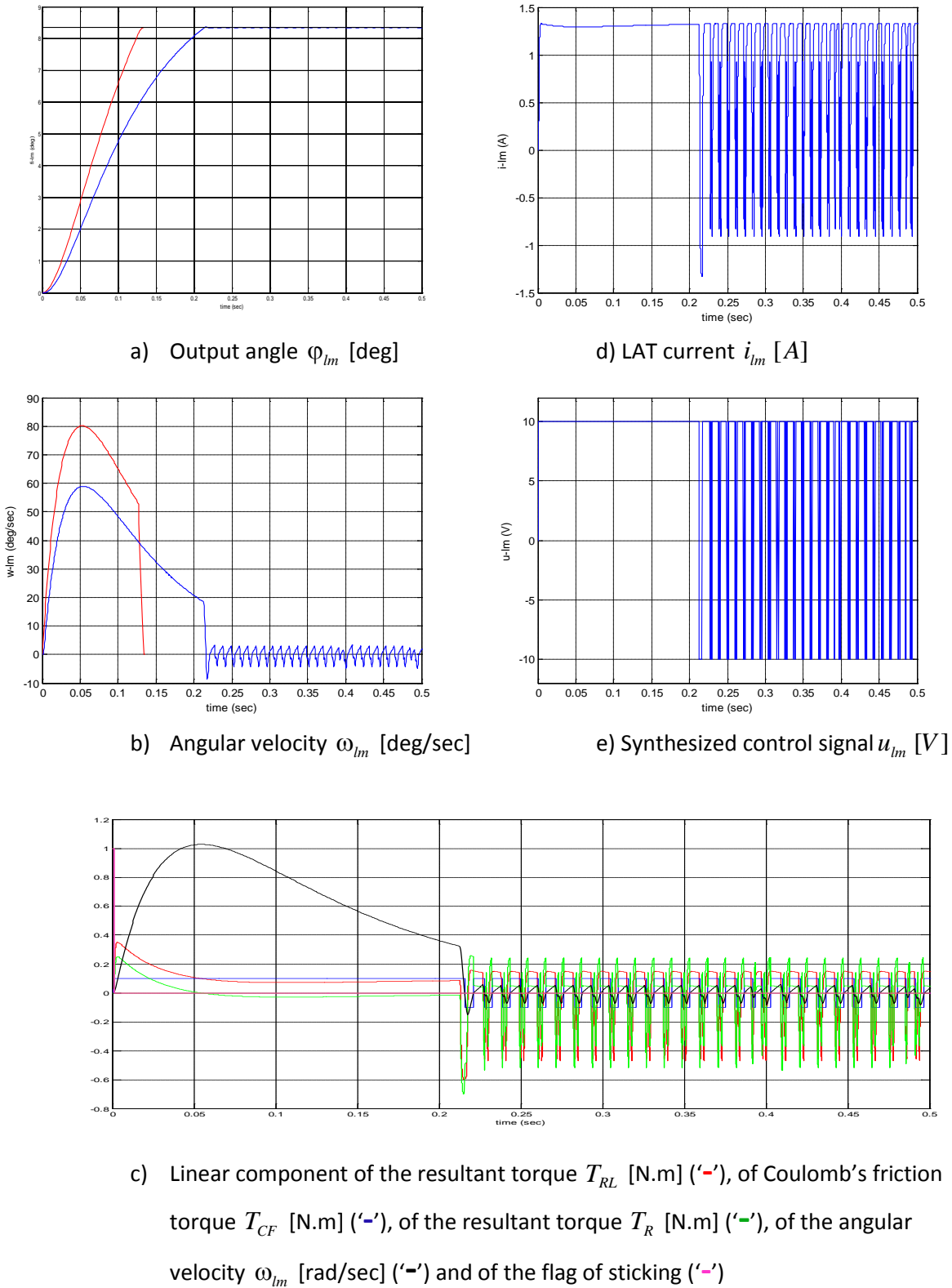

a) Output angle $\varphi_{lm}$ [deg]

d) LAT current $i_{lm}$ [A]

b) Angular velocity $\omega_{lm}$ [deg/sec]

e) Synthesized control signal $u_{lm}$ [V]

c) Linear component of the resultant torque $T_{RL}$ [N.m] ('–'), of Coulomb's friction torque $T_{CF}$ [N.m] ('–'), of the resultant torque $T_R$ [N.m] ('–'), of the angular velocity $\omega_{lm}$ [rad/sec] ('–') and of the flag of sticking ('–')

**Figure 31.** Time-diagrams in the closed loop system "Time optimal controller with sampling time of 1 ms and the prediction mechanism included – Large mirror actuator's the non-linear model of 3$^{rd}$ order with included Coulomb's friction model". The demand position on $\varphi_{lm}$ is $8.35$ [deg] and $u_0 = 10$ [V]. In a) and b) by ('–') are shown also the near time optimal solution on $\varphi_{lm}$ and $\omega_{lm}$ for the linear actuator's model.

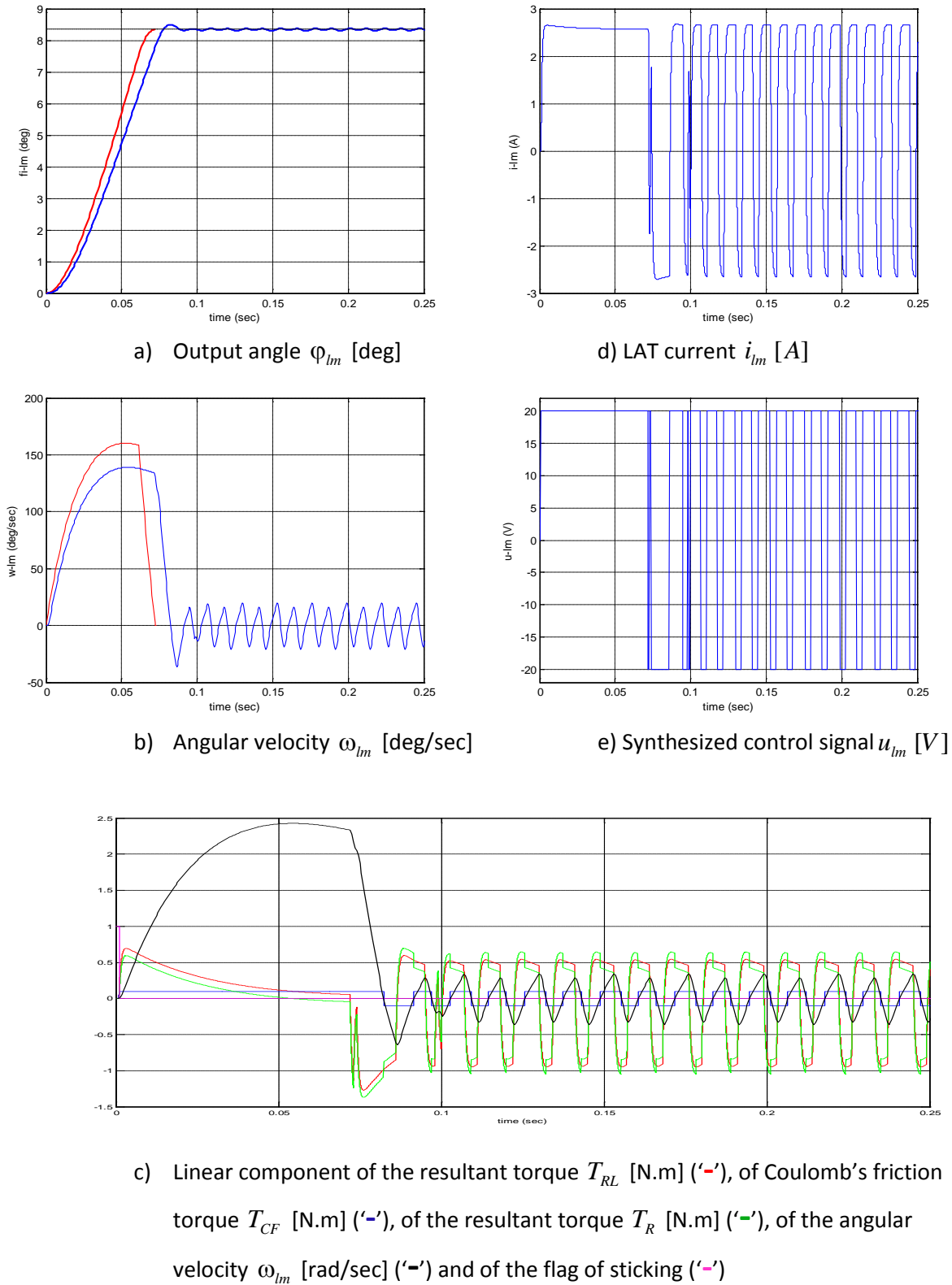

a)  Output angle $\varphi_{lm}$ [deg]

d) LAT current $i_{lm}$ [A]

b)  Angular velocity $\omega_{lm}$ [deg/sec]

e) Synthesized control signal $u_{lm}$ [V]

c)  Linear component of the resultant torque $T_{RL}$ [N.m] ('−'), of Coulomb's friction

torque $T_{CF}$ [N.m] ('−'), of the resultant torque $T_R$ [N.m] ('−'), of the angular

velocity $\omega_{lm}$ [rad/sec] ('−') and of the flag of sticking ('−')

**Figure 32. Time-diagrams in the closed loop system "Time optimal controller with sampling time of 1 ms and the prediction mechanism included − Large mirror actuator's the non-linear model of 3$^{rd}$ order with included Coulomb's friction model". The demand position on** $\varphi_{lm}$ **is** $8.35\,[\deg]$ **and** $u_0 = 20\,[V]$ **. In a) and b) by ('−')** **are shown also the near time optimal solution on** $\varphi_{lm}$ **and** $\omega_{lm}$ **for the linear actuator's model.**

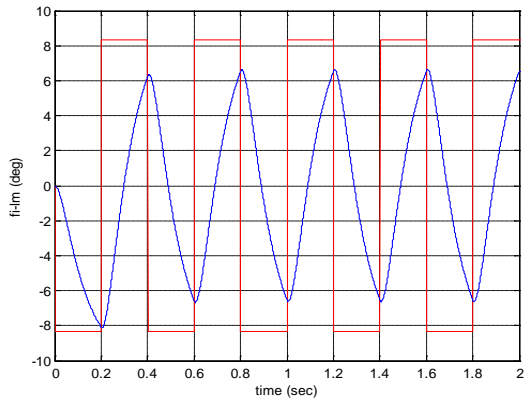

a) Output angle $\varphi_{lm}$ [deg]

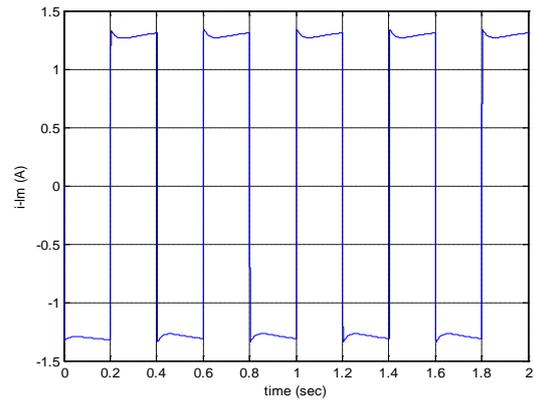

d) LAT current $i_{lm}$ [A]

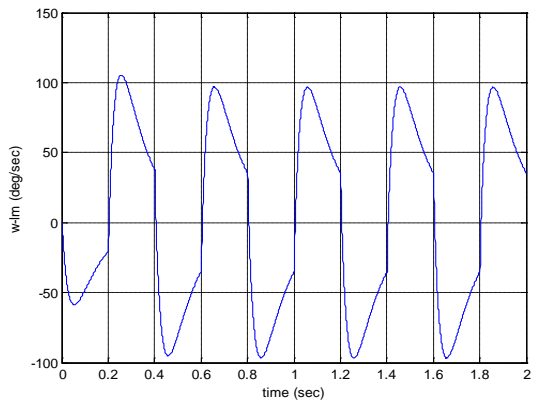

b) Angular velocity $\omega_{lm}$ [deg/sec]

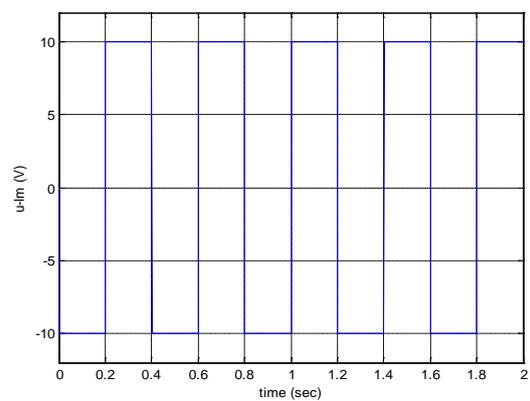

e) Synthesized control signal $u_{lm}$ [V]

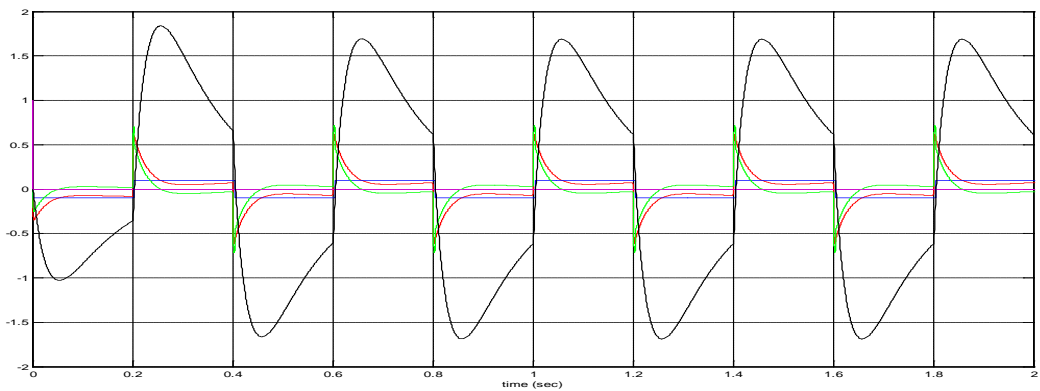

c) Linear component of the resultant torque $T_{RL}$ [N.m] ('–'), of Coulomb's friction torque $T_{CF}$ [N.m] ('–'), of the resultant torque $T_R$ [N.m] ('–'), of the angular velocity $\omega_{lm}$ [rad/sec] ('–') and of the flag of sticking ('–')

**Figure 33. Time-diagrams in the closed loop system "Time optimal controller with sampling time of 1 ms and the prediction mechanism included – Large mirror actuator's non-linear model of 3<sup>rd</sup> order with included Coulomb's friction model" at periodic demand signal, representing a square wave with an amplitude of $8.35\,[\text{deg}]$ and frequency 2.5 Hz, in case $u_0 = 10\,[V]$ ('–' in a).**

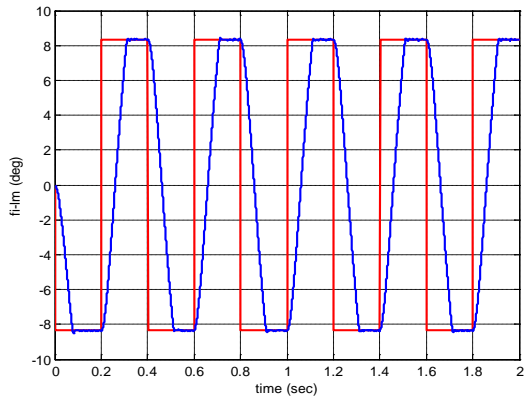

a)  Output angle $\varphi_{lm}$ [deg]

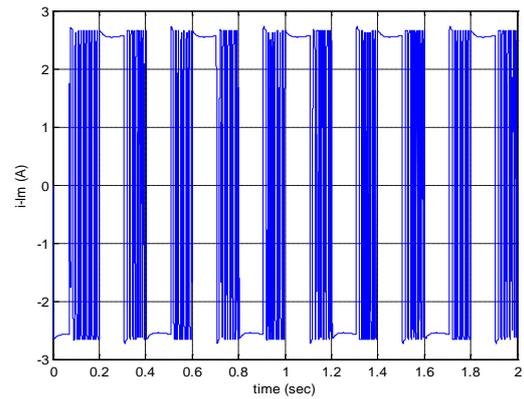

d) LAT current $i_{lm}$ [A]

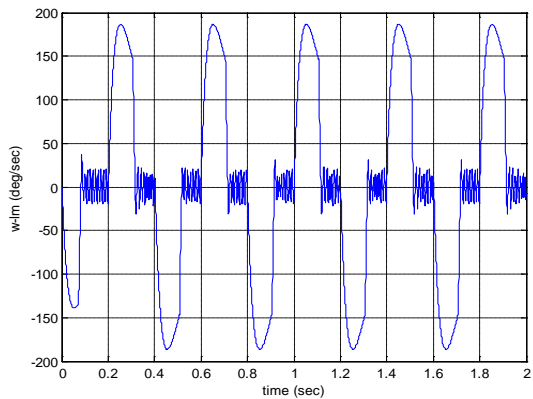

b)  Angular velocity $\omega_{lm}$ [deg/sec]

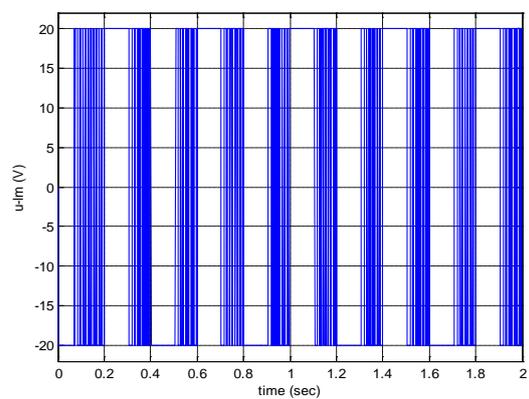

e) Synthesized control signal $u_{lm}$ [V]

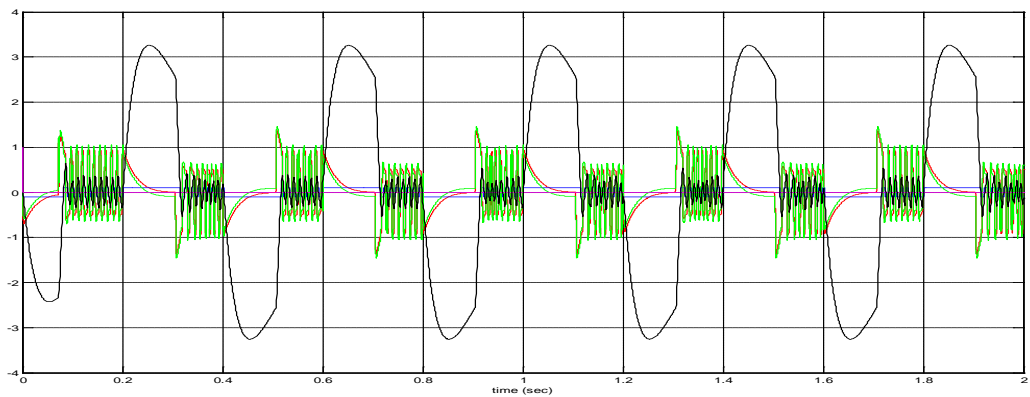

c)  Linear component of the resultant torque $T_{RL}$ [N.m] ('–'), of Coulomb's friction

    torque $T_{CF}$ [N.m] ('–'), of the resultant torque $T_R$ [N.m] ('–'), of the angular

    velocity $\omega_{lm}$ [rad/sec] ('–') and of the flag of sticking ('–')

**Figure 34. Time-diagrams in the closed loop system "Time optimal controller with sampling time of 1 ms and the prediction mechanism included – Large mirror actuator's non-linear model of 3$^{rd}$ order with included Coulomb's friction model" at periodic demand signal, representing a square wave with an amplitude of $8.35 \, [\mathrm{deg}]$ and frequency 2.5 Hz, in case $u_0 = 20 \, [V]$ ('–' in a).**

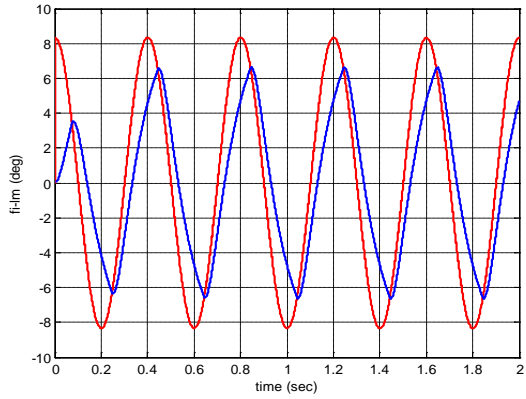

a) Output angle $\varphi_{lm}$ [deg]

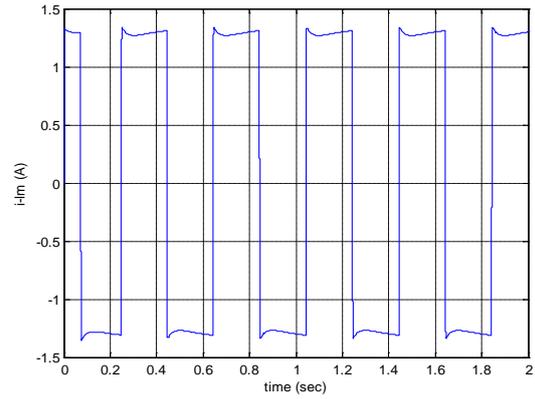

d) LAT current $i_{lm}$ [A]

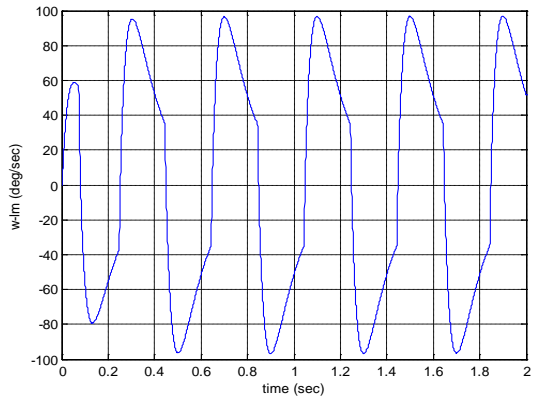

b) Angular velocity $\omega_{lm}$ [deg/sec]

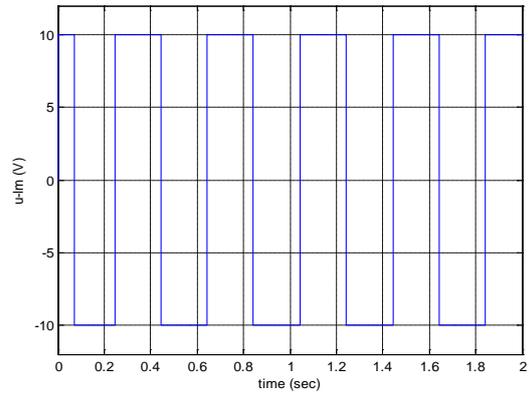

e) Synthesized control signal $u_{lm}$ [V]

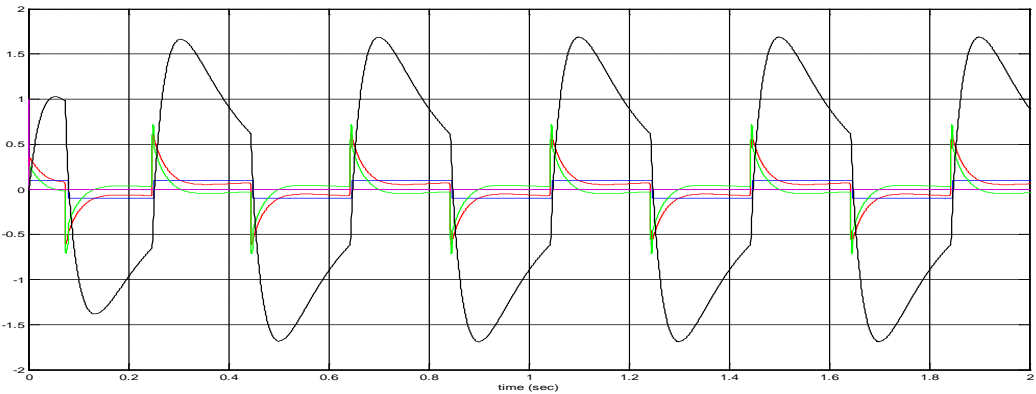

c) Linear component of the resultant torque $T_{RL}$ [N.m] ('–'), of Coulomb's friction torque $T_{CF}$ [N.m] ('–'), of the resultant torque $T_R$ [N.m] ('–'), of the angular velocity $\omega_{lm}$ [rad/sec] ('–') and of the flag of sticking ('–')

**Figure 35.** Time-diagrams in the closed loop system "Time optimal controller with sampling time of 1 ms and the prediction mechanism included – Large mirror actuator's non-linear model of 3<sup>rd</sup> order with included Coulomb's friction model" at periodic demand signal, representing a sinusoidal wave with an amplitude of $8.35\,[\deg]$ **and frequency 2.5 Hz, in case** $u_0 = 10\,[V]$ **('–' in a).**

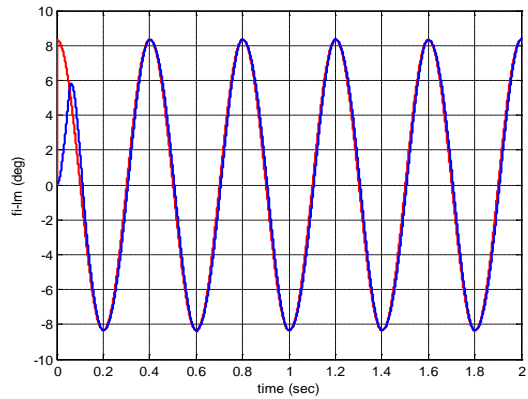

a)  Output angle $\varphi_{lm}$ [deg]

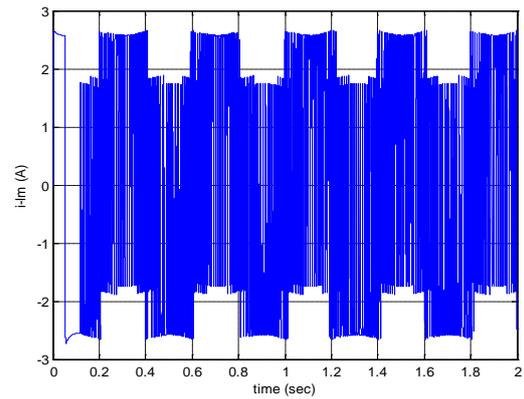

d) LAT current $i_{lm}$ [A]

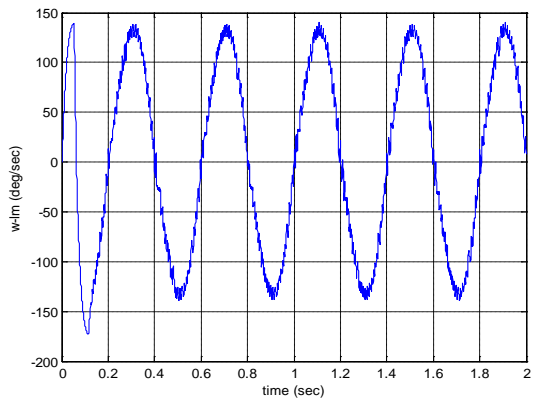

b)  Angular velocity $\omega_{lm}$ [deg/sec]

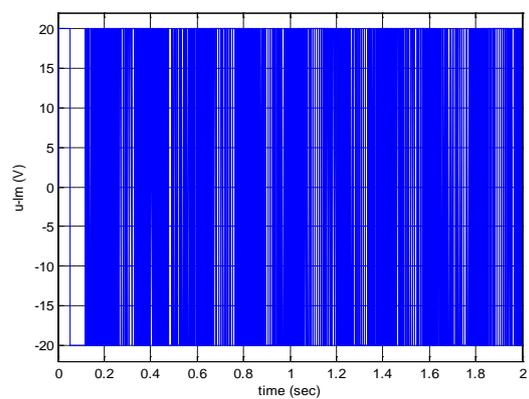

e) Synthesized control signal $u_{lm}$ [V]

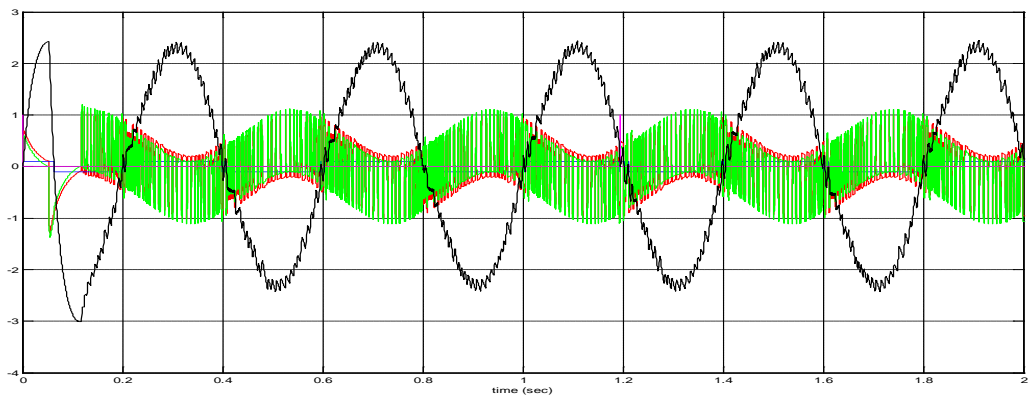

c)  Linear component of the resultant torque $T_{RL}$ [N.m] ('–'), of Coulomb's friction

torque $T_{CF}$ [N.m] ('–'), of the resultant torque $T_R$ [N.m] ('–'), of the angular

velocity $\omega_{lm}$ [rad/sec] ('–') and of the flag of sticking ('–')

**Figure 36. Time-diagrams in the closed loop system "Time optimal controller with sampling time of 1 ms and the prediction mechanism included – Large mirror actuator's non-linear model of 3$^{rd}$ order with included Coulomb's friction model" at periodic demand signal, representing a sinusoidal wave with an amplitude of $8.35\,[\mathrm{deg}]$ and frequency 2.5 Hz, in case $u_0 = 20\,[V]$ ('–' in a).**

## 4.2 Approach based on the time optimal control of the large mirror actuator with changed pivot stiffness $c_{lm}$ of the mechanical subsystem

### 4.2.1 Actuator's model at zero pivot stiffness of the mechanical subsystem

Let us investigate the case with zero pivot stiffness. Considering the linear model of the actuator presented in Figure 5, zero pivot stiffness eliminates the load position feedback by the coefficient $c_{lm}$ on the output angle $\varphi_{lm}$. At the state space model, this change $c_{lm} = 0$ resets the element $a_{21}$ of $A_{lm}$ of the linear model of the large mirror actuator of $3^{rd}$ order

$$\dot{\boldsymbol{x}}_{lm} = A_{lm}\boldsymbol{x}_{lm} + B_{lm}u_{lm},$$

$$y_{lm} = C_{lm}\boldsymbol{x}_{lm} + D_{lm}u_{lm},$$

$$\boldsymbol{x}_{lm} = \begin{bmatrix} \varphi_{lm} \\ \omega_{lm} \\ i_{lm} \end{bmatrix}, \quad A_{lm} = \begin{bmatrix} 0 & 1 & 0 \\ -c_{lm}/J_{lm} & -h_{lm}/J_{lm} & Kt/J_{lm} \\ 0 & -Kb/Lm & -Rm/Lm \end{bmatrix}, \quad B_{lm} = \begin{bmatrix} 0 \\ 0 \\ 1/Lm \end{bmatrix},$$

$$C_{lm} = \begin{bmatrix} 1 & 0 & 0 \end{bmatrix}, \quad D_{lm} = 0.$$

So, by resetting the element $a_{21}$ at the initial representation of $A_{lm}$

$$A_{lm} = \begin{bmatrix} 0 & 1 & 0 \\ -314.29 & -4.0816 & 57.755 \\ 0 & -62.889 & -1666.7 \end{bmatrix},$$

the new representation of $A_{lm}$ becomes

$$A_{lm} = \begin{bmatrix} 0 & 1 & 0 \\ 0 & -4.0816 & 57.755 \\ 0 & -62.889 & -1666.7 \end{bmatrix}$$

having eigenvalues

$$\boldsymbol{\lambda}_{lm} = \begin{bmatrix} 0 \\ -6.2692 \\ -1664.5 \end{bmatrix}.$$

The transfer function becomes

$$W_{\varphi_{lm}u_{lm}}(p) = \frac{K_{lm}}{a_{0lm}p^3 + a_{1lm}p^2 + a_{2lm}p + a_{3lm}} = \frac{K_{lm}}{p(a_{0lm}p^2 + a_{1lm}p + a_{2lm})},$$

$$K_{lm} = \left(\frac{K_t}{R_m h_{lm}}\right) \Big/ \left(1 + \frac{K_t K_b}{R_m h_{lm}}\right) \quad \left[\frac{rad}{Om.A.s} = \frac{rad}{V.s}\right],$$

$$a_{0lm} = (\frac{L_m}{R_m})(\frac{J_{lm}}{h_{lm}}) / (1 + \frac{K_t K_b}{R_m h_{lm}}) \quad [s^2], \quad a_{1lm} = \left(\left(\frac{L_m}{R_m}\right) + \left(\frac{J_{lm}}{h_{lm}}\right)\right) \Big/ \left(1 + \frac{K_t K_b}{R_m h_{lm}}\right) \quad [s],$$

$$a_{2lm} = 1 \quad [-], \quad a_{3lm} = (\frac{c_{lm}}{h_{lm}}) / (1 + \frac{K_t K_b}{R_m h_{lm}}) = 0,$$

where

$$K_{lm} = 1.23 \quad \left[\frac{rad}{V.s}\right],$$

$$a_{0lm} = 9.5832\text{e-}005 \quad [s^2], \quad a_{1lm} = 0.16011 \quad [s], \quad a_{2lm} = 1 \quad [-].$$

The correspondent step responses, $u_{lm}(t) = 1(t)$, of the above state space model, transfer function and the simulation model in Figure 5 at the initial data for the large mirror actuator, but with with zero pivot stiffness, are identical. They are shown in the following Figure 37 (on angular velocity $\omega_{lm}$ (rad/s)) and Figure 38 (on output angle $\varphi_{lm}$ (rad)). The correspondent Bode diagrams of the large mirror actuator's model are given in Figure 39 (on angular velocity $\omega_{lm}$) and Figure 40 (on output angle $\varphi_{lm}$).

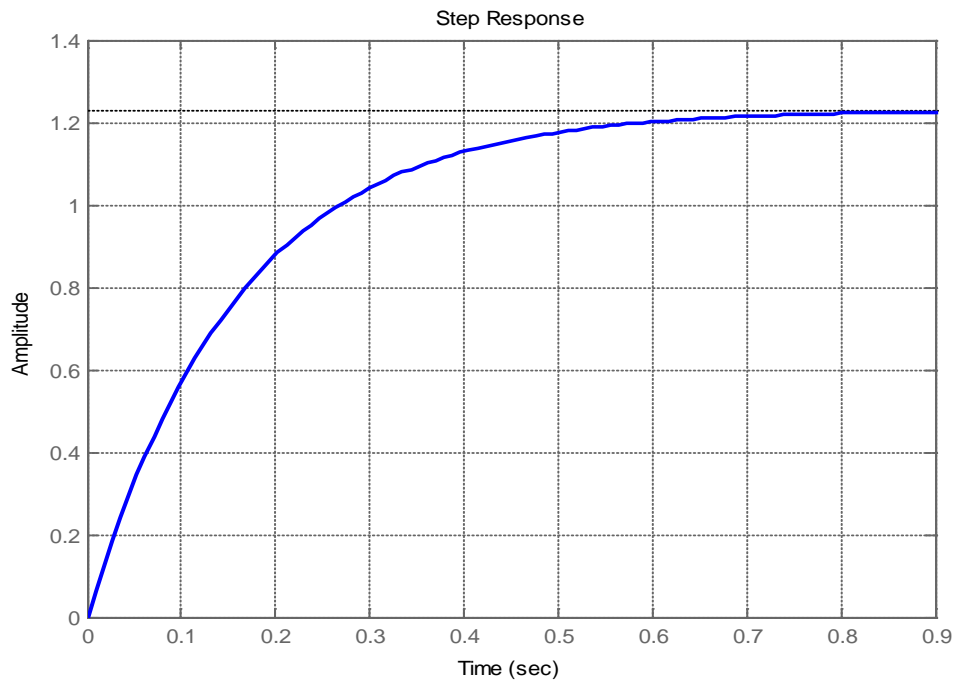

**Figure 37. Step response on angular velocity** $\omega_{lm}$ **(rad/s) of the large mirror actuator with zero pivot stiffness of the mechanical subsystem.**

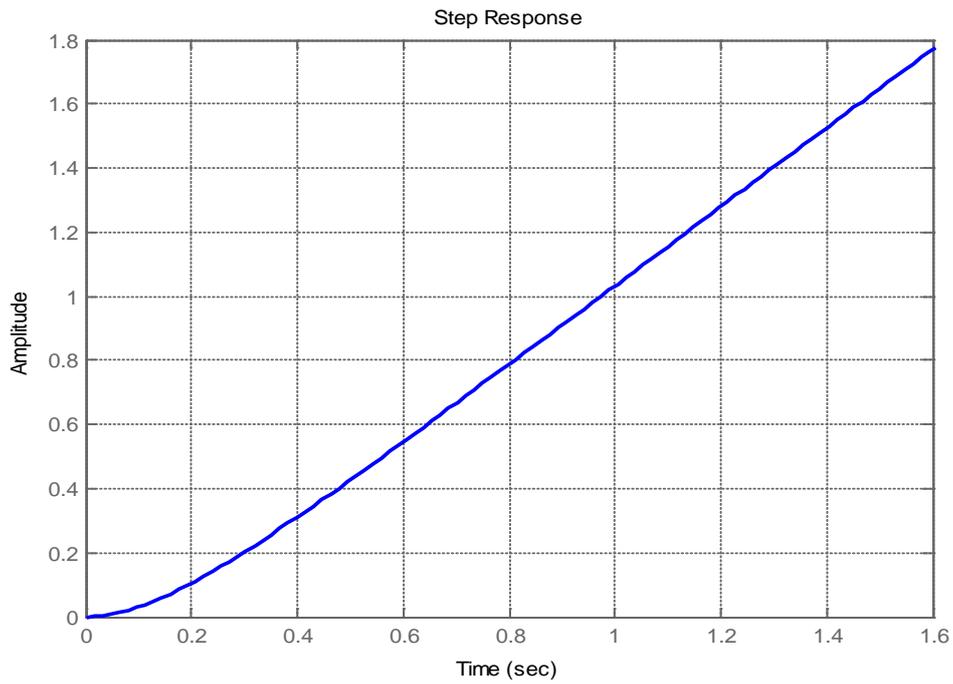

**Figure 38. Step response on output angle $\varphi_{lm}$ (rad) of the large mirror actuator with zero pivot stiffness of the mechanical subsystem.**

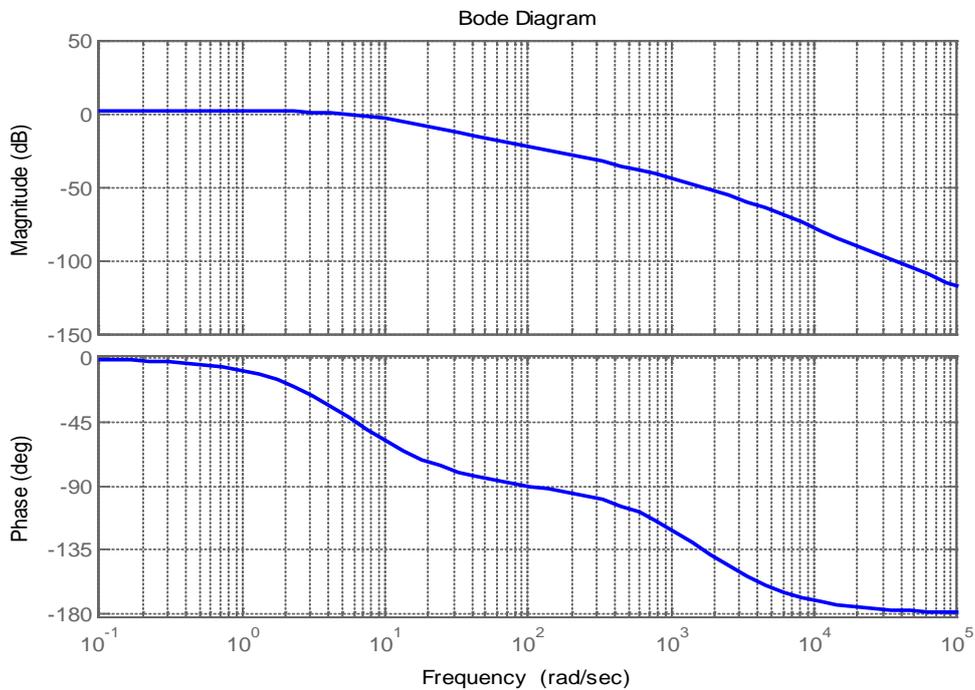

**Figure 39. Bode diagram on angular velocity of the large mirror actuator with zero pivot stiffness of the mechanical subsystem.**

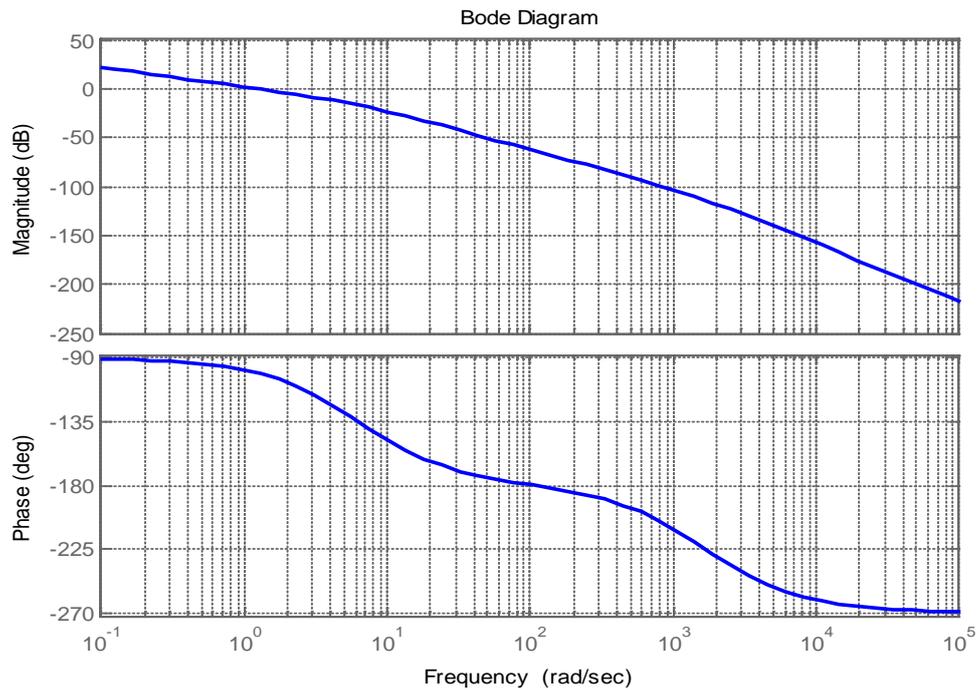

**Figure 40. Bode diagram on output angle of the large mirror actuator with zero pivot stiffness of the mechanical subsystem.**

## 4.2.2 Near time optimal control solutions for positioning

Let us deal with the above linear state space model of the actuator with zero pivot stiffness and consider the time optimal control problem for the transition from zero initial state to the final state

$$\boldsymbol{x}_{lm}(t_f) = \begin{bmatrix} \varphi_{lm}(t_f) \\ \omega_{lm}(t_f) \\ i_{lm}(t_f) \end{bmatrix} = \begin{bmatrix} 8.35 \\ 0 \\ 0 \end{bmatrix} \begin{pmatrix} \deg \\ \deg/s \\ A \end{pmatrix}.$$

Let us assume, as in the previous cases, the constraints for the control signal $u_{lm}(t)$ are

$$-u_0 \le u_{lm}(t) \le u_0.$$

We solve the problem by the author's method consequently for $u_0 = 10\,[V]$ and $u_0 = 20\,[V]$.

A near time optimal control solution for $u_0 = 10\,[V]$ is presented in Table 5 and Figure 41. The solution represents a piece-wise constant function $u_{lm}^{\tilde{o}}$ shown in Figure 41d with amplitude of $u_0 = 10\,[V]$ having three intervals of constancy with respective length given in Table 5. The near minimum time for this transition is $t_f^{\tilde{o}} = 0.088324\,[s]$ and the accuracy achieved is respectively: on $\varphi_{lm} - 1.0\mathrm{e}\text{-}7\,[\deg]$, on $\omega_{lm} - 2.0\mathrm{e}\text{-}5\,[\deg/s]$ and on $i_{lm} - 1.0\mathrm{e}\text{-}5\,[A]$.

**Table 5. A near time optimal control solution for positioning at 8.35 (deg) in case $u_0 = 10 \, [V]$.**

| Length of the first interval [sec] | 0.04967 |
|---|---|
| Length of the second interval [sec] | 0.038238 |
| Length of the third interval [sec] | 0.00041643 |
| Near minimum time for the transition $t_f^{\tilde{o}}$ [sec] | 0.088324 |

**Table 6. A near time optimal control solution for positioning at 8.35 (deg) in case $u_0 = 20 \, [V]$.**

| Length of the first interval [s] | 0.033801 |
|---|---|
| Length of the second interval [s] | 0.028293 |
| Length of the third interval [s] | 0.00041642 |
| Near minimum time for the transition $t_f^{\tilde{o}}$ [s] | 0.062511 |

A near time optimal control solution for $u_0 = 20 \, [V]$ is presented in Table 6 and Figure 42. The solution represents a piece-wise constant function $u_{lm}^{\tilde{o}}$ shown in Figure 42d with amplitude of $u_0 = 20 \, [V]$ having three intervals of constancy with respective length given in Table 6. The near minimum time for this transition is $t_f^{\tilde{o}} = 0.062511$ [s] and the accuracy achieved is respectively: on $\varphi_{lm} - 1.0e\text{-}6 \, [\deg]$, on $\omega_{lm} - 2.0e\text{-}4 \, [\deg/s]$ and on $i_{lm} - 5.0e\text{-}5 \, [A]$.

### 4.2.3  Tracking control systems

Based on the conclusions at the investigation of the closed loop digital control systems with time optimal controller for the large mirror actuator with increased damping of the mechanical subsystem, we implement here the control technique based on the synthesis of time optimal control for the linear model of the actuator with also the one-step prediction mechanism included. The control synthesis is based on a specially developed software for solving the linear time optimal control problem with given accuracy in case the eigenvalues of the controlled model of 3$^{rd}$ order are of the type as the type of eigenvalues of the above linear model of the large mirror actuator with zero pivot stiffness.

### 4.2.3.1 Response to constant demand signal

The demand signal here is the same as in the previous case

$$\boldsymbol{x}_{lm}(t_f) = \begin{bmatrix} \varphi_{lm}(t_f) \\ \omega_{lm}(t_f) \\ i_{lm}(t_f) \end{bmatrix} = \begin{bmatrix} 8.35 \\ 0 \\ 0 \end{bmatrix} \begin{pmatrix} \deg \\ \deg/s \\ A \end{pmatrix}.$$

First the model of the actuator we control represents the linear state space model, and then the model of the controlled system is changed to the non-linear model of the actuator with included Coulomb's friction model. The processes in case the model of the controlled system is the linear model of the actuator with zero pivot stiffness are shown in Figure 43 and Figure 44 for constraints on the control signal respectively $u_0 = 10\,[V]$ and $u_0 = 20\,[V]$.

The processes in the closed loop systems in case the model of the actuator is the non-linear model with included Coulomb's friction are shown in Figure 45 and Figure 46 for constraints on the control signal respectively $u_0 = 10\,[V]$ and $u_0 = 20\,[V]$. The comparison with the near time optimal solutions for constraints $u_0 = 10\,[V]$ and $u_0 = 20\,[V]$, presented also in these figures, shows that the processes in the closed loop systems on output angle, angular velocity, LAT current and control signal are very close to the respective time diagrams at the linear near time optimal solutions. The synthesized control manages well with the existence of Coulomb's friction. The steady state of the closed loop control system represents a sliding mode around the demand signals.

## 4.2.3.2 Response to sinusoidal signal with frequency 2.5 Hz and amplitude of 8.35 degrees

Here the processes in the digital tracking control system in case the demand signal on output angle $\varphi_{lm}$ is a discretized sinusoidal signal with amplitude of $8.35\,[\text{deg}]$ and frequency $2.5\,[\text{Hz}]$ are investigated. Figure 47 shows the processes in case the sampling rate of the digital part of the system is 1 ms, the constraint on the control signal is $u_0 = 10\,[V]$ and the model of the controlled system is the non-linear model of the actuator with zero pivot stiffness and Coulomb's friction model included. Figure 48 shows the processes in the control system but with changed to $u_0 = 20\,[V]$ constraint on the control signal. Both the systems manage very well with tracking - there is only one sticking phase at the beginning of the processes while the initial rising of the resultant torque $T_R$, Figure 47d and Figure 48d respectively, the movement at rest part of the process is smooth without sticking, the control system works in sliding mode.

Digital tracking control systems working simultaneously at two sampling rates, for sampling the demand signal every 4 ms and controlling the actuator with sampling rate of 1 ms, are also investigated. Figure 49 and Figure 51 show the processes in the control systems in case the constraints on the control signal are $u_0 = 10\,[V]$ and $u_0 = 20\,[V]$ respectively. Figure 50 and Figure 52 show the comparison between the processes on output angle $\varphi_{lm}$, angular velocity $\omega_{lm}$ and LAT current $i_{lm}$ in the control system with sampling rate of 1 ms and the control system working with two sampling rates of 4 ms and 1ms in case the constraints on the control signal are $u_0 = 10\,[V]$ and $u_0 = 20\,[V]$ respectively. The tracking is smoother at the system working at sampling rate of 1 ms, Figure 50b and Figure 52b. The maximum differences between the output angles at the system with two sampling rates of 4 ms and 1 ms and the system with sampling rate of 1 ms in case the constraints on the control signal are $u_0 = 10\,[V]$ and $u_0 = 20\,[V]$, Figure 53 and Figure 54, are less than $0.46\,[\text{deg}]$ and $0.35\,[\text{deg}]$ respectively.

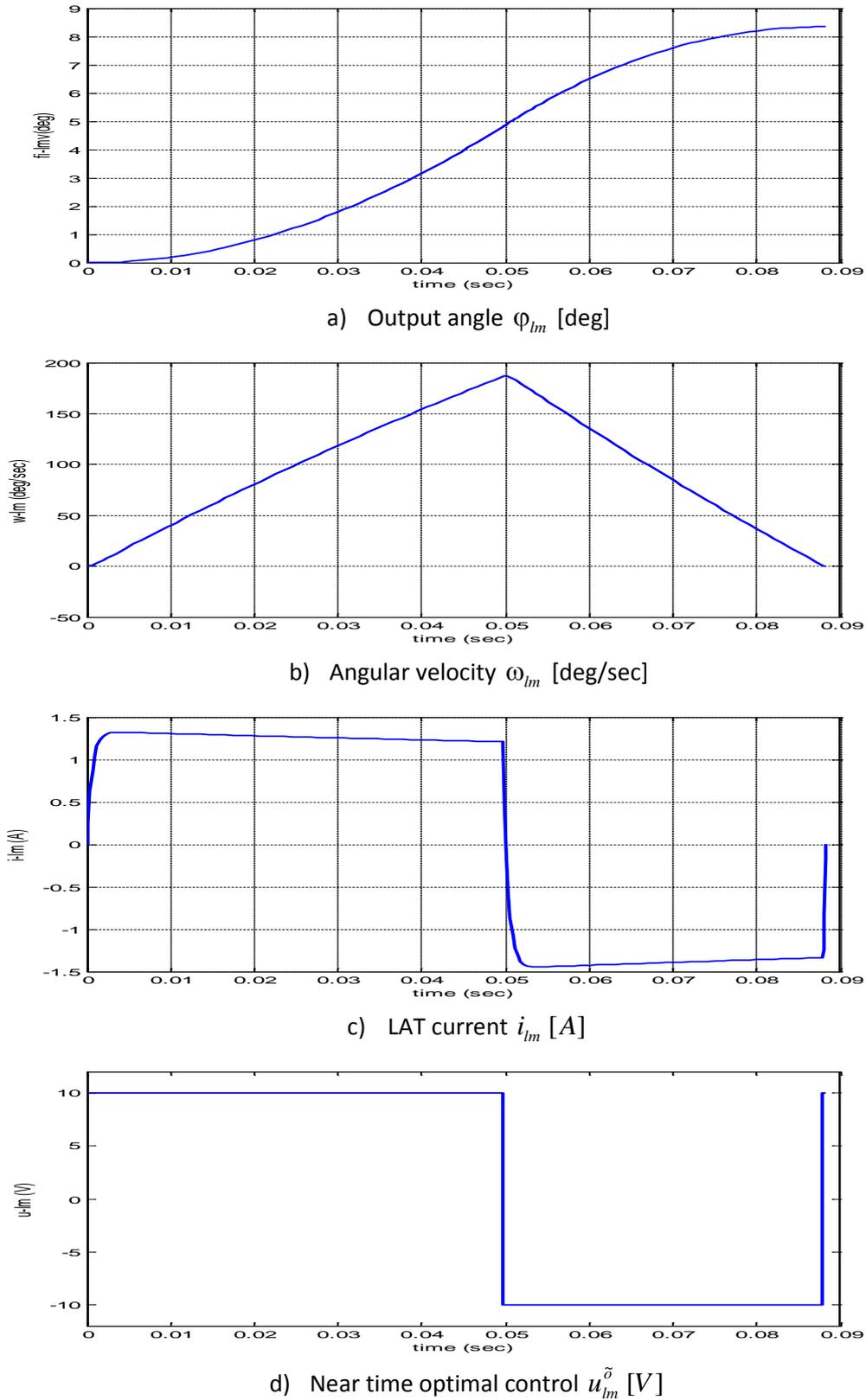

a) Output angle $\varphi_{lm}$ [deg]

b) Angular velocity $\omega_{lm}$ [deg/sec]

c) LAT current $i_{lm}$ [A]

d) Near time optimal control $u_{lm}^{\tilde{o}}$ [V]

**Figure 41. A Near time optimal control solution for the transition from zero initial state to demand position on** $\varphi_{lm}$ 8.35 [deg] **in case** $u_0 = 10 \,[V]$ **with accuracy on** $\varphi_{lm} - 1.0\mathrm{e}\text{-}7\,[\mathrm{deg}]$, **on** $\omega_{lm} - 2.0\mathrm{e}\text{-}5\,[\mathrm{deg/s}]$ **and on** $i_{lm} - 1.0\mathrm{e}\text{-}5\,[A]$.

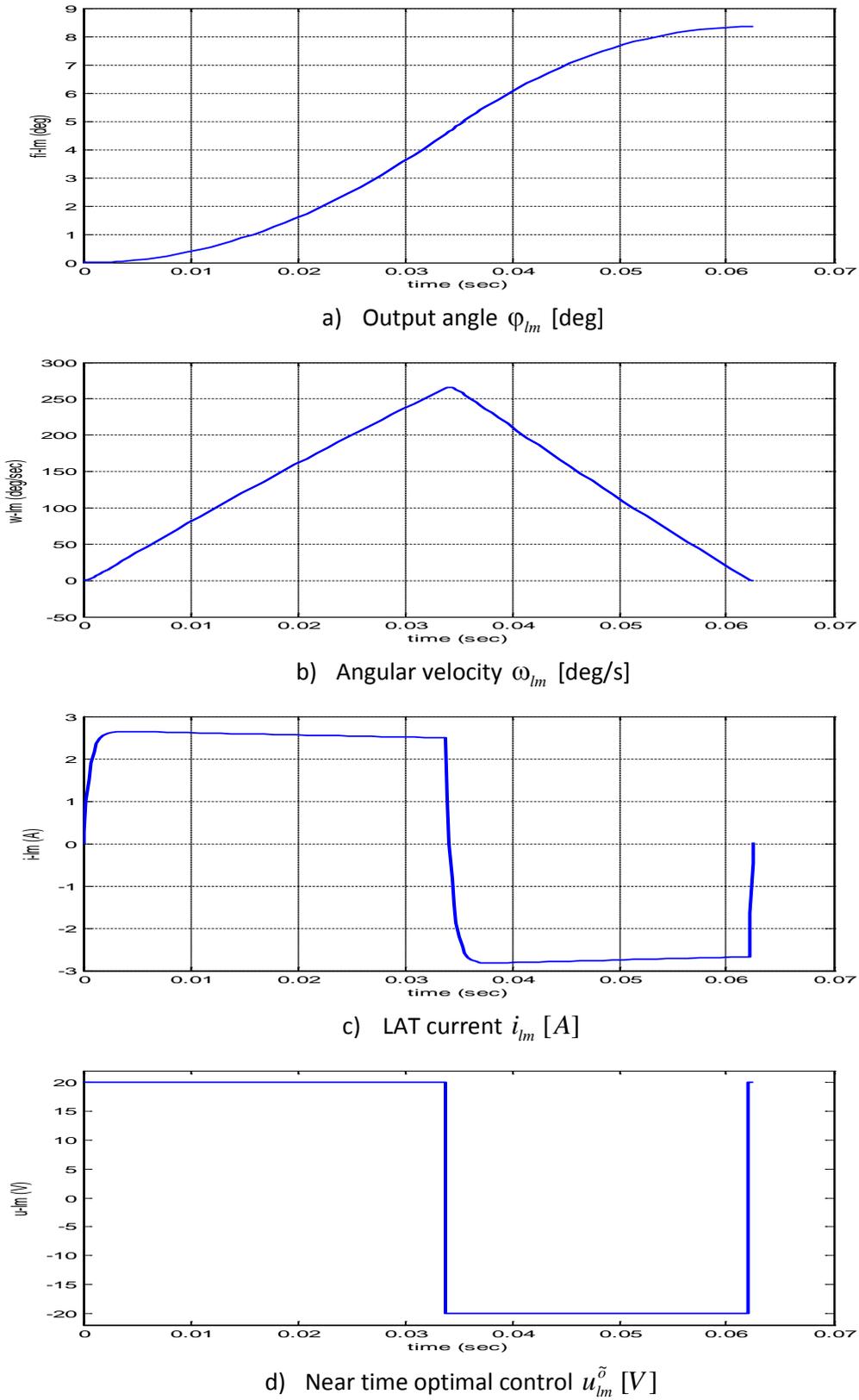

a) Output angle $\varphi_{lm}$ [deg]

b) Angular velocity $\omega_{lm}$ [deg/s]

c) LAT current $i_{lm}$ [A]

d) Near time optimal control $u_{lm}^{\tilde{o}}$ [V]

**Figure 42. A Near time optimal control solution for the transition from zero initial state to demand position on** $\varphi_{lm}$ 8.35 [deg] **in case** $u_0 = 20\,[V]$ **with accuracy on** $\varphi_{lm} - 1.0\mathrm{e}\text{-}6\,[\mathrm{deg}]$**, on** $\omega_{lm} - 2.0\mathrm{e}\text{-}4\,[\mathrm{deg/s}]$ **and on** $i_{lm} - 5.0\mathrm{e}\text{-}5\,[A]$**.**

Let us investigate the properties of the digital tracking control systems with sampling of 4 ms. Compared with the previous tracking control systems working at sampling rate of 1 ms and two sampling rates of 4 ms and 1ms, a sampling rate of 4 ms for the whole system lightens the controller's work, but some reducing of tracking quality could be expected. Figure 55 and Figure 56 show the processes in the tracking control systems with sampling rate of 4 ms in case the constraints for the control signal are $u_0 = 10\,[V]$ and $u_0 = 20\,[V]$ respectively. Both the systems manages with tracking but the comparison with the processes in the systems working simultaneously at two sampling rates of 4 ms and 1 ms, shown in Figure 57 and Figure 58 especially Figure 57b and Figure 58b, for constraints on the control signal $u_0 = 10\,[V]$ and $u_0 = 20\,[V]$ respectively, shows that tracking is more fine at the system with two sampling rates. The difference on output angle between the system with two sampling rates of 4 ms and 1 ms and the system with sampling rate of 4 ms is shown in Figure 59 for constraint on $u_0 = 10\,[V]$. The maximum difference is $0.57\,[\deg]$ but at steady sliding mode the difference is less than $0.46\,[\deg]$. For constraint $u_0 = 20\,[V]$ the difference between the system with two sampling rates of 4 ms and 1 ms and the system with sampling rate of 4 ms is shown in Figure 60. The maximum difference is $1.24\,[\deg]$ but at steady sliding mode the difference is less than $0.42\,[\deg]$.

### 4.3 Conclusions on controlling the large mirror actuator

The main idea at controlling the large mirror actuator applying techniques based on solving of time optimal control problems consists of transforming first the oscillating properties of the actuator as a controlled system having one negative eigenvalue and other two conjugate complex eigenvalues with negative real part into a controlled system with all negative and different eigenvalues.

Two types of correcting the actuator's properties are investigated: by increasing damping and by changing the pivot stiffness of the mechanical subsystem respectively. These approaches lead to solving of many time optimal control problems and synthesis of time optimal control with given accuracy while the specially synthesized digital tracking control systems are running based on a specially developed software.

Digital tracking control systems working at different sampling rates and solving time optimal control problems for constraints on the control signal $u_0 = 10\,[V]$ and $u_0 = 20\,[V]$ respectively are investigated. An inclusion of a further special one-step prediction mechanism allows achieving better results of tracking. The models of the controlled systems are refined by inclusion of Coulomb's friction model.

The approach based on zero pivot stiffness of the mechanical subsystem allows good results at tracking at a lower sampling rate of 4 ms and lower constraint on the control signal $u_0 = 10\,[V]$, while controlling the large mirror actuator with increased damping of the mechanical subsystem requires higher constraint on the control signal, $u_0 = 20\,[V]$, and higher rate of sampling, for example 1 ms.

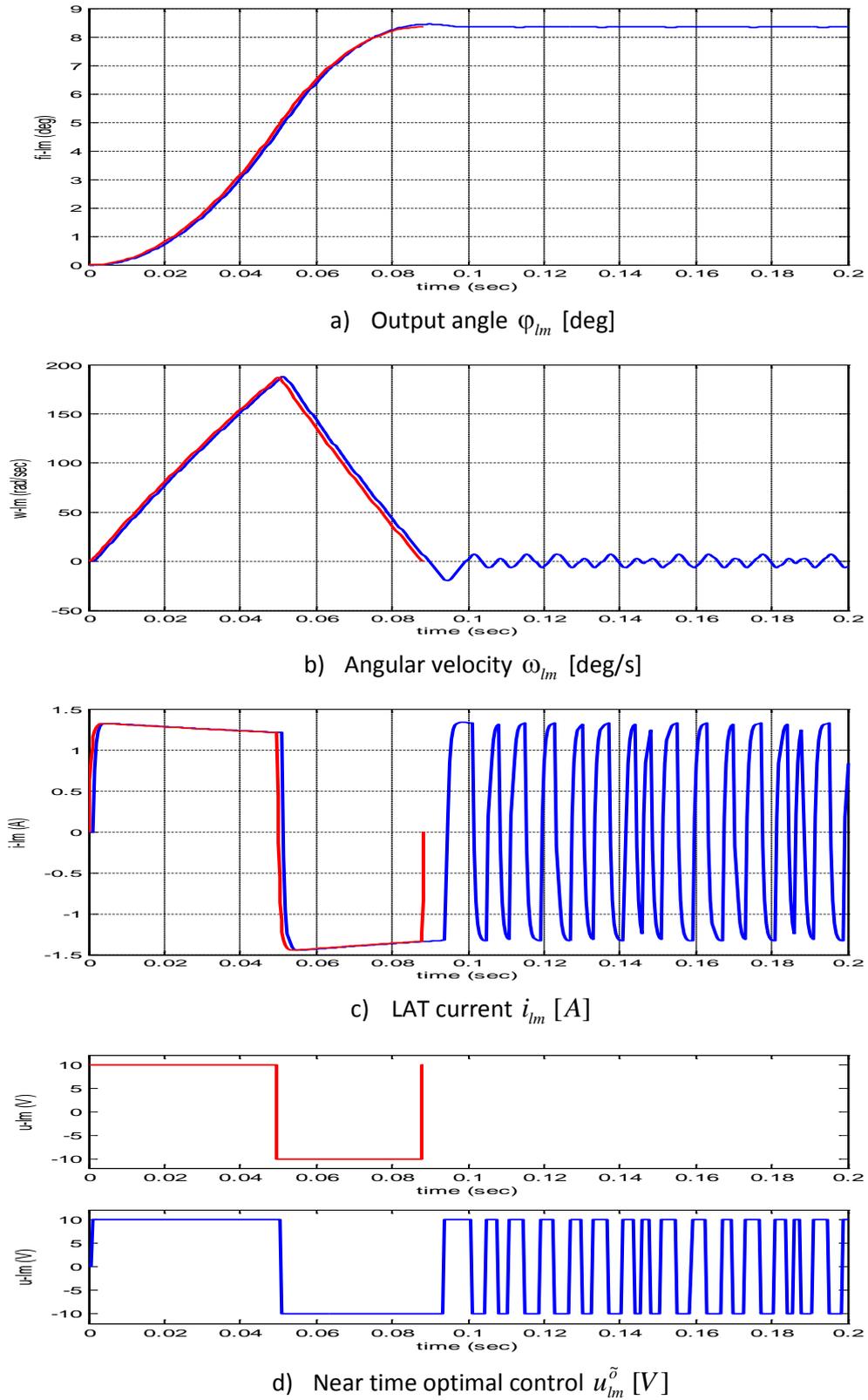

a) Output angle $\varphi_{lm}$ [deg]

b) Angular velocity $\omega_{lm}$ [deg/s]

c) LAT current $i_{lm}$ [A]

d) Near time optimal control $u_{lm}^{\tilde{o}}$ [V]

**Figure 43. Comparison between the near linear time optimal control solution ('–') and the processes in the tracking control system with the linear model of the large mirror actuator with zero pivot stiffness as a model of the controlled system in case the demand position on** $\varphi_{lm}$ **is** $8.35\,[\deg]$**, the constraint on** $u_0 = 10\,[V]$ **and sampling time of 1ms.**

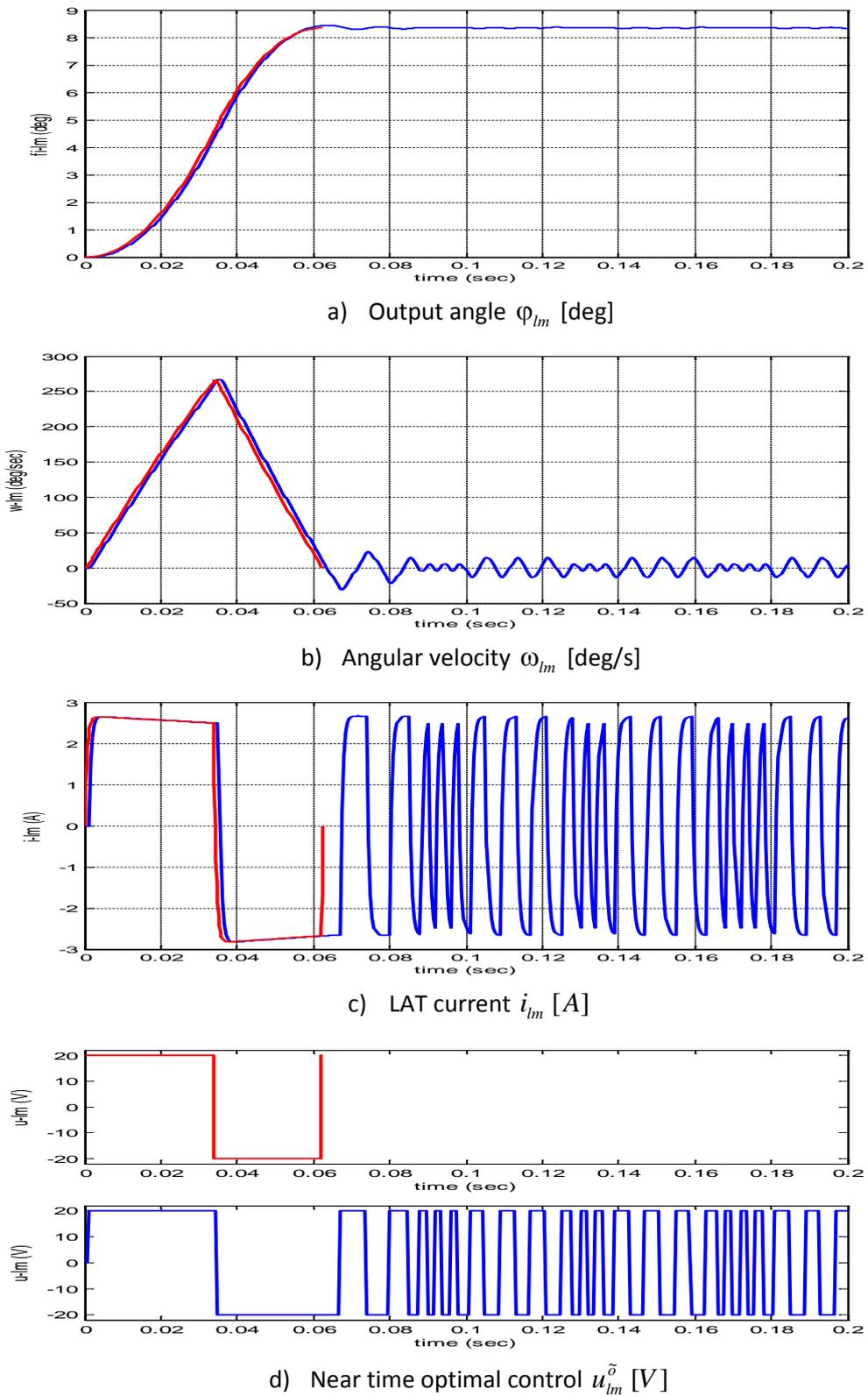

a) Output angle $\varphi_{lm}$ [deg]

b) Angular velocity $\omega_{lm}$ [deg/s]

c) LAT current $i_{lm}$ [A]

d) Near time optimal control $u_{lm}^{\tilde{o}}$ [V]

**Figure 44. Comparison between the near linear time optimal control solution ('–') and the processes in the digital tracking control system with the linear model of the large mirror actuator with zero pivot stiffness as a model of the controlled system in case the demand position on $\varphi_{lm}$ is $8.35\,[\text{deg}]$, the constraint on $u_0 = 20\,[V]$ and sampling time of 1ms.**

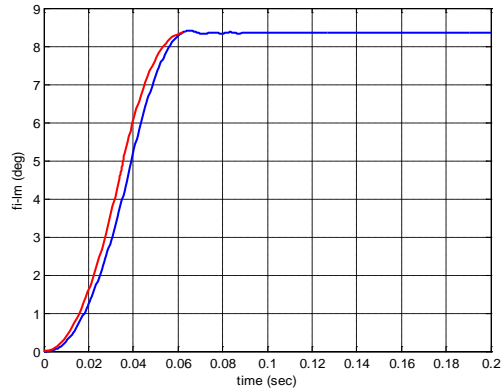

a) Output angle $\varphi_{lm}$ [deg]

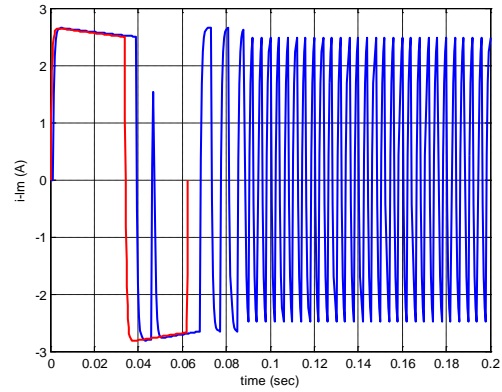

d) LAT current $i_{lm}$ $[A]$

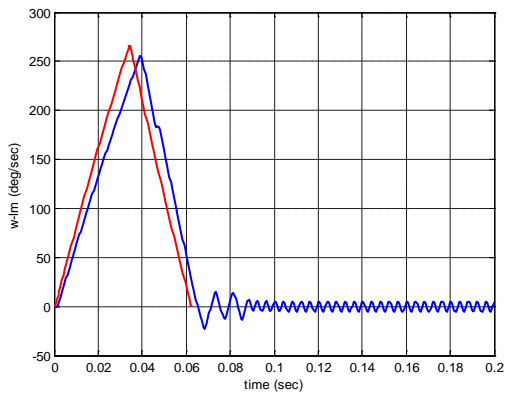

b) Angular velocity $\omega_{lm}$ [deg/sec]

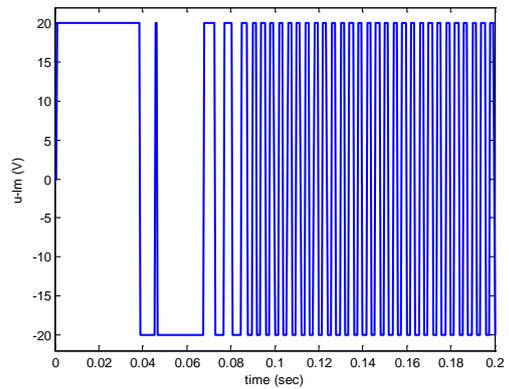

e) Synthesized control signal $u_{lm}$ $[V]$

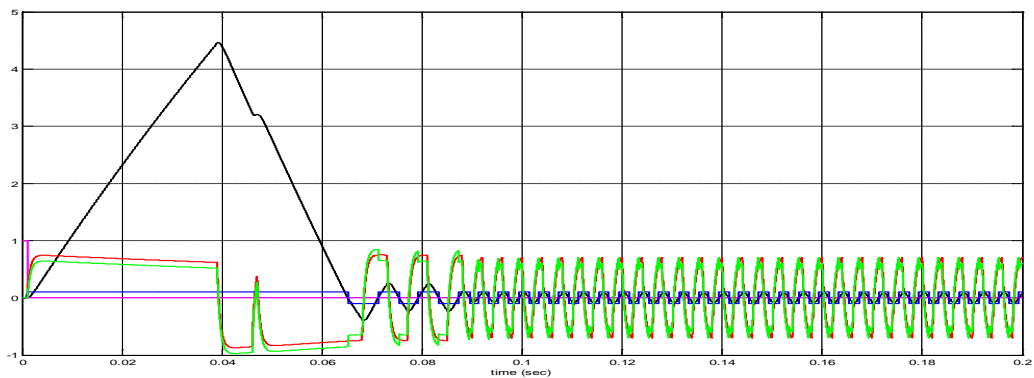

c) Linear component of the resultant torque $T_{RL}$ [N.m] ('–'), of Coulomb's friction

torque $T_{CF}$ [N.m] ('–'), of the resultant torque $T_R$ [N.m] ('–'), of the angular

velocity $\omega_{lm}$ [rad/sec] ('–') and of the flag of sticking ('–')

**Figure 45. Time-diagrams of the processes in the digital tracking control system with the non-linear model of the large mirror actuator with zero pivot stiffness and Coulomb's friction model included as a model of the controlled system in case the demand position on $\varphi_{lm}$ is $8.35\,[\text{deg}]$, the constraint on $u_0 = 10\,[V]$ and sampling time of 1ms. In a), b) and d) the processes are compared with the linear near time optimal solution ('–') on $\varphi_{lm}$ and $\omega_{lm}$ and $i_{lm}$.**

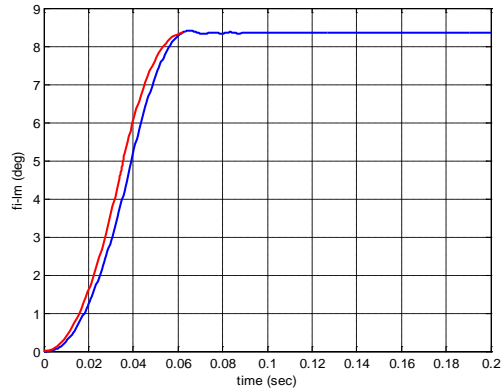

a)  Output angle $\varphi_{lm}$ [deg]

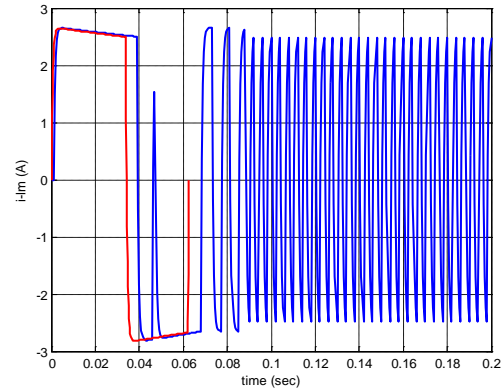

d) LAT current $i_{lm}$ [A]

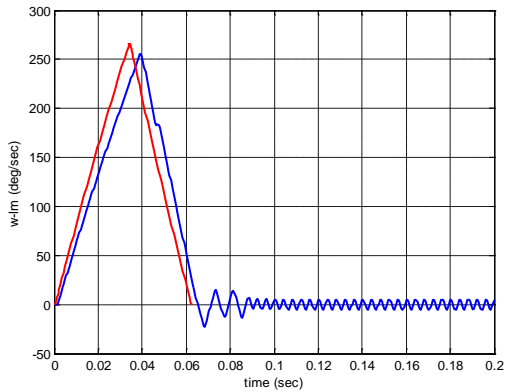

b)  Angular velocity $\omega_{lm}$ [deg/sec]

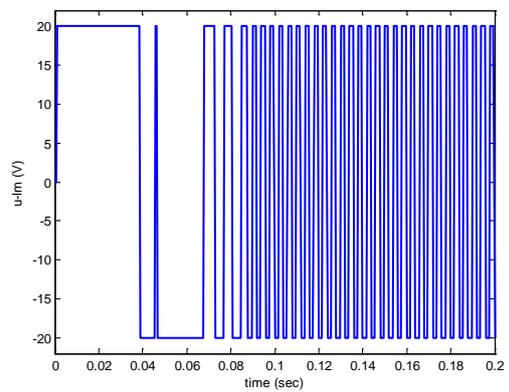

e) Synthesized control signal $u_{lm}$ [V]

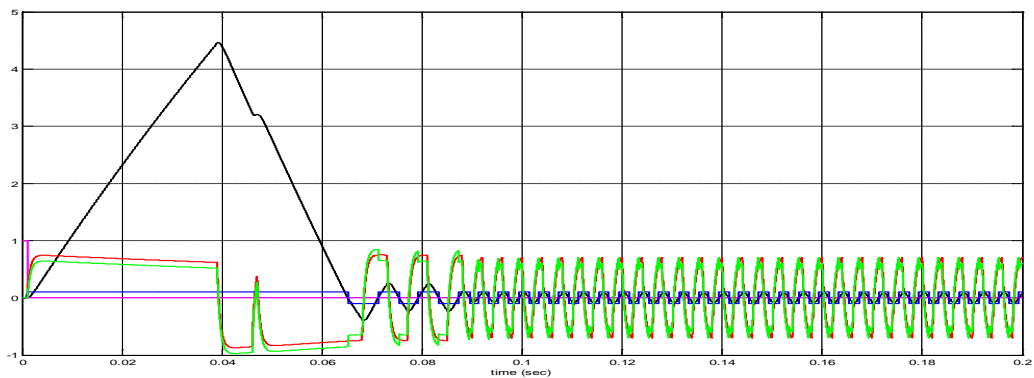

c)  Linear component of the resultant torque $T_{RL}$ [N.m] ('–'), of Coulomb's friction
    torque $T_{CF}$ [N.m] ('–'), of the resultant torque $T_R$ [N.m] ('–'), of the angular
    velocity $\omega_{lm}$ [rad/sec] ('–') and of the flag of sticking ('–')

**Figure 46. Time-diagrams of the processes in the digital tracking control system with the non-linear model of the large mirror actuator with zero pivot stiffness and Coulomb's friction model included as a model of the controlled system in case the demand position on $\varphi_{lm}$ is $8.35$ [deg], the constraint on $u_0 = 20$ [V] and sampling time of 1ms. In a), b) and d) the processes are compared with the linear near time optimal solution ('–') on $\varphi_{lm}$ and $\omega_{lm}$ and $i_{lm}$.**

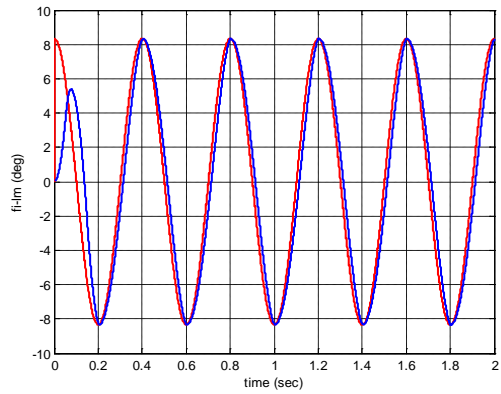

a) Output angle $\varphi_{lm}$ [deg]

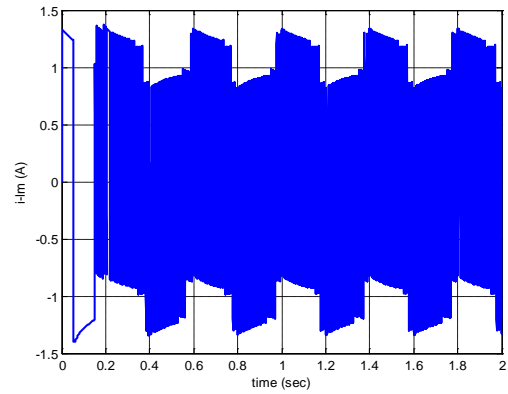

d) LAT current $i_{lm}$ $[A]$

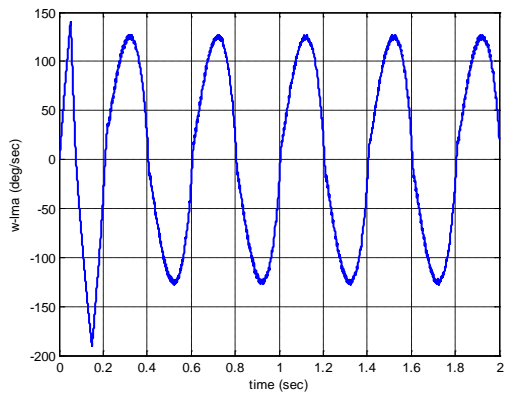

b) Angular velocity $\omega_{lm}$ [deg/sec]

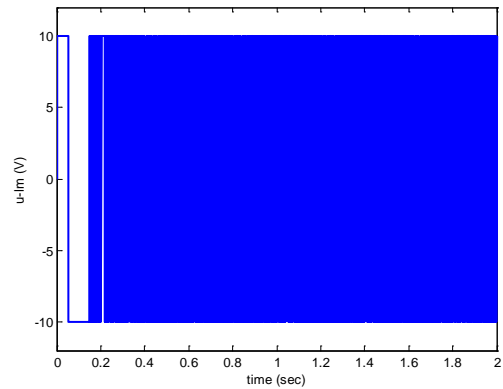

e) Synthesized control signal $u_{lm}$ $[V]$

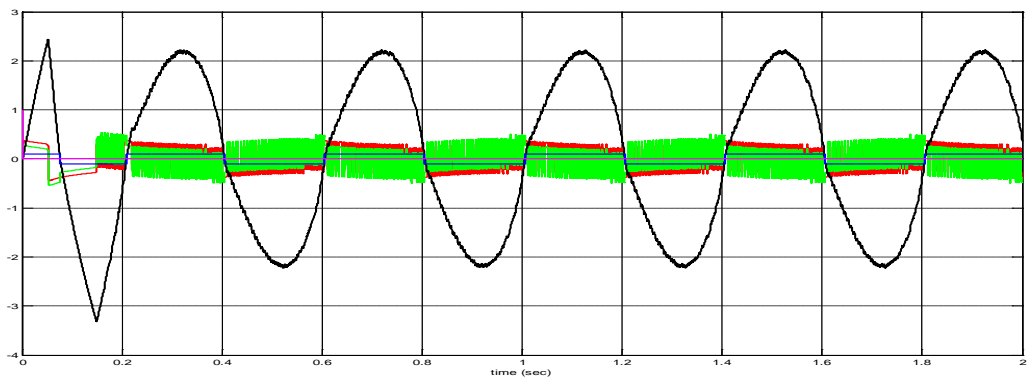

c) Linear component of the resultant torque $T_{RL}$ [N.m] ('–'), of Coulomb's friction

torque $T_{CF}$ [N.m] ('–'), of the resultant torque $T_R$ [N.m] ('–'), of the angular

velocity $\omega_{lm}$ [rad/sec] ('–') and of the flag of sticking ('–')

**Figure 47. Time-diagrams of the processes in the digital tracking control system, working at sampling rate of 1 ms, with the non-linear model of the large mirror actuator with zero pivot stiffness and Coulomb's friction model included as a model of the controlled system in case the demand position on $\varphi_{lm}$ is periodic signal with amplitude of $8.35\,[\deg]$ and frequency 2.5 Hz. The constraint on $u_0 = 10\,[V]$.**

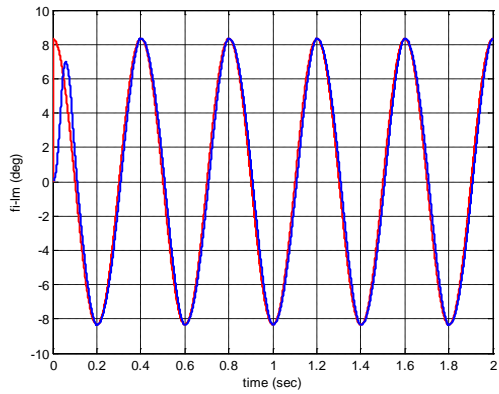

a)   Output angle $\varphi_{lm}$ [deg]

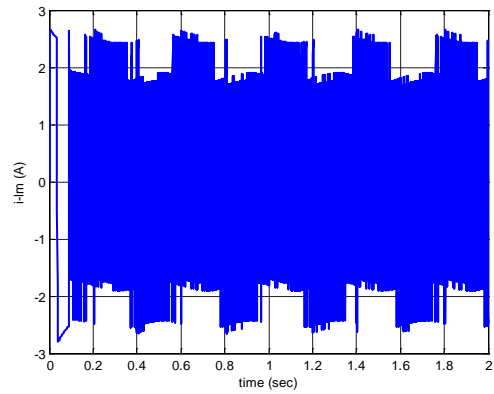

d) LAT current $i_{lm}$ $[A]$

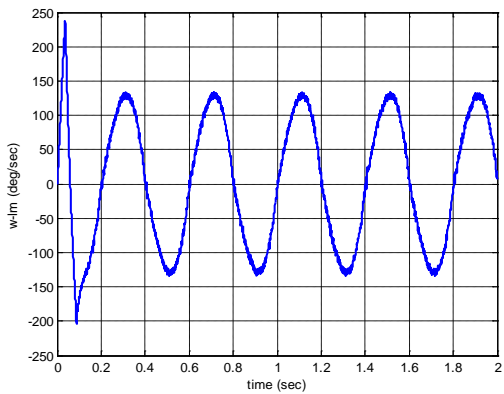

b)   Angular velocity $\omega_{lm}$ [deg/sec]

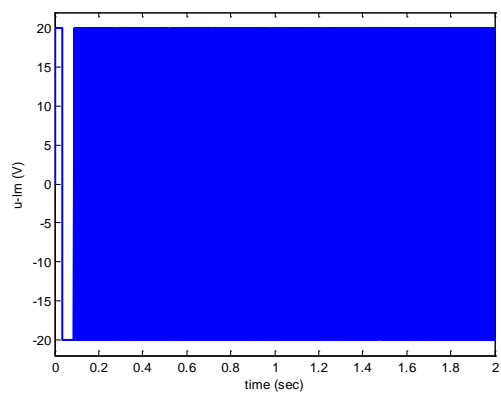

e) Synthesized control signal $u_{lm}$ $[V]$

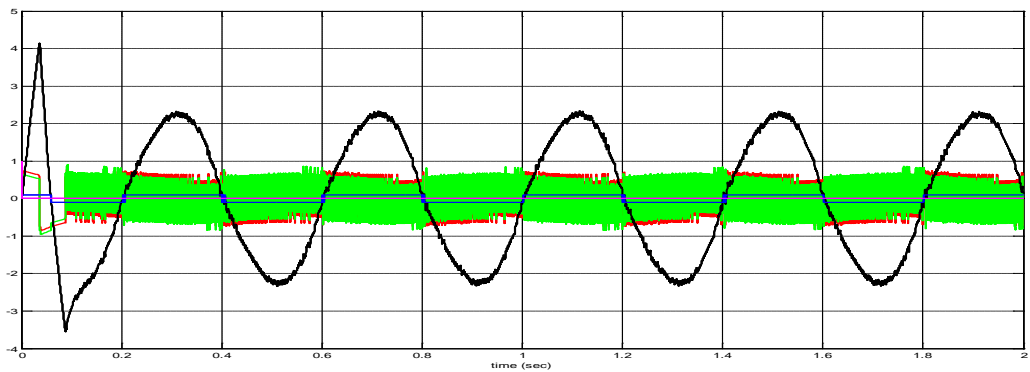

c)   Linear component of the resultant torque $T_{RL}$ [N.m] ('‒'), of Coulomb's friction

torque $T_{CF}$ [N.m] ('‒'), of the resultant torque $T_R$ [N.m] ('‒'), of the angular

velocity $\omega_{lm}$ [rad/sec] ('‒') and of the flag of sticking ('‒')

**Figure 48. Time-diagrams of the processes in the digital tracking control system, working at sampling rate of 1 ms, with the non-linear model of the large mirror actuator with zero pivot stiffness and Coulomb's friction model included as a model of the controlled system in case the demand position on $\varphi_{lm}$ is periodic signal with amplitude of $8.35\,[\mathrm{deg}]$ and frequency 2.5 Hz. The constraint on $u_0 = 20\,[V]$.**

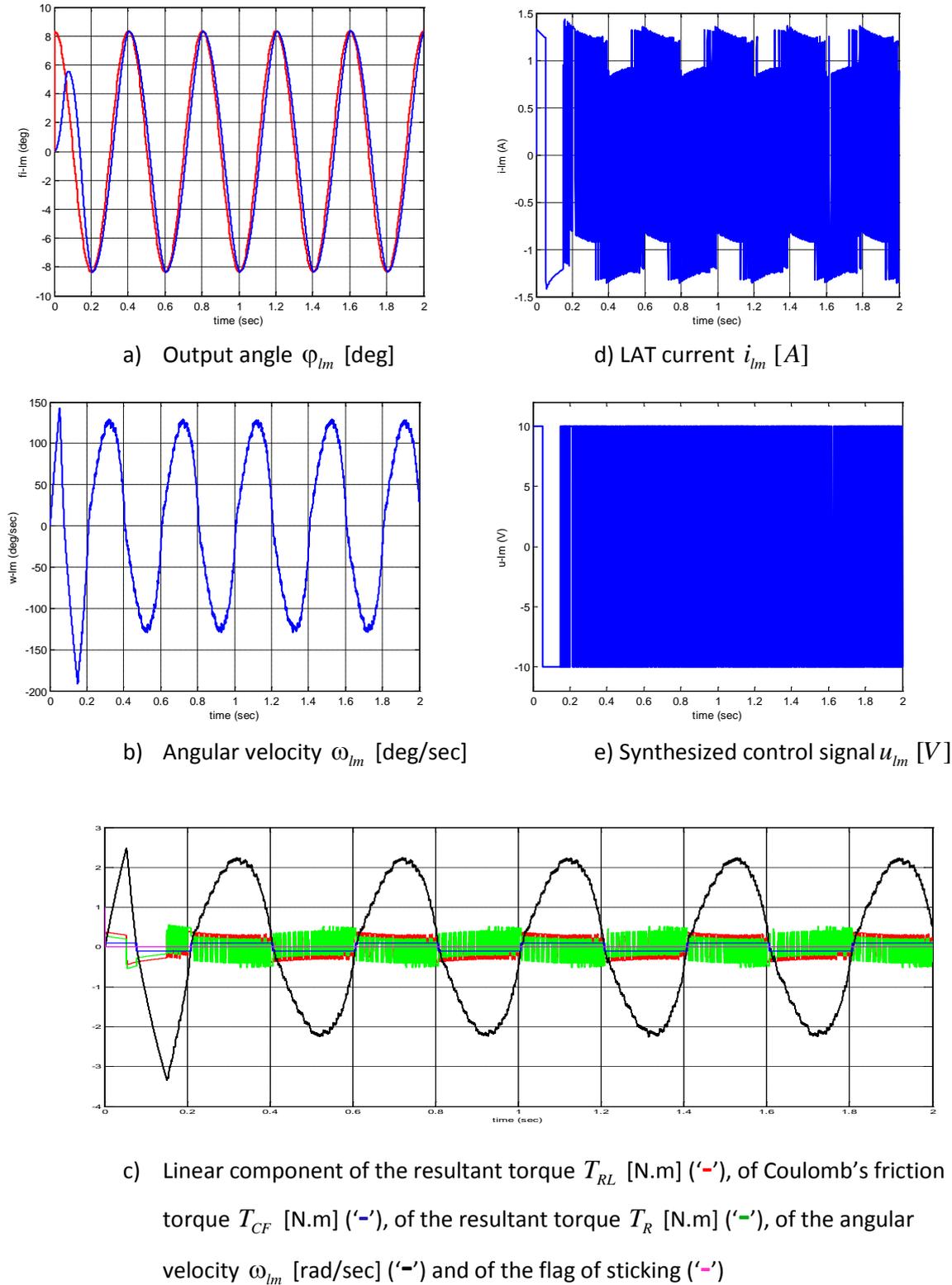

a) Output angle $\varphi_{lm}$ [deg]

d) LAT current $i_{lm}$ [A]

b) Angular velocity $\omega_{lm}$ [deg/sec]

e) Synthesized control signal $u_{lm}$ [V]

c) Linear component of the resultant torque $T_{RL}$ [N.m] ('–'), of Coulomb's friction

torque $T_{CF}$ [N.m] ('–'), of the resultant torque $T_R$ [N.m] ('–'), of the angular

velocity $\omega_{lm}$ [rad/sec] ('–') and of the flag of sticking ('–')

**Figure 49. Time-diagrams of the processes in the digital tracking control system with the non-linear model of the large mirror actuator with zero pivot stiffness and Coulomb's friction model included as a model of the controlled system in case the demand position on $\varphi_{lm}$ is periodic signal with amplitude of $8.35\,[\text{deg}]$ and frequency 2.5 Hz discretized by sampling time of 4 ms (by ('–') in a)), while the rest part works at sampling rate of 1ms. The constraint on $u_0 = 10\,[V]$.**

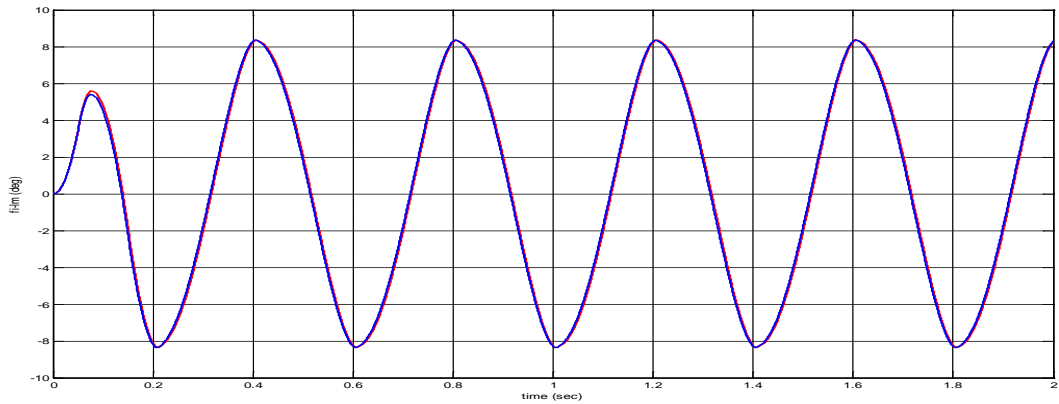

a)  Output angle  $\varphi_{lm}$  [deg]

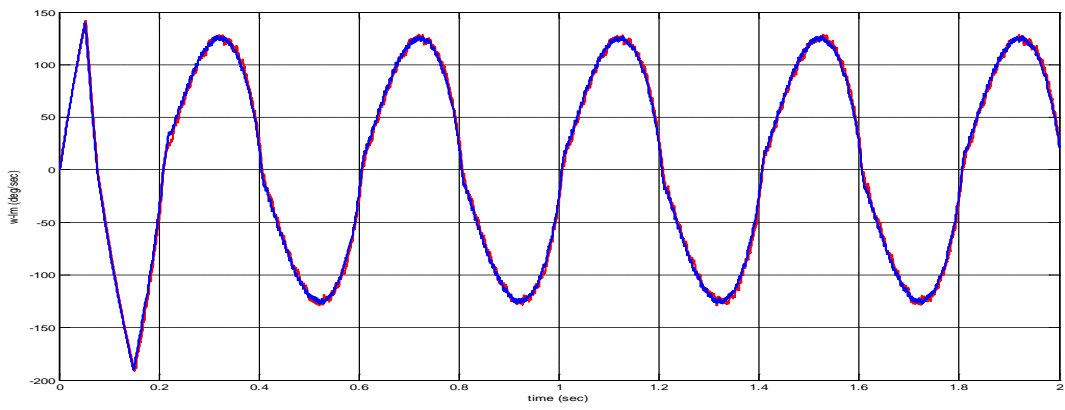

b)  Angular velocity  $\omega_{lm}$  [deg/sec]

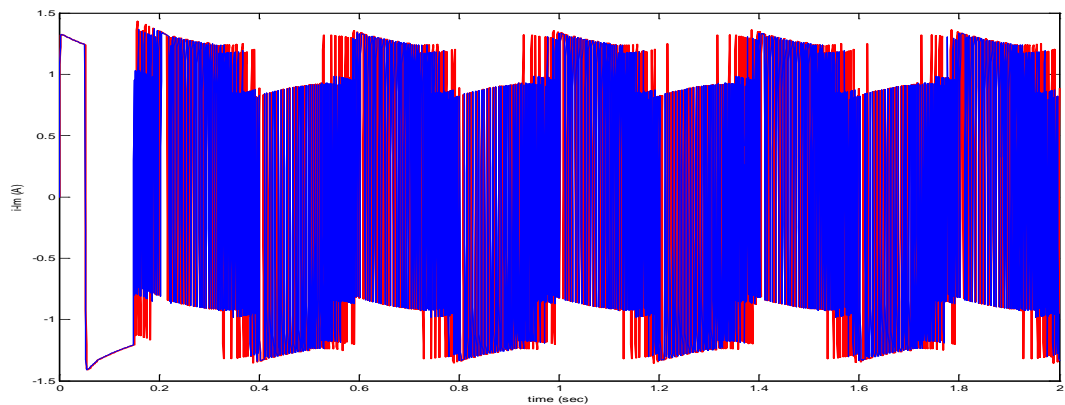

c)  LAT current  $i_{lm}$  [A]

**Figure 50. Comparison between the processes on output angle  $\varphi_{lm}$ , angular velocity  $\omega_{lm}$  and LAT current  $i_{lm}$  in the digital tracking control systems for the large mirror actuator controlling the non-linear model of the actuator with zero pivot stiffness and Coulomb's friction model included. The first one works at sampling rate of 1 ms ('−') but the second one works at two sampling rates of 4 ms and 1 ms ('−'). The constraint on  $u_0 = 10 \, [V]$ .**

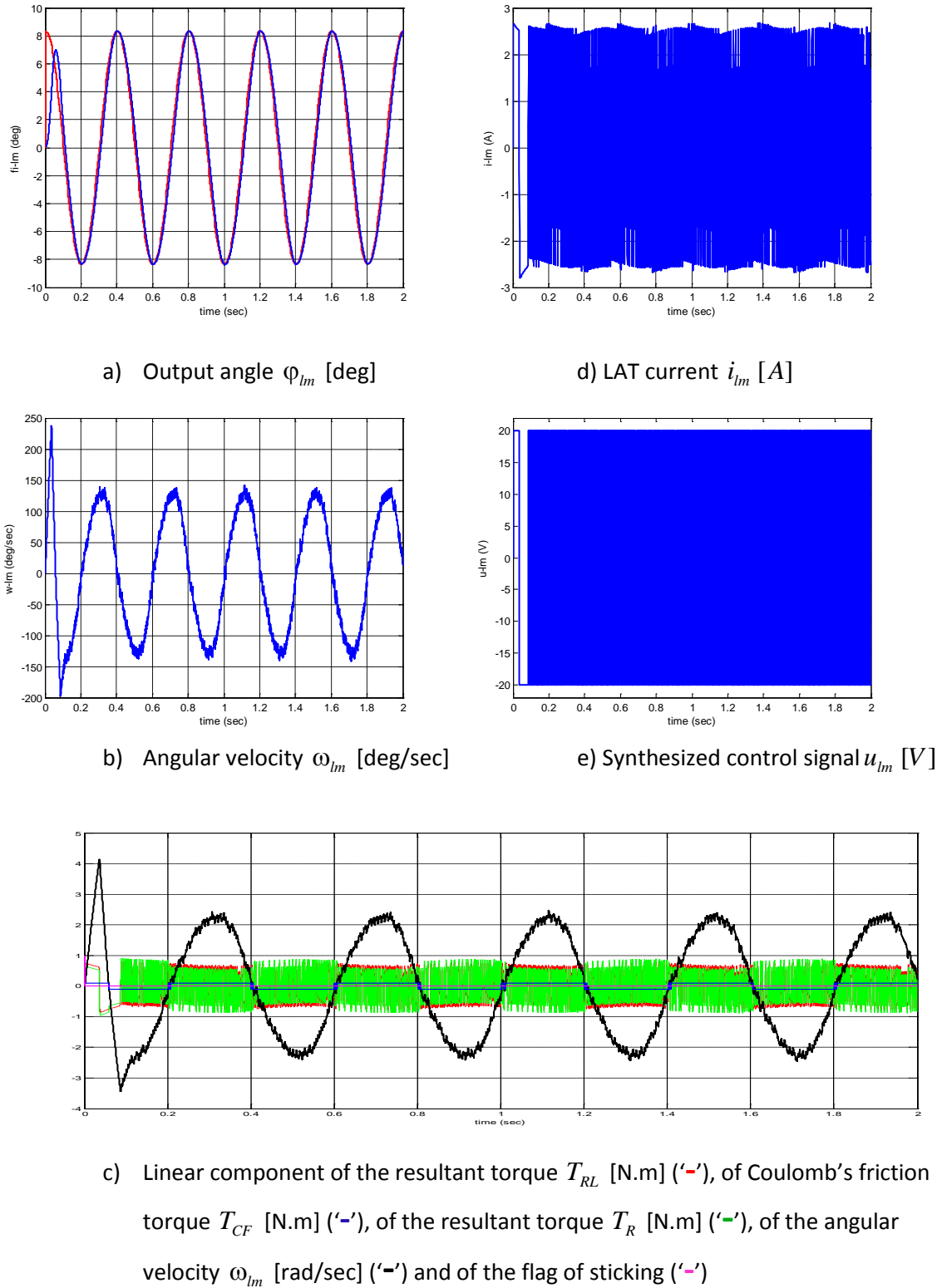

a)   Output angle $\varphi_{lm}$ [deg]

d) LAT current $i_{lm}$ $[A]$

b)   Angular velocity $\omega_{lm}$ [deg/sec]

e) Synthesized control signal $u_{lm}$ $[V]$

c)   Linear component of the resultant torque $T_{RL}$ [N.m] ('–'), of Coulomb's friction

torque $T_{CF}$ [N.m] ('–'), of the resultant torque $T_R$ [N.m] ('–'), of the angular

velocity $\omega_{lm}$ [rad/sec] ('–') and of the flag of sticking ('–')

**Figure 51. Time-diagrams of the processes in the digital tracking control system with the non-linear model of the large mirror actuator with zero pivot stiffness and Coulomb's friction model included as a model of the controlled system in case the demand position on** $\varphi_{lm}$ **is periodic signal with amplitude of** $8.35\,[\text{deg}]$ **and frequency 2.5 Hz discretized by sampling time of 4 ms (by ('–') in a)), while the rest part works at sampling rate of 1ms. The constraint on** $u_0 = 20\,[V]$ **.**

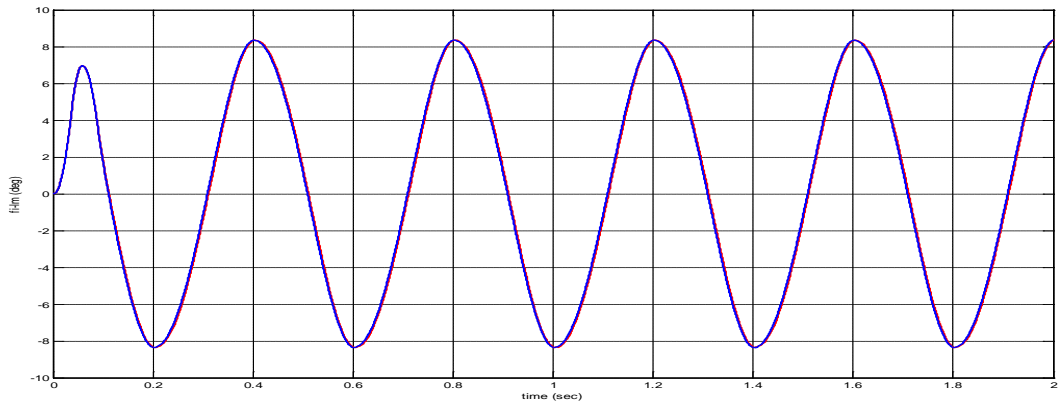

a) Output angle $\varphi_{lm}$ [deg]

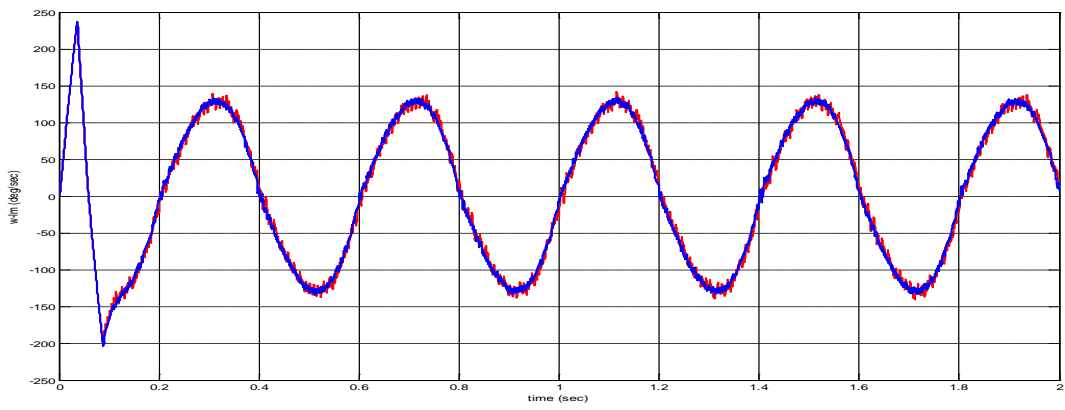

b) Angular velocity $\omega_{lm}$ [deg/sec]

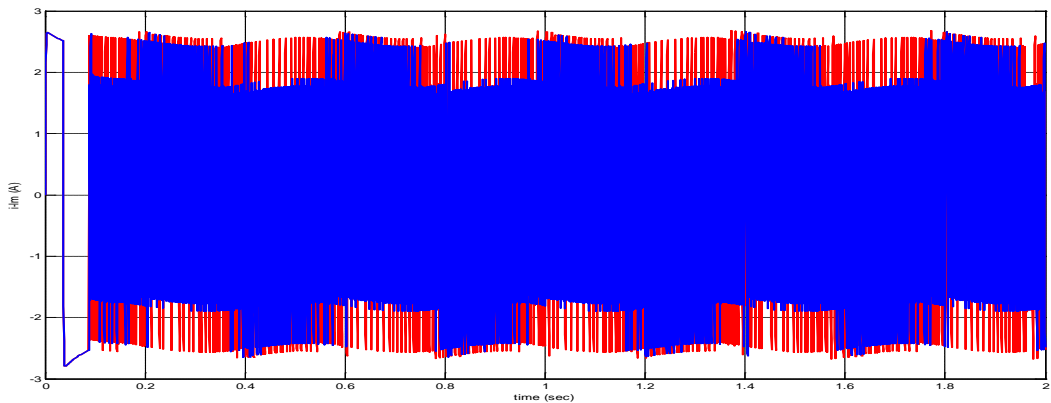

c) LAT current $i_{lm}$ [A]

**Figure 52. Comparison between the processes on output angle $\varphi_{lm}$, angular velocity $\omega_{lm}$ and LAT current $i_{lm}$ in the digital tracking control systems for the large mirror actuator controlling the non-linear model of the actuator with zero pivot stiffness and Coulomb's friction model included. The first one works at sampling rate of 1 ms ('–') but the second one works at two sampling rates of 4 ms and 1 ms ('–'). The constraint on $u_0 = 20\,[V]$.**

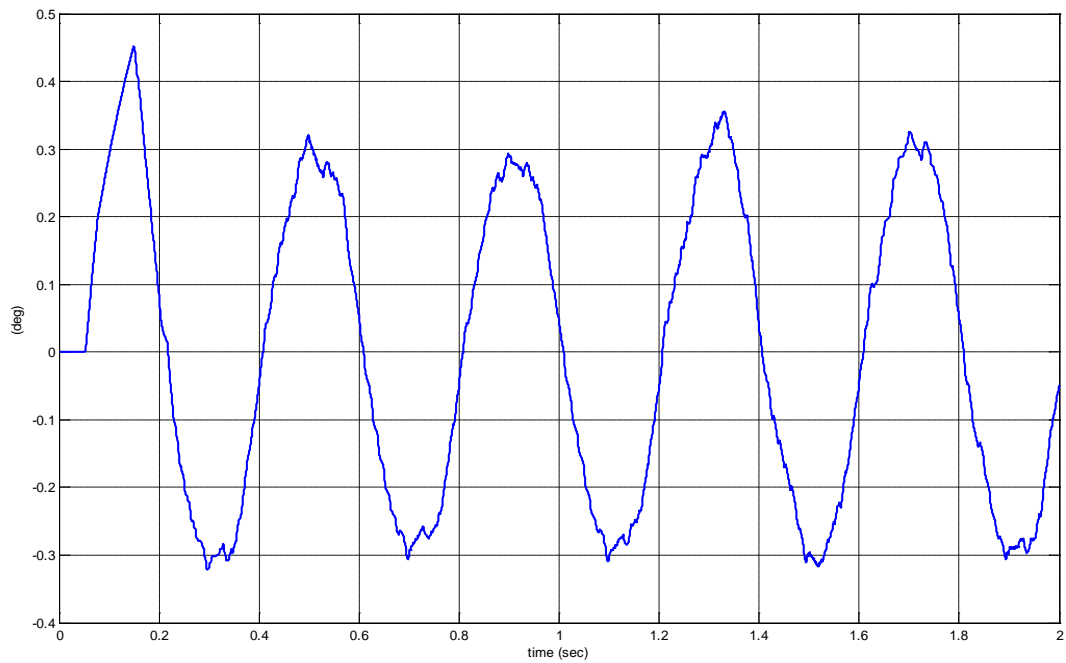

**Figure 53. Difference between the output angles at the system with two sampling rates of 4 ms and 1 ms and the system with sampling rate of 1 ms in case the constraint on control signal is** $u_0 = 10\,[V]$**.**

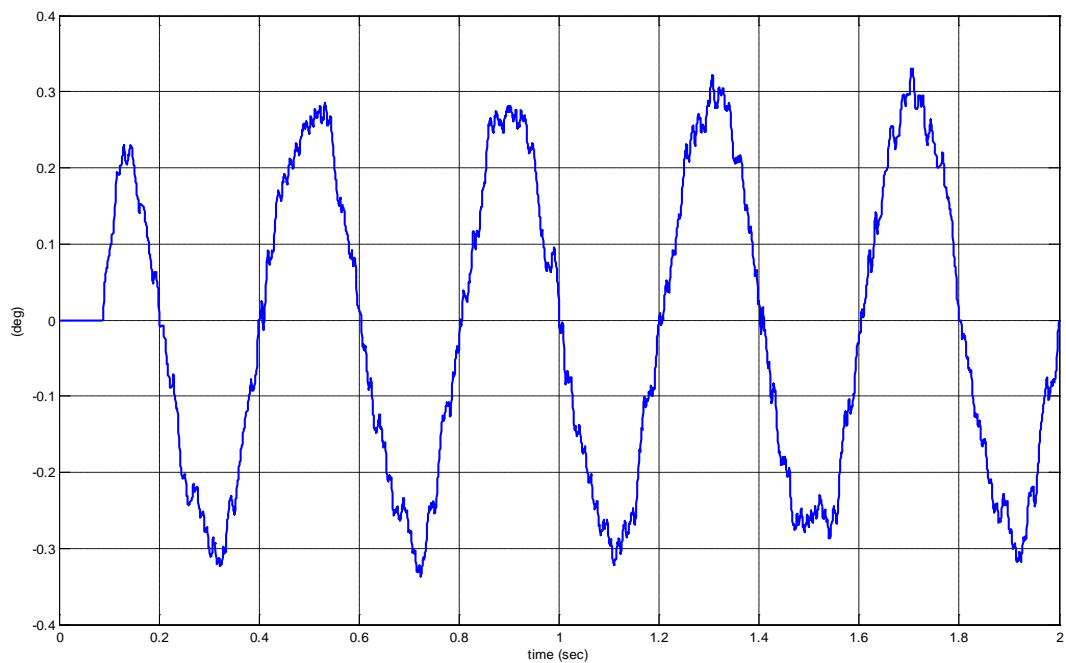

**Figure 54. Difference between the output angles at the system with sampling rate of 1 ms and the system with two sampling rates of 4 ms and 1 ms in case the constraint on control signal is** $u_0 = 20\,[V]$**.**

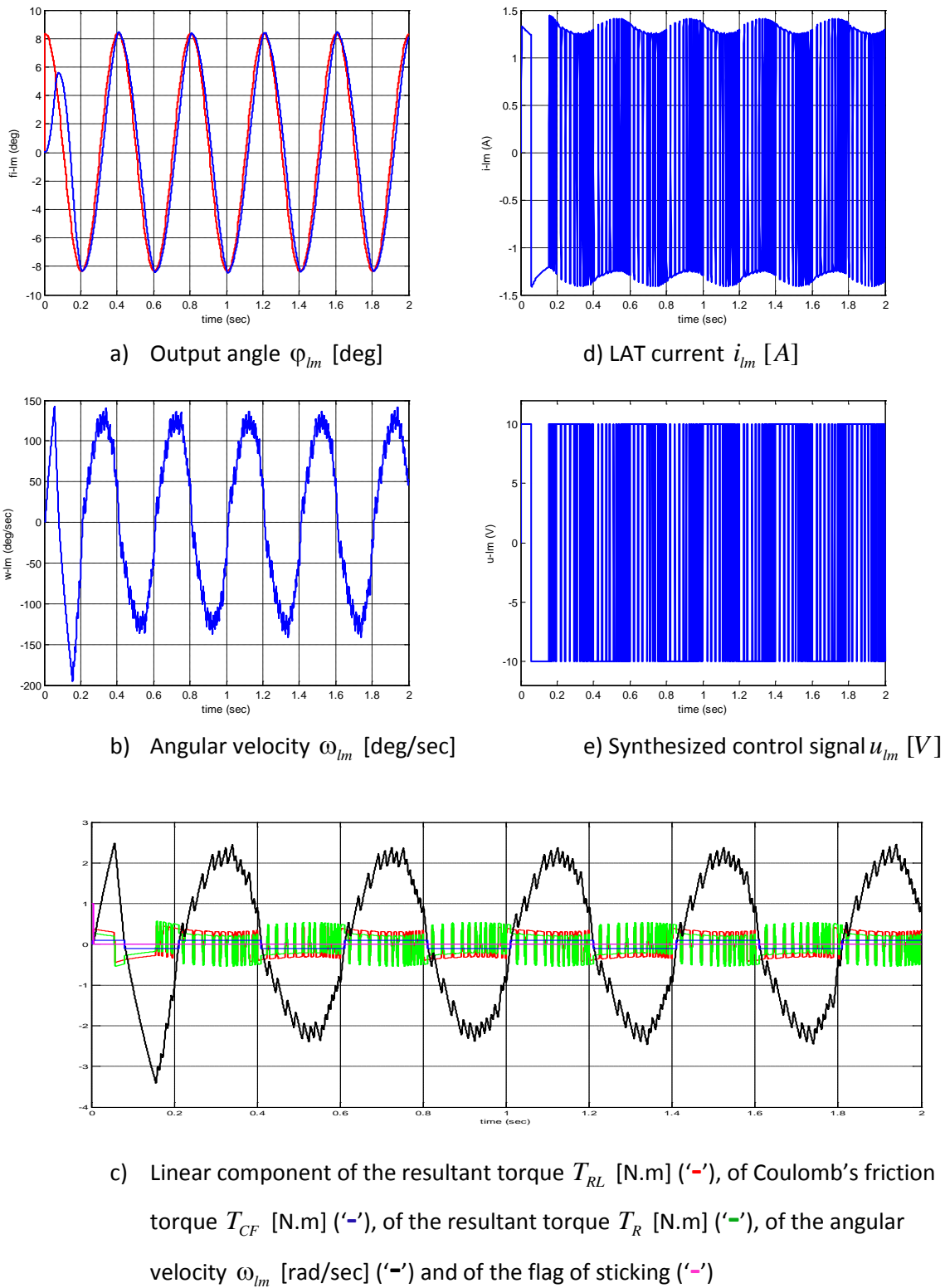

a)   Output angle $\varphi_{lm}$ [deg]

d) LAT current $i_{lm}$ $[A]$

b)   Angular velocity $\omega_{lm}$ [deg/sec]

e) Synthesized control signal $u_{lm}$ $[V]$

c)   Linear component of the resultant torque $T_{RL}$ [N.m] ('**–**'), of Coulomb's friction

torque $T_{CF}$ [N.m] ('**–**'), of the resultant torque $T_R$ [N.m] ('**–**'), of the angular

velocity $\omega_{lm}$ [rad/sec] ('**–**') and of the flag of sticking ('**–**')

**Figure 55. Time-diagrams of the processes in the digital tracking control system, working at sampling rate of 4 ms, with the non-linear model of the large mirror actuator with zero pivot stiffness and Coulomb's friction model included as a model of the controlled system in case the demand position on** $\varphi_{lm}$ **is periodic signal with amplitude of** $8.35\,[\text{deg}]$ **and frequency 2.5 Hz. The constraint on** $u_0 = 10\,[V]$ **.**

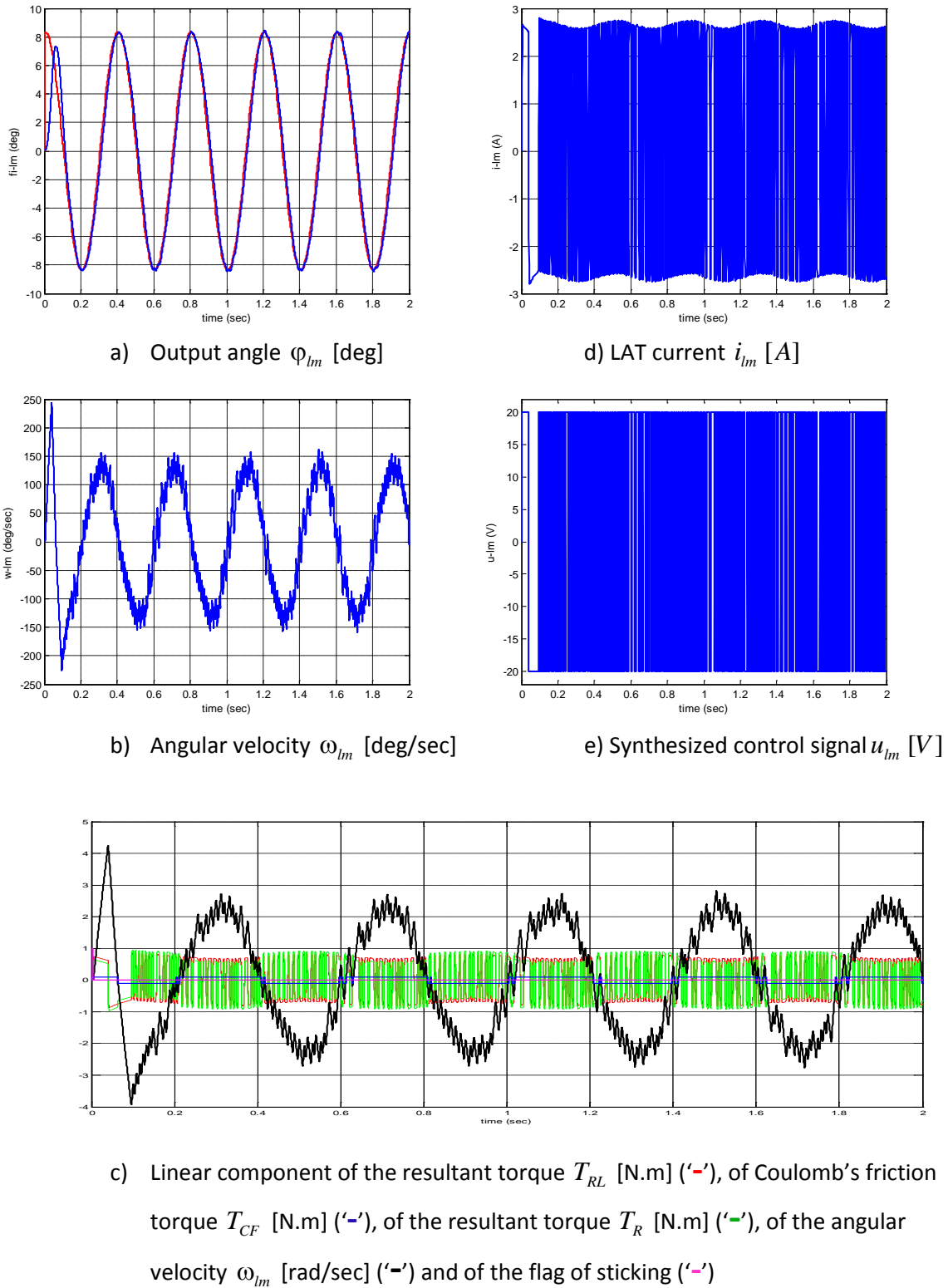

a) Output angle $\varphi_{lm}$ [deg]

d) LAT current $i_{lm}$ $[A]$

b) Angular velocity $\omega_{lm}$ [deg/sec]

e) Synthesized control signal $u_{lm}$ $[V]$

c) Linear component of the resultant torque $T_{RL}$ [N.m] ('—'), of Coulomb's friction torque $T_{CF}$ [N.m] ('—'), of the resultant torque $T_R$ [N.m] ('—'), of the angular velocity $\omega_{lm}$ [rad/sec] ('—') and of the flag of sticking ('—')

**Figure 56. Time-diagrams of the processes in the digital tracking control system, working at sampling rate of 4 ms, with the non-linear model of the large mirror actuator with zero pivot stiffness and Coulomb's friction model included as a model of the controlled system in case the demand position on $\varphi_{lm}$ is periodic signal with amplitude of $8.35\,[\text{deg}]$ and frequency 2.5 Hz. The constraint on $u_0 = 20\,[V]$.**

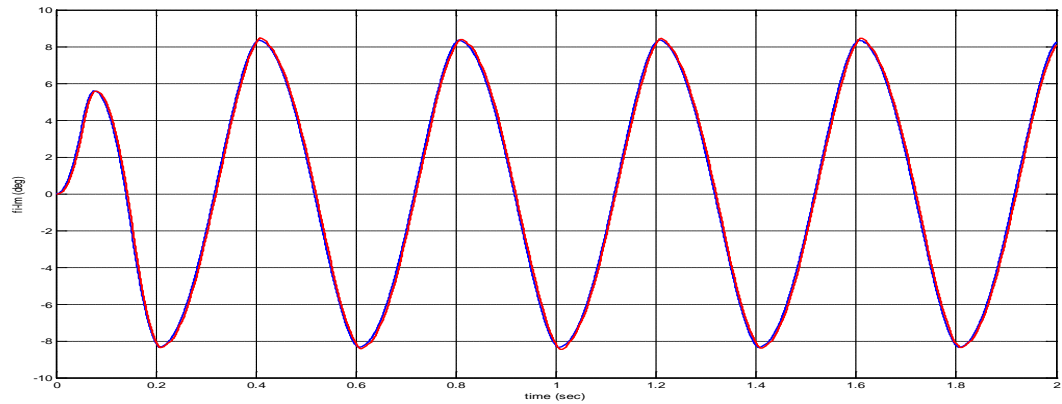

a) Output angle $\varphi_{lm}$ [deg]

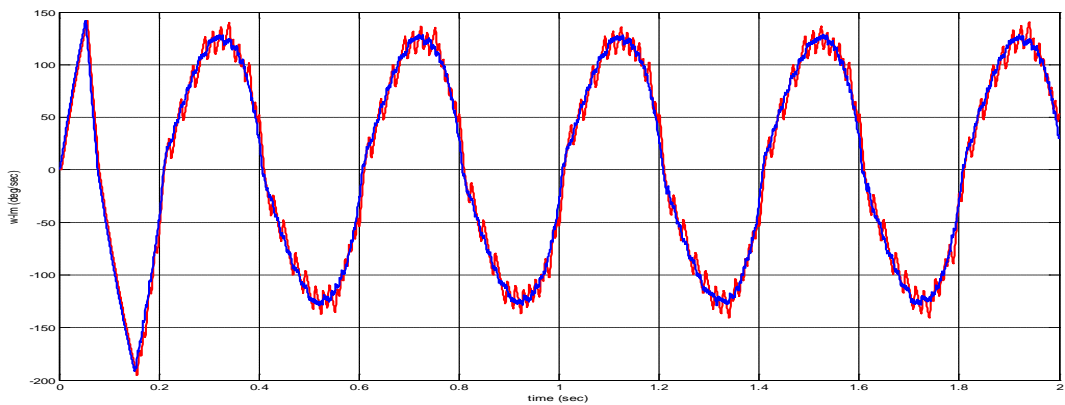

b) Angular velocity $\omega_{lm}$ [deg/sec]

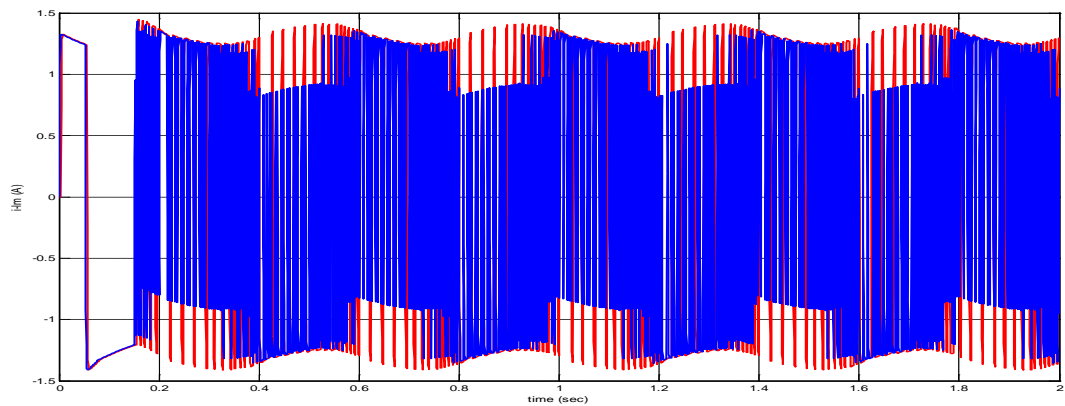

c) LAT current $i_{lm}$ [A]

**Figure 57. Comparison between the processes on output angle $\varphi_{lm}$, angular velocity $\omega_{lm}$ and LAT current $i_{lm}$ in the digital tracking control systems for the large mirror actuator controlling the non-non-linear model of the actuator with zero pivot stiffness and Coulomb's friction model included. The first one works at two sampling rates of 4 ms and 1 ms ('−') but the second one works at sampling rate of 4 ms ('~'). The constraint on $u_0 = 10\,[V]$.**

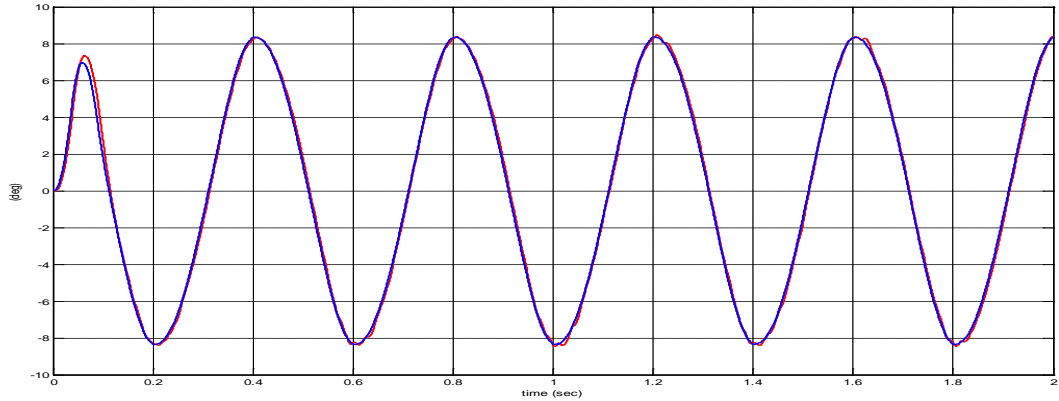

a)  Output angle  $\varphi_{lm}$  [deg]

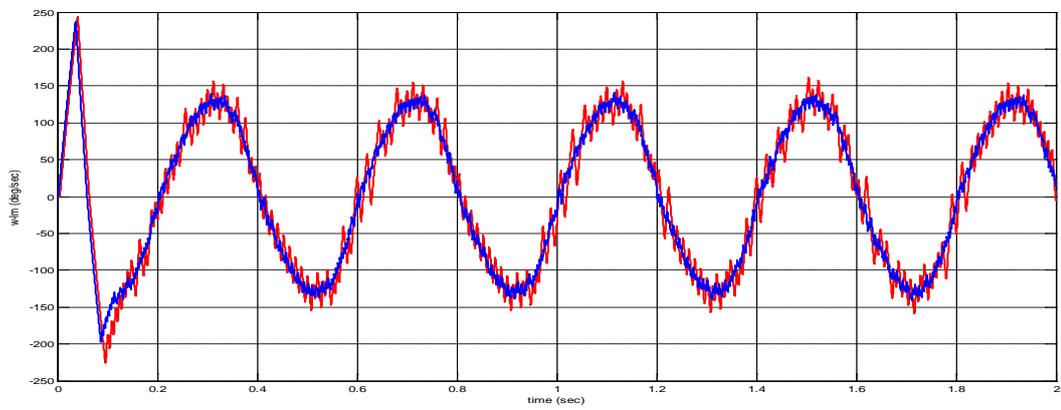

b)  Angular velocity  $\omega_{lm}$  [deg/sec]

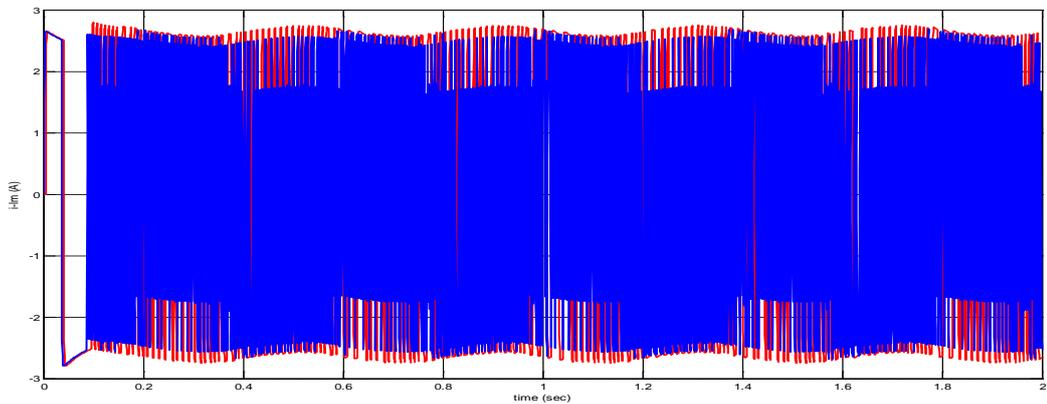

c)  LAT current  $i_{lm}$  [A]

**Figure 58. Comparison between the processes on output angle  $\varphi_{lm}$ , angular velocity  $\omega_{lm}$  and LAT current  $i_{lm}$  in the digital tracking control systems for the large mirror actuator controlling the non-linear model of the actuator with zero pivot stiffness and Coulomb's friction model included. The first one works at two sampling rates of 4 ms and 1 ms ('~') but the second one works at sampling rate of 4 ms ('~'). The constraint on  $u_0 = 20\,[V]$ .**

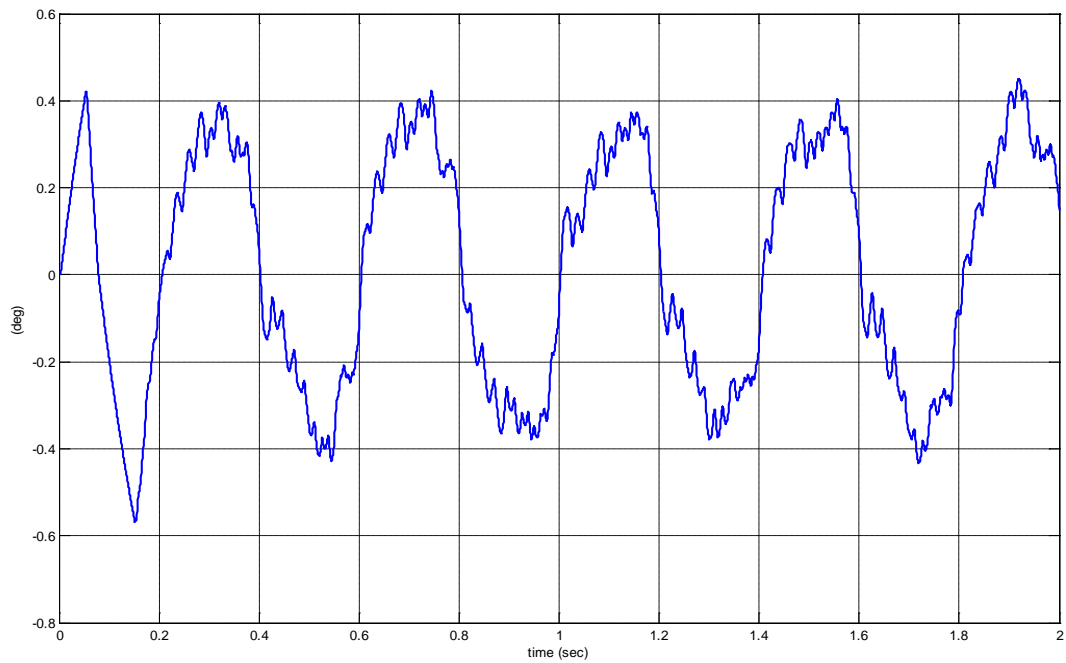

**Figure 59. Difference between the output angles at the system with two sampling rates of 4 ms and 1 ms and the system with sampling rate of 4 ms in case the constraint on control signal is $u_0 = 10\,[V]$.**

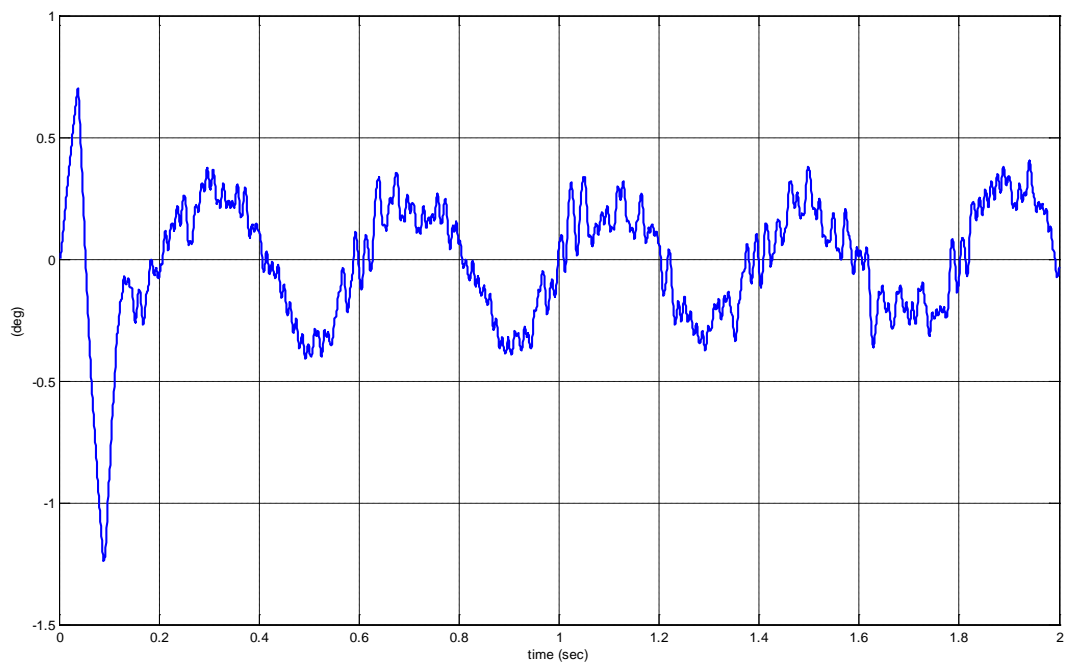

**Figure 60. Difference between the output angles at the system with two sampling rates of 4 ms and 1 ms and the system with sampling rate of 4 ms in case the constraint on control signal is $u_0 = 20\,[V]$.**

# 5. Closed loop systems for the small mirror actuator based on the synthesis of time optimal control

Let us consider the state space model of the small mirror actuator obtained at the modelling of the mirror actuators.

$$\dot{\boldsymbol{x}}_{sm} = A_{sm}\boldsymbol{x}_{sm} + B_{sm}u_{sm},$$

where

$$A_{sm} = \begin{bmatrix} 0 & 1 & 0 \\ -17571 & -42.857 & 404.29 \\ 0 & -62.889 & -1666.7 \end{bmatrix}, \quad B_{sm} = \begin{bmatrix} 0 \\ 0 \\ 222.22 \end{bmatrix}.$$

The eigenvalues of this system

$$\boldsymbol{\lambda}_{sm} = \begin{bmatrix} -29.282 + 129.93\,j \\ -29.282 - 129.93\,j \\ -1651 \end{bmatrix}$$

are of the same type as the eigenvalues of the large mirror actuator's model, one negative eigenvalue and two conjugate complex eigenvalues with negative real part, which determine the oscillating properties of the actuator presented at modelling. So this is a precondition for applying here the same approaches as at the controlling the large mirror actuator. On the experience and conclusions made at controlling the large mirror actuator, we shall focus on the approach based on zero pivot stiffness of the mechanical subsystem.

## 5.1 Approach based on the time optimal control of the small mirror actuator with zero pivot stiffness of the mechanical subsystem

As mentioned at considering the linear state space model of the large mirror actuator with zero pivot stiffness, zero pivot stiffness of the mechanical subsystem causes resetting the element $a_{21}$ of the system matrix of the model. So for the linear model of the small mirror actuator

$$\dot{\boldsymbol{x}}_{sm} = A_{sm}\boldsymbol{x}_{sm} + B_{sm}u_{sm},$$

$$y_{sm} = C_{sm}\boldsymbol{x}_{sm} + D_{sm}u_{sm},$$

$$\boldsymbol{x}_{sm} = \begin{bmatrix} \varphi_{sm} \\ \omega_{sm} \\ i_{sm} \end{bmatrix}, \quad A_{sm} = \begin{bmatrix} 0 & 1 & 0 \\ -c_{sm}/J_{sm} & -h_{sm}/J_{sm} & Kt/J_{sm} \\ 0 & -Kb/Lm & -Rm/Lm \end{bmatrix}, \quad B_{sm} = \begin{bmatrix} 0 \\ 0 \\ 1/Lm \end{bmatrix},$$

$$C_{sm} = \begin{bmatrix} 1 & 0 & 0 \end{bmatrix}, \quad D_{sm} = 0,$$

by resetting the element $a_{21}$ ($c_{sm} = 0$) at the initial representation of $A_{sm}$, presented above, the new representation of $A_{sm}$ becomes

$$A_{sm} = \begin{bmatrix} 0 & 1 & 0 \\ 0 & -42.857 & 404.29 \\ 0 & -62.889 & -1666.7 \end{bmatrix}$$

having eigenvalues

$$\lambda_{sm} = \begin{bmatrix} 0 \\ -58.669 \\ -1650.9 \end{bmatrix}.$$

Now all the eigenvalues of the actuator with zero pivot stiffness are negative and different. So by this correction the model of the actuator is suitable for applying author's method for solving of time optimal control problems for a specified class of linear systems.

### 5.1.1 Near time optimal control solutions for positioning

Let us deal with the above linear state space model of the actuator with zero pivot stiffness and consider the time optimal control problem for the transition from zero initial state to the final state

$$\boldsymbol{x}_{sm}(t_f) = \begin{bmatrix} \varphi_{sm}(t_f) \\ \omega_{sm}(t_f) \\ i_{sm}(t_f) \end{bmatrix} = \begin{bmatrix} 3.57 \\ 0 \\ 0 \end{bmatrix} \begin{pmatrix} \deg \\ \deg/s \\ A \end{pmatrix}.$$

Let us assume, as in the previous cases, the constraints for the control signal $u_{sm}(t)$ are

$$-u_0 \leq u_{sm}(t) \leq u_0.$$

We solve the problem by the author's method consequently for $u_0 = 10\,[V]$ and $u_0 = 20\,[V]$.

A near time optimal control solution for $u_0 = 10\,[V]$ is presented in Table 7 and Figure 61. The solution represents a piece-wise constant function $u_{sm}^{\bar{o}}$ shown in Figure 61d with amplitude of $u_0 = 10\,[V]$ having three intervals of constancy with respective length given in Table 7. The near minimum time for this transition is $t_f^{\bar{o}} = 0.022978$ [s] and the accuracy achieved is respectively: on $\varphi_{sm} - 7.0\text{e-}10\,[\deg]$, on $\omega_{sm} - 1.5\text{e-}6\,[\deg/s]$ and on $i_{sm} - 8.5\text{e-}8\,[A]$.

**Table 7. A near time optimal control solution for positioning at 3.57 (deg) in case $u_0 = 10\,[V]$.**

| | |
|---|---|
| Length of the first interval [s] | 0.014428 |
| Length of the second interval [s] | 0.008131 |
| Length of the third interval [s] | 0.000420 |
| Near minimum time for the transition $t_f^{\tilde{o}}$ [s] | 0.022978 |

**Table 8. A near time optimal control solution for positioning at 3.57 (deg) in case $u_0 = 20\,[V]$.**

| | |
|---|---|
| Length of the first interval [s] | 0.009388 |
| Length of the second interval [s] | 0.006449 |
| Length of the third interval [s] | 0.000420 |
| Near minimum time for the transition $t_f^{\tilde{o}}$ [s] | 0.016257 |

A near time optimal control solution for $u_0 = 20\,[V]$ is presented in Table 8 and Figure 62. The solution represents a piece-wise constant function $u_{sm}^{\tilde{o}}$ shown in Figure 62d with amplitude of $u_0 = 20\,[V]$ having three intervals of constancy with respective length given in Table 8. The near minimum time for this transition is $t_f^{\tilde{o}} = 0.016257$ [s] and the accuracy achieved is respectively: on $\varphi_{sm} - 3.5\text{e-}10\,[\text{deg}]$, on $\omega_{sm} - 9.0\text{e-}7\,[\text{deg/s}]$ and on $i_{sm} - 6.5\text{e-}8\,[A]$.

### 5.1.2  Tracking control systems

We investigate the synthesized control systems for the small mirror actuator in the same manner we investigate the control systems for the large mirror actuator. The control synthesis is based on a specially developed software for solving the linear time optimal control problem with given accuracy in case the eigenvalues of the controlled model of 3rd order are of the type as the type of eigenvalues of the above corrected linear model of the small mirror actuator with zero pivot stiffness. The systems include also an one-step prediction mechanism, which reduces the amplitude of steady oscillations in sliding mode and improves the control reaction caused by the discretization and time delay of one step at synthesizing the respective control signal in the tracking system.

### 5.1.2.1  Response to constant demand signal

The demand signal here is the same as in the previous case

$$\boldsymbol{x}_{sm}(t_f) = \begin{bmatrix} \varphi_{sm}(t_f) \\ \omega_{sm}(t_f) \\ i_{sm}(t_f) \end{bmatrix} = \begin{bmatrix} 3.57 \\ 0 \\ 0 \end{bmatrix} \begin{pmatrix} \deg \\ \deg/s \\ A \end{pmatrix}.$$

First the model of the actuator we control represents the linear state space model, and then the model of the controlled system is changed to the non-linear model of the actuator with included Coulomb's friction model. The processes in case the model of the controlled system is the linear model of the actuator with zero pivot stiffness are shown in Figure 63 and Figure 64 for constraints on the control signal $u_0 = 10\,[V]$ and $u_0 = 20\,[V]$ respectively. The processes in the closed loop systems in case the model of the actuator is the non-linear model with included Coulomb's friction are shown in Figure 65 and Figure 66 for constraints on the control signal $u_0 = 10\,[V]$ and $u_0 = 20\,[V]$ respectively. The comparison with the near time optimal solutions for constraints $u_0 = 10\,[V]$ and $u_0 = 20\,[V]$, presented also in these figures, shows that the processes in the closed loop systems on output angle, angular velocity, LAT current and control signal are close to the respective time diagrams at the linear near time optimal solutions at the transition to the demand position, but the steady state of the closed systems represents periodic oscillations with amplitude around 0.5 deg on output angle and 50 and 100 deg/s on angular velocity for constraints $u_0 = 10\,[V]$ and $u_0 = 20\,[V]$ respectively. In case the controlled system represents the non-linear model of the actuator with Coulomb's friction included, Figure 65 and Figure 66 for $u_0 = 10\,[V]$ and $u_0 = 20\,[V]$ respectively, the behavior of the control systems at steady state around the demand position is with significantly less amplitude of oscillations. There is only one sticking phase in the very beginning of the processes while the initial rising of the resultant torque, Figure 65c and Figure 66c, but after that initial phase the movement of the actuator's load is smooth without sticking.

## 5.1.2.2 Response to sinusoidal demand signal with frequency 20 Hz and amplitude of 3.57 degrees

Let the demand signal be a sinusoidal signal with frequency $20\,\text{Hz}$ and amplitude of $3.57\,\text{deg}$ as required for the lidar's scanning system. Time-diagrams of the processes in the digital tracking control systems working at sampling rate of $0.1\,\text{ms}$ and controlling the linear model of the small mirror actuator with zero pivot stiffness are presented in Figure 67 and Figure 68 for constraints on $u_0 = 10\,[V]$ and $u_0 = 20\,[V]$ respectively. The time-diagrams show that unlike the system with constraint on $u_0 = 10\,[V]$, Figure 67, the system with constraint on $u_0 = 20\,[V]$, Figure 68, manages very well with tracking the demand periodic signal. The achieved accuracy in this case is $0.0217\,\text{deg}$.

Let us now increase the sampling time ten times from $0.1\,\text{ms}$ to $1\,\text{ms}$. Figure 69 shows the time-diagrams of the processes in the control system with constraint on $u_0 = 20\,[V]$. The achieved accuracy at this sampling rate is $0.1181\,\text{deg}$.

The processes in the digital tracking control system with constraint on $u_0 = 20\,[V]$ and sampling rate of $1\,\text{ms}$ but controlling the non-linear model of the actuator with Coulomb's friction model included are presented in Figure 70. The achieved accuracy in this case is $0.0775\deg$.

Decreasing the sampling rate to $4\,\mathrm{ms}$ causes poor tracking behaviour. It can be seen in Figure 71, especially in Figure 71a, where there are samplings from the output angle with difference of more than 2 degrees compared with amplitude of the demand periodic signal.

A compromise is a digital tracking control system working simultaneously at two sampling rates of $4\,\mathrm{ms}$ and $1\,\mathrm{ms}$. The demand signal is discretized by sampling rate of $4\,\mathrm{ms}$, but the rest part of the control system works at sampling rate of $1\,\mathrm{ms}$. The processes in this system are shown in Figure 72. In Figure 72a the discretized demand periodic signal ('-'), the output angle $\varphi_{sm}$ and the samplings from it ('*') at rate of $4\,\mathrm{ms}$ are presented together. The achieved in this system accuracy on tracking is $0.1759\deg$.

### 5.1.3 Conclusions on controlling the small mirror actuator

The approach of applying techniques for controlling the small mirror actuator on the basis of synthesis of time optimal control shows the possibility of achieving high dynamic performance at the control of the small mirror actuator. At the chosen approach a correction of the actuator's design is needed. Here the case with zero pivot stiffness of the mechanical subsystem is investigated. A sampling rate of $4\,\mathrm{ms}$ required for the demand signal is insufficient for controlling effectively the non-linear model of the actuator with zero pivot stiffness. At sampling times of $1\,\mathrm{ms}$ or less the effectiveness of tracking is much better. So by keeping the sampling time of $4\,\mathrm{ms}$ for the demand signal but controlling the actuator with sampling rate of $1\,\mathrm{ms}$ the digital tracking control system synthesizing a near time optimal control with constraints of it $u_0 = 20\,[V]$ manages well the tracking the periodic sinusoidal demand signal of $20\,\mathrm{Hz}$ and amplitude of $3.57\deg$ as required for the lidar's scanning system with accuracy of $0.1759\deg$.

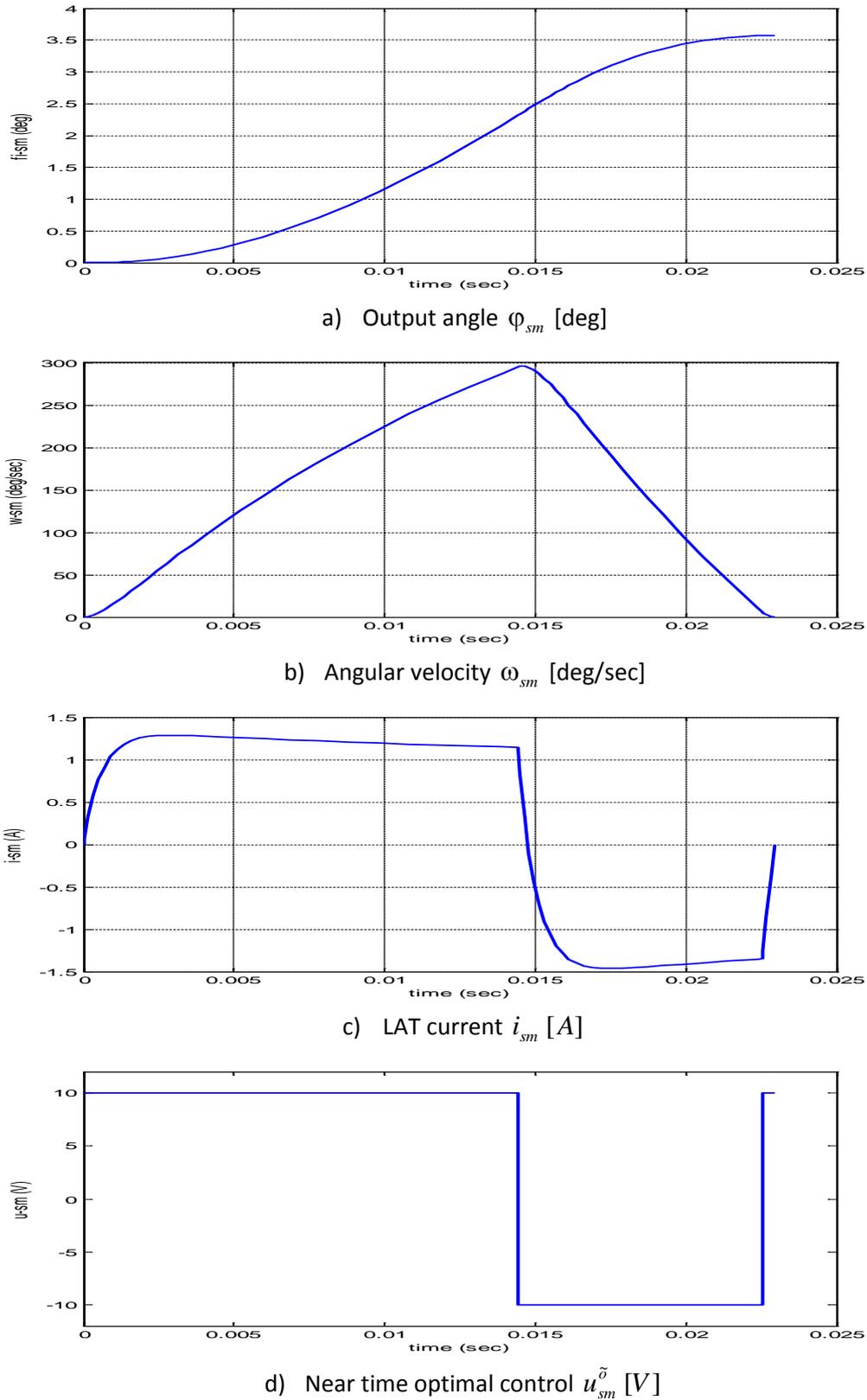

a)  Output angle $\varphi_{sm}$ [deg]

b)  Angular velocity $\omega_{sm}$ [deg/sec]

c)  LAT current $i_{sm}$ [A]

d)  Near time optimal control $u_{sm}^{\tilde{o}}$ [V]

**Figure 61. A Near time optimal control solution for the transition from zero initial state to demand position on** $\varphi_{sm}$ 3.57 [deg] **in case** $u_0 = 10$ [V] **with accuracy on** $\varphi_{sm} - 7.0\text{e-}10$ [deg]**, on** $\omega_{sm} - 1.5\text{e-}6$ [deg/s] **and on** $i_{sm} - 8.5\text{e-}8$ [A]**.**

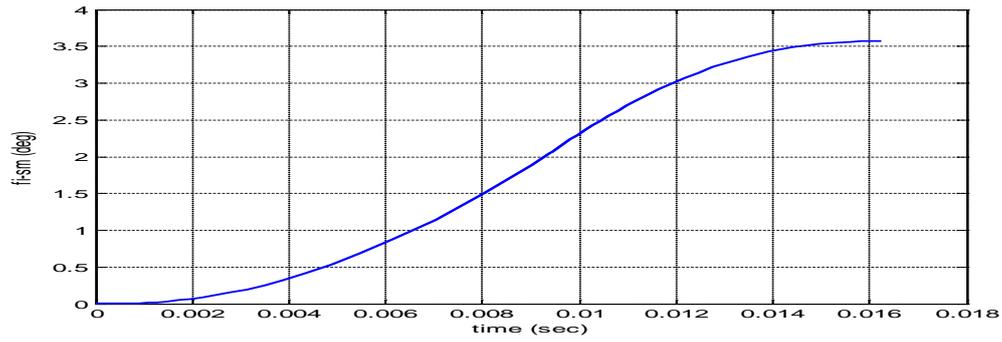

a) Output angle $\varphi_{lm}$ [deg]

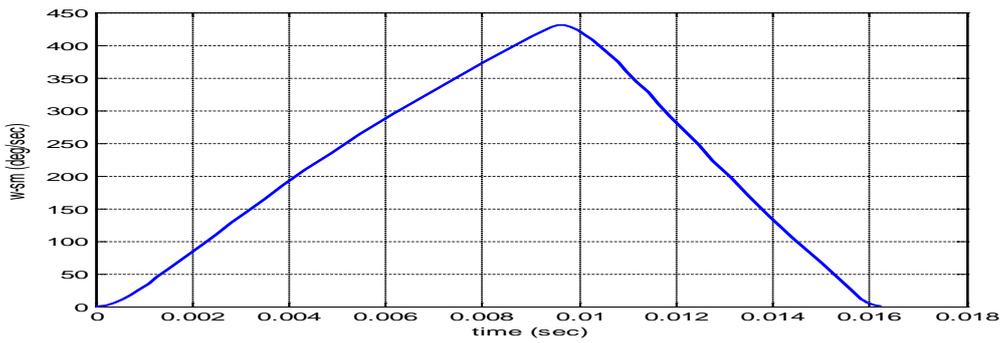

b) Angular velocity $\omega_{lm}$ [deg/s]

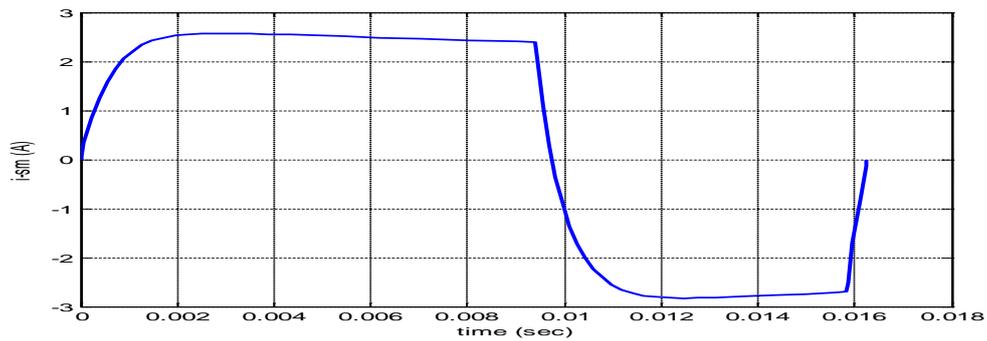

c) LAT current $i_{lm}$ [A]

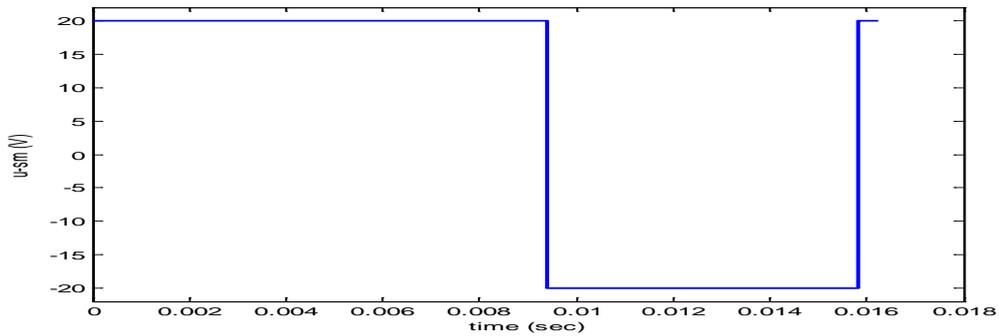

d) Near time optimal control $u_{sm}^{\tilde{o}}$ [V]

**Figure 62. A Near time optimal control solution for the transition from zero initial state to demand position on** $\varphi_{sm}$ 3.57 [deg] **in case** $u_0 = 20\,[V]$ **with accuracy on** $\varphi_{sm} - 3.5\mathrm{e}\text{-}10\,[\text{deg}]$, **on** $\omega_{sm} - 9.0\mathrm{e}\text{-}7\,[\text{deg/s}]$ **and on** $i_{sm} - 6.5\mathrm{e}\text{-}8\,[A]$.

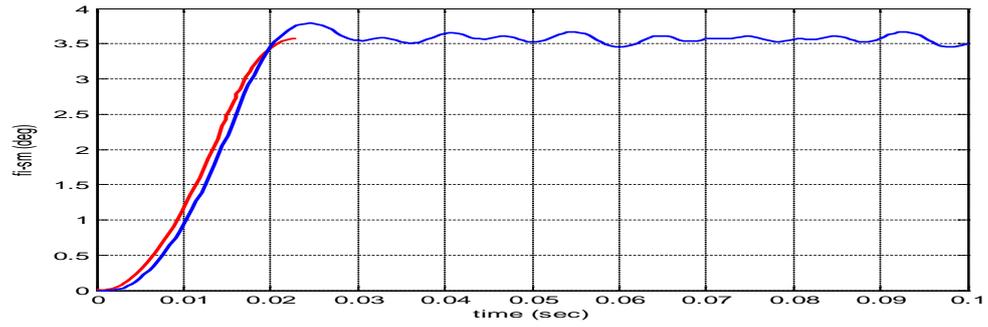

a)   Output angle $\varphi_{sm}$ [deg]

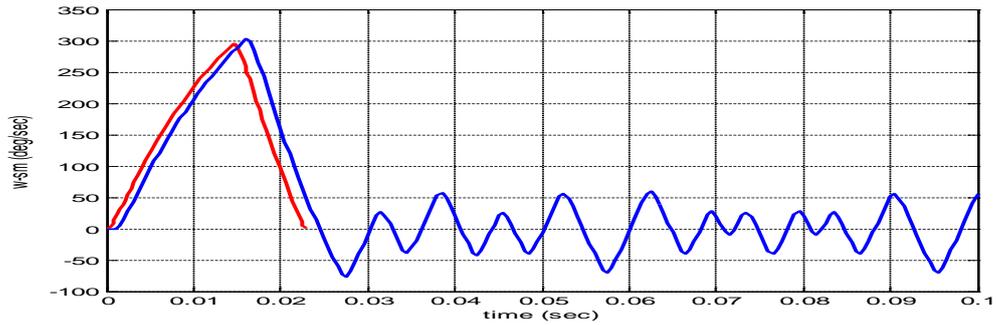

b)   Angular velocity $\omega_{sm}$ [deg/s]

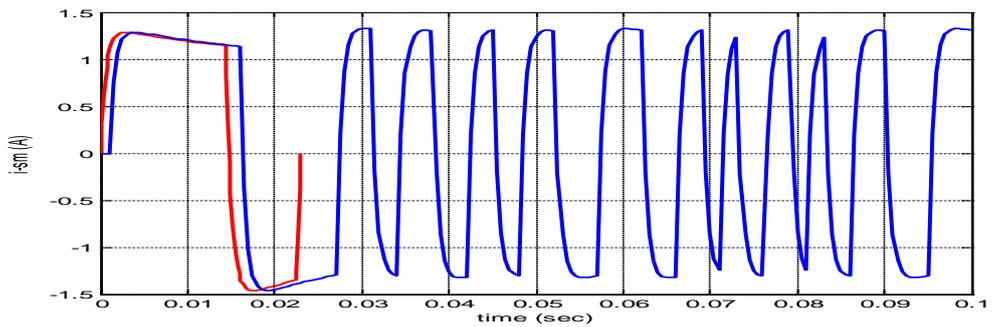

c)   LAT current $i_{sm}$ [A]

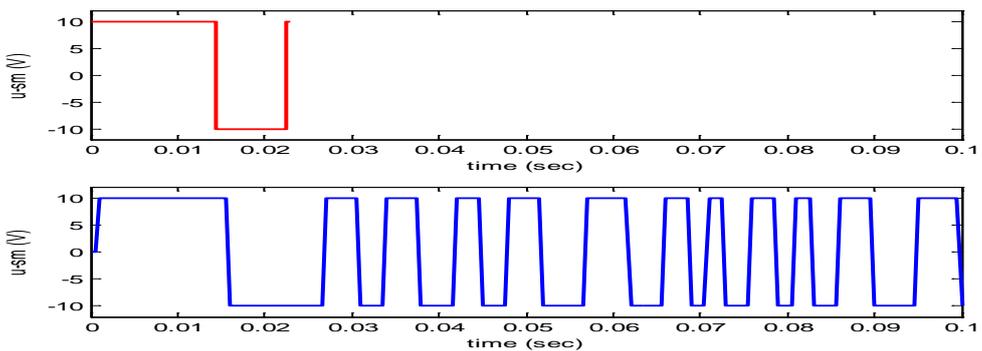

d)   Near time optimal control $u_{sm}^{\tilde{o}}$ [V]

**Figure 63. Comparison between the near linear time optimal control solution ('–') and the processes in the tracking control system with the linear model of the small mirror actuator with zero pivot stiffness as a model of the controlled system in case the demand position on $\varphi_{sm}$ is $3.57\,[\mathrm{deg}]$, the constraint on $u_0 = 10\,[V]$ and sampling time of 1ms.**

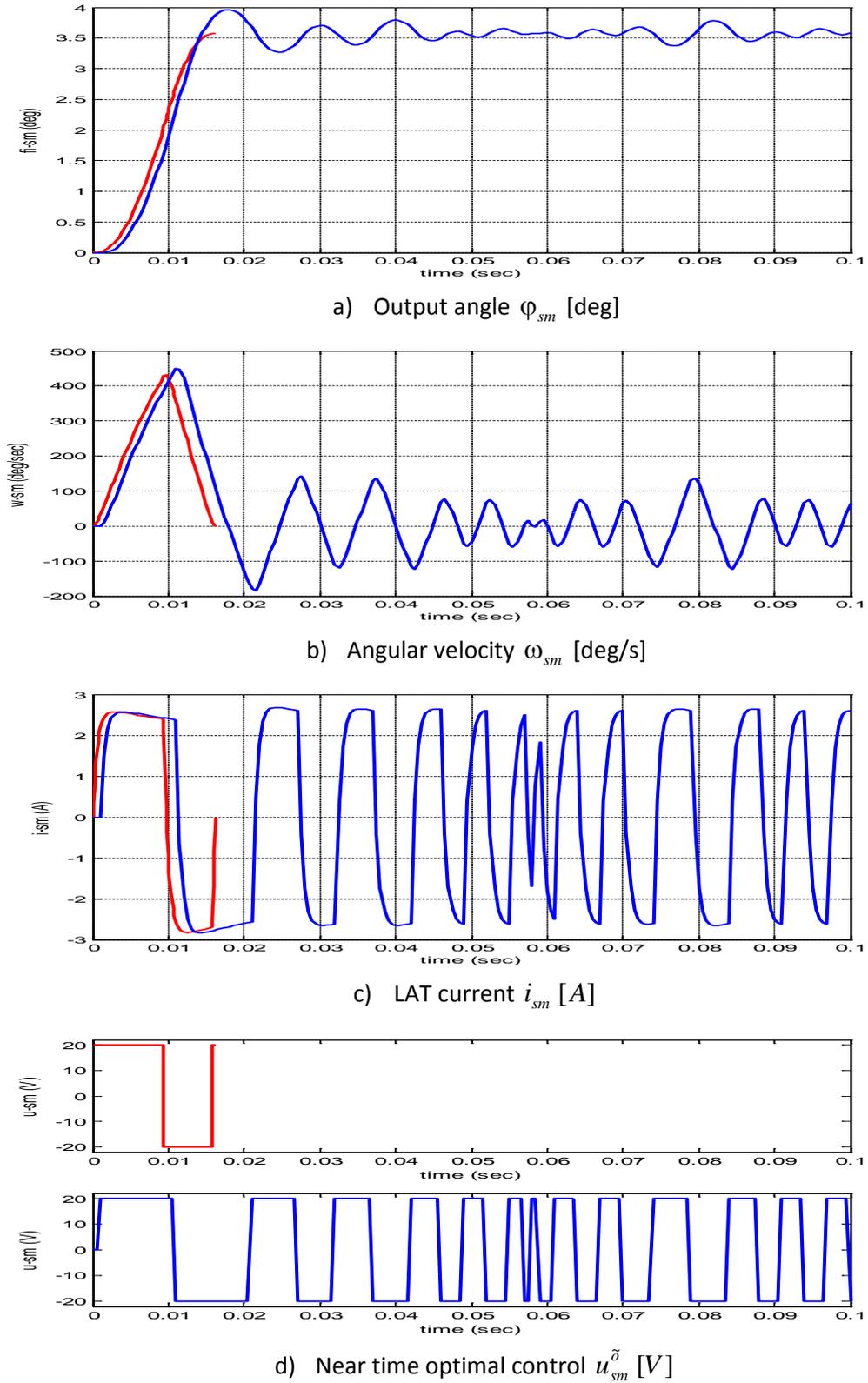

a)   Output angle $\varphi_{sm}$ [deg]

b)   Angular velocity $\omega_{sm}$ [deg/s]

c)   LAT current $i_{sm}$ [A]

d)   Near time optimal control $u_{sm}^{\tilde{o}}$ [V]

**Figure 64. Comparison between the near linear time optimal control solution ('–') and the processes in the tracking control system with the linear model of the small mirror actuator with zero pivot stiffness as a model of the controlled system in case the demand position on** $\varphi_{sm}$ **is** $3.57\,[\mathrm{deg}]$**, the constraint on** $u_0 = 20\,[V]$ **and sampling time of 1ms.**

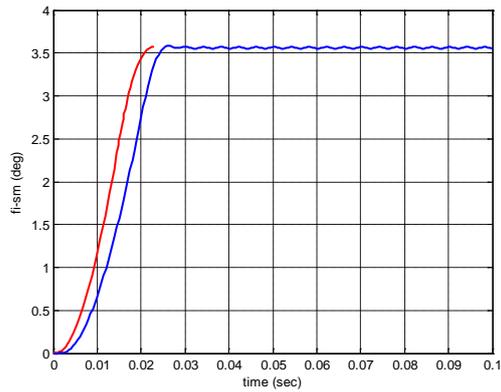

a)   Output angle $\varphi_{sm}$ [deg]

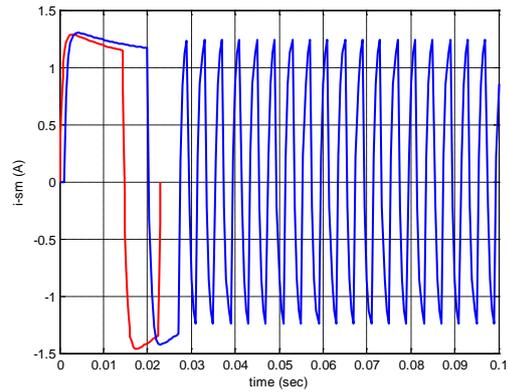

d) LAT current $i_{lm}$ $[A]$

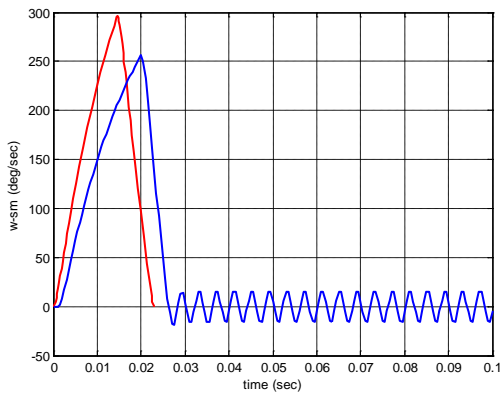

b)   Angular velocity $\omega_{lm}$ [deg/sec]

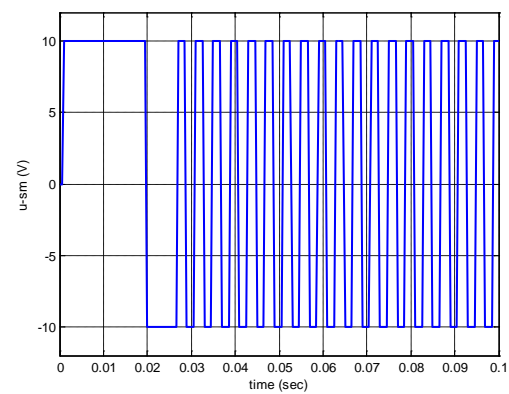

e) Synthesized control signal $u_{lm}$ $[V]$

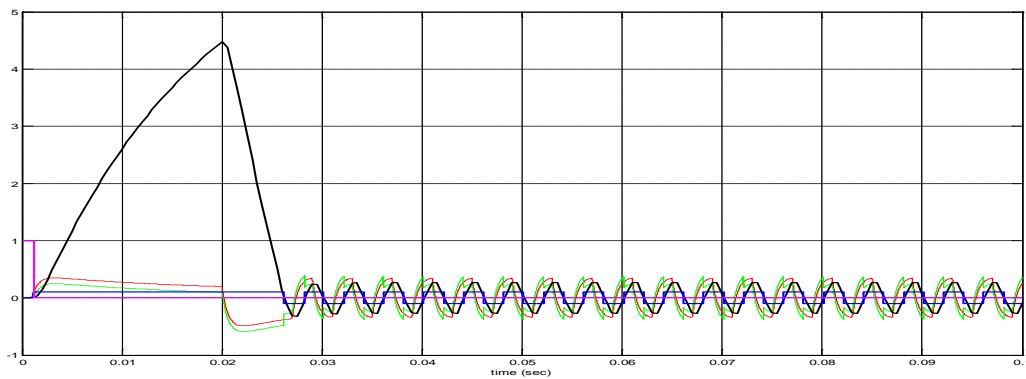

c)   Linear component of the resultant torque $T_{RL}$ [N.m] ('–'), of Coulomb's friction

torque $T_{CF}$ [N.m] ('–'), of the resultant torque $T_R$ [N.m] ('–'), of the angular

velocity $\omega_{sm}$ [rad/sec] ('–') and of the flag of sticking ('–')

**Figure 65. Time-diagrams of the processes in the digital tracking control system with the non-linear model of the small mirror actuator with zero pivot stiffness and Coulomb's friction model included as a model of the controlled system in case the demand position on $\varphi_{sm}$ is $3.57$ [deg], the constraint on $u_0 = 10$ [V] and sampling time of 1ms. In a), b) and d) the processes are compared with the linear near time optimal solution ('–') on $\varphi_{sm}$ and $\omega_{sm}$ and $i_{sm}$.**

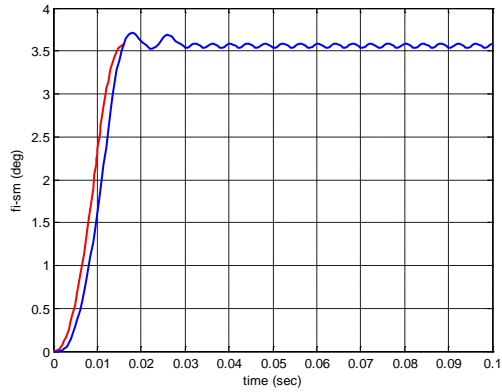

a) Output angle $\varphi_{sm}$ [deg]

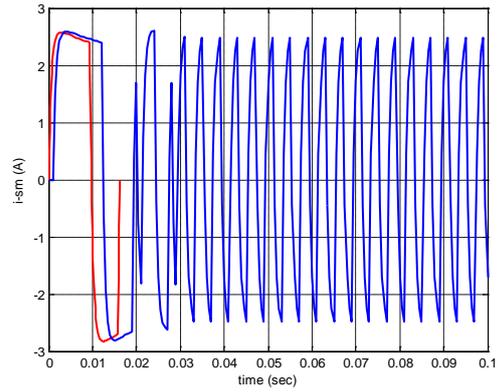

d) LAT current $i_{sm}$ [A]

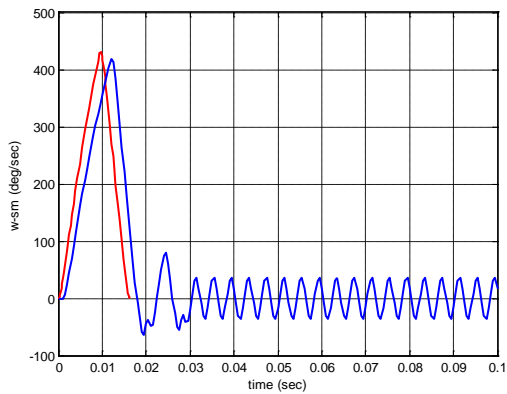

b) Angular velocity $\omega_{sm}$ [deg/sec]

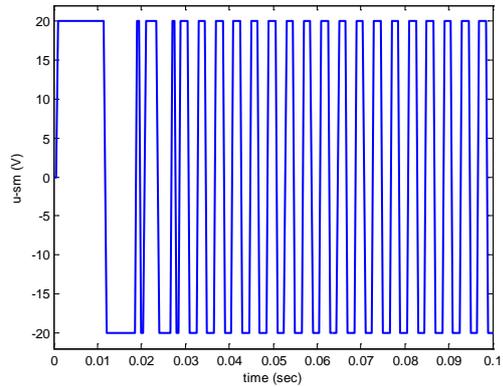

e) Synthesized control signal $u_{sm}$ [V]

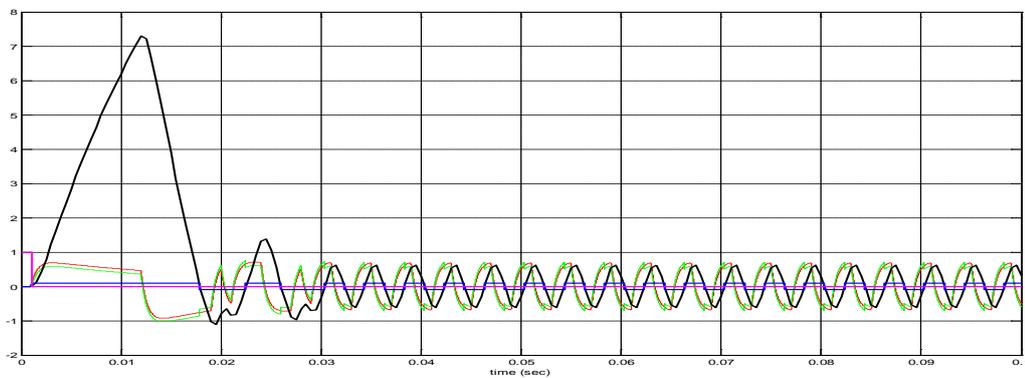

c) Linear component of the resultant torque $T_{RL}$ [N.m] ('—'), of Coulomb's friction torque $T_{CF}$ [N.m] ('—'), of the resultant torque $T_R$ [N.m] ('—'), of the angular velocity $\omega_{sm}$ [rad/sec] ('—') and of the flag of sticking ('—')

**Figure 66. Time-diagrams of the processes in the digital tracking control system with the non-linear model of the small mirror actuator with zero pivot stiffness and Coulomb's friction model included as a model of the controlled system in case the demand position on $\varphi_{sm}$ is $3.57\,[\text{deg}]$, the constraint on $u_0 = 20\,[V]$ and sampling time of 1ms. In a), b) and d) the processes are compared with the linear near time optimal solution ('—') on $\varphi_{sm}$ and $\omega_{sm}$ and $i_{sm}$.**

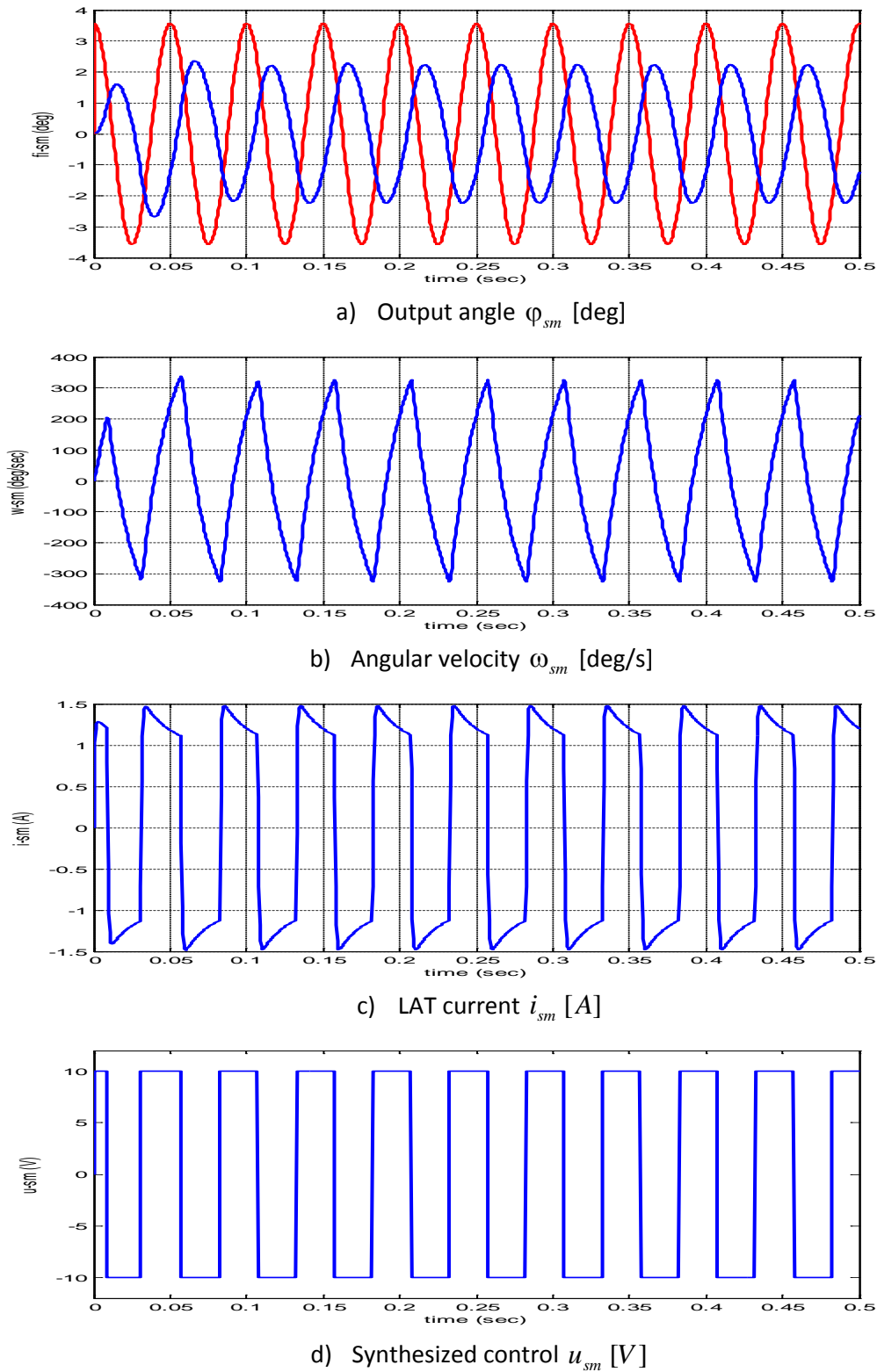

a)  Output angle $\varphi_{sm}$ [deg]

b)  Angular velocity $\omega_{sm}$ [deg/s]

c)  LAT current $i_{sm}$ [A]

d)  Synthesized control $u_{sm}$ [V]

**Figure 67. Time-diagrams of the processes in the digital tracking control system working at sampling rate of** $0.1\,\mathrm{ms}$ **and controlling the linear model of the small mirror actuator with zero pivot stiffness in case the demand position on** $\varphi_{sm}$ **is periodic signal with amplitude of** $3.57\,\mathrm{deg}$ **and frequency** $20\,\mathrm{Hz}$. **The constraint on** $u_0 = 10\,[V]$.

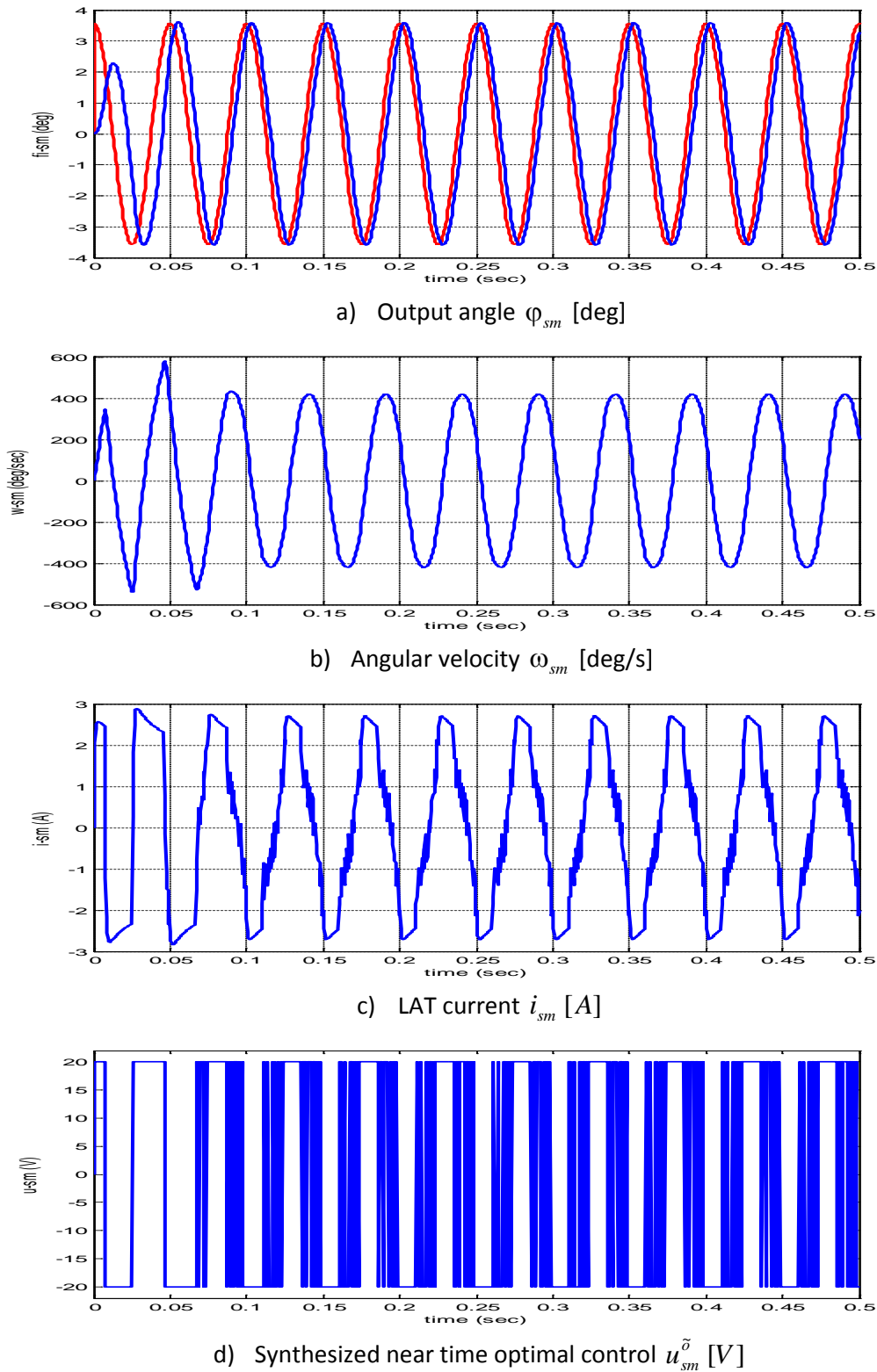

a)  Output angle $\varphi_{sm}$ [deg]

b)  Angular velocity $\omega_{sm}$ [deg/s]

c)  LAT current $i_{sm}$ [A]

d)  Synthesized near time optimal control $u_{sm}^{\tilde{o}}$ [V]

**Figure 68. Time-diagrams of the processes in the digital tracking control system working at sampling rate of** $0.1\,\mathrm{ms}$ **and controlling the linear model of the small mirror actuator with zero pivot stiffness in case the demand position on** $\varphi_{sm}$ **is periodic signal with amplitude of** $3.57\,\mathrm{deg}$ **and frequency** $20\,\mathrm{Hz}$**. The constraint on** $u_0 = 20\,[V]$**.**

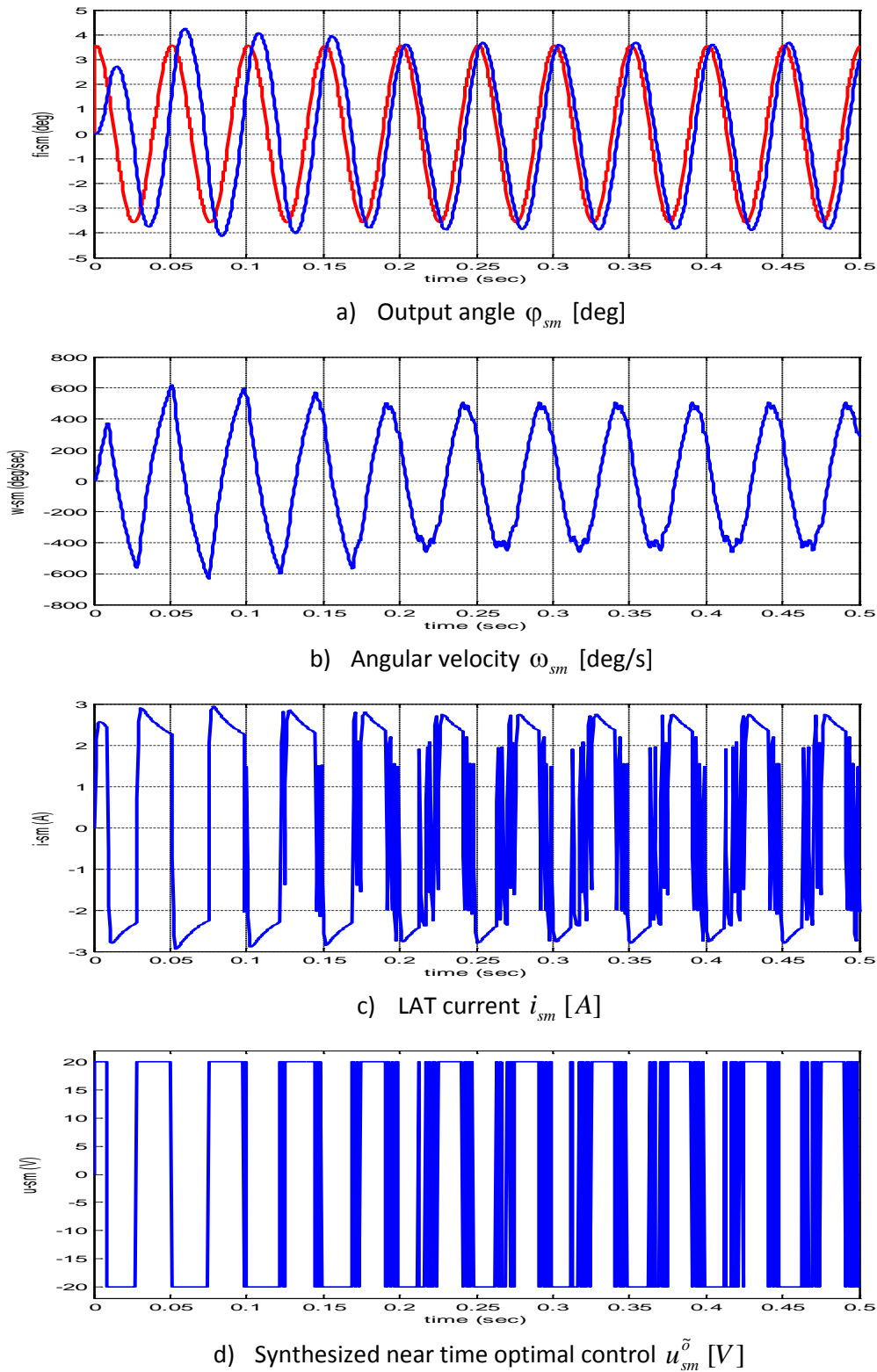

a)   Output angle $\varphi_{sm}$ [deg]

b)   Angular velocity $\omega_{sm}$ [deg/s]

c)   LAT current $i_{sm}$ [$A$]

d)   Synthesized near time optimal control $u_{sm}^{\tilde{o}}$ [$V$]

**Figure 69. Time-diagrams of the processes in the digital tracking control system working at sampling rate of** $1\,\mathrm{ms}$ **and controlling the linear model of the small mirror actuator with zero pivot stiffness in case the demand position on** $\varphi_{sm}$ **is periodic signal with amplitude of** $3.57\,\mathrm{deg}$ **and frequency** $20\,\mathrm{Hz}$ **. The constraint on** $u_0 = 20\,[V]$ **.**

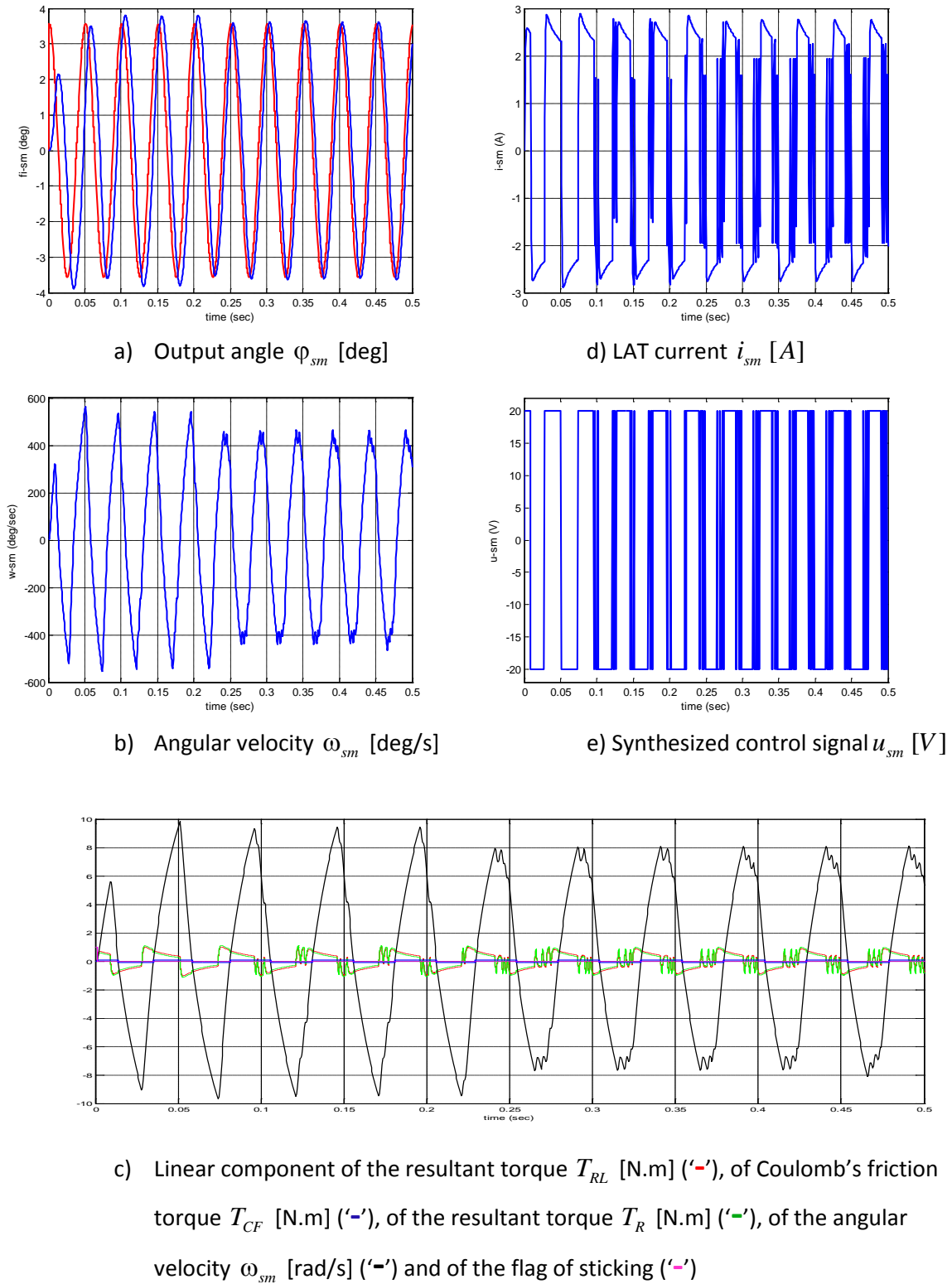

a)  Output angle $\varphi_{sm}$ [deg]

d) LAT current $i_{sm}$ [A]

b)  Angular velocity $\omega_{sm}$ [deg/s]

e) Synthesized control signal $u_{sm}$ [V]

c)  Linear component of the resultant torque $T_{RL}$ [N.m] ('–'), of Coulomb's friction

torque $T_{CF}$ [N.m] ('–'), of the resultant torque $T_R$ [N.m] ('–'), of the angular

velocity $\omega_{sm}$ [rad/s] ('–') and of the flag of sticking ('–')

**Figure 70. Time-diagrams of the processes in the digital tracking control system working at sampling rate of** $1\,\text{ms}$ **controlling the non-linear model of the small mirror actuator with zero pivot stiffness and Coulomb's friction model included in case the demand position on** $\varphi_{sm}$**, ('–') in a), is periodic signal with amplitude of** $3.57\,\text{deg}$ **and frequency** $20\,\text{Hz}$**. The constraint on** $u_0 = 20\,[V]$**.**

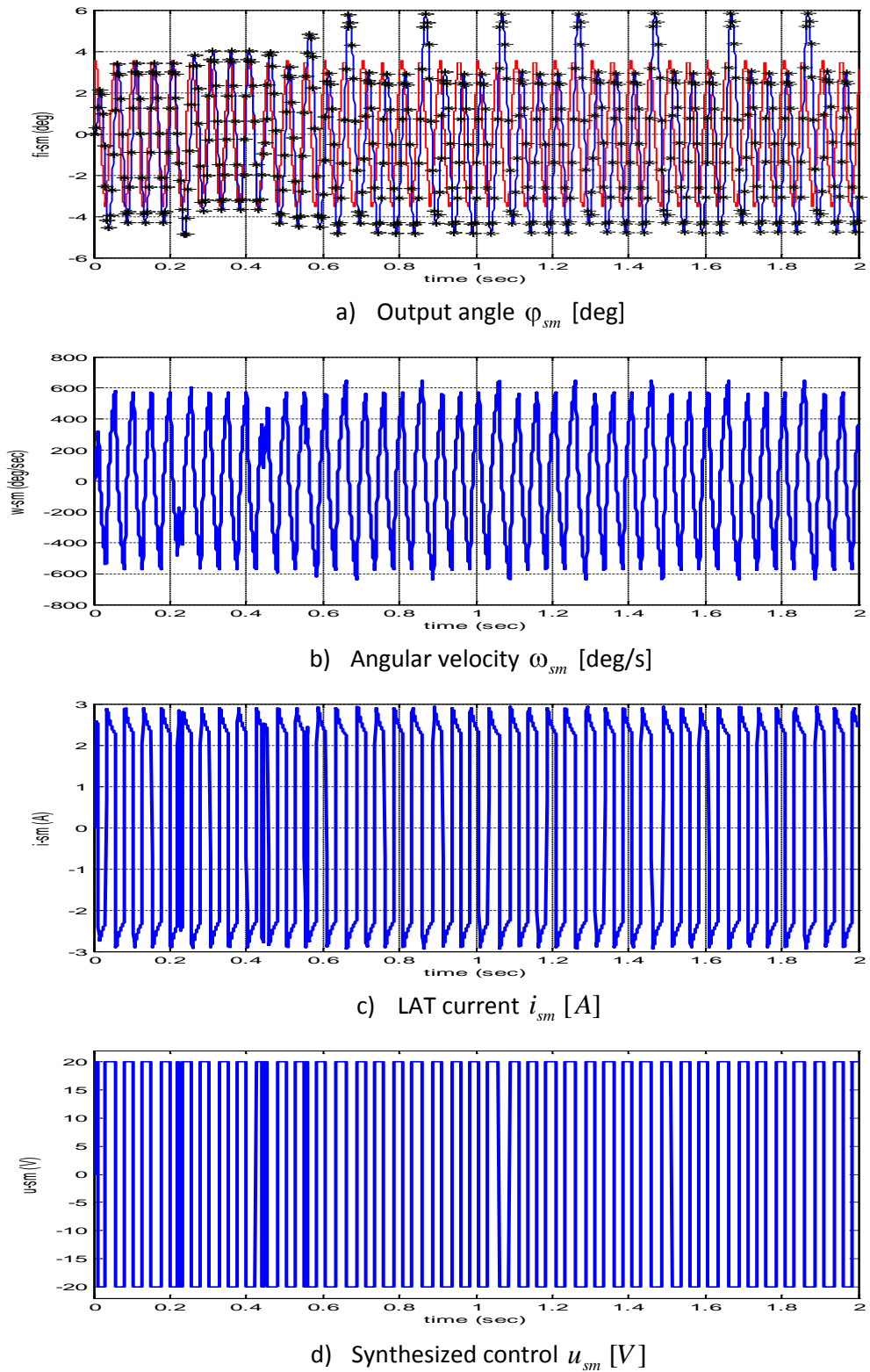

a)   Output angle $\varphi_{sm}$ [deg]

b)   Angular velocity $\omega_{sm}$ [deg/s]

c)   LAT current $i_{sm}$ [A]

d)   Synthesized control $u_{sm}$ [V]

**Figure 71. Time-diagrams of the processes in the digital tracking control system working at sampling rate of** $4\,\mathrm{ms}$ **controlling the non-linear model of the small mirror actuator with zero pivot stiffness and Coulomb's friction model included in case the demand position on** $\varphi_{sm}$ **, '–' in a), is periodic signal with amplitude of** $3.57\,\mathrm{deg}$ **and frequency** $20\,\mathrm{Hz}$ **. By '\*' in a) are presented also the samplings from the output angle at rate** $4\,\mathrm{ms}$ **. The constraint on** $u_0 = 20\,[V]$ **.**

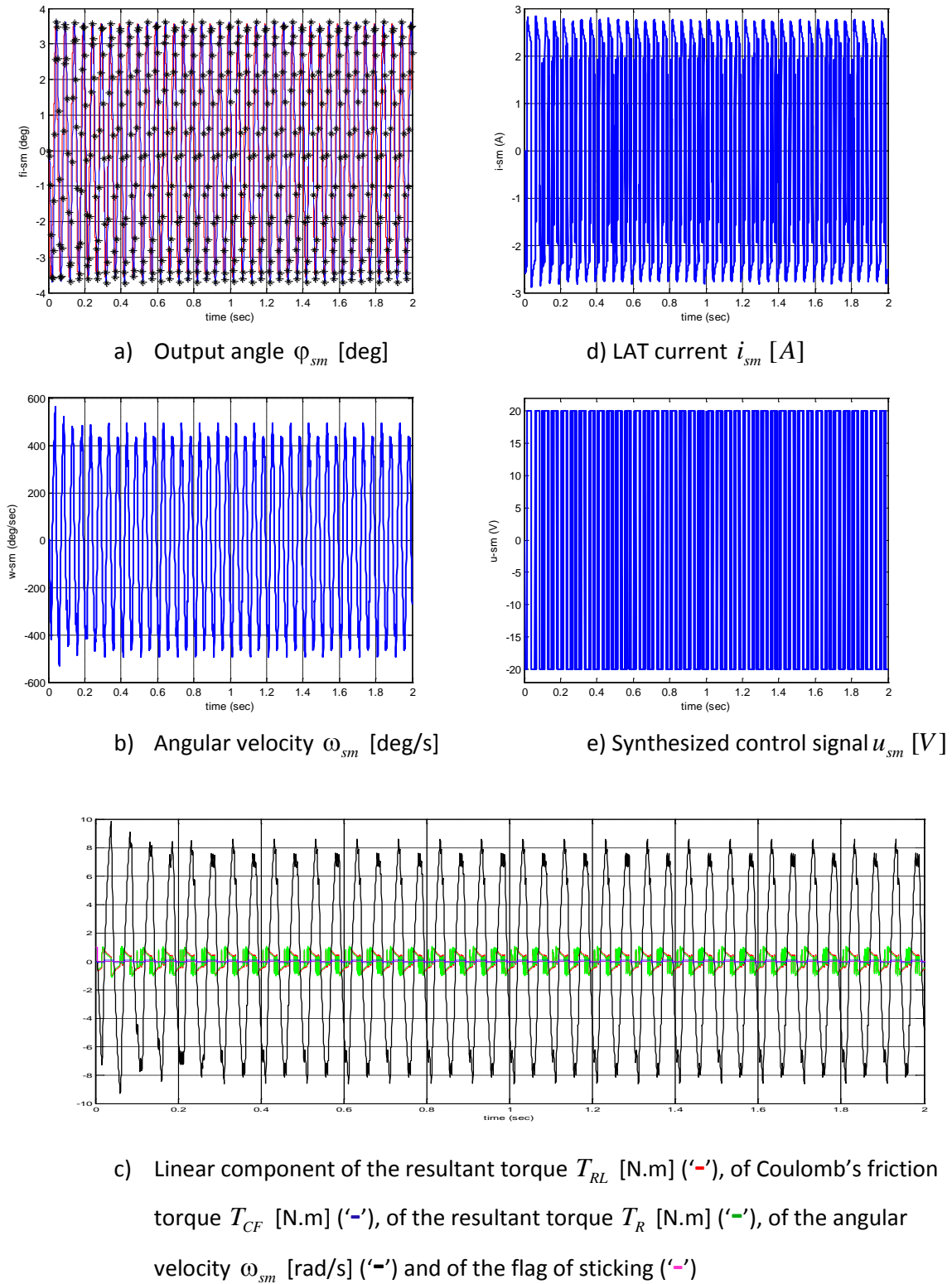

a)  Output angle $\varphi_{sm}$ [deg]

d)  LAT current $i_{sm}$ $[A]$

b)  Angular velocity $\omega_{sm}$ [deg/s]

e)  Synthesized control signal $u_{sm}$ $[V]$

c)  Linear component of the resultant torque $T_{RL}$ [N.m] ('–'), of Coulomb's friction

torque $T_{CF}$ [N.m] ('–'), of the resultant torque $T_R$ [N.m] ('–'), of the angular

velocity $\omega_{sm}$ [rad/s] ('–') and of the flag of sticking ('–')

**Figure 72. Time-diagrams of the processes in the digital tracking control system working simultaneously at two sampling rates of** $4\,\mathrm{ms}$ **and** $1\,\mathrm{ms}$ **controlling the non-linear model of the small mirror actuator with zero pivot stiffness and Coulomb's friction model included in case the demand position on** $\varphi_{sm}$**, '–' in a), is periodic signal with amplitude of** $3.57\,\mathrm{deg}$ **and frequency** $20\,\mathrm{Hz}$**. By '\*' in a) are presented also the samplings from the output angle at rate** $4\,\mathrm{ms}$**. The constraint on** $u_0 = 20\,[V]$**.**

# 6. Spatial modelling of the scanning system taking into account the dynamics of the actuators' control systems included

Here we show results on the spatial modeling of the scanning system taking into account the dynamic behavior of the synthesized digital control systems for controlling the non-linear models of the large and the small mirror actuators with Coulomb's friction model included on the basis of synthesizing a near time optimal control for each one linear model of the actuator. The way we model this behavior includes a replacement of the blocks Sine Wave and Sine Wave1 in Figure 2 by the outputs of the digital tracking control systems for the small and large mirror actuators respectively. The systems we choose are the tracking control system for the small mirror actuator with zero pivot stiffness of the mechanical subsystem working simultaneously at two sampling rates of $4\,\mathrm{ms}$ and $1\,\mathrm{ms}$ and the tracking control system for the large mirror actuator with zero pivot stiffness of the mechanical subsystem working at sampling rate of $4\,\mathrm{ms}$ and having constraints on control signal $u_0 = 20\,[V]$. The first five passes, $200\,\mathrm{ms}$ each, of the scanned area with scanned points ('o') for the mirror actuators and respective scanning in the vertical plane at range of 200 m in front of the aircraft are shown in Figure 73. The next group of five passes, from Pass 6 till Pass 10 respectively, is shown in Figure 74. The first five odd passes 1, 3, 5, 7 and 9 of the mirror actuators and scanned area are presented together in Figure 75 and Figure 77 respectively. The even passes 2, 4, 6, 8 and 10 are shown in Figure 76 and Figure 78. The positions of each one mirror actuator and respective scanned points in the vertical plane at range of 200 m in front of the aircraft in the first time interval 0-2 seconds at sampling rate of $4\,\mathrm{ms}$ are presented in Table 9.

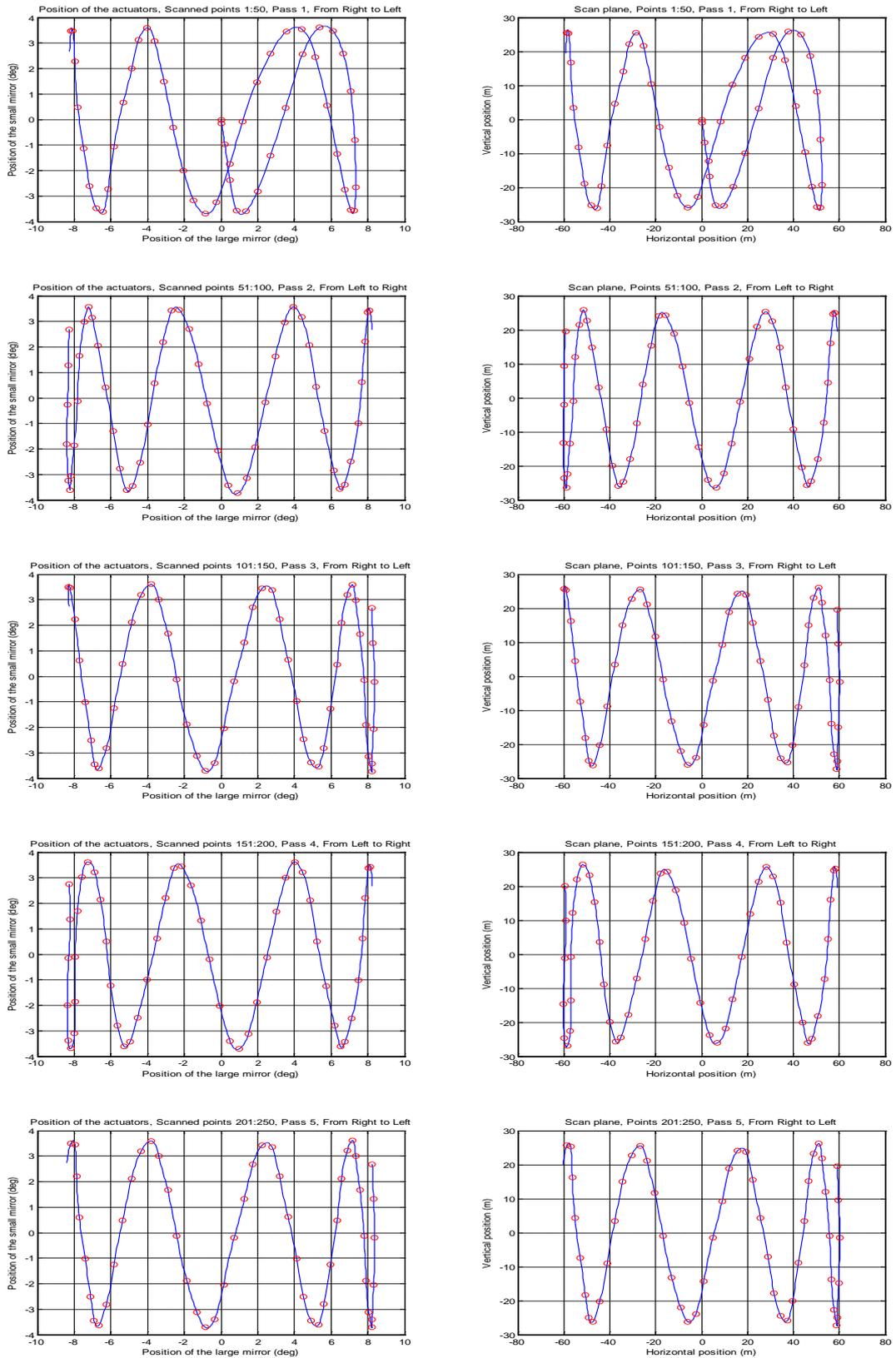

**Figure 73. First five passes of the scanned area with scanned points ('o') for the mirror actuators and respective scanning in the vertical plane at range of 200 m in front of the aircraft.**

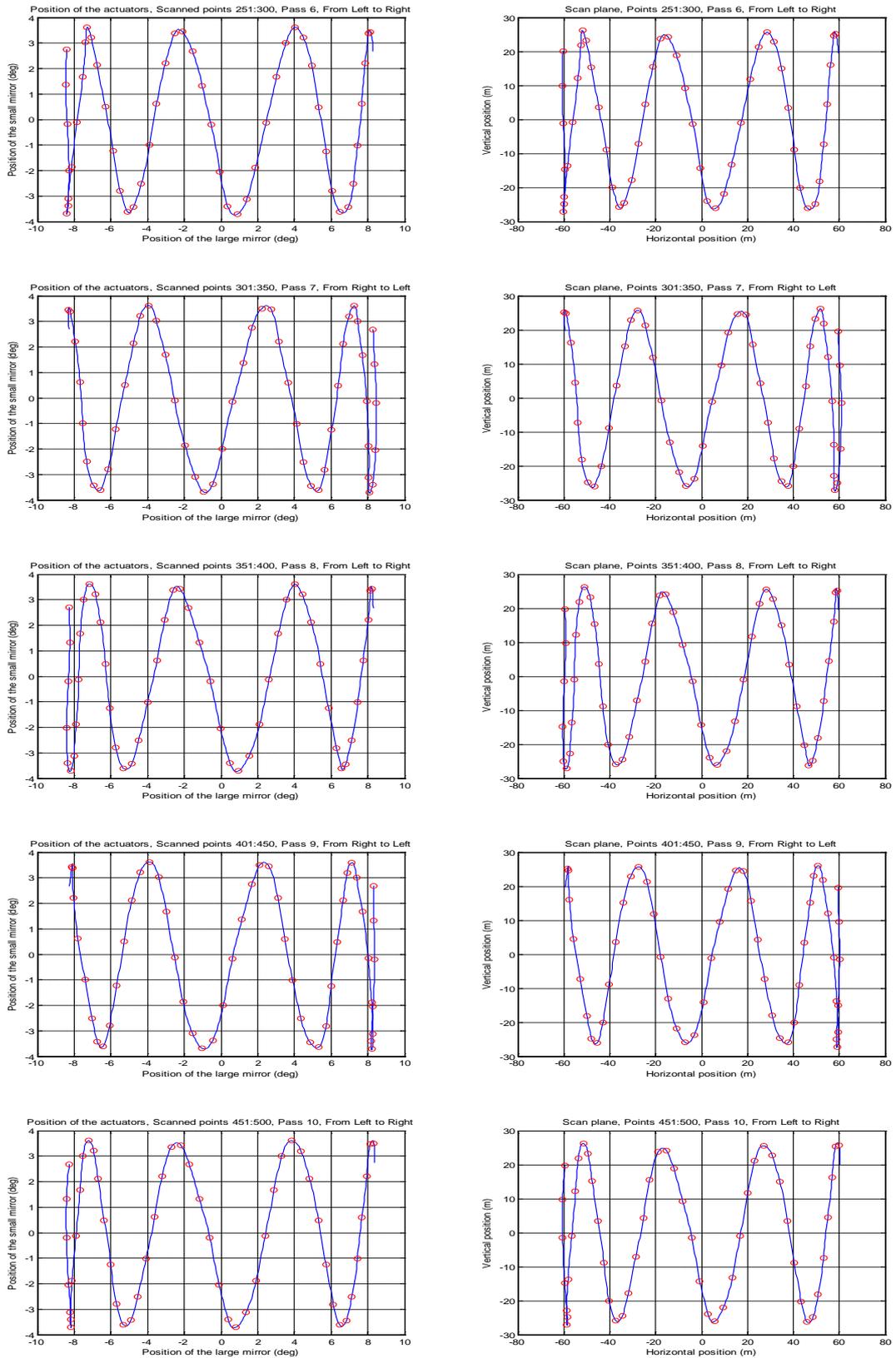

**Figure 74. Second five passes of the scanned area with scanned points ('o') for the mirror actuators and respective scanning in the vertical plane at range of 200 m in front of the aircraft.**

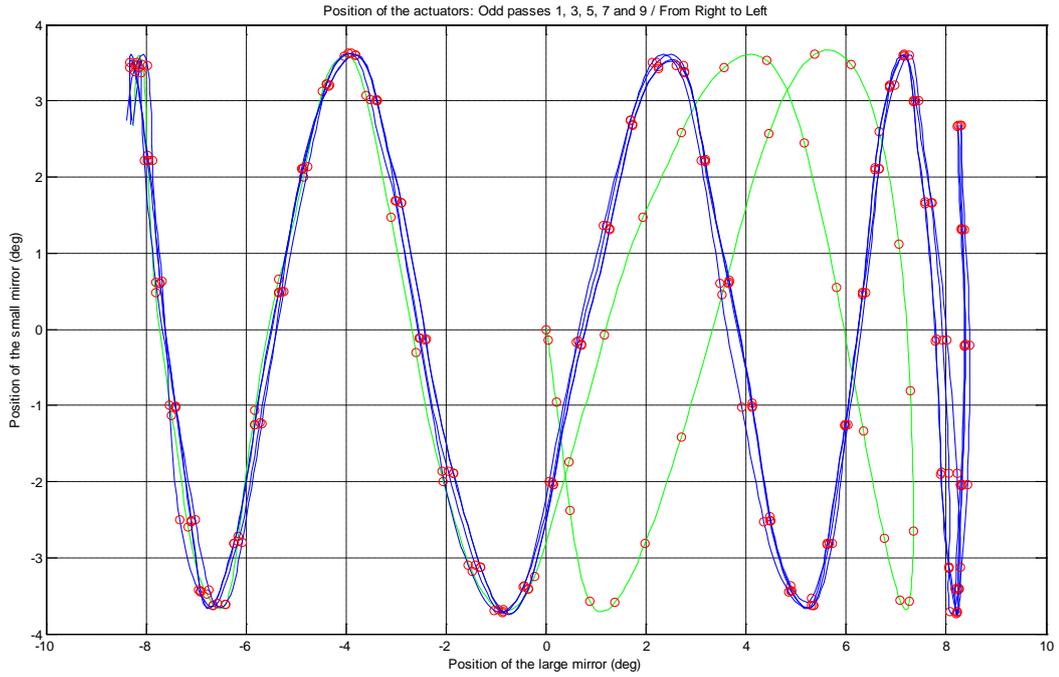

**Figure 75. Positions of the actuators at odd passes 1, 3, 5, 7 and 9. The first pass is shown by ('–'). The scanned points are presented by ('o').**

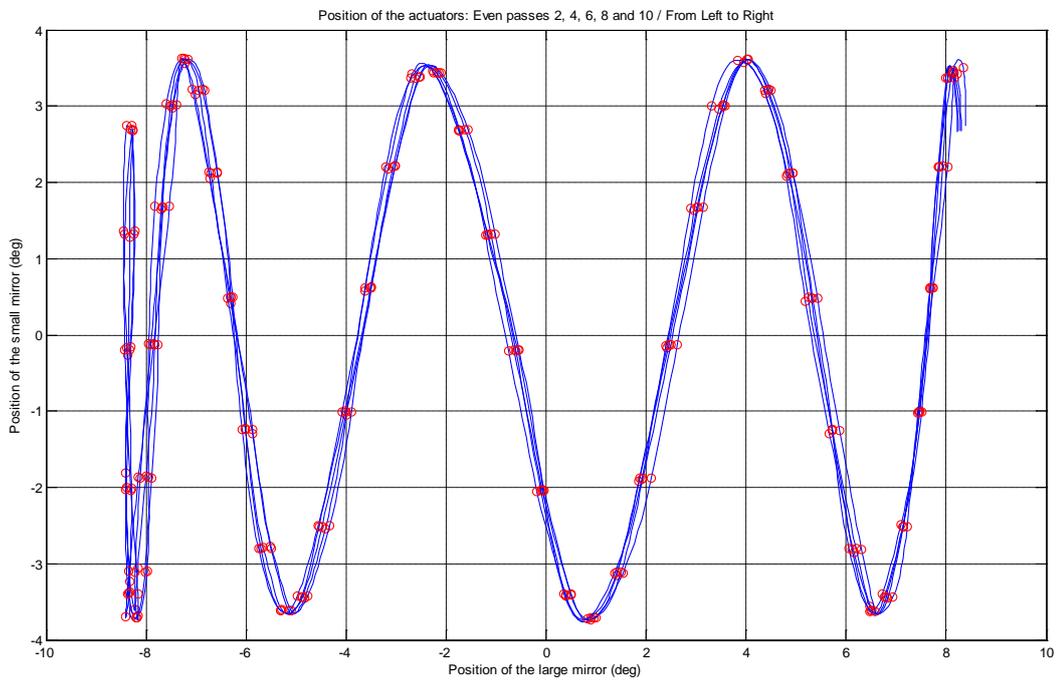

**Figure 76. Positions of the actuators at even passes 2, 4, 6, 8 and 10. The scanned points are shown by ('o').**

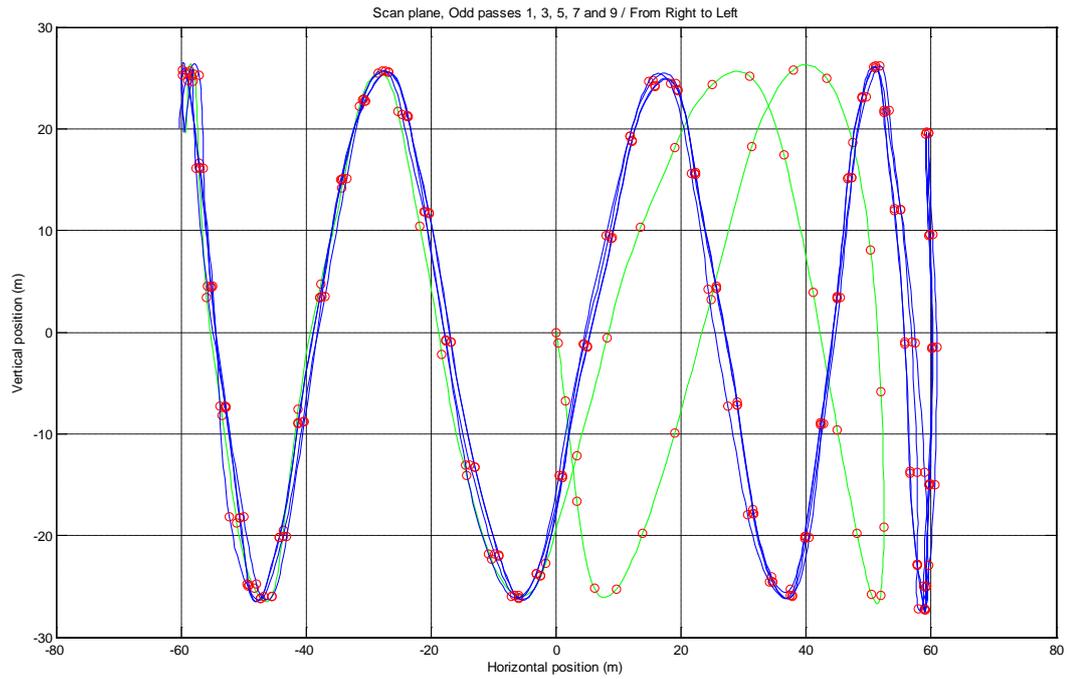

**Figure 77. Scanning in the vertical plane at range of 200 m at odd passes 1, 3, 5, 7 and 9. The scanned points are presented by ('o'), the first pass is shown by ('–').**

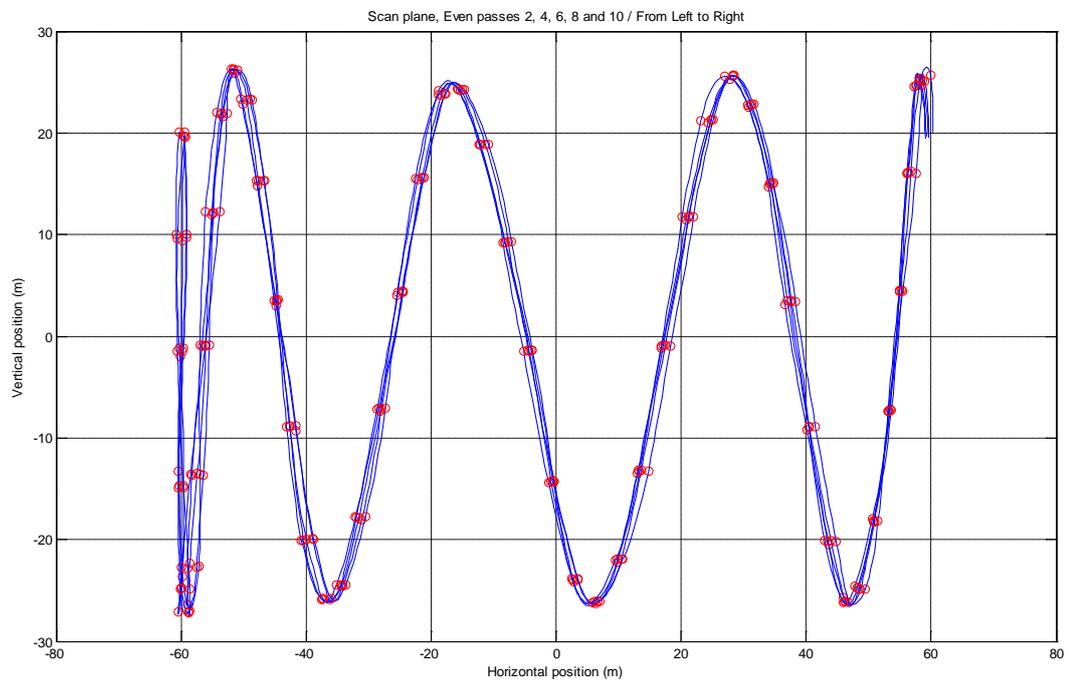

**Figure 78. Scanning in the vertical plane at range of 200 m at even passes 2, 4, 6, 8 and 10. The scanned points are shown by ('o').**

**Table 9. Positions of the mirror actuators and respective scanned points in the vertical plane at range of 200 m in front of the aircraft in the first time interval 0-2 seconds at sampling rate of 4 ms.**

| Time (s) | Large mirror Position (deg) | Small mirror Position (deg) | Scan plane at range 200 m. Horizontal position (m) | Scan plane at range 200 m. Vertical position (m) | Time (s) | Large mirror Position (deg) | Small mirror Position (deg) | Scan plane at range 200 m. Horizontal position (m) | Scan plane at range 200 m. Vertical position (m) |
|---|---|---|---|---|---|---|---|---|---|
| 0.000 | 0.000 | 0.000 | 0.000 | 0.000 | 1.000 | -8.394 | 2.740 | -60.340 | 20.070 |
| 0.004 | 0.000 | 0.000 | 0.000 | 0.000 | 1.004 | -8.456 | 1.364 | -60.812 | 9.974 |
| 0.008 | 0.043 | -0.147 | 0.301 | -1.025 | 1.008 | -8.381 | -0.165 | -60.240 | -1.204 |
| 0.012 | 0.203 | -0.962 | 1.419 | -6.731 | 1.012 | -8.291 | -2.008 | -59.557 | -14.675 |
| 0.016 | 0.480 | -2.379 | 3.355 | -16.672 | 1.016 | -8.341 | -3.378 | -59.931 | -24.769 |
| 0.020 | 0.872 | -3.578 | 6.089 | -25.160 | 1.020 | -8.405 | -3.689 | -60.424 | -27.092 |
| 0.024 | 1.375 | -3.592 | 9.606 | -25.277 | 1.024 | -8.337 | -3.102 | -59.906 | -22.725 |
| 0.028 | 1.986 | -2.808 | 13.890 | -19.740 | 1.028 | -8.150 | -1.866 | -58.482 | -13.609 |
| 0.032 | 2.704 | -1.415 | 18.934 | -9.949 | 1.032 | -7.846 | -0.117 | -56.186 | -0.847 |
| 0.036 | 3.525 | 0.457 | 24.733 | 3.219 | 1.036 | -7.533 | 1.684 | -53.835 | 12.205 |
| 0.040 | 4.447 | 2.571 | 31.295 | 18.245 | 1.040 | -7.379 | 3.020 | -52.689 | 21.917 |
| 0.044 | 5.362 | 3.607 | 37.877 | 25.805 | 1.044 | -7.283 | 3.621 | -51.973 | 26.300 |
| 0.048 | 6.103 | 3.478 | 43.263 | 25.004 | 1.048 | -7.075 | 3.219 | -50.422 | 23.308 |
| 0.052 | 6.667 | 2.593 | 47.406 | 18.685 | 1.052 | -6.751 | 2.129 | -48.020 | 15.338 |
| 0.056 | 7.060 | 1.122 | 50.309 | 8.094 | 1.056 | -6.313 | 0.496 | -44.801 | 3.554 |
| 0.060 | 7.284 | -0.813 | 51.979 | -5.873 | 1.060 | -5.870 | -1.236 | -41.562 | -8.834 |
| 0.064 | 7.345 | -2.646 | 52.435 | -19.179 | 1.064 | -5.485 | -2.795 | -38.766 | -19.972 |
| 0.068 | 7.270 | -3.569 | 51.870 | -25.915 | 1.068 | -5.099 | -3.609 | -35.979 | -25.777 |
| 0.072 | 7.075 | -3.553 | 50.421 | -25.750 | 1.072 | -4.770 | -3.428 | -33.614 | -24.421 |
| 0.076 | 6.764 | -2.748 | 48.116 | -19.824 | 1.076 | -4.335 | -2.506 | -30.494 | -17.767 |
| 0.080 | 6.339 | -1.340 | 44.991 | -9.613 | 1.080 | -3.893 | -1.005 | -27.348 | -7.094 |
| 0.084 | 5.804 | 0.544 | 41.082 | 3.886 | 1.084 | -3.510 | 0.623 | -24.629 | 4.387 |
| 0.088 | 5.161 | 2.452 | 36.423 | 17.470 | 1.088 | -3.022 | 2.213 | -21.173 | 15.593 |
| 0.092 | 4.412 | 3.536 | 31.051 | 25.144 | 1.092 | -2.529 | 3.377 | -17.699 | 23.812 |
| 0.096 | 3.562 | 3.443 | 24.993 | 24.380 | 1.096 | -2.095 | 3.439 | -14.653 | 24.229 |
| 0.100 | 2.715 | 2.585 | 19.013 | 18.203 | 1.100 | -1.558 | 2.688 | -10.885 | 18.877 |
| 0.104 | 1.937 | 1.465 | 13.544 | 10.276 | 1.104 | -1.017 | 1.322 | -7.101 | 9.257 |
| 0.108 | 1.168 | -0.078 | 8.158 | -0.548 | 1.108 | -0.537 | -0.198 | -3.747 | -1.388 |
| 0.112 | 0.465 | -1.739 | 3.248 | -12.178 | 1.112 | -0.058 | -2.036 | -0.405 | -14.256 |
| 0.116 | -0.231 | -3.242 | -1.609 | -22.766 | 1.116 | 0.361 | -3.400 | 2.522 | -23.886 |
| 0.120 | -0.862 | -3.683 | -6.017 | -25.906 | 1.120 | 0.885 | -3.707 | 6.181 | -26.079 |
| 0.124 | -1.487 | -3.178 | -10.394 | -22.340 | 1.124 | 1.413 | -3.117 | 9.870 | -21.910 |
| 0.128 | -2.050 | -2.007 | -14.339 | -14.089 | 1.128 | 1.880 | -1.880 | 13.141 | -13.189 |
| 0.132 | -2.610 | -0.309 | -18.271 | -2.169 | 1.132 | 2.450 | -0.129 | 17.146 | -0.906 |
| 0.136 | -3.108 | 1.474 | -21.784 | 10.378 | 1.136 | 3.023 | 1.673 | 21.182 | 11.773 |
| 0.140 | -3.604 | 3.074 | -25.296 | 21.745 | 1.140 | 3.534 | 3.010 | 24.799 | 21.284 |
| 0.144 | -4.041 | 3.589 | -28.399 | 25.481 | 1.144 | 4.043 | 3.612 | 28.416 | 25.642 |
| 0.148 | -4.477 | 3.130 | -31.512 | 22.245 | 1.148 | 4.492 | 3.210 | 31.621 | 22.817 |
| 0.152 | -4.855 | 1.996 | -34.222 | 14.180 | 1.152 | 4.940 | 2.121 | 34.837 | 15.077 |
| 0.156 | -5.338 | 0.658 | -37.706 | 4.682 | 1.156 | 5.330 | 0.488 | 37.648 | 3.469 |
| 0.160 | -5.826 | -1.058 | -41.247 | -7.556 | 1.160 | 5.721 | -1.245 | 40.478 | -8.883 |
| 0.164 | -6.151 | -2.718 | -43.611 | -19.505 | 1.164 | 6.054 | -2.803 | 42.907 | -20.111 |
| 0.168 | -6.413 | -3.611 | -45.535 | -26.031 | 1.168 | 6.494 | -3.617 | 46.132 | -26.093 |
| 0.172 | -6.784 | -3.483 | -48.266 | -25.173 | 1.172 | 6.940 | -3.436 | 49.422 | -24.864 |
| 0.176 | -7.162 | -2.599 | -51.070 | -18.802 | 1.176 | 7.223 | -2.514 | 51.523 | -18.194 |
| 0.180 | -7.484 | -1.128 | -53.469 | -8.166 | 1.180 | 7.445 | -1.013 | 53.181 | -7.328 |
| 0.184 | -7.808 | 0.476 | -55.899 | 3.455 | 1.184 | 7.672 | 0.615 | 54.882 | 4.460 |
| 0.188 | -7.972 | 2.288 | -57.136 | 16.669 | 1.188 | 7.847 | 2.206 | 56.191 | 16.051 |
| 0.192 | -8.078 | 3.475 | -57.940 | 25.419 | 1.192 | 8.027 | 3.369 | 57.551 | 24.625 |
| 0.196 | -8.192 | 3.476 | -58.805 | 25.453 | 1.196 | 8.155 | 3.432 | 58.523 | 25.120 |
| 0.200 | -8.256 | 2.682 | -59.290 | 19.613 | 1.200 | 8.291 | 2.680 | 59.552 | 19.611 |
| 0.204 | -8.329 | 1.282 | -59.843 | 9.364 | 1.204 | 8.375 | 1.315 | 60.197 | 9.606 |
| 0.208 | -8.360 | -0.265 | -60.077 | -1.934 | 1.208 | 8.469 | -0.206 | 60.907 | -1.505 |
| 0.212 | -8.402 | -1.816 | -60.396 | -13.278 | 1.212 | 8.422 | -2.043 | 60.552 | -14.950 |
| 0.216 | -8.316 | -3.231 | -59.745 | -23.676 | 1.216 | 8.255 | -3.408 | 59.283 | -24.966 |
| 0.220 | -8.216 | -3.605 | -58.985 | -26.418 | 1.220 | 8.076 | -3.715 | 57.924 | -27.191 |
| 0.224 | -8.166 | -3.058 | -58.604 | -22.363 | 1.224 | 8.051 | -3.125 | 57.733 | -22.831 |
| 0.228 | -8.002 | -1.854 | -57.364 | -13.501 | 1.228 | 8.054 | -1.887 | 57.760 | -13.752 |
| 0.232 | -7.826 | -0.130 | -56.033 | -0.943 | 1.232 | 7.938 | -0.137 | 56.881 | -0.992 |
| 0.236 | -7.701 | 1.651 | -55.096 | 11.984 | 1.236 | 7.703 | 1.665 | 55.114 | 12.089 |

| | | | | | | | | |
|---|---|---|---|---|---|---|---|---|
| 0.240 | -7.464 | 2.971 | -53.324 | 21.578 | 1.240 | 7.458 | 3.002 | 53.275 | 21.803 |
| 0.244 | -7.217 | 3.560 | -51.479 | 25.835 | 1.244 | 7.266 | 3.604 | 51.842 | 26.171 |
| 0.248 | -7.023 | 3.150 | -50.040 | 22.796 | 1.248 | 6.964 | 3.202 | 49.597 | 23.165 |
| 0.252 | -6.720 | 2.055 | -47.792 | 14.795 | 1.252 | 6.652 | 2.113 | 47.296 | 15.209 |
| 0.256 | -6.302 | 0.417 | -44.722 | 2.985 | 1.256 | 6.396 | 0.480 | 45.412 | 3.442 |
| 0.260 | -5.879 | -1.296 | -41.627 | -9.261 | 1.260 | 6.032 | -1.252 | 42.742 | -8.957 |
| 0.264 | -5.513 | -2.779 | -38.969 | -19.860 | 1.264 | 5.659 | -2.811 | 40.031 | -20.108 |
| 0.268 | -5.146 | -3.617 | -36.315 | -25.838 | 1.268 | 5.344 | -3.625 | 37.744 | -25.929 |
| 0.272 | -4.835 | -3.451 | -34.080 | -24.596 | 1.272 | 4.921 | -3.443 | 34.696 | -24.552 |
| 0.276 | -4.417 | -2.540 | -31.085 | -18.021 | 1.276 | 4.492 | -2.521 | 31.619 | -17.891 |
| 0.280 | -3.993 | -1.048 | -28.060 | -7.405 | 1.280 | 4.121 | -1.020 | 28.970 | -7.210 |
| 0.284 | -3.627 | 0.572 | -25.457 | 4.033 | 1.284 | 3.644 | 0.608 | 25.581 | 4.283 |
| 0.288 | -3.155 | 2.181 | -22.115 | 15.369 | 1.288 | 3.163 | 2.222 | 22.172 | 15.661 |
| 0.292 | -2.678 | 3.418 | -18.750 | 24.119 | 1.292 | 2.741 | 3.464 | 19.194 | 24.449 |
| 0.296 | -2.260 | 3.459 | -15.812 | 24.378 | 1.296 | 2.214 | 3.508 | 15.490 | 24.721 |
| 0.300 | -1.738 | 2.692 | -12.148 | 18.915 | 1.300 | 1.684 | 2.743 | 11.772 | 19.272 |
| 0.304 | -1.212 | 1.315 | -8.466 | 9.207 | 1.304 | 1.215 | 1.367 | 8.485 | 9.575 |
| 0.308 | -0.746 | -0.215 | -5.212 | -1.507 | 1.308 | 0.642 | -0.162 | 4.481 | -1.130 |
| 0.312 | -0.177 | -2.060 | -1.239 | -14.429 | 1.312 | 0.067 | -2.005 | 0.464 | -14.042 |
| 0.316 | 0.394 | -3.430 | 2.751 | -24.102 | 1.316 | -0.447 | -3.375 | -3.122 | -23.707 |
| 0.320 | 0.904 | -3.742 | 6.312 | -26.327 | 1.320 | -0.958 | -3.686 | -6.694 | -25.927 |
| 0.324 | 1.412 | -3.155 | 9.863 | -22.176 | 1.324 | -1.410 | -3.098 | -9.850 | -21.774 |
| 0.328 | 1.859 | -1.919 | 12.998 | -13.469 | 1.328 | -1.965 | -1.862 | -13.738 | -13.070 |
| 0.332 | 2.411 | -0.171 | 16.870 | -1.198 | 1.332 | -2.523 | -0.113 | -17.656 | -0.795 |
| 0.336 | 2.965 | 1.629 | 20.776 | 11.467 | 1.336 | -3.019 | 1.687 | -21.156 | 11.875 |
| 0.340 | 3.459 | 2.966 | 24.263 | 20.963 | 1.340 | -3.514 | 3.023 | -24.656 | 21.379 |
| 0.344 | 3.950 | 3.567 | 27.753 | 25.308 | 1.344 | -3.949 | 3.625 | -27.745 | 25.722 |
| 0.348 | 4.382 | 3.164 | 30.833 | 22.477 | 1.348 | -4.384 | 3.222 | -30.845 | 22.892 |
| 0.352 | 4.814 | 2.075 | 33.926 | 14.738 | 1.352 | -4.760 | 2.133 | -33.542 | 15.147 |
| 0.356 | 5.187 | 0.441 | 36.615 | 3.137 | 1.356 | -5.242 | 0.499 | -37.013 | 3.551 |
| 0.360 | 5.666 | -1.291 | 40.084 | -9.211 | 1.360 | -5.729 | -1.233 | -40.539 | -8.801 |
| 0.364 | 6.150 | -2.850 | 43.610 | -20.464 | 1.364 | -6.157 | -2.792 | -43.655 | -20.045 |
| 0.368 | 6.471 | -3.557 | 45.957 | -25.648 | 1.368 | -6.584 | -3.606 | -46.789 | -26.030 |
| 0.372 | 6.729 | -3.394 | 47.861 | -24.511 | 1.372 | -6.953 | -3.425 | -49.516 | -24.785 |
| 0.376 | 7.096 | -2.485 | 50.579 | -17.968 | 1.376 | -7.323 | -2.502 | -52.270 | -18.130 |
| 0.380 | 7.471 | -0.995 | 53.370 | -7.206 | 1.380 | -7.533 | -1.002 | -53.834 | -7.255 |
| 0.384 | 7.684 | 0.623 | 54.969 | 4.521 | 1.384 | -7.683 | 0.626 | -54.963 | 4.539 |
| 0.388 | 7.838 | 2.207 | 56.131 | 16.060 | 1.388 | -7.945 | 2.217 | -56.934 | 16.146 |
| 0.392 | 8.000 | 3.365 | 57.346 | 24.589 | 1.392 | -8.217 | 3.380 | -58.991 | 24.753 |
| 0.396 | 8.109 | 3.424 | 58.175 | 25.050 | 1.396 | -8.330 | 3.443 | -59.851 | 25.246 |
| 0.400 | 8.227 | 2.670 | 59.065 | 19.520 | 1.400 | -8.294 | 2.691 | -59.579 | 19.691 |
| 0.404 | 8.294 | 1.302 | 59.574 | 9.506 | 1.404 | -8.242 | 1.326 | -59.180 | 9.672 |
| 0.408 | 8.369 | -0.220 | 60.150 | -1.608 | 1.408 | -8.319 | -0.195 | -59.767 | -1.424 |
| 0.412 | 8.310 | -2.059 | 59.696 | -15.046 | 1.412 | -8.411 | -2.032 | -60.467 | -14.868 |
| 0.416 | 8.235 | -3.424 | 59.127 | -25.083 | 1.416 | -8.364 | -3.397 | -60.106 | -24.914 |
| 0.420 | 8.203 | -3.732 | 58.886 | -27.355 | 1.420 | -8.196 | -3.704 | -58.833 | -27.144 |
| 0.424 | 8.057 | -3.143 | 57.776 | -22.963 | 1.424 | -8.016 | -3.114 | -57.470 | -22.743 |
| 0.428 | 7.897 | -1.905 | 56.572 | -13.863 | 1.428 | -7.888 | -1.876 | -56.503 | -13.651 |
| 0.432 | 7.789 | -0.155 | 55.758 | -1.125 | 1.432 | -7.752 | -0.126 | -55.483 | -0.912 |
| 0.436 | 7.569 | 1.646 | 54.103 | 11.937 | 1.436 | -7.668 | 1.676 | -54.847 | 12.163 |
| 0.440 | 7.337 | 2.983 | 52.372 | 21.641 | 1.440 | -7.470 | 3.013 | -53.368 | 21.885 |
| 0.444 | 7.159 | 3.585 | 51.044 | 26.005 | 1.444 | -7.157 | 3.615 | -51.029 | 26.224 |
| 0.448 | 6.870 | 3.183 | 48.901 | 23.006 | 1.448 | -6.834 | 3.213 | -48.638 | 23.218 |
| 0.452 | 6.572 | 2.094 | 46.700 | 15.060 | 1.452 | -6.567 | 2.124 | -46.667 | 15.276 |
| 0.456 | 6.328 | 0.461 | 44.910 | 3.302 | 1.456 | -6.296 | 0.491 | -44.677 | 3.517 |
| 0.460 | 5.976 | -1.271 | 42.333 | -9.092 | 1.460 | -6.079 | -1.241 | -43.090 | -8.883 |
| 0.464 | 5.615 | -2.830 | 39.712 | -20.241 | 1.464 | -5.753 | -2.800 | -40.712 | -20.043 |
| 0.468 | 5.311 | -3.537 | 37.511 | -25.288 | 1.468 | -5.314 | -3.614 | -37.527 | -25.846 |
| 0.472 | 4.900 | -3.373 | 34.546 | -24.046 | 1.472 | -4.868 | -3.433 | -34.319 | -24.467 |
| 0.476 | 4.482 | -2.465 | 31.550 | -17.491 | 1.476 | -4.482 | -2.510 | -31.547 | -17.813 |
| 0.480 | 4.122 | -0.975 | 28.978 | -6.892 | 1.480 | -3.990 | -1.009 | -28.035 | -7.129 |
| 0.484 | 3.656 | 0.644 | 25.664 | 4.538 | 1.484 | -3.493 | 0.618 | -24.510 | 4.356 |
| 0.488 | 3.185 | 2.228 | 22.328 | 15.703 | 1.488 | -3.057 | 2.209 | -21.420 | 15.565 |
| 0.492 | 2.773 | 3.386 | 19.420 | 23.894 | 1.492 | -2.620 | 3.373 | -18.344 | 23.789 |
| 0.496 | 2.256 | 3.444 | 15.785 | 24.274 | 1.496 | -2.242 | 3.435 | -15.685 | 24.208 |
| 0.500 | 1.736 | 2.690 | 12.133 | 18.900 | 1.500 | -1.758 | 2.684 | -12.292 | 18.855 |
| 0.504 | 1.276 | 1.323 | 8.911 | 9.263 | 1.504 | -1.166 | 1.318 | -8.142 | 9.230 |
| 0.508 | 0.712 | -0.200 | 4.971 | -1.398 | 1.508 | -0.571 | -0.203 | -3.986 | -1.417 |
| 0.512 | 0.145 | -2.038 | 1.015 | -14.275 | 1.512 | -0.038 | -2.040 | -0.268 | -14.286 |
| 0.516 | -0.360 | -3.404 | -2.510 | -23.912 | 1.516 | 0.491 | -3.404 | 3.432 | -23.918 |
| 0.520 | -0.862 | -3.712 | -6.022 | -26.109 | 1.520 | 0.961 | -3.711 | 6.710 | -26.111 |
| 0.524 | -1.305 | -3.122 | -9.120 | -21.941 | 1.524 | 1.533 | -3.122 | 10.715 | -21.944 |
| 0.528 | -1.852 | -1.885 | -12.950 | -13.225 | 1.528 | 2.108 | -1.884 | 14.746 | -13.225 |
| 0.532 | -2.402 | -0.135 | -16.811 | -0.945 | 1.532 | 2.622 | -0.133 | 18.354 | -0.935 |
| 0.536 | -2.891 | 1.667 | -20.254 | 11.726 | 1.536 | 3.133 | 1.668 | 21.959 | 11.748 |
| 0.540 | -3.379 | 3.004 | -23.697 | 21.227 | 1.540 | 3.584 | 3.006 | 25.151 | 21.259 |
| 0.544 | -3.806 | 3.606 | -26.731 | 25.569 | 1.544 | 4.034 | 3.608 | 28.350 | 25.611 |
| 0.548 | -4.339 | 3.204 | -30.523 | 22.753 | 1.548 | 4.426 | 3.206 | 31.145 | 22.779 |
| 0.552 | -4.874 | 2.114 | -34.360 | 15.026 | 1.552 | 4.923 | 2.116 | 34.708 | 15.046 |
| 0.556 | -5.349 | 0.481 | -37.783 | 3.424 | 1.556 | 5.424 | 0.483 | 38.324 | 3.441 |
| 0.560 | -5.823 | -1.251 | -41.218 | -8.936 | 1.560 | 5.865 | -1.249 | 41.528 | -8.922 |
| 0.564 | -6.237 | -2.810 | -44.243 | -20.186 | 1.564 | 6.306 | -2.808 | 44.749 | -20.180 |
| 0.568 | -6.651 | -3.624 | -47.289 | -26.174 | 1.568 | 6.584 | -3.621 | 46.792 | -26.143 |
| 0.572 | -6.904 | -3.442 | -49.153 | -24.904 | 1.572 | 6.802 | -3.440 | 48.396 | -24.865 |
| 0.576 | -7.096 | -2.520 | -50.581 | -18.221 | 1.576 | 7.129 | -2.518 | 50.820 | -18.209 |
| 0.580 | -7.399 | -1.019 | -52.835 | -7.373 | 1.580 | 7.464 | -1.017 | 53.322 | -7.360 |
| 0.584 | -7.711 | 0.608 | -55.170 | 4.414 | 1.584 | 7.744 | 0.611 | 55.420 | 4.432 |
| 0.588 | -7.967 | 2.223 | -57.103 | 16.193 | 1.588 | 8.027 | 2.202 | 57.555 | 16.049 |
| 0.592 | -8.228 | 3.465 | -59.076 | 25.382 | 1.592 | 8.152 | 3.365 | 58.497 | 24.625 |
| 0.596 | -8.330 | 3.508 | -59.854 | 25.733 | 1.596 | 8.219 | 3.428 | 59.010 | 25.106 |
| 0.600 | -8.286 | 2.744 | -59.517 | 20.078 | 1.600 | 8.296 | 2.676 | 59.591 | 19.581 |
| 0.604 | -8.226 | 1.368 | -59.059 | 9.981 | 1.604 | 8.328 | 1.311 | 59.837 | 9.571 |
| 0.608 | -8.297 | -0.161 | -59.601 | -1.172 | 1.608 | 8.371 | -0.210 | 60.166 | -1.534 |
| 0.612 | -8.383 | -2.004 | -60.257 | -14.658 | 1.612 | 8.287 | -2.047 | 59.523 | -14.959 |
| 0.616 | -8.332 | -3.374 | -59.862 | -24.735 | 1.616 | 8.188 | -3.412 | 58.771 | -24.979 |
| 0.620 | -8.160 | -3.685 | -58.558 | -26.992 | 1.620 | 8.230 | -3.719 | 59.089 | -27.265 |
| 0.624 | -7.976 | -3.097 | -57.165 | -22.611 | 1.624 | 8.288 | -3.129 | 59.531 | -22.917 |
| 0.628 | -7.946 | -1.861 | -56.941 | -13.550 | 1.628 | 8.215 | -1.891 | 58.976 | -13.805 |
| 0.632 | -7.946 | -0.112 | -56.945 | -0.817 | 1.632 | 8.022 | -0.141 | 57.519 | -1.024 |
| 0.636 | -7.827 | 1.688 | -56.047 | 12.270 | 1.636 | 7.714 | 1.661 | 55.192 | 12.059 |
| 0.640 | -7.590 | 3.024 | -54.265 | 21.991 | 1.640 | 7.396 | 2.998 | 52.813 | 21.760 |
| 0.644 | -7.238 | 3.626 | -51.632 | 26.319 | 1.644 | 7.134 | 3.600 | 50.858 | 26.109 |
| 0.648 | -6.877 | 3.223 | -48.957 | 23.299 | 1.648 | 6.867 | 3.198 | 48.883 | 23.115 |
| 0.652 | -6.573 | 2.133 | -46.713 | 15.346 | 1.652 | 6.655 | 2.109 | 47.314 | 15.179 |
| 0.656 | -6.266 | 0.500 | -44.460 | 3.583 | 1.656 | 6.333 | 0.476 | 44.947 | 3.411 |
| 0.660 | -6.015 | -1.232 | -42.618 | -8.813 | 1.660 | 6.002 | -1.256 | 42.528 | -8.985 |
| 0.664 | -5.654 | -2.791 | -39.993 | -19.965 | 1.664 | 5.727 | -2.815 | 40.526 | -20.147 |
| 0.668 | -5.286 | -3.605 | -37.326 | -25.778 | 1.668 | 5.344 | -3.629 | 37.748 | -25.959 |
| 0.672 | -4.974 | -3.424 | -35.079 | -24.420 | 1.672 | 4.849 | -3.448 | 34.183 | -24.572 |
| 0.676 | -4.555 | -2.502 | -32.072 | -17.759 | 1.676 | 4.350 | -2.525 | 30.607 | -17.907 |
| 0.680 | -4.026 | -1.001 | -28.290 | -7.069 | 1.680 | 3.911 | -1.024 | 27.475 | -7.232 |
| 0.684 | -3.492 | 0.627 | -24.503 | 4.416 | 1.684 | 3.472 | 0.603 | 24.359 | 4.250 |
| 0.688 | -3.020 | 2.218 | -21.161 | 15.623 | 1.688 | 3.092 | 2.218 | 21.667 | 15.627 |
| 0.692 | -2.549 | 3.381 | -17.840 | 23.844 | 1.692 | 2.605 | 3.460 | 18.240 | 24.408 |
| 0.696 | -2.136 | 3.444 | -14.943 | 24.261 | 1.696 | 2.115 | 3.503 | 14.791 | 24.685 |
| 0.700 | -1.620 | 2.692 | -11.319 | 18.909 | 1.700 | 1.684 | 2.739 | 11.768 | 19.242 |
| 0.704 | -1.099 | 1.326 | -7.676 | 9.288 | 1.704 | 1.148 | 1.363 | 8.021 | 9.545 |
| 0.708 | -0.639 | -0.194 | -4.459 | -1.359 | 1.708 | 0.610 | -0.166 | 4.257 | -1.159 |
| 0.712 | -0.075 | -2.031 | -0.521 | -14.227 | 1.712 | 0.132 | -2.009 | 0.920 | -14.072 |
| 0.716 | 0.492 | -3.396 | 3.434 | -23.858 | 1.716 | -0.449 | -3.379 | -3.137 | -23.737 |
| 0.720 | 0.997 | -3.703 | 6.963 | -26.052 | 1.720 | -1.033 | -3.690 | -7.211 | -25.959 |
| 0.724 | 1.500 | -3.113 | 10.481 | -21.884 | 1.724 | -1.554 | -3.103 | -10.859 | -21.809 |
| 0.728 | 1.943 | -1.875 | 13.586 | -13.161 | 1.728 | -2.073 | -1.866 | -14.496 | -13.103 |
| 0.732 | 2.490 | -0.125 | 17.428 | -0.876 | 1.732 | -2.531 | -0.118 | -17.719 | -0.825 |
| 0.736 | 3.040 | 1.677 | 21.305 | 11.803 | 1.736 | -2.989 | 1.683 | -20.944 | 11.844 |
| 0.740 | 3.529 | 3.014 | 24.765 | 21.313 | 1.740 | -3.388 | 3.019 | -23.764 | 21.338 |
| 0.744 | 4.017 | 3.616 | 28.228 | 25.669 | 1.744 | -3.892 | 3.620 | -27.340 | 25.685 |
| 0.748 | 4.445 | 3.214 | 31.281 | 22.841 | 1.748 | -4.400 | 3.218 | -30.962 | 22.864 |
| 0.752 | 4.872 | 2.125 | 34.348 | 15.101 | 1.752 | -4.848 | 2.128 | -34.174 | 15.125 |
| 0.756 | 5.242 | 0.492 | 37.011 | 3.497 | 1.756 | -5.296 | 0.495 | -37.397 | 3.522 |
| 0.760 | 5.718 | -1.240 | 40.455 | -8.853 | 1.760 | -5.685 | -1.237 | -40.215 | -8.829 |
| 0.764 | 6.198 | -2.799 | 43.958 | -20.103 | 1.764 | -6.074 | -2.796 | -43.053 | -20.063 |
| 0.768 | 6.514 | -3.613 | 46.280 | -26.067 | 1.768 | -6.407 | -3.610 | -45.489 | -26.024 |
| 0.772 | 6.770 | -3.432 | 48.160 | -24.797 | 1.772 | -6.742 | -3.429 | -47.955 | -24.771 |
| 0.776 | 7.133 | -2.509 | 50.854 | -18.149 | 1.776 | -7.021 | -2.507 | -50.020 | -18.111 |
| 0.780 | 7.504 | -1.008 | 53.623 | -7.302 | 1.780 | -7.408 | -1.006 | -52.901 | -7.277 |
| 0.784 | 7.715 | 0.619 | 55.199 | 4.492 | 1.784 | -7.802 | 0.622 | -55.856 | 4.514 |
| 0.788 | 7.866 | 2.210 | 56.338 | 16.084 | 1.788 | -8.035 | 2.213 | -57.610 | 16.130 |
| 0.792 | 8.024 | 3.373 | 57.530 | 24.655 | 1.792 | -8.103 | 3.376 | -58.130 | 24.693 |
| 0.796 | 8.131 | 3.436 | 58.337 | 25.145 | 1.796 | -8.128 | 3.439 | -58.319 | 25.162 |
| 0.800 | 8.245 | 2.685 | 59.205 | 19.632 | 1.800 | -8.269 | 2.687 | -59.387 | 19.655 |
| 0.804 | 8.309 | 1.319 | 59.693 | 9.630 | 1.804 | -8.423 | 1.321 | -60.558 | 9.669 |
| 0.808 | 8.382 | -0.202 | 60.249 | -1.473 | 1.808 | -8.426 | -0.199 | -60.581 | -1.456 |
| 0.812 | 8.320 | -2.039 | 59.778 | -14.903 | 1.812 | -8.303 | -2.036 | -59.649 | -14.882 |
| 0.816 | 8.244 | -3.403 | 59.194 | -24.932 | 1.816 | -8.167 | -3.401 | -58.616 | -24.894 |
| 0.820 | 8.211 | -3.711 | 58.947 | -27.198 | 1.820 | -8.179 | -3.708 | -58.707 | -27.171 |
| 0.824 | 8.064 | -3.121 | 57.832 | -22.803 | 1.824 | -8.212 | -3.118 | -58.952 | -22.819 |
| 0.828 | 7.904 | -1.883 | 56.622 | -13.701 | 1.828 | -8.119 | -1.880 | -58.246 | -13.713 |
| 0.832 | 7.795 | -0.132 | 55.803 | -0.961 | 1.832 | -7.907 | -0.130 | -56.644 | -0.944 |
| 0.836 | 7.574 | 1.669 | 54.144 | 12.104 | 1.836 | -7.683 | 1.672 | -54.963 | 12.134 |
| 0.840 | 7.342 | 3.007 | 52.408 | 21.810 | 1.840 | -7.513 | 3.009 | -53.686 | 21.863 |
| 0.844 | 7.163 | 3.609 | 51.075 | 26.177 | 1.844 | -7.232 | 3.611 | -51.587 | 26.211 |
| 0.848 | 6.873 | 3.207 | 48.928 | 23.177 | 1.848 | -6.941 | 3.209 | -49.428 | 23.208 |
| 0.852 | 6.575 | 2.117 | 46.722 | 15.229 | 1.852 | -6.705 | 2.120 | -47.683 | 15.263 |
| 0.856 | 6.330 | 0.484 | 44.938 | 3.471 | 1.856 | -6.360 | 0.487 | -45.143 | 3.489 |
| 0.860 | 5.977 | -1.248 | 42.347 | -8.923 | 1.860 | -6.006 | -1.245 | -42.558 | -8.908 |
| 0.864 | 5.617 | -2.807 | 39.722 | -20.071 | 1.864 | -5.709 | -2.804 | -40.395 | -20.067 |
| 0.868 | 5.312 | -3.621 | 37.517 | -25.893 | 1.868 | -5.305 | -3.618 | -37.463 | -25.874 |
| 0.872 | 4.900 | -3.439 | 34.549 | -24.519 | 1.872 | -4.893 | -3.437 | -34.499 | -24.501 |
| 0.876 | 4.482 | -2.517 | 31.549 | -17.860 | 1.876 | -4.540 | -2.514 | -31.961 | -17.849 |
| 0.880 | 4.122 | -1.016 | 28.974 | -7.180 | 1.880 | -4.080 | -1.013 | -28.678 | -7.161 |
| 0.884 | 3.655 | 0.612 | 25.657 | 4.313 | 1.884 | -3.615 | 0.614 | -25.372 | 4.329 |
| 0.888 | 3.184 | 2.203 | 22.318 | 15.526 | 1.888 | -3.209 | 2.205 | -22.496 | 15.544 |
| 0.892 | 2.771 | 3.366 | 19.407 | 23.754 | 1.892 | -2.698 | 3.368 | -18.890 | 23.765 |
| 0.896 | 2.254 | 3.429 | 15.768 | 24.162 | 1.896 | -2.183 | 3.431 | -15.269 | 24.174 |
| 0.900 | 1.733 | 2.677 | 12.114 | 18.807 | 1.900 | -1.728 | 2.680 | -12.079 | 18.824 |
| 0.904 | 1.272 | 1.311 | 8.889 | 9.185 | 1.904 | -1.170 | 1.314 | -8.171 | 9.200 |
| 0.908 | 0.708 | -0.209 | 4.945 | -1.464 | 1.908 | -0.609 | -0.207 | -4.249 | -1.447 |
| 0.912 | 0.141 | -2.046 | 0.988 | -14.332 | 1.912 | -0.109 | -2.044 | -0.759 | -14.315 |
| 0.916 | -0.364 | -3.411 | -2.541 | -23.963 | 1.916 | 0.389 | -3.408 | 2.718 | -23.946 |
| 0.920 | -0.867 | -3.718 | -6.056 | -26.155 | 1.920 | 0.827 | -3.716 | 5.778 | -26.137 |
| 0.924 | -1.311 | -3.128 | -9.155 | -21.983 | 1.924 | 1.370 | -3.626 | 9.569 | -21.968 |
| 0.928 | -1.858 | -1.890 | -12.988 | -13.264 | 1.928 | 1.915 | -1.888 | 13.390 | -13.249 |
| 0.932 | -2.408 | -0.140 | -16.852 | -0.981 | 1.932 | 2.400 | -0.137 | 16.792 | -0.964 |
| 0.936 | -2.897 | 1.662 | -20.298 | 11.691 | 1.936 | 2.883 | 1.664 | 20.192 | 11.708 |
| 0.940 | -3.385 | 2.999 | -23.743 | 21.194 | 1.940 | 3.306 | 3.002 | 23.184 | 21.204 |
| 0.944 | -3.813 | 3.601 | -26.780 | 25.537 | 1.944 | 3.834 | 3.603 | 26.928 | 25.556 |
| 0.948 | -4.346 | 3.199 | -30.573 | 22.722 | 1.948 | 4.366 | 3.202 | 30.715 | 22.742 |
| 0.952 | -4.882 | 2.110 | -34.413 | 14.995 | 1.952 | 4.836 | 2.112 | 34.089 | 15.008 |
| 0.956 | -5.357 | 0.477 | -37.839 | 3.393 | 1.956 | 5.306 | 0.479 | 37.474 | 3.409 |
| 0.960 | -5.831 | -1.255 | -41.276 | -8.967 | 1.960 | 5.717 | -1.253 | 40.449 | -8.943 |
| 0.964 | -6.245 | -2.814 | -44.304 | -20.218 | 1.964 | 6.128 | -2.812 | 43.444 | -20.183 |
| 0.968 | -6.660 | -3.628 | -47.352 | -26.207 | 1.968 | 6.481 | -3.626 | 46.034 | -26.151 |
| 0.972 | -6.913 | -3.447 | -49.219 | -24.937 | 1.972 | 6.836 | -3.444 | 48.651 | -24.903 |
| 0.976 | -7.105 | -2.524 | -50.648 | -18.253 | 1.976 | 7.135 | -2.522 | 50.866 | -18.240 |
| 0.980 | -7.408 | -1.023 | -52.905 | -7.404 | 1.980 | 7.436 | -1.021 | 53.115 | -7.389 |
| 0.984 | -7.720 | 0.604 | -55.242 | 4.383 | 1.984 | 7.683 | 0.607 | 54.964 | 4.399 |
| 0.988 | -7.873 | 2.218 | -56.389 | 16.147 | 1.988 | 7.935 | 2.221 | 56.855 | 16.175 |
| 0.992 | -7.968 | 3.461 | -57.105 | 25.284 | 1.992 | 8.132 | 3.463 | 58.348 | 25.343 |
| 0.996 | -8.175 | 3.504 | -58.676 | 25.660 | 1.996 | 8.335 | 3.507 | 59.890 | 25.721 |
| 1.000 | -8.394 | 2.740 | -60.340 | 20.070 | 2.000 | 8.382 | 2.742 | 60.248 | 20.085 |

**End of Table 9.**

# 7. Ways for improvement of the real scan picture

## 7.1 Introducing positive phase shifts into demand sinusoidal signals of the actuators' tracking control systems

A further adjustment of both the actuator's control systems could improve the scan picture. By introducing a positive phase shift $\Delta\varphi_{lm}$ into demand sinusoidal signal for the large mirror actuator and a respective phase shift $\Delta\varphi_{sm}$ into demand sinusoidal signal for the small mirror actuator, which correspond to the phase delay at the driven frequency, the phases of the outputs of the actuators' control systems are synchronized with the phases of the sinusoidal signals forming the ideal work of the scanning system. The effect of this adjustment is presented in Figure 79 for the control system of the large mirror actuator and in Figure 80, Figure 81 and Figure 82 for the control system of the small mirror actuator. It can be seen also in these figures that time of $0.2\,\mathrm{s}$ is sufficient for the transition to steady synchronized oscillations for both the control systems. The numerical results of the scanning in this case in presented in Table 10.

So after this phase adjustment the first five passes of the scanned area with scanned points ('o') for the mirror actuators and respective scanning in the vertical plane at range of 200 m in front of the aircraft are shown in Figure 83. The second group of five passes, from Pass 6 till Pass 10 respectively, is presented in Figure 84. Figure 85 shows the ideal mirror actuators' odd pass and ideal scanned points together with the positions of the mirror actuators at odd passes 1, 3, 5, 7 and 9. Figure 86 shows the ideal mirror actuators' even pass and ideal scanned points together with the positions of the mirror actuators at even passes 2, 4, 6, 8 and 10. The scanning in the vertical plane at range of 200 m in front of the aircraft for odd passes is shown in Figure 87. The respective scanning at even passes is shown in Figure 88. Now the symmetry of the scanning with respect to the initial point with coordinates $(0,0)$ is better: Figure 85 versus Figure 75, Figure 86 versus Figure 76, Figure 87 versus Figure 77 and Figure 88 versus Figure 78.

## 7.2 Increasing the sampling rate of the actuators' tracking control systems

The sampling rate of $4\,\mathrm{ms}$ for the demand signals for both the tracking control systems of the large and the small mirror actuators is a request of the design of the scanning/imaging system as a whole. As we have shown at the investigation of the control systems of the large and the small mirror actuators based on synthesis of time optimal control, a sampling rate of $4\,\mathrm{ms}$ for the rest part of each one system, except the discretizing of the demand signal at sampling rate of $4\,\mathrm{ms}$, could be considered as a compromise with the imposed request and quality of the synthesized system for the tracking control system of the large mirror actuator and is unacceptable for the tracking control system of the small mirror actuator. So an approach for improvement the quality of tracking the periodic sinusoidal demand signal is the increase of the sampling rate of each one control system except the sampling rate for the demand signals. Combining that with an further phase adjustment for the sampling rate chosen the scan picture as a product of the dynamic behavior of both control systems could be additionally improved.

Here we show the effect of increasing the sampling rate to $0.1\,\text{ms}$ combined with the phase adjustment with respect to the scan picture pattern. Figure 89 shows the time-diagrams of the processes in the large mirror actuator digital tracking control system in the time interval $[0, 1]\,\text{s}$. Figure 90 shows the output angle $\varphi_{lm}$ presented together with the discretized sinusoidal signal ('-') and the ideal sinusoidal signal with amplitude of $8.35\,[\text{deg}]$ and frequency $2.5\,[\text{Hz}]$ for the scan pattern ('~') in the time intervals $[0, 0.6]\,\text{s}$ and $[0.6, 1.2]\,\text{s}$. The introduced positive phase shift $\Delta\varphi_{lm} = 9.58\,[\text{deg}]$ provides the phase synchronisation shown in Figure 90a and Figure 90b. Figure 89 and Figure 90a show that the transition time to steady tracking lasts practically less than $0.2\,\text{s}$.

Dealing with the small mirror actuator's control system Figure 91 shows the time-diagrams of the processes in the system in the time interval $[0, 0.25]\,\text{s}$. Figure 92 shows the output angle $\varphi_{sm}$ together with the discretized sinusoidal signal ('-') and the ideal sinusoidal signal with amplitude of $3.57\,[\text{deg}]$ and frequency $20\,[\text{Hz}]$ for the scan pattern ('~') in the time intervals $[0, 0.15]\,\text{s}$ and $[0.15, 0.3]\,\text{s}$. The introduced positive phase shift here for providing the phase synchronisation shown in Figure 92a and Figure 92b is $\Delta\varphi_{sm} = 49.11\,[\text{deg}]$. Figure 91 and Figure 92a show that the transition time to steady tracking lasts practically less than $0.1\,\text{s}$.

The effect of the above further adjustment of the actuators' control system is shown in next figures. Figure 93 and Figure 94 present the first 15 old and 15 even passes by the actuators of the scan area with scanned points. Figure 95 and Figure 96 show the scan picture at range of 200 m in front of the aircraft at these odd and even passes. Alongside these passes the ideal odd and even passes forming the scan pattern are also shown for comparison. Excluding the first odd pass of scan area by actuators with respective scanned points, shown by ('~') and ('o') in Figure 93 and Figure 95, there is a very good repetition and clearness of the rest of the real odd and even passes matching the scan pattern.

Alongside with the increase of the sampling rate the amount of near time optimal control synthesis problems solved drastically increases.

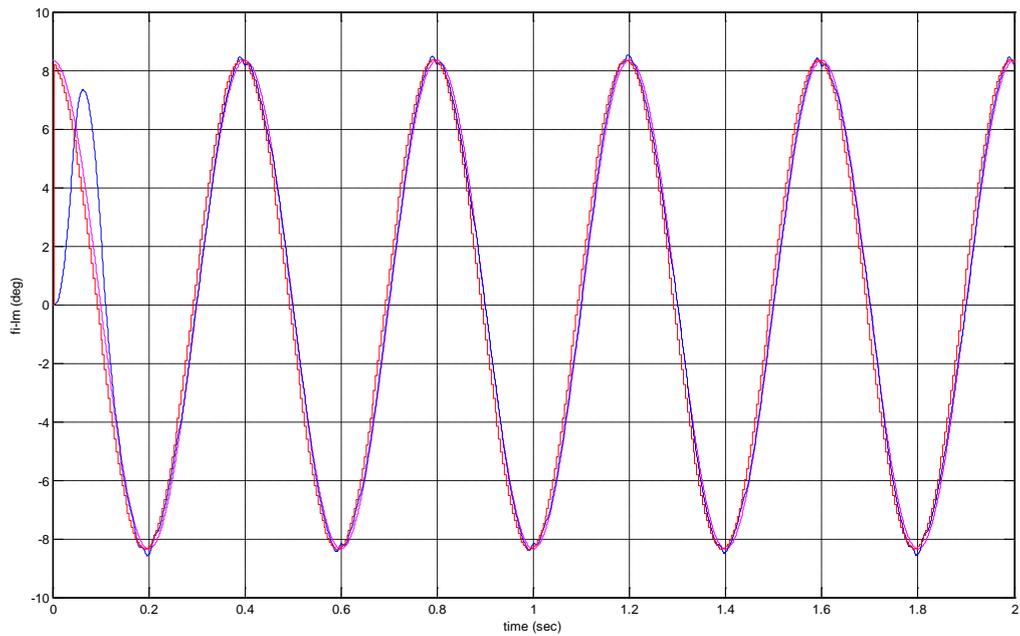

**Figure 79. Phase synchronisation between the output signal ('–') of the large mirror actuator's control system and sinusoidal signal ('~') forming the ideal work of the scanning system. The discretized demand periodic signal for the actuator's control system is presented by ('~').**

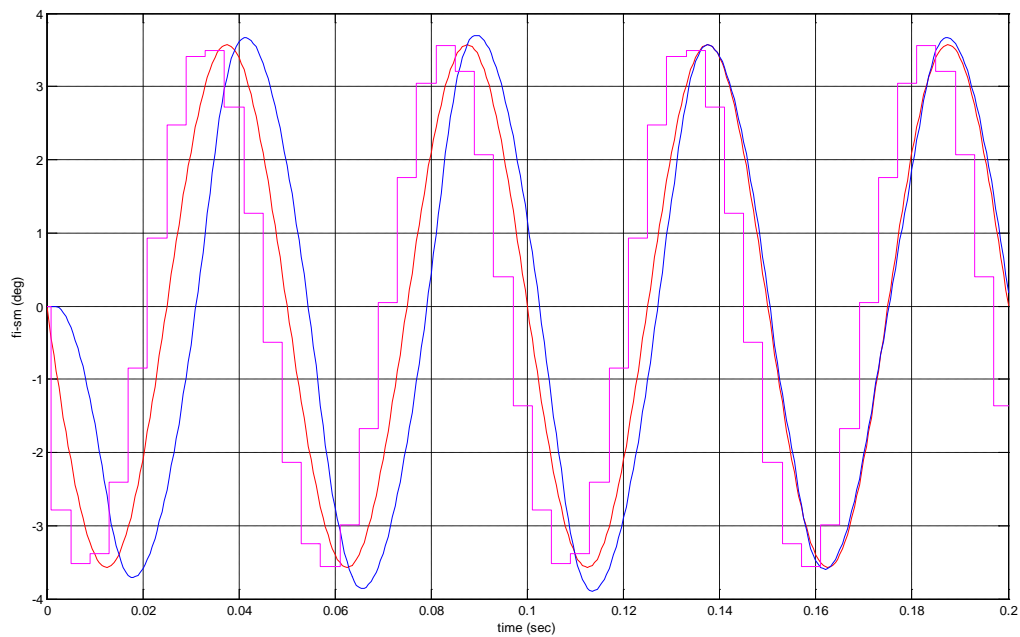

**Figure 80. Phase synchronization, shown in the time interval 0-0.2 s, between the output signal ('–') of the small mirror actuator's control system and sinusoidal signal ('~') forming the ideal work of the scanning system. The discretized demand periodic signal for the actuator's control system is presented by ('~').**

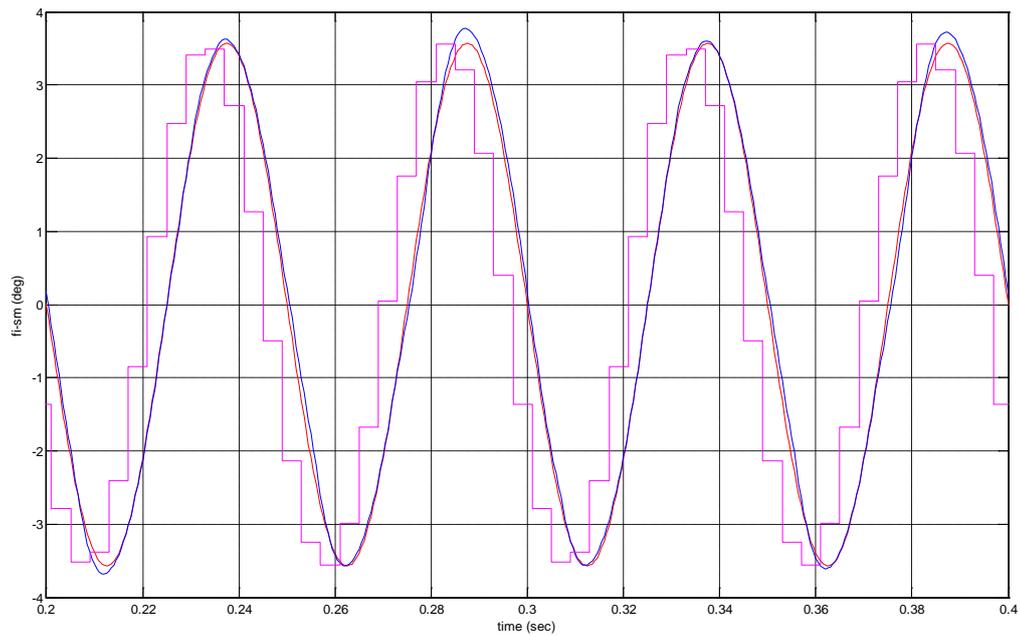

**Figure 81. Phase synchronization, shown in the time interval 0.2-0.4 s, between the output signal ('~') of the small mirror actuator's control system and sinusoidal signal ('~') forming the ideal work of the scanning system. The discretized demand periodic signal for the actuator's control system is presented by ('~').**

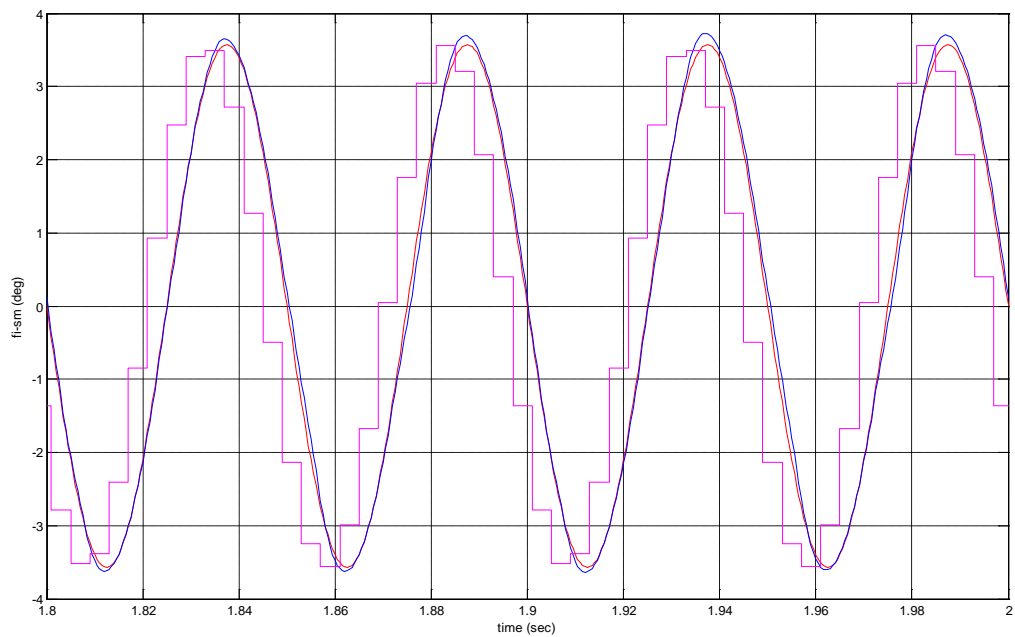

**Figure 82. Phase synchronization in the time interval 1.8-2 s between the output signal ('~') of the small mirror actuator's control system and sinusoidal signal ('~') forming the ideal work of the scanning system. The discretized demand periodic signal for the actuator's control system is presented by ('~').**

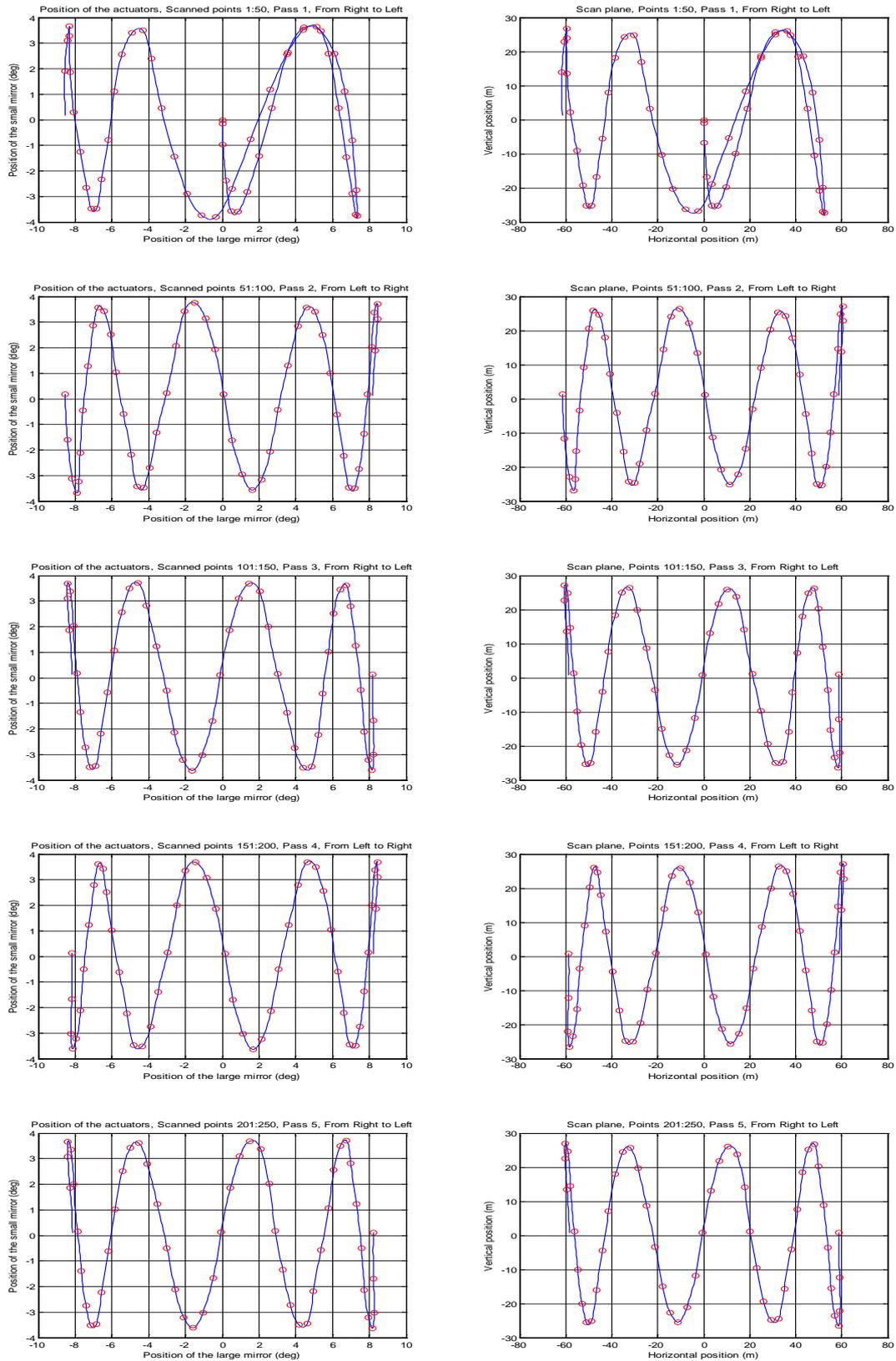

**Figure 83. First five passes of the scanned area with scanned points ('o') for the mirror actuators and respective scanning in the vertical plane at range of 200 m in front of the aircraft after the further phase adjustment of the actuators' control systems.**

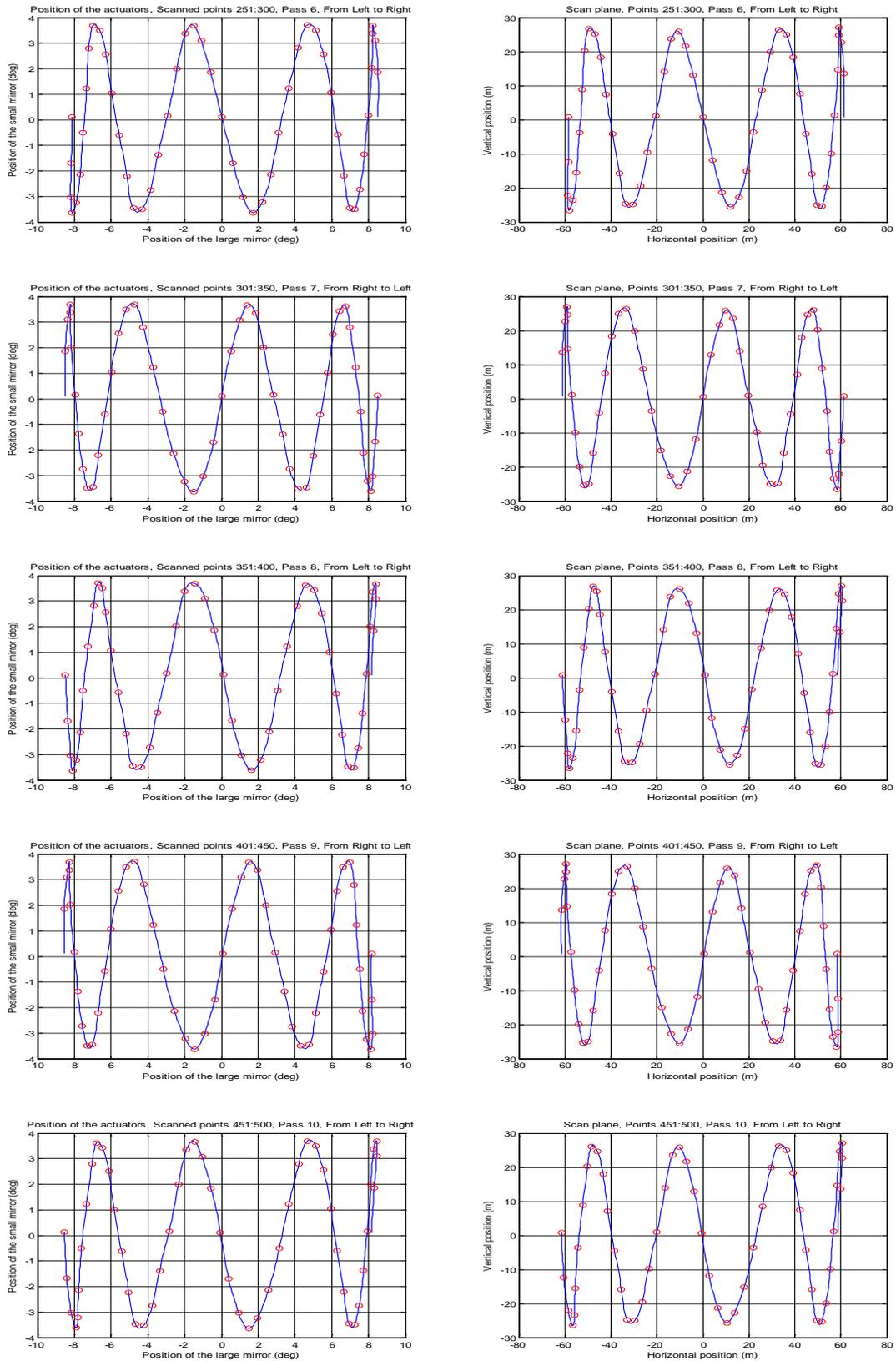

**Figure 84. Second group of five passes of the scanned area with scanned points ('o') for the mirror actuators and respective scanning in the vertical plane at range of 200 m in front of the aircraft after the further phase adjustment of the actuators' control systems.**

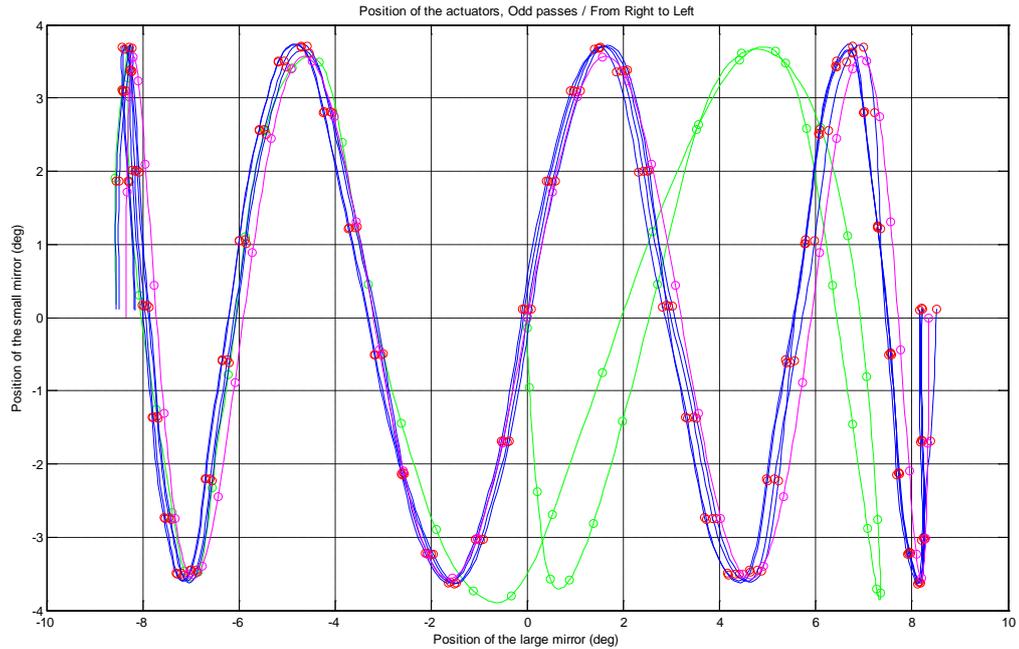

**Figure 85. The ideal mirror actuators' odd pass ('–') and ideal scanned points ('o') shown together with the positions of the mirror actuators at odd passes 1, 3, 5, 7 and 9 after the further phase adjustment of the actuators' control systems. The first pass with scanned points are shown by ('–') and ('o'). The scanned points are presented by ('o').**

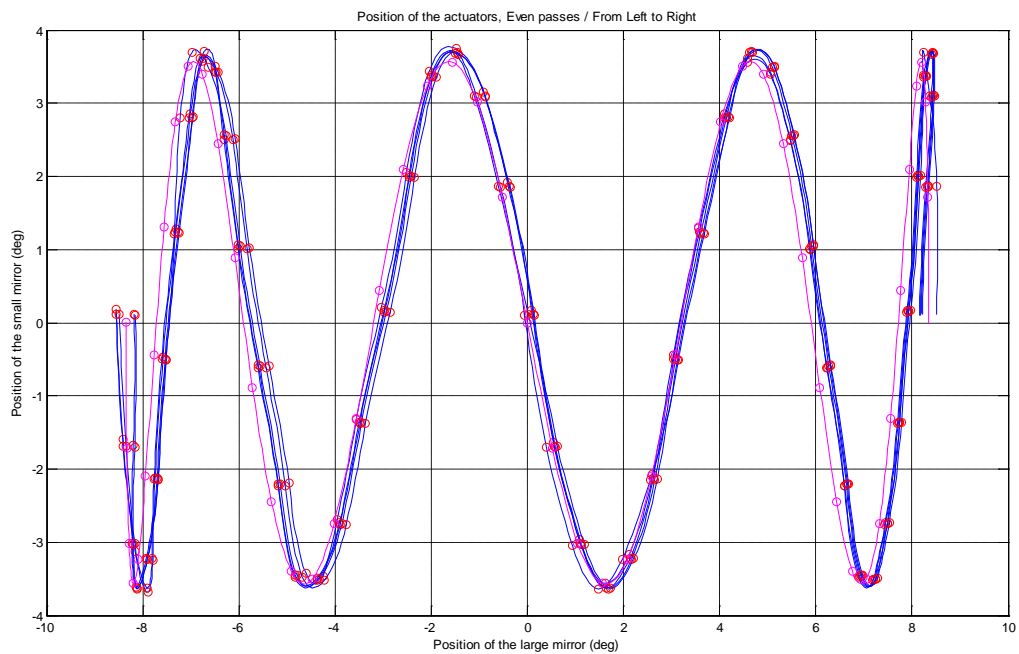

**Figure 86. The ideal mirror actuators' even pass ('–') and ideal scanned points ('o') shown together with the positions of the mirror actuators at even passes 2, 4, 6, 8 and 10 after the further phase adjustment of the actuators' control systems. The scanned points are presented by ('o').**

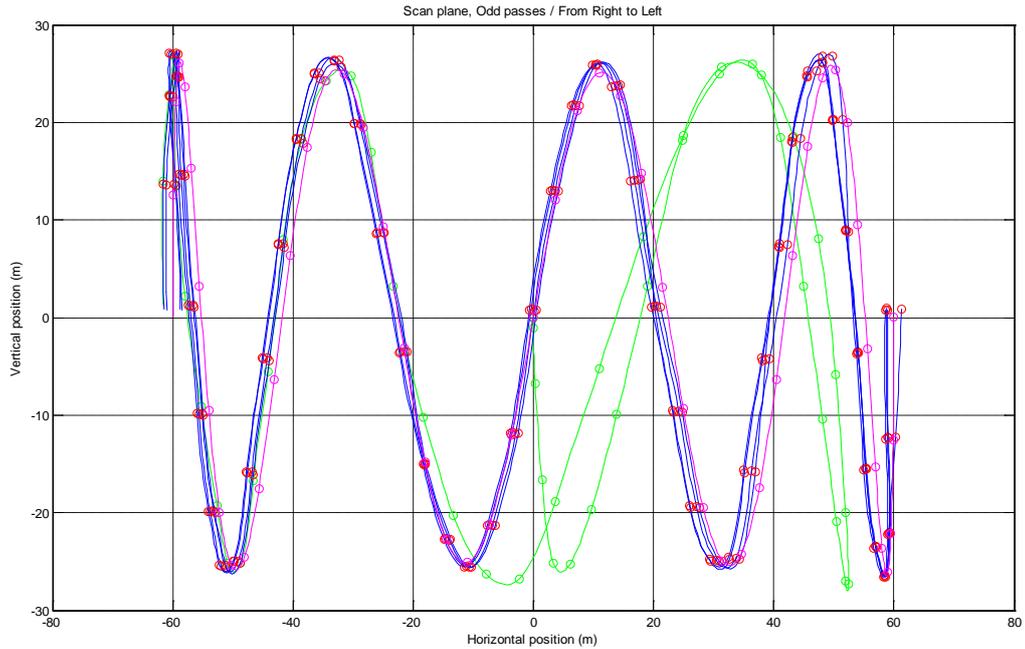

**Figure 87. Scanning in the vertical plane at range of 200 m in front of the aircraft at the ideal odd pass ('–') with ideal scanned points ('o') shown together with the scanning at odd passes 1, 3, 5, 7 and 9 after the further phase adjustment of the actuators' control systems. The scanned points are presented by ('o'). The first pass with scanned points are shown by ('–') and ('o').**

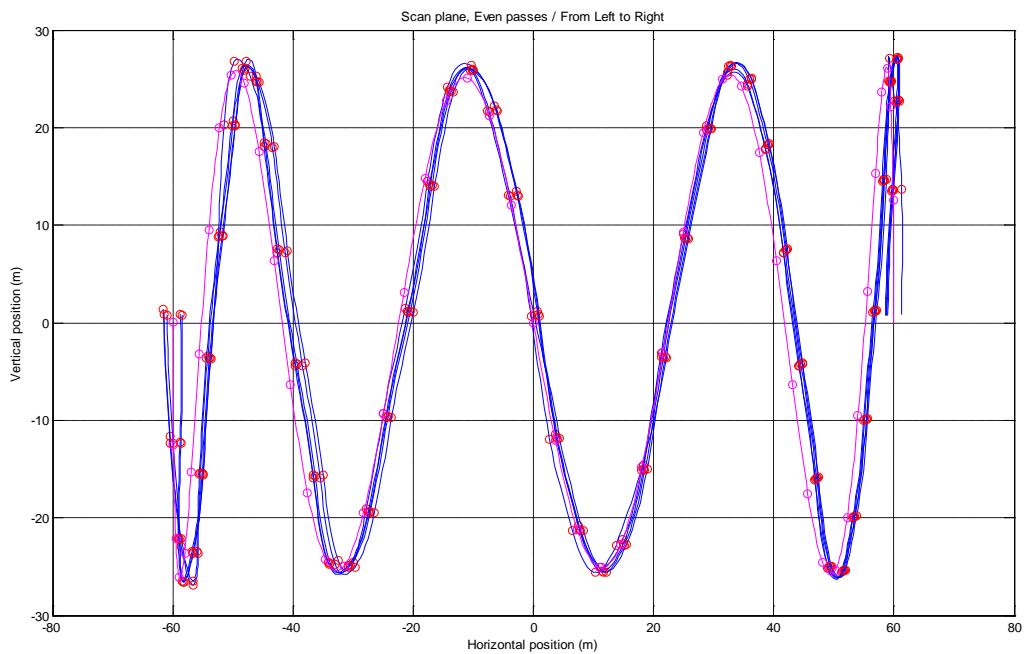

**Figure 88. Scanning in the vertical plane at range of 200 m in front of the aircraft at the ideal even pass ('–') with ideal scanned points ('o') shown together with the even passes 2, 4, 6, 8 and 10 ('–') after the further phase adjustment of the actuators' control systems. The scanned points are presented by ('o').**

**Table 10. Positions of the mirror actuators and respective scanned points in the vertical plane at range of 200 m in front of the aircraft in the first time interval 0-2 seconds at sampling rate of 4 ms after the further phase adjustment of the actuators' control systems.**

| Time (s) | Large mirror Position (deg) | Small mirror Position (deg) | Scan plane at range 200 m. Horizontal position (m) | Scan plane at range 200 m. Vertical position (m) | Time (s) | Large mirror Position (deg) | Small mirror Position (deg) | Scan plane at range 200 m. Horizontal position (m) | Scan plane at range 200 m. Vertical position (m) |
|---|---|---|---|---|---|---|---|---|---|
| 0.000 | 0.000 | 0.000 | 0.000 | 0.000 | 1.000 | -8.152 | 0.100 | -58.496 | 0.727 |
| 0.004 | 0.000 | -0.147 | 0.000 | -1.025 | 1.004 | -8.167 | -1.701 | -58.615 | -12.404 |
| 0.008 | 0.043 | -0.962 | 0.301 | -6.730 | 1.008 | -8.203 | -3.037 | -58.884 | -22.218 |
| 0.012 | 0.203 | -2.379 | 1.419 | -16.670 | 1.012 | -8.112 | -3.638 | -58.196 | -26.634 |
| 0.016 | 0.480 | -3.578 | 3.355 | -25.152 | 1.016 | -7.902 | -3.236 | -56.612 | -23.613 |
| 0.020 | 0.872 | -3.592 | 6.089 | -25.260 | 1.020 | -7.681 | -2.146 | -54.949 | -15.590 |
| 0.024 | 1.375 | -2.808 | 9.606 | -19.715 | 1.024 | -7.513 | -0.513 | -53.688 | -3.713 |
| 0.028 | 1.986 | -1.415 | 13.890 | -9.929 | 1.028 | -7.339 | 1.219 | -52.385 | 8.819 |
| 0.032 | 2.704 | 0.457 | 18.934 | 3.209 | 1.032 | -7.216 | 2.802 | -51.471 | 20.294 |
| 0.036 | 3.525 | 2.571 | 24.733 | 18.163 | 1.036 | -6.982 | 3.695 | -49.729 | 26.766 |
| 0.040 | 4.447 | 3.607 | 31.295 | 25.663 | 1.040 | -6.632 | 3.496 | -47.143 | 25.235 |
| 0.044 | 5.362 | 3.478 | 37.877 | 24.873 | 1.044 | -6.274 | 2.560 | -44.514 | 18.388 |
| 0.048 | 6.103 | 2.593 | 43.263 | 18.602 | 1.048 | -5.972 | 1.049 | -42.310 | 7.502 |
| 0.052 | 6.667 | 1.122 | 47.406 | 8.067 | 1.052 | -5.563 | -0.586 | -39.336 | -4.179 |
| 0.056 | 7.060 | -0.813 | 50.309 | -5.862 | 1.056 | -5.148 | -2.207 | -36.332 | -15.715 |
| 0.060 | 7.284 | -2.753 | 51.979 | -19.950 | 1.060 | -4.790 | -3.454 | -33.757 | -24.611 |
| 0.064 | 7.345 | -3.762 | 52.435 | -27.351 | 1.064 | -4.326 | -3.502 | -30.436 | -24.887 |
| 0.068 | 7.270 | -3.712 | 51.870 | -26.960 | 1.068 | -3.857 | -2.740 | -27.094 | -19.388 |
| 0.072 | 7.075 | -2.882 | 50.421 | -20.853 | 1.072 | -3.448 | -1.365 | -24.185 | -9.624 |
| 0.076 | 6.764 | -1.455 | 48.116 | -10.472 | 1.076 | -2.933 | 0.162 | -20.547 | 1.137 |
| 0.080 | 6.339 | 0.445 | 44.991 | 3.191 | 1.080 | -2.414 | 2.004 | -16.895 | 14.085 |
| 0.084 | 5.804 | 2.582 | 41.082 | 18.477 | 1.084 | -1.956 | 3.373 | -13.677 | 23.746 |
| 0.088 | 5.161 | 3.635 | 36.423 | 25.973 | 1.088 | -1.499 | 3.683 | -10.472 | 25.929 |
| 0.092 | 4.412 | 3.518 | 31.051 | 25.016 | 1.092 | -1.100 | 3.095 | -7.684 | 21.742 |
| 0.096 | 3.562 | 2.641 | 24.993 | 18.665 | 1.096 | -0.596 | 1.859 | -4.164 | 13.018 |
| 0.100 | 2.611 | 1.177 | 18.277 | 8.267 | 1.100 | 0.016 | 0.110 | 0.111 | 0.766 |
| 0.104 | 1.562 | -0.753 | 10.917 | -5.274 | 1.104 | 0.630 | -1.691 | 4.396 | -11.840 |
| 0.108 | 0.523 | -2.696 | 3.652 | -18.911 | 1.108 | 1.181 | -3.028 | 8.248 | -21.266 |
| 0.112 | -0.338 | -3.808 | -2.363 | -26.784 | 1.112 | 1.729 | -3.629 | 12.083 | -25.556 |
| 0.116 | -1.125 | -3.735 | -7.858 | -26.282 | 1.116 | 2.216 | -3.227 | 15.499 | -22.724 |
| 0.120 | -1.902 | -2.890 | -13.301 | -20.317 | 1.120 | 2.701 | -2.137 | 18.912 | -15.036 |
| 0.124 | -2.613 | -1.450 | -18.294 | -10.186 | 1.124 | 3.127 | -0.504 | 21.916 | -3.544 |
| 0.128 | -3.317 | 0.461 | -23.260 | 3.243 | 1.128 | 3.657 | 1.228 | 25.670 | 8.665 |
| 0.132 | -3.851 | 2.389 | -27.049 | 16.891 | 1.132 | 4.191 | 2.811 | 29.469 | 19.930 |
| 0.136 | -4.319 | 3.488 | -30.380 | 24.789 | 1.136 | 4.664 | 3.704 | 32.849 | 26.388 |
| 0.140 | -4.889 | 3.406 | -34.469 | 24.282 | 1.140 | 5.135 | 3.505 | 36.241 | 25.027 |
| 0.144 | -5.463 | 2.555 | -38.605 | 18.244 | 1.144 | 5.548 | 2.569 | 39.225 | 18.355 |
| 0.148 | -5.870 | 1.111 | -41.561 | 7.936 | 1.148 | 5.961 | 1.058 | 42.227 | 7.565 |
| 0.152 | -6.213 | -0.780 | -44.069 | -5.583 | 1.152 | 6.316 | -0.578 | 44.824 | -4.139 |
| 0.156 | -6.558 | -2.324 | -46.604 | -16.723 | 1.156 | 6.673 | -2.198 | 47.448 | -15.826 |
| 0.160 | -6.848 | -3.481 | -48.739 | -25.177 | 1.160 | 6.974 | -3.445 | 49.670 | -24.940 |
| 0.164 | -7.140 | -3.466 | -50.908 | -25.127 | 1.164 | 7.277 | -3.493 | 51.926 | -25.354 |
| 0.168 | -7.378 | -2.661 | -52.681 | -19.292 | 1.168 | 7.526 | -2.731 | 53.783 | -19.830 |
| 0.172 | -7.725 | -1.253 | -55.280 | -9.091 | 1.172 | 7.779 | -1.357 | 55.681 | -9.853 |
| 0.176 | -8.080 | 0.302 | -57.957 | 2.197 | 1.176 | 7.978 | 0.171 | 57.182 | 1.241 |
| 0.180 | -8.275 | 2.013 | -59.430 | 13.571 | 1.180 | 8.183 | 2.013 | 58.731 | 14.692 |
| 0.184 | -8.308 | 3.278 | -59.682 | 24.019 | 1.184 | 8.231 | 3.382 | 59.099 | 24.768 |
| 0.188 | -8.310 | 3.655 | -59.696 | 26.815 | 1.188 | 8.243 | 3.692 | 59.193 | 27.068 |
| 0.192 | -8.431 | 3.110 | -60.618 | 22.810 | 1.192 | 8.373 | 3.104 | 60.180 | 22.752 |
| 0.196 | -8.565 | 1.907 | -61.646 | 13.976 | 1.196 | 8.517 | 1.868 | 61.273 | 13.677 |
| 0.200 | -8.552 | 0.185 | -61.546 | 1.353 | 1.200 | 8.511 | 0.118 | 61.228 | 0.866 |
| 0.204 | -8.416 | -1.595 | -60.504 | -11.659 | 1.204 | 8.380 | -1.682 | 60.235 | -12.296 |
| 0.208 | -8.162 | -3.131 | -58.572 | -22.900 | 1.208 | 8.237 | -3.019 | 59.193 | -22.090 |
| 0.212 | -7.897 | -3.680 | -56.572 | -26.883 | 1.212 | 8.144 | -3.620 | 58.441 | -26.509 |
| 0.216 | -7.791 | -3.241 | -55.773 | -23.625 | 1.216 | 7.939 | -3.218 | 56.889 | -23.488 |
| 0.220 | -7.741 | -2.123 | -55.399 | -15.427 | 1.220 | 7.722 | -2.128 | 55.256 | -15.465 |
| 0.224 | -7.578 | -0.467 | -54.171 | -3.381 | 1.224 | 7.558 | -0.495 | 54.026 | -3.584 |
| 0.228 | -7.297 | 1.284 | -52.075 | 9.279 | 1.228 | 7.283 | 1.237 | 51.972 | 8.944 |
| 0.232 | -7.007 | 2.857 | -49.917 | 20.655 | 1.232 | 6.999 | 2.797 | 49.856 | 20.214 |
| 0.236 | -6.771 | 3.575 | -48.173 | 25.841 | 1.236 | 6.769 | 3.610 | 48.153 | 26.102 |
| 0.240 | -6.427 | 3.419 | -45.634 | 24.634 | 1.240 | 6.429 | 3.429 | 45.653 | 24.709 |
| 0.244 | -6.074 | 2.516 | -43.050 | 18.039 | 1.244 | 6.082 | 2.507 | 43.107 | 17.975 |
| 0.248 | -5.777 | 1.030 | -40.887 | 7.355 | 1.248 | 5.790 | 1.006 | 40.981 | 7.182 |
| 0.252 | -5.373 | -0.585 | -37.956 | -4.166 | 1.252 | 5.391 | -0.622 | 38.084 | -4.426 |
| 0.256 | -4.962 | -2.190 | -34.992 | -15.574 | 1.256 | 4.985 | -2.236 | 35.155 | -15.905 |
| 0.260 | -4.609 | -3.424 | -32.456 | -24.372 | 1.260 | 4.636 | -3.478 | 32.651 | -24.760 |
| 0.264 | -4.254 | -3.463 | -29.918 | -24.601 | 1.264 | 4.181 | -3.522 | 29.400 | -25.013 |
| 0.268 | -3.955 | -2.695 | -27.790 | -19.079 | 1.268 | 3.721 | -2.757 | 26.125 | -19.502 |
| 0.272 | -3.549 | -1.316 | -24.904 | -9.279 | 1.272 | 3.320 | -1.381 | 23.280 | -9.730 |
| 0.276 | -3.032 | 0.215 | -21.245 | 1.512 | 1.276 | 2.813 | 0.148 | 19.704 | 1.037 |
| 0.280 | -2.511 | 2.060 | -17.573 | 14.486 | 1.280 | 2.303 | 1.991 | 16.112 | 13.989 |
| 0.284 | -2.050 | 3.431 | -14.336 | 24.168 | 1.284 | 1.853 | 3.361 | 12.951 | 23.654 |
| 0.288 | -1.486 | 3.743 | -10.381 | 26.359 | 1.288 | 1.403 | 3.672 | 9.802 | 25.844 |
| 0.292 | -0.919 | 3.157 | -6.417 | 22.173 | 1.292 | 1.012 | 3.084 | 7.067 | 21.662 |
| 0.296 | -0.413 | 1.921 | -2.886 | 13.455 | 1.296 | 0.516 | 1.848 | 3.600 | 12.943 |
| 0.300 | 0.090 | 0.173 | 0.628 | 1.208 | 1.300 | 0.015 | 0.099 | 0.104 | 0.694 |
| 0.304 | 0.533 | -1.627 | 3.725 | -11.392 | 1.304 | -0.426 | -1.701 | -2.974 | -11.909 |
| 0.308 | 1.081 | -2.963 | 7.549 | -20.809 | 1.308 | -0.971 | -3.038 | -6.779 | -21.330 |
| 0.312 | 1.631 | -3.564 | 11.401 | -25.092 | 1.312 | -1.519 | -3.639 | -10.612 | -25.615 |
| 0.316 | 2.121 | -3.162 | 14.832 | -22.259 | 1.316 | -2.006 | -3.236 | -14.025 | -22.780 |
| 0.320 | 2.608 | -2.072 | 18.261 | -14.574 | 1.320 | -2.595 | -2.147 | -18.169 | -15.099 |
| 0.324 | 3.037 | -0.439 | 21.279 | -3.085 | 1.324 | -3.187 | -0.513 | -22.343 | -3.612 |
| 0.328 | 3.569 | 1.294 | 25.047 | 9.121 | 1.328 | -3.717 | 1.219 | -26.096 | 8.600 |
| 0.332 | 4.105 | 2.853 | 28.857 | 20.219 | 1.332 | -4.244 | 2.802 | -29.848 | 19.868 |
| 0.336 | 4.580 | 3.559 | 32.252 | 25.338 | 1.336 | -4.711 | 3.694 | -33.186 | 26.327 |
| 0.340 | 5.054 | 3.396 | 35.656 | 24.233 | 1.340 | -5.176 | 3.495 | -36.534 | 24.965 |
| 0.344 | 5.469 | 2.488 | 38.652 | 17.760 | 1.344 | -5.583 | 2.560 | -39.475 | 18.292 |
| 0.348 | 5.884 | 0.998 | 41.665 | 7.130 | 1.348 | -5.989 | 1.049 | -42.434 | 7.499 |
| 0.352 | 6.241 | -0.621 | 44.274 | -4.447 | 1.352 | -6.339 | -0.587 | -44.989 | -4.207 |
| 0.356 | 6.600 | -2.228 | 46.909 | -16.033 | 1.356 | -6.690 | -2.208 | -47.572 | -15.896 |
| 0.360 | 6.902 | -3.465 | 49.142 | -25.066 | 1.360 | -6.985 | -3.455 | -49.754 | -25.011 |
| 0.364 | 7.208 | -3.505 | 51.409 | -25.426 | 1.364 | -7.283 | -3.502 | -51.969 | -25.424 |
| 0.368 | 7.458 | -2.738 | 53.277 | -19.868 | 1.368 | -7.526 | -2.740 | -53.787 | -19.898 |
| 0.372 | 7.713 | -1.360 | 55.186 | -9.868 | 1.372 | -7.774 | -1.366 | -55.646 | -9.920 |
| 0.376 | 7.914 | 0.171 | 56.698 | 1.242 | 1.376 | -7.968 | 0.161 | -57.109 | 1.173 |
| 0.380 | 8.120 | 2.016 | 58.258 | 14.703 | 1.380 | -8.168 | 2.004 | -58.620 | 14.621 |
| 0.384 | 8.274 | 3.386 | 59.424 | 24.813 | 1.384 | -8.212 | 3.372 | -58.953 | 24.693 |
| 0.388 | 8.434 | 3.698 | 60.645 | 27.168 | 1.388 | -8.221 | 3.683 | -59.024 | 26.993 |
| 0.392 | 8.443 | 3.111 | 60.712 | 22.822 | 1.392 | -8.349 | 3.095 | -59.991 | 22.677 |
| 0.396 | 8.325 | 1.876 | 59.815 | 13.707 | 1.396 | -8.489 | 1.858 | -61.064 | 13.604 |
| 0.400 | 8.194 | 0.127 | 58.817 | 0.925 | 1.400 | -8.481 | 0.109 | -61.003 | 0.797 |
| 0.404 | 8.210 | -1.673 | 58.938 | -12.209 | 1.404 | -8.349 | -1.692 | -59.997 | -12.361 |
| 0.408 | 8.245 | -3.009 | 59.207 | -22.024 | 1.408 | -8.204 | -3.028 | -58.892 | -22.152 |
| 0.412 | 8.155 | -3.610 | 58.520 | -26.440 | 1.412 | -8.109 | -3.629 | -58.177 | -26.569 |
| 0.416 | 7.945 | -3.208 | 56.935 | -23.418 | 1.416 | -7.903 | -3.227 | -56.613 | -23.549 |
| 0.420 | 7.724 | -2.118 | 55.271 | -15.393 | 1.420 | -7.684 | -2.138 | -54.970 | -15.527 |
| 0.424 | 7.556 | -0.485 | 54.010 | -3.512 | 1.424 | -7.518 | -0.504 | -53.728 | -3.651 |
| 0.428 | 7.277 | 1.247 | 51.927 | 9.017 | 1.428 | -7.242 | 1.228 | -51.664 | 8.873 |
| 0.432 | 6.989 | 2.807 | 49.783 | 20.287 | 1.432 | -6.956 | 2.811 | -49.538 | 20.309 |
| 0.436 | 6.755 | 3.621 | 48.052 | 26.173 | 1.436 | -6.724 | 3.703 | -47.825 | 26.769 |
| 0.440 | 6.412 | 3.439 | 45.526 | 24.780 | 1.440 | -6.488 | 3.504 | -46.084 | 25.267 |
| 0.444 | 6.061 | 2.517 | 42.954 | 18.046 | 1.444 | -6.305 | 2.569 | -44.740 | 18.454 |
| 0.448 | 5.766 | 1.016 | 40.803 | 7.254 | 1.448 | -6.011 | 1.058 | -42.593 | 7.564 |
| 0.452 | 5.467 | -0.611 | 38.640 | -4.354 | 1.452 | -5.604 | -0.578 | -39.629 | -4.121 |
| 0.456 | 5.224 | -2.226 | 36.880 | -15.855 | 1.456 | -5.190 | -2.199 | -36.635 | -15.659 |
| 0.460 | 4.872 | -3.468 | 34.342 | -24.720 | 1.460 | -4.834 | -3.446 | -34.070 | -24.557 |
| 0.464 | 4.407 | -3.511 | 31.012 | -24.969 | 1.464 | -4.372 | -3.493 | -30.758 | -24.833 |
| 0.468 | 3.937 | -2.747 | 27.662 | -19.448 | 1.468 | -3.904 | -2.731 | -27.426 | -19.333 |
| 0.472 | 3.527 | -1.371 | 24.745 | -9.666 | 1.472 | -3.496 | -1.357 | -24.526 | -9.567 |
| 0.476 | 3.011 | 0.158 | 21.099 | 1.110 | 1.476 | -2.982 | 0.170 | -20.896 | 1.197 |
| 0.480 | 2.492 | 2.001 | 17.439 | 14.070 | 1.480 | -2.465 | 2.013 | -17.253 | 14.147 |
| 0.484 | 2.033 | 3.371 | 14.214 | 23.738 | 1.484 | -2.008 | 3.381 | -14.043 | 23.809 |
| 0.488 | 1.470 | 3.682 | 10.272 | 25.921 | 1.488 | -1.448 | 3.691 | -10.116 | 25.987 |
| 0.492 | 0.905 | 3.095 | 6.319 | 21.733 | 1.492 | -0.884 | 3.104 | -6.177 | 21.796 |
| 0.496 | 0.401 | 1.859 | 2.799 | 13.015 | 1.496 | -0.383 | 1.867 | -2.671 | 13.075 |
| 0.500 | -0.101 | 0.110 | -0.704 | 0.767 | 1.500 | 0.117 | 0.118 | 0.819 | 0.825 |
| 0.504 | -0.543 | -1.691 | -3.790 | -11.837 | 1.504 | 0.557 | -1.683 | 3.892 | -11.780 |
| 0.508 | -1.089 | -3.027 | -7.605 | -21.260 | 1.508 | 1.102 | -3.019 | 7.694 | -21.204 |
| 0.512 | -1.638 | -3.628 | -11.446 | -25.547 | 1.512 | 1.649 | -3.621 | 11.524 | -25.492 |
| 0.516 | -2.126 | -3.226 | -14.868 | -22.714 | 1.516 | 2.135 | -3.218 | 14.934 | -22.659 |
| 0.520 | -2.612 | -2.136 | -18.287 | -15.026 | 1.520 | 2.620 | -2.129 | 18.341 | -14.973 |
| 0.524 | -3.039 | -0.503 | -21.296 | -3.537 | 1.524 | 3.045 | -0.495 | 21.339 | -3.484 |
| 0.528 | -3.570 | 1.229 | -25.055 | 8.668 | 1.528 | 3.575 | 1.237 | 25.086 | 8.721 |
| 0.532 | -4.105 | 2.812 | -28.856 | 19.928 | 1.532 | 4.108 | 2.796 | 28.876 | 19.814 |
| 0.536 | -4.579 | 3.705 | -32.242 | 26.382 | 1.536 | 4.580 | 3.610 | 32.252 | 25.701 |
| 0.540 | -5.052 | 3.429 | -35.638 | 25.021 | 1.540 | 5.052 | 3.429 | 35.636 | 24.466 |
| 0.544 | -5.465 | 2.570 | -38.625 | 18.352 | 1.544 | 5.464 | 2.506 | 38.613 | 17.892 |
| 0.548 | -5.879 | 1.059 | -41.629 | 7.567 | 1.548 | 5.876 | 1.005 | 41.608 | 7.183 |
| 0.552 | -6.235 | -0.577 | -44.230 | -4.129 | 1.552 | 6.231 | -0.622 | 44.198 | -4.456 |
| 0.556 | -6.593 | -2.197 | -46.857 | -15.809 | 1.556 | 6.587 | -2.236 | 46.815 | -16.091 |
| 0.560 | -6.894 | -3.444 | -49.082 | -24.915 | 1.560 | 6.887 | -3.478 | 49.030 | -25.163 |
| 0.564 | -7.199 | -3.492 | -51.341 | -25.329 | 1.564 | 7.190 | -3.522 | 51.279 | -25.551 |
| 0.568 | -7.448 | -2.730 | -53.201 | -19.808 | 1.568 | 7.438 | -2.758 | 53.130 | -20.012 |
| 0.572 | -7.702 | -1.356 | -55.102 | -9.838 | 1.572 | 7.691 | -1.382 | 55.021 | -10.027 |
| 0.576 | -7.902 | 0.172 | -56.607 | 1.248 | 1.576 | 7.890 | 0.147 | 56.517 | 1.068 |
| 0.580 | -8.107 | 2.014 | -58.159 | 14.688 | 1.580 | 8.094 | 1.991 | 58.060 | 14.515 |
| 0.584 | -8.260 | 3.383 | -59.318 | 24.782 | 1.584 | 8.246 | 3.360 | 59.210 | 24.612 |
| 0.588 | -8.420 | 3.693 | -60.532 | 27.126 | 1.588 | 8.404 | 3.671 | 60.416 | 26.959 |
| 0.592 | -8.427 | 3.105 | -60.593 | 22.773 | 1.592 | 8.411 | 3.084 | 60.469 | 22.611 |
| 0.596 | -8.309 | 1.869 | -59.692 | 13.654 | 1.596 | 8.292 | 1.848 | 59.562 | 13.498 |
| 0.600 | -8.177 | 0.119 | -58.689 | 0.870 | 1.600 | 8.159 | 0.099 | 58.554 | 0.720 |
| 0.604 | -8.192 | -1.681 | -58.805 | -12.265 | 1.604 | 8.174 | -1.702 | 58.665 | -12.412 |
| 0.608 | -8.227 | -3.018 | -59.072 | -22.081 | 1.608 | 8.208 | -3.038 | 58.927 | -22.226 |
| 0.612 | -8.136 | -3.619 | -58.382 | -26.499 | 1.612 | 8.117 | -3.639 | 58.234 | -26.643 |
| 0.616 | -7.927 | -3.217 | -56.795 | -23.478 | 1.616 | 7.907 | -3.237 | 56.646 | -23.621 |
| 0.620 | -7.705 | -2.127 | -55.130 | -15.455 | 1.620 | 7.685 | -2.147 | 54.978 | -15.598 |
| 0.624 | -7.537 | -0.494 | -53.867 | -3.576 | 1.624 | 7.516 | -0.514 | 53.713 | -3.720 |
| 0.628 | -7.258 | 1.238 | -51.782 | 8.949 | 1.628 | 7.341 | 1.218 | 52.405 | 8.812 |
| 0.632 | -6.969 | 2.797 | -49.636 | 20.216 | 1.632 | 7.218 | 2.801 | 51.486 | 20.287 |
| 0.636 | -6.735 | 3.611 | -47.904 | 26.102 | 1.636 | 6.983 | 3.694 | 49.740 | 26.759 |
| 0.640 | -6.496 | 3.430 | -46.144 | 24.730 | 1.640 | 6.633 | 3.495 | 47.150 | 25.228 |
| 0.644 | -6.310 | 2.508 | -44.782 | 18.013 | 1.644 | 6.274 | 2.559 | 44.517 | 18.381 |
| 0.648 | -6.015 | 1.007 | -42.618 | 7.201 | 1.648 | 5.972 | 1.048 | 42.309 | 7.495 |
| 0.652 | -5.605 | -0.621 | -39.637 | -4.425 | 1.652 | 5.563 | -0.587 | 39.331 | -4.186 |
| 0.656 | -5.189 | -2.235 | -36.626 | -15.918 | 1.656 | 5.147 | -2.208 | 36.323 | -15.722 |
| 0.660 | -4.830 | -3.477 | -34.046 | -24.781 | 1.660 | 4.789 | -3.455 | 33.745 | -24.618 |
| 0.664 | -4.366 | -3.521 | -30.719 | -25.030 | 1.664 | 4.324 | -3.503 | 30.420 | -24.894 |
| 0.668 | -3.896 | -2.756 | -27.371 | -19.511 | 1.668 | 3.855 | -2.741 | 27.075 | -19.395 |
| 0.672 | -3.486 | -1.380 | -24.457 | -9.730 | 1.672 | 3.444 | -1.366 | 24.163 | -9.631 |
| 0.676 | -2.971 | 0.149 | -20.814 | 1.044 | 1.676 | 2.929 | 0.161 | 20.522 | 1.130 |
| 0.680 | -2.452 | 1.992 | -17.157 | 14.002 | 1.680 | 2.410 | 2.003 | 16.867 | 14.078 |
| 0.684 | -1.993 | 3.362 | -13.934 | 23.669 | 1.684 | 1.952 | 3.372 | 13.646 | 23.739 |
| 0.688 | -1.430 | 3.673 | -9.994 | 25.852 | 1.688 | 1.494 | 3.682 | 10.439 | 25.922 |
| 0.692 | -0.865 | 3.085 | -6.043 | 21.665 | 1.692 | 1.095 | 3.094 | 7.647 | 21.735 |
| 0.696 | -0.362 | 1.849 | -2.525 | 12.949 | 1.696 | 0.591 | 1.858 | 4.125 | 13.011 |
| 0.700 | 0.140 | 0.100 | 0.977 | 0.701 | 1.700 | 0.083 | 0.109 | 0.577 | 0.759 |
| 0.704 | 0.582 | -1.700 | 4.062 | -11.903 | 1.704 | -0.366 | -1.692 | -2.552 | -11.845 |
| 0.708 | 1.128 | -3.037 | 7.876 | -21.327 | 1.708 | -0.918 | -3.029 | -6.408 | -21.266 |
| 0.712 | 1.676 | -3.638 | 11.717 | -25.615 | 1.712 | -1.472 | -3.630 | -10.289 | -25.550 |
| 0.716 | 2.164 | -3.235 | 15.138 | -22.783 | 1.716 | -1.966 | -3.228 | -13.749 | -22.717 |
| 0.720 | 2.650 | -2.146 | 18.556 | -15.094 | 1.720 | -2.563 | -2.138 | -17.939 | -15.036 |
| 0.724 | 3.077 | -0.512 | 21.565 | -3.603 | 1.724 | -3.161 | -0.505 | -22.158 | -3.551 |
| 0.728 | 3.608 | 1.220 | 25.324 | 8.603 | 1.728 | -3.697 | 1.227 | -25.955 | 8.660 |
| 0.732 | 4.143 | 2.803 | 29.125 | 19.865 | 1.732 | -4.230 | 2.810 | -29.750 | 19.927 |
| 0.736 | 4.617 | 3.695 | 32.512 | 26.320 | 1.736 | -4.703 | 3.703 | -33.131 | 26.387 |
| 0.740 | 5.089 | 3.496 | 35.907 | 24.959 | 1.740 | -5.174 | 3.504 | -36.521 | 25.026 |
| 0.744 | 5.503 | 2.561 | 38.895 | 18.289 | 1.744 | -5.586 | 2.568 | -39.503 | 18.353 |
| 0.748 | 5.916 | 1.050 | 41.900 | 7.502 | 1.748 | -5.999 | 1.057 | -42.503 | 7.560 |
| 0.752 | 6.272 | -0.586 | 44.501 | -4.198 | 1.752 | -6.354 | -0.579 | -45.099 | -4.147 |
| 0.756 | 6.630 | -2.207 | 47.129 | -15.881 | 1.756 | -6.710 | -2.199 | -47.722 | -15.838 |
| 0.760 | 6.931 | -3.454 | 49.354 | -24.992 | 1.760 | -7.010 | -3.446 | -49.942 | -24.955 |
| 0.764 | 7.235 | -3.501 | 51.614 | -25.406 | 1.764 | -7.314 | -3.494 | -52.197 | -25.369 |
| 0.768 | 7.485 | -2.739 | 53.474 | -19.883 | 1.768 | -7.562 | -2.732 | -54.052 | -19.844 |
| 0.772 | 7.738 | -1.365 | 55.375 | -9.910 | 1.772 | -7.814 | -1.358 | -55.949 | -9.863 |
| 0.776 | 7.938 | 0.162 | 56.880 | 1.180 | 1.776 | -8.013 | 0.170 | -57.448 | 1.235 |
| 0.780 | 8.143 | 2.005 | 58.433 | 14.625 | 1.780 | -8.217 | 2.012 | -58.996 | 14.690 |
| 0.784 | 8.296 | 3.373 | 59.592 | 24.722 | 1.784 | -8.266 | 3.381 | -59.361 | 24.769 |
| 0.788 | 8.455 | 3.684 | 60.806 | 27.066 | 1.788 | -8.278 | 3.691 | -59.454 | 27.071 |
| 0.792 | 8.463 | 3.096 | 60.866 | 22.712 | 1.792 | -8.407 | 3.103 | -60.441 | 22.753 |
| 0.796 | 8.345 | 1.859 | 59.964 | 13.590 | 1.796 | -8.551 | 1.867 | -61.533 | 13.674 |
| 0.800 | 8.213 | 0.110 | 58.960 | 0.802 | 1.800 | -8.544 | 0.117 | -61.486 | 0.859 |
| 0.804 | 8.228 | -1.691 | 59.076 | -12.338 | 1.804 | -8.414 | -1.683 | -60.490 | -12.308 |
| 0.808 | 8.263 | -3.027 | 59.342 | -22.158 | 1.808 | -8.166 | -3.020 | -58.605 | -22.082 |
| 0.812 | 8.172 | -3.628 | 58.651 | -26.578 | 1.812 | -7.907 | -3.621 | -56.649 | -26.453 |
| 0.816 | 7.962 | -3.226 | 57.063 | -23.555 | 1.816 | -7.807 | -3.219 | -55.893 | -23.465 |
| 0.820 | 7.741 | -2.137 | 55.396 | -15.529 | 1.820 | -7.763 | -2.129 | -55.560 | -15.478 |
| 0.824 | 7.572 | -0.503 | 54.132 | -3.646 | 1.824 | -7.605 | -0.496 | -54.373 | -3.593 |
| 0.828 | 7.293 | 1.229 | 52.046 | 8.884 | 1.828 | -7.329 | 1.236 | -52.314 | 8.940 |
| 0.832 | 7.004 | 2.812 | 49.899 | 20.325 | 1.832 | -7.044 | 2.796 | -50.170 | 20.215 |
| 0.836 | 6.770 | 3.704 | 48.165 | 26.787 | 1.836 | -6.814 | 3.609 | -48.485 | 26.104 |
| 0.840 | 6.427 | 3.505 | 45.635 | 25.262 | 1.840 | -6.474 | 3.428 | -45.981 | 24.711 |
| 0.844 | 6.075 | 2.570 | 43.061 | 18.428 | 1.844 | -6.126 | 2.506 | -43.430 | 17.973 |
| 0.848 | 5.780 | 1.059 | 40.907 | 7.559 | 1.848 | -5.834 | 1.005 | -41.300 | 7.177 |
| 0.852 | 5.377 | -0.577 | 37.985 | -4.107 | 1.852 | -5.434 | -0.623 | -38.399 | -4.434 |
| 0.856 | 4.967 | -2.198 | 35.030 | -15.630 | 1.856 | -5.028 | -2.237 | -35.465 | -15.916 |
| 0.860 | 4.615 | -3.445 | 32.502 | -24.519 | 1.860 | -4.679 | -3.479 | -32.958 | -24.773 |
| 0.864 | 4.157 | -3.492 | 29.226 | -24.798 | 1.864 | -4.224 | -3.523 | -29.702 | -25.026 |
| 0.868 | 3.693 | -2.730 | 25.928 | -19.307 | 1.868 | -3.763 | -2.758 | -26.423 | -19.513 |
| 0.872 | 3.289 | -1.356 | 23.061 | -9.551 | 1.872 | -3.361 | -1.382 | -23.575 | -9.739 |
| 0.876 | 2.884 | 0.171 | 20.200 | 1.203 | 1.876 | -2.855 | 0.147 | -19.996 | 1.030 |
| 0.880 | 2.536 | 2.014 | 17.752 | 14.157 | 1.880 | -2.344 | 1.990 | -16.401 | 13.983 |
| 0.884 | 2.082 | 3.382 | 14.562 | 23.820 | 1.884 | -1.893 | 3.360 | -13.237 | 23.649 |
| 0.888 | 1.518 | 3.692 | 10.611 | 25.997 | 1.888 | -1.443 | 3.671 | -10.085 | 25.838 |
| 0.892 | 0.952 | 3.105 | 6.649 | 21.804 | 1.892 | -1.052 | 3.083 | -7.348 | 21.656 |
| 0.896 | 0.447 | 1.868 | 3.121 | 13.083 | 1.896 | -0.555 | 1.847 | -3.878 | 12.936 |
| 0.900 | -0.056 | 0.119 | -0.390 | 0.832 | 1.900 | -0.055 | 0.098 | -0.380 | 0.687 |
| 0.904 | -0.499 | -1.682 | -3.483 | -11.773 | 1.904 | 0.387 | -1.702 | 2.699 | -11.916 |
| 0.908 | -1.046 | -3.018 | -7.304 | -21.195 | 1.908 | 0.932 | -3.039 | 6.507 | -21.336 |
| 0.912 | -1.596 | -3.620 | -11.153 | -25.482 | 1.912 | 1.480 | -3.640 | 10.341 | -25.620 |
| 0.916 | -2.085 | -3.217 | -14.581 | -22.650 | 1.916 | 1.967 | -3.237 | 13.754 | -22.785 |
| 0.920 | -2.572 | -2.128 | -18.005 | -14.964 | 1.920 | 2.557 | -2.148 | 17.900 | -15.104 |
| 0.924 | -3.000 | -0.494 | -21.021 | -3.476 | 1.924 | 3.149 | -0.514 | 22.074 | -3.618 |
| 0.928 | -3.532 | 1.238 | -24.784 | 8.726 | 1.928 | 3.679 | 1.218 | 25.828 | 8.592 |
| 0.932 | -4.068 | 2.797 | -28.590 | 19.818 | 1.932 | 4.206 | 2.801 | 29.579 | 19.857 |
| 0.936 | -4.543 | 3.611 | -31.982 | 25.703 | 1.936 | 4.673 | 3.693 | 32.918 | 26.314 |
| 0.940 | -5.016 | 3.430 | -35.381 | 24.467 | 1.940 | 5.139 | 3.494 | 36.266 | 24.952 |
| 0.944 | -5.431 | 2.507 | -38.373 | 17.895 | 1.944 | 5.546 | 2.559 | 39.207 | 18.280 |
| 0.948 | -5.845 | 1.006 | -41.382 | 7.188 | 1.948 | 5.953 | 1.048 | 42.166 | 7.490 |
| 0.952 | -6.202 | -0.621 | -43.987 | -4.448 | 1.952 | 6.302 | -0.588 | 44.722 | -4.213 |
| 0.956 | -6.560 | -2.235 | -46.618 | -16.080 | 1.956 | 6.654 | -2.209 | 47.304 | -15.899 |
| 0.960 | -6.863 | -3.477 | -48.848 | -25.151 | 1.960 | 6.949 | -3.456 | 49.486 | -25.010 |
| 0.964 | -7.168 | -3.521 | -51.111 | -25.538 | 1.964 | 7.247 | -3.503 | 51.702 | -25.423 |
| 0.968 | -7.418 | -2.757 | -52.975 | -20.000 | 1.968 | 7.491 | -2.741 | 53.519 | -19.899 |
| 0.972 | -7.672 | -1.381 | -54.880 | -10.018 | 1.972 | 7.738 | -1.367 | 55.378 | -9.924 |
| 0.976 | -7.873 | 0.148 | -56.389 | 1.075 | 1.976 | 7.933 | 0.160 | 56.842 | 1.165 |
| 0.980 | -8.079 | 1.992 | -57.946 | 14.520 | 1.980 | 8.133 | 2.003 | 58.353 | 14.609 |
| 0.984 | -8.232 | 3.361 | -59.109 | 24.616 | 1.984 | 8.280 | 3.371 | 59.473 | 24.704 |
| 0.988 | -8.393 | 3.672 | -60.327 | 26.963 | 1.988 | 8.435 | 3.682 | 60.647 | 27.045 |
| 0.992 | -8.401 | 3.085 | -60.392 | 22.616 | 1.992 | 8.438 | 3.094 | 60.673 | 22.691 |
| 0.996 | -8.283 | 1.849 | -59.495 | 13.504 | 1.996 | 8.316 | 1.857 | 59.743 | 13.572 |
| 1.000 | -8.152 | 0.100 | -58.496 | 0.727 | 2.000 | 8.180 | 0.108 | 58.712 | 0.788 |

**End of Table 10.**

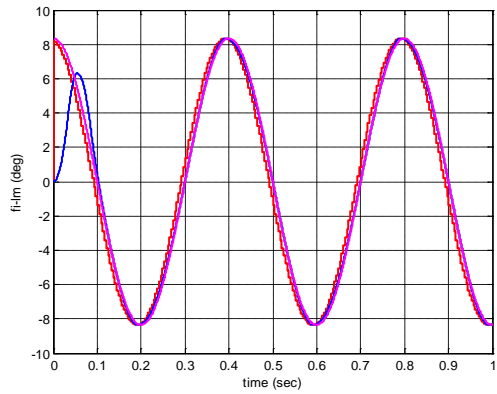

a)   Output angle $\varphi_{lm}$ [deg]

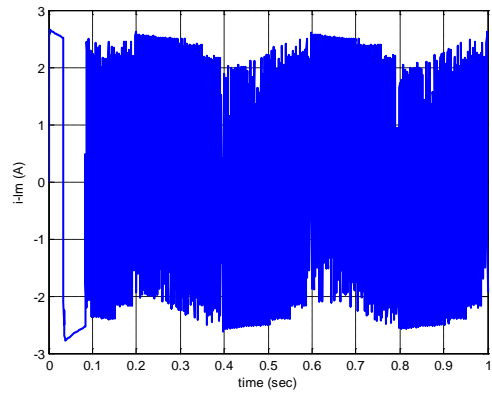

d) LAT current $i_{lm}$ [A]

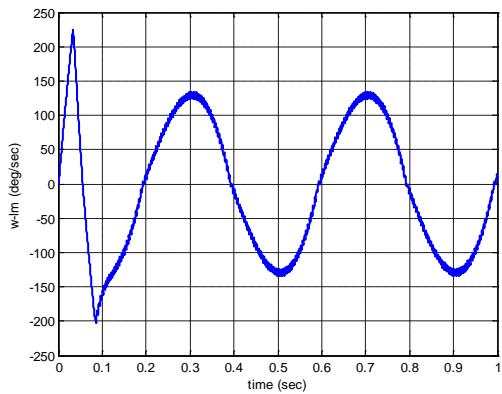

b)   Angular velocity $\omega_{lm}$ [deg/sec]

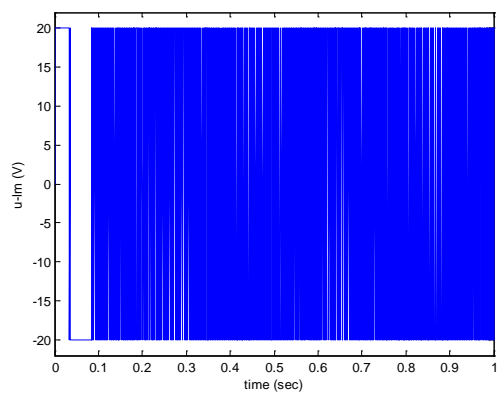

e) Synthesized control signal $u_{lm}$ [V]

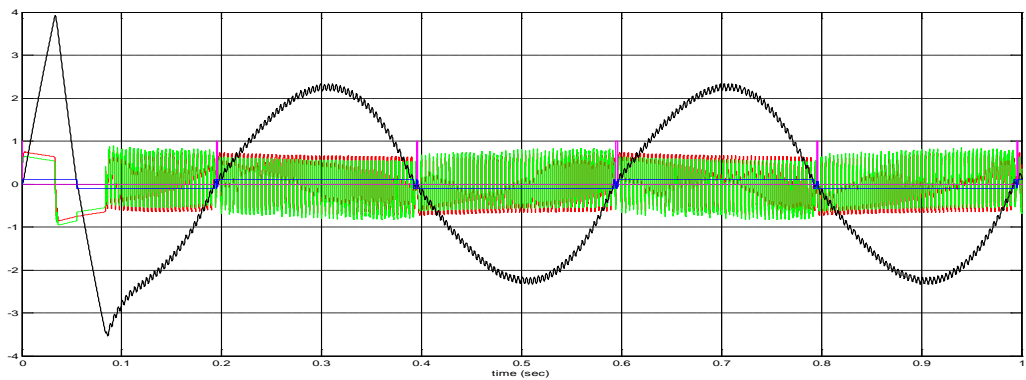

c)   Linear component of the resultant torque $T_{RL}$ [N.m] ('–'), of Coulomb's friction

torque $T_{CF}$ [N.m] ('–'), of the resultant torque $T_R$ [N.m] ('–'), of the angular

velocity $\omega_{lm}$ [rad/sec] ('–') and of the flag of sticking ('–')

**Figure 89. Time-diagrams of the processes in the large mirror actuator digital tracking control system in the time interval [0, 1] s working at sampling rate of 0.1 ms, except the demand sinusoidal signal, ('–') in a), discretized at sampling rate of 4 ms. The ideal sinusoidal signal for the scan pattern is presented in a) by ('–').**

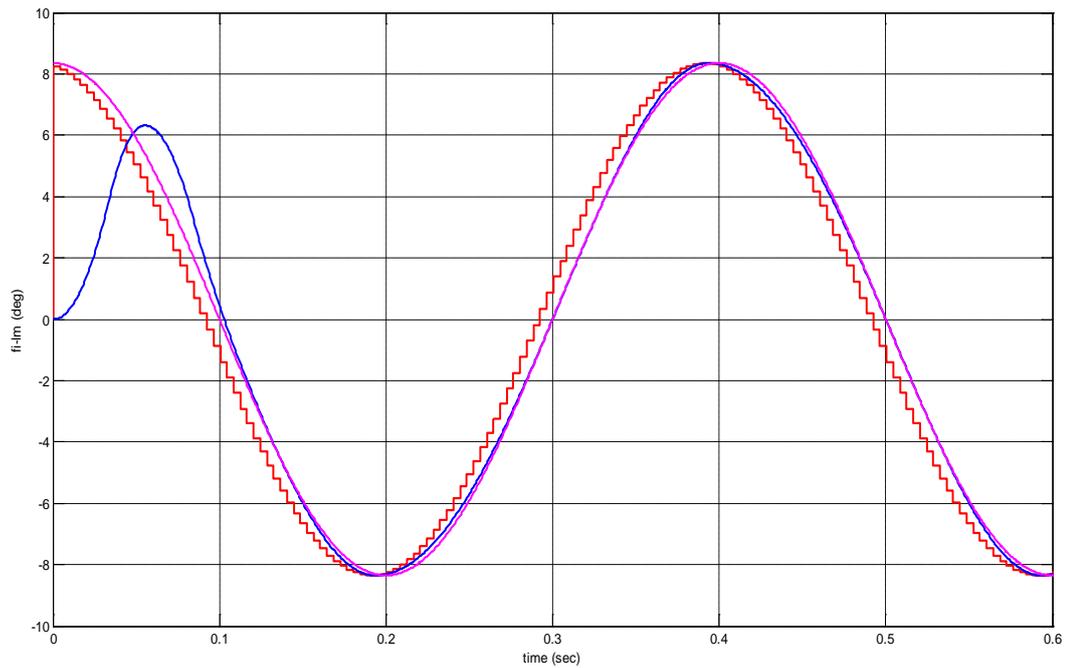

a)

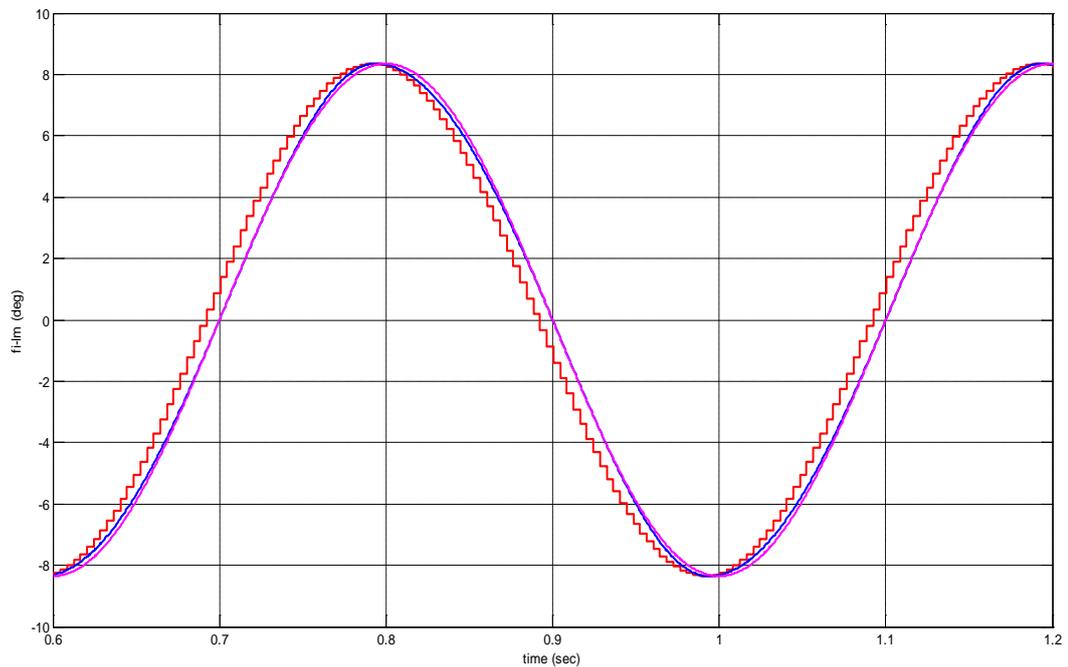

b)

**Figure 90. Output angle of the large mirror actuator digital tracking control system ('−') presented together with the discretized sinusoidal signal ('−') and the ideal sinusoidal signal with amplitude of** $8.35\,[\text{deg}]$ **and frequency** $2.5\,\text{Hz}$ **for the scan pattern ('∼') in the time intervals** $[0, 0.6]\,\text{s}$ **and** $[0.6, 1.2]\,\text{s}$.

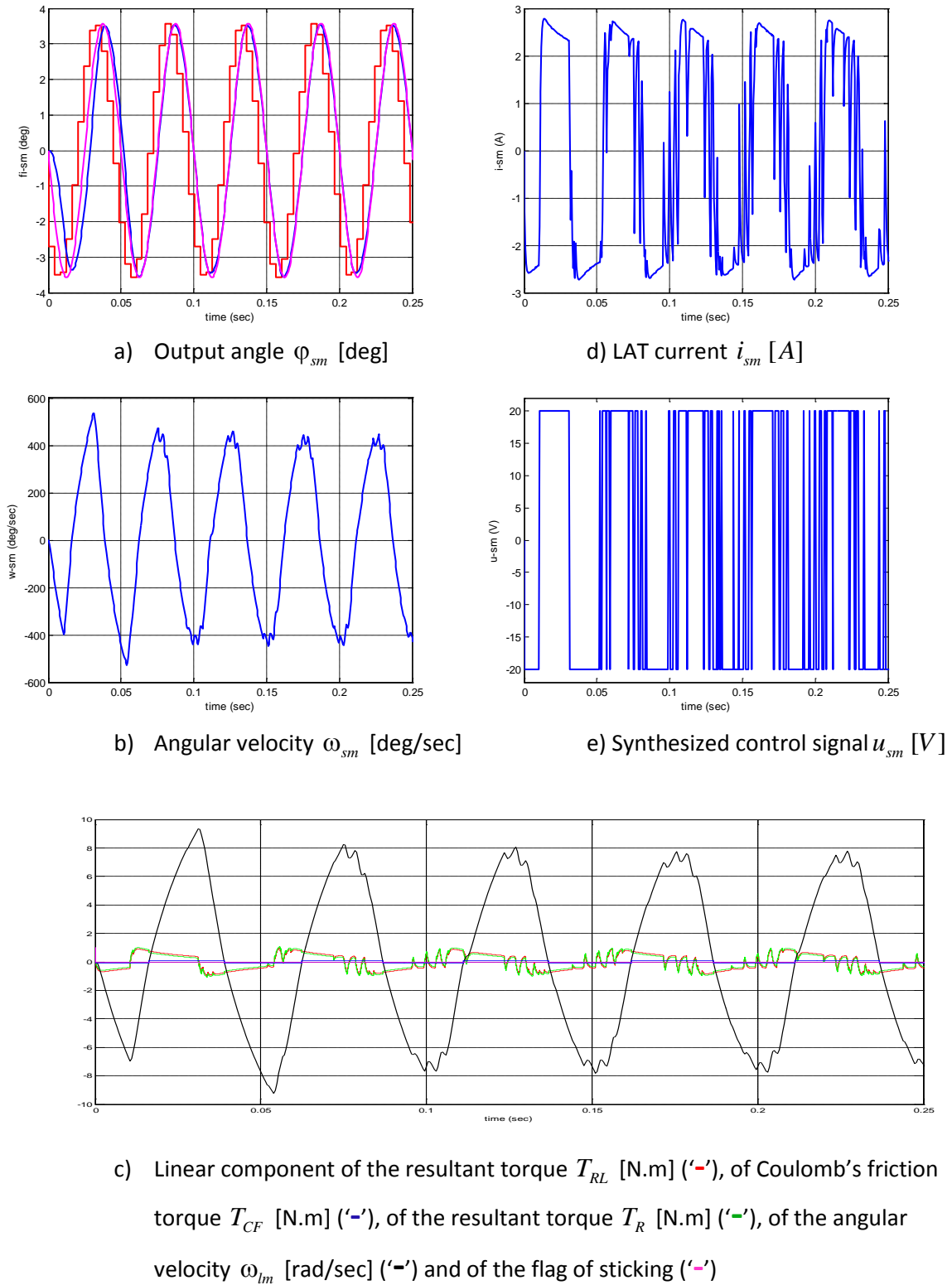

a)  Output angle $\varphi_{sm}$ [deg]

d) LAT current $i_{sm}$ [A]

b)  Angular velocity $\omega_{sm}$ [deg/sec]

e) Synthesized control signal $u_{sm}$ [V]

c)  Linear component of the resultant torque $T_{RL}$ [N.m] ('–'), of Coulomb's friction

torque $T_{CF}$ [N.m] ('–'), of the resultant torque $T_R$ [N.m] ('–'), of the angular

velocity $\omega_{lm}$ [rad/sec] ('–') and of the flag of sticking ('–')

**Figure 91. Time-diagrams of the processes in the small mirror actuator digital tracking control system in the time interval [0, 0.25] s working at sampling rate of 0.1 ms, except the demand sinusoidal signal, ('–') in a), discretized at sampling rate of 4 ms. The ideal sinusoidal signal for the scan pattern is presented in a) by ('–').**

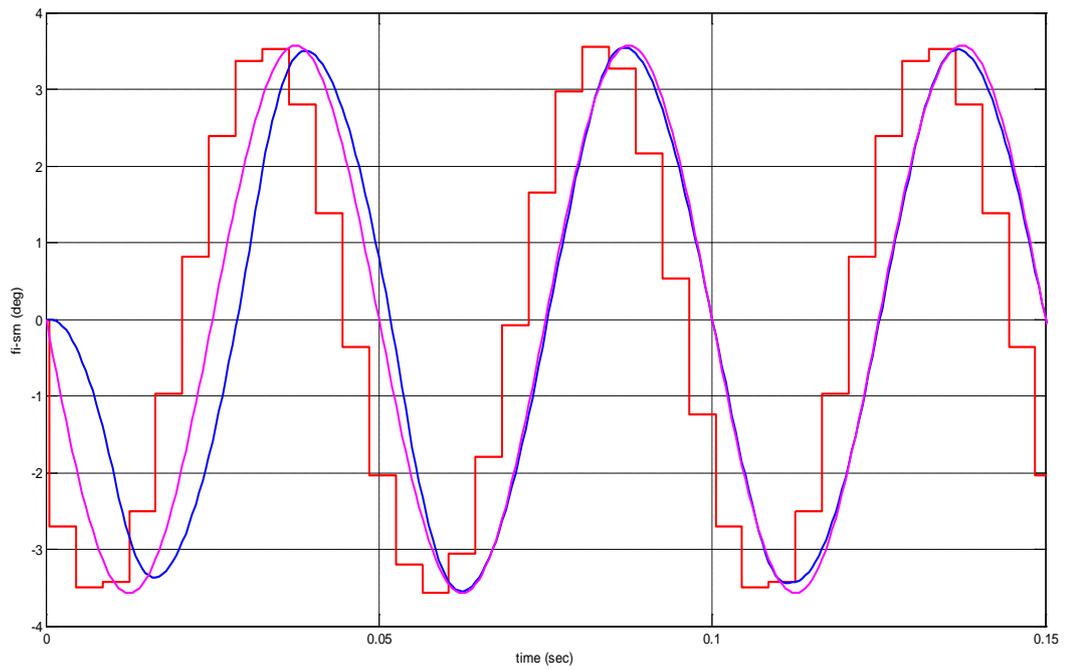

a)

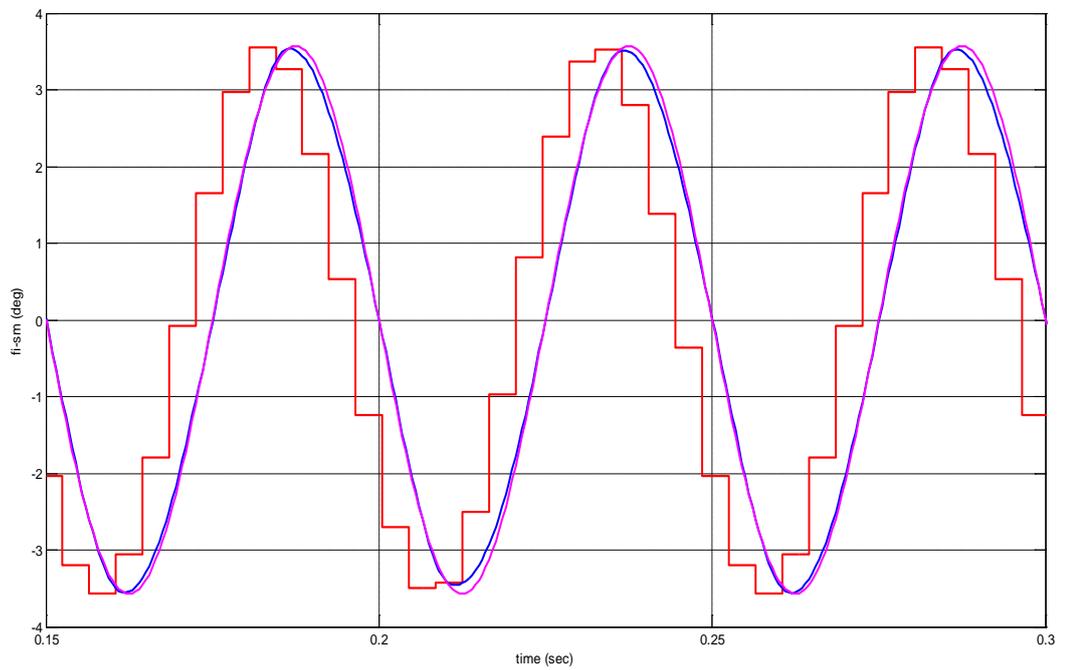

b)

**Figure 92. Output angle of the small mirror actuator digital tracking control system ('–') presented together with the discretized sinusoidal signal ('–') and the ideal sinusoidal signal with amplitude of $3.57\,\mathrm{deg}$ and frequency $20\,\mathrm{Hz}$ for the scan pattern ('–') in the time intervals $[0, 0.15]\,\mathrm{s}$ and $[0.15, 0.3]\,\mathrm{s}$.**

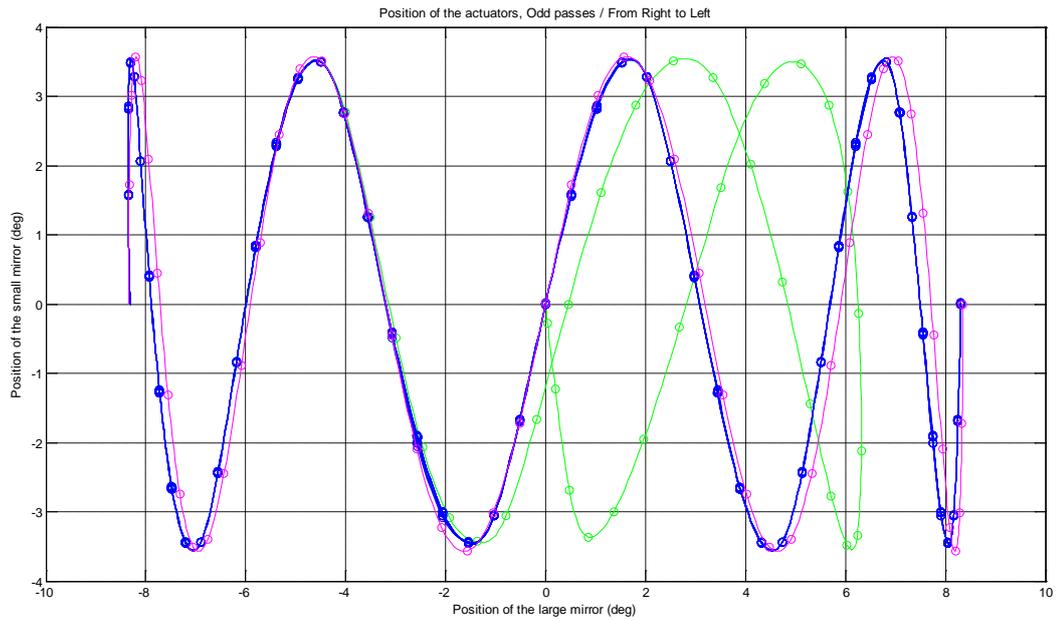

**Figure 93. The ideal mirror actuators' odd pass ('–') and ideal scanned points ('o') shown together with the positions of the mirror actuators at odd passes ('–') with scanned points ('o') after improvement of the actuators' control systems by increased sampling rate of 0.1 ms combined with further phase adjustment. The first pass with scanned points are shown by ('–') and ('o').**

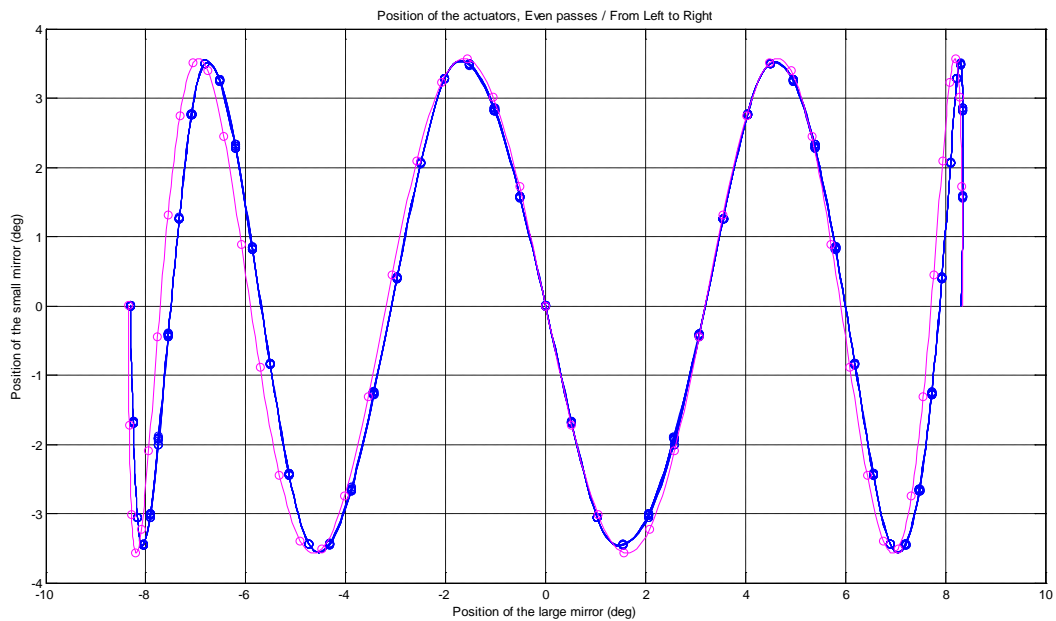

**Figure 94. The ideal mirror actuators' even pass ('–') and ideal scanned points ('o') shown together with the positions of the mirror actuators at even passes ('–') with scanned points ('o') after improvement of the actuators' control systems by increased sampling rate of 0.1 ms and further phase adjustment.**

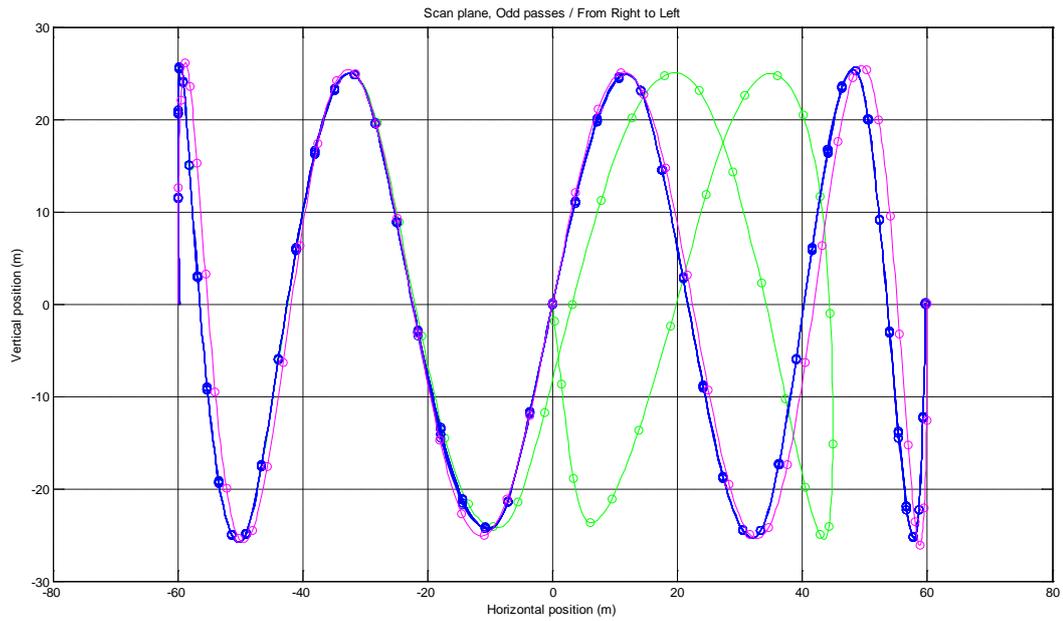

**Figure 95. Scanning in the vertical plane at range of 200 m in front of the aircraft at the ideal odd pass ('–') with ideal scanned points ('o') shown together with the scanning at odd passes ('–') with scanned points ('o') after improvement of the actuators' control systems by increased sampling rate of 0.1 ms and further phase adjustment. The first pass with scanned points are shown by ('–') and ('o').**

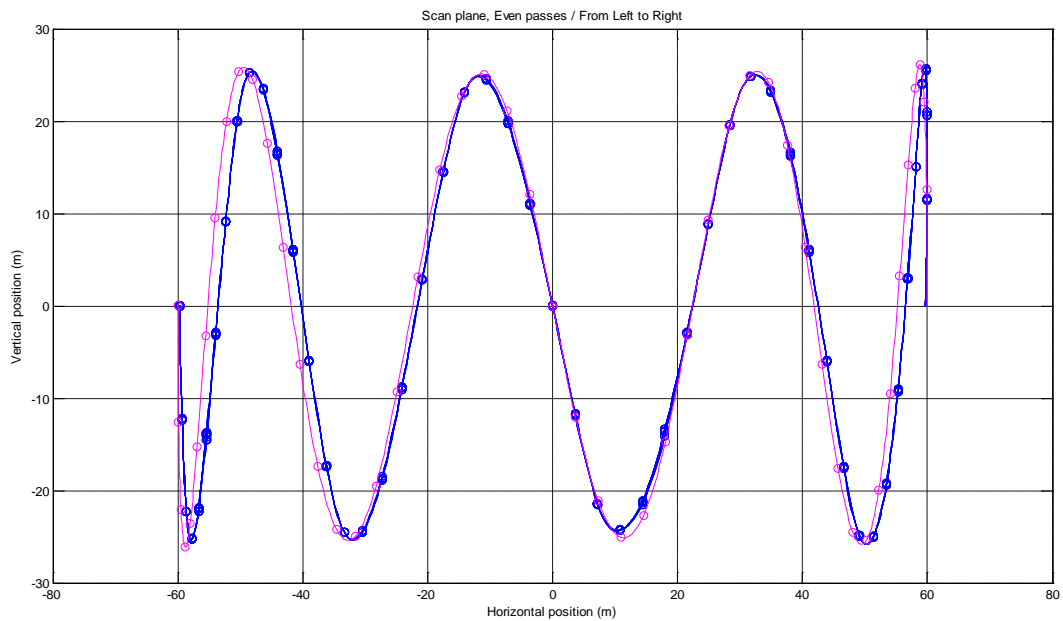

**Figure 96. Scanning in the vertical plane at range of 200 m in front of the aircraft at the ideal even pass ('–') with ideal scanned points ('o') shown together with the scanning at even passes ('–') with scanned points ('o') after improvement of the actuators' control systems by increased sampling rate of 0.1 ms and further phase adjustment.**

# 8. Conclusions

Summarizing, the research done here answers positively the question regarding the possibility for time optimal control of the scanning system of the Green-Wake lidar.

The analysis is decomposed into several group of problems solved consecutively.

First a spatial mathematical model of the scanning system is developed. The inputs of the model are the angular positions of the large and the small mirrors while the outputs are the XY coordinates in the vertical plane in front of the aircraft at given range, where that range and the distance between the axis of the mirrors actuators present model's parameters. By simulation of the model in case the range is 200 m and the inputs are pure sinusoidal signals with frequencies 2.5 Hz and 20 Hz, imitating the ideal work of the large and the small mirror actuators without any inertia properties, the model's outputs form the scan pattern of the scanning system. The obtained simulation data match the data for the scan pattern of the project partner Sula Systems Ltd.

The second group of problems are connected with the mathematical modelling and simulation of the dynamic properties of the large and small mirror actuators. Linear models of the actuators are developed and simulated. They show the oscillating properties of each one actuator when it is driven by sinusoidal signal with frequency 2.5 Hz for the large mirror and 20 Hz for the small mirror actuator respectively, which frequencies are very close to the respective resonant frequency of each one actuator. At that design approach the actuators' control systems are considered as open loop systems. Next a model of Coulomb's friction is developed and introduced into the linear models of the actuators, so by this more precise and truthful non-linear models of the actuators are obtained.

The third group of problems deals with the control synthesis of the actuators with emphasis on the design of closed loop tracking control systems of both the large and the small mirror actuators. The design is based on the method for synthesis of time optimal control of any order for a class of linear systems and a special software developed by the author. For implementation of that approach the oscillating properties of the actuators are first corrected by a change of the parameters of their mechanical subsystems, so that all the eigenvalues of the actuators' linear models become different and negative. Two approaches to achieve this are investigated: by changing of the damping and by eliminating the pivot stiffness of the actuator's mechanical subsystem. At each one approach digital tracking control systems controlling consecutively the respective linear and non-linear models of each one actuator are investigated. The sampling rate for discretizing the demand signal imposed in advance by the design of the scanning system as a whole has been fixed to 4 ms, so the systems' behaviour at various sampling rates for the rest part of the control systems is examined. Results regarding the eligible sampling rates and constraints on the control signal for each one actuator's tracking control system are obtained so that the developed tracking control systems could accomplish the tracking of the sinusoidal demand signals of given amplitude and frequency. Finally the digital tracking control systems of the large and the small mirror actuators based on the synthesis of near time optimal control for their linear models with given accuracy synthesized here have the best dynamic performance and provide the best tracking of the demand signals within the given constraints on the control signals.

Having already developed digital tracking control systems of both mirror actuators the fourth group of problems is connected with modelling the real scan picture considering the actuators' control systems dynamic. The impact of the dynamic behaviour of the control systems is investigated on the basis of the tracking control systems of the actuators with zero pivot stiffness of their mechanical subsystems. The control system of the large mirror actuator works at sampling rate of 4 ms while the control system of the small mirror actuator works at two sampling rates: 4 ms for discretizing the demand sinusoidal signal and 1 ms for the rest part of the control system. The simulated scanning in the vertical plane at range of 200 m in front of the aircraft shows good covering and repetition at both odd and even passes of the scanning area but no symmetry of the scan picture with respect to the central point with coordinates (0,0). The transition time to steady tracking is less than 0.2 s at constraints on the control signals of 20 V and practically there is a steady tracking after the second half of the very first pass of the scan area.

So the next fifth group of problems investigate the improvement of the real scan picture. The first and main approach proposed and implemented here is introducing a respective positive phase shift into demand sinusoidal signal for each one mirror actuator's tracking control system corresponding to the phase delay of the system at the frequency of the demand signal providing by this technique the symmetry of the scanning and synchronization with the scan pattern. The second approach is based on the increase of the sampling rates of both the control systems, except the sampling rate for the demand sinusoidal signals of 4 ms, combined with phase adjustment. By sampling rate of 0.1 ms chosen here the aim is rather to show how the almost ideal tracking control systems based on the synthesis of time optimal control within the given constraints on the control signals could serve the scanning. Here an excellent repetition and clearness of the scanning alongside with symmetry and matching the scan pattern could be seen.

Finally, here the way to a real scanning system based on the time optimal control synthesis in closed loop is started and the first very important steps are done.

## *Acknowledgments*

The author would like to thank the colleagues from Sula Systems Ltd. Stephen Bowater for support and cooperation as a project manager and Martin Humphries and Dr David Bamford for providing the initial data regarding the large and the small mirror actuators and helpful comments at discussing the scanning system's issues.

# 9. References

**[ 1 ]**    **Атанс М., Фалб П.** Оптимальное управление. Москва, Машиностроение, 1968. / **Athans, M., P. L. Falb,** *Optimal Control*, McGraw-Hill, New York, 1966.

**[ 2 ]**    **Болтянский В. Г**. Математические методы оптимального управления. Москва, Наука, 1969. / **Boltyanskii, V. G.**, *Mathematical Methods of Optimal Control*, Holt, Rinehart and Winston, New York, 1971.

**[ 3 ]**    **Габасов Р., Кириллова Ф. М., Прищепова С. В.** Синтез оптимальной по быстродействию дискретной системой. Автоматика и телемеханика, 1991, № 12, стр. 92-99.

**[ 4 ]**    **Иванов В. А., Фалдин Н. В.** Теория оптимальных систем автоматического управления. Москва, Наука, 1981.

**[ 5 ]**    **Павлов А. А.** Синтез релейных систем, оптимальных по быстродействию (метод фазового пространства). Москва, Наука, 1966.

**[ 6 ]**    **Понтрягин Л. С., Болтянский В. Г., Гамкрелидзе Р. В., Мищенко Е. Ф.** Математическая теория оптимальных процессов. 4-е изд., Москва, Наука, 1983. / **Pontryagin, L. S., V. G. Boltyanskii, R. V. Gamkrelidze and E. F. Mischenko**, *The Mathematical Theory of Optimal Processes*, Pergamon Press, Oxford, 1964.

**[ 7 ]**    **Фельдбаум А. А.** Простейшие релейные системы автоматического регулирования. Автоматика и телемеханика, т. X, № 4, 1949.

**[ 8 ]**    **Фельдбаум А. А.** Оптимальные процессы в системах автоматического регулирования. Автоматика и телемеханика, т. XIV, № 6, 1953.

**[ 9 ]**    **Фельдбаум А. А.** О синтезе оптимальных систем с помощью фазового пространства. Автоматика и телемеханика, т. XVI, № 2, 1955.

**[ 10 ]**    **Фельдбаум А. А.** Вычислительные устройства в автоматических системах. Москва, Физматгиз, 1959.

**[ 11 ]**    **Фельдбаум А. А.** Основы теории оптимальных автоматических систем, Москва, Физматгиз, 1963.

**[ 12 ]**    **Фельдбаум А. А., Бутковский А. Г.** Методы теории автоматического управления. Москва, Наука, 1971.

**[ 13 ]**    **Чаки Ф.** Современная теория управления. Нелинейные, оптимальные и адаптивные системы. Москва, Мир, 1975.

**[ 14 ]**    **Leitmann, G.**, *The calculus of variations and optimal control*, Plenum Press, 1981.

**[ 15 ]**    **Penev, B. G.**, "A Method for Synthesis of Time-Optimal Control of Any Order for a Class of Linear Problems for Time-Optimal Control", Ph.D. Dissertation, Technical University of Sofia, 1999 (in Bulgarian).

**[ 16 ]**    **Penev, B. G. and N. D. Christov**, "On the Synthesis of Time Optimal Control for a Class of Linear Systems", in Proc. 2002 American Control Conference, Anchorage, May 8-10, 2002, pp. 316-321.

[ 17]    **Penev, B. G. and N. D. Christov**, "On the State-Space Analysis in the Synthesis of Time-Optimal Control for a Class of Linear Systems", in Proc. 2004 American Control Conference, Boston, June 30 - July 2, 2004, Vol. 1, pp. 40-45.

[ 18]    **Penev, B. G. and N. D. Christov**, "A fast time-optimal control synthesis algorithm for a class of linear systems", in Proc. 2005 American Control Conference, Portland, 8-10 June 2005, Vol. 2, pp. 883-888.

[ 19]    **Penev, B.**, "Linear modelling in time-optimal control of a rotary pneumatic actuator", Journal of the Technical University at Plovdiv, Vol. 14 (1), 2009, pp. 155-160.

[ 20]    **Pinch, E. R.**, *Optimal Control and the Calculus of Variations*, Oxford University Press, Oxford, 1993.

# 10. Tables of figures

## 10.1 Figures

## 10.2 Tables